\PassOptionsToPackage{table}{xcolor}
\documentclass[acmsmall,dvipsnames,nonacm,screen]{acmart}
\setcopyright{none}

\usepackage[capitalise,noabbrev]{cleveref}
\usepackage{csquotes}
\usepackage{enumitem}
\usepackage{mathtools}
\usepackage{mathpartir}
\usepackage{multicol}
\usepackage{tabularx}
\usepackage{thmtools}
\usepackage{thm-restate}
\usepackage{xspace}

\usepackage[skins]{tcolorbox}
\AtBeginDocument{\tcbset{enhanced jigsaw,colback=white!95!black,arc=0pt,boxrule=0pt,frame hidden}}

\def\OPTIONConf{1}

\usepackage{mathpartir}
\usepackage{amsmath,amsthm}

\usepackage{srcltx}

\usepackage{jdunfield}
\usepackage{rulelinks}

\newcommand{\reviewcomment}[3]{\textit{\textcolor{#3}{#2: #1}}\typeout{LaTeX Warning: #1 on input line \the\inputlineno.}}

\newcommand{\rulename}[1]{\textsf{#1}}
\newcommand{\labtyperulename}[1]{\rulename{#1}}

\newcommand{\termname}[1]{\mathsf{#1}}

\newcommand{\lemmaref}[1]{\cref{#1} (\nameref{#1})}

\newcommand{\alt}{\ | \ }

\newcommand{\PSystemF}{Implicit Polarized F\xspace}

\newcommand{\Cbpv}{Call-by-push-value\xspace}
\newcommand{\cbpv}{call-by-push-value\xspace}

\newcommand{\lhs}{left-hand side\xspace}
\newcommand{\rhs}{right-hand side\xspace}

\newcommand{\mlf}{$\text{ML}^\text{F}$\xspace}
\newcommand{\systemfsub}{System F$_{<:}$\xspace}

\newcommand{\shiftdown}{\mathord{\downarrow}}
\newcommand{\shiftup}{\mathord{\uparrow}}
\newcommand{\funarrow}{\rightarrow}

\newcommand{\tyvar}[1]{#1}
\newcommand{\shiftd}[1]{\shiftdown #1}

\newcommand{\arrowtype}[2]{#1 \funarrow #2}
\newcommand{\shiftu}[1]{\shiftup #1}
\newcommand{\foralltype}[2]{\forall \tyvar{#1} \ldotp #2}

\newcommand{\possubtype}{\leq^+}
\newcommand{\negsubtype}{\leq^-}
\newcommand{\pnsubtype}{\leq^\pm}

\newcommand{\postype}{\mathop{}\mathopen{}\mathrm{type}^+}
\newcommand{\negtype}{\mathop{}\mathopen{}\mathrm{type}^-}
\newcommand{\posnegtype}{\mathop{}\mathopen{}\mathrm{type}^\pm}

\newcommand{\conwf}{\mathop{}\mathopen{}\mathrm{ctx}}

\DeclareMathOperator{\EV}{EV}

\DeclareMathOperator{\FreeEV}{FEV}
\DeclareMathOperator{\FreeUV}{FUV}

\newcommand{\dcontext}{\Theta}
\newcommand{\dconwithhole}[1]{\dcontext_L, #1, \dcontext_R}

\newcommand{\makedec}[1]{\left\lVert#1\right\rVert}

\newcommand{\wfpostypeJudg}[2]{#1 \entails #2 \postype}
\newcommand{\wfnegtypeJudg}[2]{#1 \entails #2 \negtype}
\newcommand{\wfposnegtypeJudg}[2]{#1 \entails #2 \posnegtype}

\newcommand{\DArbJudge}[4]{#2 \entails #3 \begingroup \color{dDigPurple} #1 \endgroup #4}
\newcommand{\DPosSubtypeJudge}[3]{\DArbJudge{\possubtype} {#1}{#2}{#3}}
\newcommand{\DNegSubtypeJudge}[3]{\DArbJudge{\negsubtype} {#1}{#2}{#3}}
\newcommand{\DPnSubtypeJudge}[3]{\DArbJudge{\pnsubtype} {#1}{#2}{#3}}

\newcommand{\CnewArbConPosJudge}[4]{\DPosSubtypeJudge{\makedec{#2}}{#3}{\subcon{#1} #4}}
\newcommand{\CnewArbConNegJudge}[4]{\DNegSubtypeJudge{\makedec{#2}}{\subcon{#1} #3}{#4}}
\newcommand{\CnewPosJudge}[3]{\CnewArbConPosJudge{\ccontext}{#1}{#2}{#3}}
\newcommand{\CnewNegJudge}[3]{\CnewArbConNegJudge{\ccontext}{#1}{#2}{#3}}

\newcommand{\DPosSubtypeJudg}[3]{#1 \entails #2 \begingroup \color{dDigPurple} \possubtype \endgroup #3}
\newcommand{\DNegSubtypeJudg}[3]{#1 \entails #2 \begingroup \color{dDigPurple} \negsubtype \endgroup #3}

\newcommand{\APosSubtypeJudg}[4]{\DPosSubtypeJudg{#1}{#2}{#3} \prodcon #4}
\newcommand{\ANegSubtypeJudg}[4]{\DNegSubtypeJudg{#1}{#2}{#3} \prodcon #4}

\newrulecommand{drefl}{\labtyperulename{$\pnsubtype$Drefl}}
\newrulecommand{dshiftdown}{\labtyperulename{$\pnsubtype$Dshift$\shiftdown$}}
\newrulecommand{dforalll}{\labtyperulename{$\pnsubtype$Dforalll}}
\newrulecommand{dforallr}{\labtyperulename{$\pnsubtype$Dforallr}}
\newrulecommand{darrow}{\labtyperulename{$\pnsubtype$Darrow}}
\newrulecommand{dshiftup}{\labtyperulename{$\pnsubtype$Dshift$\shiftup$}}

\newcommand{\emptyacontext}{\cdot}
\newcommand{\acontext}{\Theta}
\newcommand{\guess}[1]{\hat{#1}}
\newcommand{\groundJudge}[1]{#1 \text{ ground}}

\newcommand{\aconwithhole}[1]{\acontext_L, #1, \acontext_R}

\newcommand{\prodcon}{\dashv}
\newcommand{\subcon}[1]{[#1]}
\newcommand{\subterm}[1]{[#1]}

\newrulecommand{arefl}{\labtyperulename{$\pnsubtype$Arefl}}
\newrulecommand{ainst}{\labtyperulename{$\pnsubtype$Ainst}}
\newrulecommand{ashiftdown}{\labtyperulename{$\pnsubtype$Ashift$\shiftdown$}}
\newrulecommand{aforallr}{\labtyperulename{$\pnsubtype$Aforallr}}
\newrulecommand{aforalll}{\labtyperulename{$\pnsubtype$Aforalll}}
\newrulecommand{aarrow}{\labtyperulename{$\pnsubtype$Aarrow}}
\newrulecommand{ashiftup}{\labtyperulename{$\pnsubtype$Ashift$\shiftup$}}

\newcommand{\ccontext}{\Omega}
\newcommand{\emptyccontext}{\cdot}

\newcommand{\congoes}{\,\longrightarrow\,}

\newcommand{\congoesJudg}[2]{#1 \congoes #2}

\newcommand{\congoesweak}{\,\Longrightarrow\,}
\newcommand{\congoesWeakJudg}[2]{#1 \congoesweak #2}

\newrulecommand{cempty}{\labtyperulename{Cempty}}
\newrulecommand{cuvar}{\labtyperulename{Cuvar}}
\newrulecommand{cunsolvedguess}{\labtyperulename{Cunsolvedguess}}
\newrulecommand{csolveguess}{\labtyperulename{Csolveguess}}
\newrulecommand{csolvedguess}{\labtyperulename{Csolvedguess}}

\newcommand{\mutualSubtypePos}{\begingroup \color{dDigPurple} \cong^+ \endgroup}
\newcommand{\mutualSubtypeNeg}{\begingroup \color{dDigPurple} \cong^- \endgroup}
\newcommand{\mutualSubtypePosNeg}{\begingroup \color{dDigPurple} \cong^\pm \endgroup}
\newcommand{\MutualSubtypeJudgeArb}[4]{#2 \entails #3 #1 #4}
\newcommand{\MutualSubtypePosJudge}[3]{\MutualSubtypeJudgeArb{\mutualSubtypePos}{#1}{#2}{#3}}
\newcommand{\MutualSubtypeNegJudge}[3]{\MutualSubtypeJudgeArb{\mutualSubtypeNeg}{#1}{#2}{#3}}
\newcommand{\MutualSubtypePosNegJudge}[3]{\MutualSubtypeJudgeArb{\mutualSubtypePosNeg}{#1}{#2}{#3}}

\newrulecommand{Cempty}{\labtyperulename{Cempty}}
\newrulecommand{Cuvar}{\labtyperulename{Cuvar}}
\newrulecommand{Cunsolvedguess}{\labtyperulename{Cunsolvedguess}}
\newrulecommand{Csolveguess}{\labtyperulename{Csolveguess}}
\newrulecommand{Csolvedguess}{\labtyperulename{Csolvedguess}}

\newrulecommand{twfuvar}{\labtyperulename{Twfuvar}}
\newrulecommand{twfguess}{\labtyperulename{Twfguess}}
\newrulecommand{twfshiftdown}{\labtyperulename{Twfshift$\shiftdown$}}
\newrulecommand{twfforall}{\labtyperulename{Twfforall}}
\newrulecommand{twfarrow}{\labtyperulename{Twfarrow}}
\newrulecommand{twfshiftup}{\labtyperulename{Twfshift$\shiftup$}}

\newrulecommand{cwfempty}{\labtyperulename{Cwfempty}}
\newrulecommand{cwfuvar}{\labtyperulename{Cwfuvar}}
\newrulecommand{cwfunsolvedguess}{\labtyperulename{Cwfunsolvedguess}}
\newrulecommand{cwfsolvedguess}{\labtyperulename{Cwfsolvedguess}}

\newcommand{\termsize}[1]{\abs{#1}_{\textsc{nq}}}

\DeclareMathOperator{\NumPrenexQuantifiers}{NPQ}
\newcommand{\numprenex}[1]{\NumPrenexQuantifiers(#1)}

\newcommand{\wellformed}{w.f.}

\newcommand{\var}[1]{#1}
\newcommand{\thunk}[1]{\{#1\}}

\newcommand{\lamterm}[3]{\lambda #1 : #2 \ldotp #3}
\newcommand{\gen}[2]{\mathrm{\Lambda} #1 \ldotp #2}
\newcommand{\return}[1]{\termname{return} \ #1}
\newcommand{\letplain}[4]{\termname{let} \ #1 = #2 (#3); #4 }
\newcommand{\letplainnobrackets}[4]{\termname{let} \ #1 = #2 \ #3; #4 }
\newcommand{\letvar}[3]{\termname{let} \ #1 = #2; #3 }

\newcommand{\letanno}[5]{\termname{let} \ #1 : #2 = #3 (#4); #5 }

\newcommand{\emptyspine}{\epsilon}
\newcommand{\spine}[2]{{#1} \begingroup \color{dDigPurple} \gg \endgroup {#2}}

\newcommand{\typecontext}{\Theta}
\newcommand{\emptytypecontext}{\cdot}

\newcommand{\conwfJudg}[1]{#1 \conwf}

\newcommand{\typeenv}{\Gamma}
\newcommand{\emptytypeenv}{\cdot}
\newcommand{\envitem}[2]{#1 : #2}

\newcommand{\envwfJudg}[2]{#1 \entails #2 \mathop{}\mathopen{}\mathrm{env}}

\newrulecommand{Ewfempty}{\labtyperulename{Ewfempty}}
\newrulecommand{Ewfvar}{\labtyperulename{Ewfvar}}

\newrulecommand{Eisoempty}{\labtyperulename{Eisoempty}}
\newrulecommand{Eisovar}{\labtyperulename{Eisovar}}

\newcommand{\synths}{:}

\newcommand{\declsynjudg}[3]{{#1} \entails {#2} \synths {#3}}

\newcommand{\algosynjudg}[4]{{#1} \entails {#2} \synths {#3} \dashv {#4}}

\newcommand{\declspinejudg}[3]{{#1} \entails {#2} :{#3}}

\newcommand{\algospinejudg}[4]{{#1} \entails {#2} : {#3} \dashv {#4}}

\newcommand{\negisotype}{\begingroup \color{dDigPurple} \cong^- \endgroup}

\newcommand{\declnisotypejudg}[3]{{#1} \entails {#2} \negisotype {#3}}

\newcommand{\posisotype}{\begingroup \color{dDigPurple} \cong^+ \endgroup}

\newcommand{\declpisotypejudg}[3]{{#1}~\entails~{#2}~\posisotype~{#3}}

\newcommand{\isovarctx}{\begingroup \color{dDigPurple} \cong \endgroup}
\newcommand{\declisovarctxjudg}[3]{{#1}~\entails~{#2}~\isovarctx~{#3}}

\newrulecommand{Wcempty}{\labtyperulename{Wcempty}}
\newrulecommand{Wcuvar}{\labtyperulename{Wcuvar}}
\newrulecommand{Wcunsolvedguess}{\labtyperulename{Wcunsolvedguess}}
\newrulecommand{Wcsolvedguess}{\labtyperulename{Wcsolvedguess}}
\newrulecommand{Wcsolveguess}{\labtyperulename{Wcsolveguess}}
\newrulecommand{Wcunsolvedextend}{\labtyperulename{Wcnewunsolvedguess}}
\newrulecommand{Wcsolvedextend}{\labtyperulename{Wcnewsolvedguess}}

\newrulecommand{Dfunabs}{\labtyperulename{D$\lambda$abs}}
\newrulecommand{Dtypeabs}{\labtyperulename{Dgen}}
\newrulecommand{Dvar}{\labtyperulename{Dvar}}
\newrulecommand{Dthunk}{\labtyperulename{Dthunk}}
\newrulecommand{Dreturn}{\labtyperulename{Dreturn}}
\newrulecommand{Dspinenil}{\labtyperulename{Dspinenil}}
\newrulecommand{Dspinecons}{\labtyperulename{Dspinecons}}
\newrulecommand{Dspinetypeabs}{\labtyperulename{Dspinetypeabs}}

\newrulecommand{Dambiguouslet}{\labtyperulename{Dambiguouslet}}
\newrulecommand{Dunambiguouslet}{\labtyperulename{Dunambiguouslet}}

\newrulecommand{Restrictempty}{\labtyperulename{$\restriction$empty}}
\newrulecommand{Restrictuvar}{\labtyperulename{$\restriction$uvar}}
\newrulecommand{Restrictguessin}{\labtyperulename{$\restriction$guessin}}
\newrulecommand{Restrictguessnotin}{\labtyperulename{$\restriction$guessnotin}}

\newcommand{\restrictcontext}[2]{{#1} \! \restriction {#2}}

\newrulecommand{Afunabs}{\labtyperulename{A$\lambda$abs}}
\newrulecommand{Atypeabs}{\labtyperulename{Agen}}
\newrulecommand{Avar}{\labtyperulename{Avar}}
\newrulecommand{Athunk}{\labtyperulename{Athunk}}
\newrulecommand{Areturn}{\labtyperulename{Areturn}}
\newrulecommand{Aspinenil}{\labtyperulename{Aspinenil}}
\newrulecommand{Aspinecons}{\labtyperulename{Aspinecons}}
\newrulecommand{Aspinetypeabsin}{\labtyperulename{Aspinetypeabsin}}
\newrulecommand{Aspinetypeabsnotin}{\labtyperulename{Aspinetypeabsnotin}}

\newrulecommand{Aambiguouslet}{\labtyperulename{Aambiguouslet}}
\newrulecommand{Aunambiguouslet}{\labtyperulename{Aunambiguouslet}}

\begin{document}

\title{\PSystemF: local type inference for impredicativity}

\author{Henry Mercer}
\orcid{0000-0003-3844-9818}
\affiliation[obeypunctuation=true]{%
  \country{United Kingdom}
}
\email{henry@henrymercer.name}

\author{Cameron Ramsay}
\affiliation[obeypunctuation=true]{%
  \country{United Kingdom}
}
\email{cfr26@cantab.ac.uk}

\author{Neel Krishnaswami}
\orcid{0000-0003-2838-5865}
\affiliation{
  \department{Department of Computer Science and Technology}
  \institution{University of Cambridge}
  \country{United Kingdom}
}
\email{nk480@cl.cam.ac.uk}

\begin{abstract}
  System F, the polymorphic lambda calculus, features the principle of \emph{impredicativity}: polymorphic types may be (explicitly)
  instantiated at other types, enabling many powerful idioms such as Church encoding and data abstraction. Unfortunately, type applications
  need to be implicit for a language to be human-usable, and the problem of inferring all type applications in System F is undecidable. As a
  result, language designers have historically avoided impredicative type inference.

  We reformulate System F in terms of \cbpv, and study type inference for it. Surprisingly, this new perspective yields
  a novel type inference algorithm which is extremely simple to implement (not even requiring unification), infers many types,
  and has a simple declarative specification. Furthermore, our approach offers type theoretic explanations of how many
  of the heuristics used in existing algorithms for impredicative polymorphism arise. 
\end{abstract}

\ccsdesc[500]{Theory of computation~Type theory}
\ccsdesc[500]{Software and its engineering~Polymorphism}
\ccsdesc[300]{Software and its engineering~Functional languages}

\keywords{type systems, impredicative polymorphism, local type inference}

\maketitle

\newcommand{\listterm}{[]}
\newcommand{\datatype}[1]{#1}
\newcommand{\inttype}{\datatype{Int}}
\newcommand{\stringtype}{\datatype{String}}
\newcommand{\prodtype}[2]{#1 \times #2}

\section{Introduction}

System F, the polymorphic lambda calculus, was invented in the early
1970s by John Reynolds~\citep{jcr-system-f} and
Jean-Yves Girard~\citep{girard-system-f}, and has remained
one of the fundamental objects of study in the theory of the lambda
calculus ever since.

Syntactically, it is a tiny extension of the simply-typed lambda
calculus with type variables and quantification over them, with only
five typing rules for the whole calculus. Despite its small size, it
is capable of modeling inductive data types via Church
encodings\footnote{Indeed, any first-order function provably total in
  second-order arithmetic can be expressed in System F.}, and supports
reasoning about data abstraction and modularity via the theory of
parametricity~\citep{reynolds-parametricity,wadler-free-theorems}.

Offering a combination of parsimony and expressive power, System F has
been important in the study of semantics, and has also served as
inspiration for much work on language
design~\citep{haskell-fc}. However, practical languages have
historically shied away from adopting the distinctive feature of
System F --- full impredicative polymorphism.  This is because the
natural specification for type inference for full System F is
undecidable.

To understand this specification, consider the following two variables:
\begin{align*}
\mathsf{fst} &: \mathsf{\forall \alpha, \beta \ldotp (\alpha \times \beta) \to \alpha} \\
\mathsf{pair} &: \mathsf{\inttype \times \stringtype}
\end{align*}
Here, \textsf{fst} has the polymorphic type of
the first projection for pairs, and \textsf{pair} is a pair with a concrete
type $\mathsf{\inttype \times \stringtype}$.
To project out the first component in System F, we would write:
\begin{align*}
\mathsf{fst\ [\inttype]\ [\stringtype]\ pair}
\end{align*}
Note that each type parameter must be given explicitly.  This is a
syntactically heavy discipline: even in this tiny example, the number
of tokens needed for type arguments equals the number of tokens
which compute anything!
Instead, we would prefer to write:
\begin{align*}
\mathsf{fst\ pair}
\end{align*}
Leaving the redundant type arguments \emph{implicit} makes
the program more readable. But how can we specify this?
In the application $\mathsf{fst\ pair}$, the function
\textsf{fst} has the type $\forall \alpha, \beta \ldotp (\alpha \times \beta) \to \alpha$,
but we wish to use it at a function type $\mathsf{(\inttype \times \stringtype) \to \inttype}$.

The standard technical device for using one type in the place of another is \emph{subtyping}.
With subtyping, the application rule is:
\begin{displaymath}
  \newcommand{\judge}[3]{{#1} \vdash {#2} : {#3}}
  \inferrule*[]
             { \judge{\typecontext, \Gamma}{\mathsf{f}}{A} \\
               \typecontext \vdash A \leq B \to C \\
               \judge{\typecontext, \Gamma}{\mathsf{v}}{B} }
             { \judge{\typecontext, \Gamma}{\mathsf{f\ v}}{C} }
\end{displaymath}
As long our subtype relation shows that
$\mathsf{\forall \alpha, \beta \ldotp (\alpha \times \beta) \to \alpha \leq (\inttype \times \stringtype) \to \inttype}$,
we are in the clear! Since we want subtyping to instantiate
quantifiers, we can introduce subtyping rules for the universal type which
reflect the \emph{specialization order}:
\begin{mathpar}
  \newcommand{\judge}[3]{{#1} \vdash {#2} \, {#3}}
  \newcommand{\subt}[3]{{#1} \vdash {#2} \leq {#3}}
  \inferrule*[right=\textsc{$\forall$L}]
             { \judge{\Theta}{A}{\mathsf{type}} \\ \subt{\Theta}{[A/\alpha]B}{C} }
             { \subt{\Theta}{\forall \alpha \ldotp B}{C} }
  \and
  \inferrule*[right=\textsc{$\forall$R}]
             { \subt{\Theta, \alpha}{B}{C} }
             { \subt{\Theta}{B}{\forall \alpha \ldotp C} }
  \\
  \inferrule*[]
             { \subt{\Theta}{X}{A} \\ \subt{\Theta}{B}{Y} }
             { \subt{\Theta}{A \to B}{X \to Y} }
  \and
  \inferrule*[]
             { \alpha \in \Theta }
             { \subt{\Theta}{\alpha}{\alpha} }
\end{mathpar}
Since a more general type like $\foralltype{\alpha, \beta}{\arrowtype{(\alpha \times \beta)}{\alpha}}$ can be used in the place of a more specific
type like $\mathsf{\arrowtype{(\inttype \times \stringtype)}{\inttype}}$, we take the standard subtyping rules for functions and base types,
and add rules reflecting the principle that a universal type is a
subtype of all of its instantiations.

This subtype relation is small, natural, and expressive. Thus, it has
been the focus of a great deal of research in subtyping.
Unfortunately, that research has revealed that this relation is
\emph{undecidable}~\citep{tiuryn-urzczyn-96,chrzaszcz-98}.  This has
prompted many efforts to understand the causes of undecidability and
to develop strategies which work around it.  To contextualize our
contributions, we briefly describe some of these efforts, but delay
extended discussion of the related work until the end of the paper.

\paragraph{Impredicativity} One line of research locates the source of
the undecidability in the fact that the \textsc{$\forall$L} rule is
\emph{impredicative} --- it permits a quantifier to be instantiated at
any type, including types with quantifiers in them.

For example, the identity function $\mathsf{id}$ can be
typed in System F at $\forall \alpha \ldotp \alpha \to \alpha$, and impredicativity
means that self-application is typeable with a polymorphic type:
\begin{displaymath}
  \mathsf{id}\ [\forall \alpha \ldotp \alpha \to \alpha]\ \mathsf{id} : \forall \alpha \ldotp \alpha \to \alpha
\end{displaymath}

A perhaps more useful example is the low-precedence application function in Haskell:
\begin{displaymath}
  \begin{array}{l}
  \mathsf{(\$)} : \forall \alpha \ldotp \forall \beta \ldotp (\alpha \funarrow \beta) \funarrow \alpha \funarrow \beta \\
  \mathsf{(\$) \ f \ x = f \ x}
  \end{array}
\end{displaymath}
This function simply applies a function to an argument, and is used in
languages like Haskell and OCaml to avoid parentheses.

In Haskell, impredicative instantiation in the apply operator, for
example in expressions like \textsf{runST \$ (return True)}
\footnote{
  Here, \textsf{runST} has type $\forall \alpha \ldotp (\forall \beta \ldotp \textsf{ST}\ \beta\ \alpha) \to \alpha$, which ensures that the internal state $\beta$ is confined within the \textsf{ST} monad and remains inaccessible to the rest of the program.
  This polymorphic parameter of \textsf{runST} necessitates impredicative instantiation of \textsf{\$}.
}, is
important enough that GHC, the primary Haskell compiler, has a special
case typing rule for it
\citep{serranoGuardedImpredicativePolymorphism2018}.
However, this
special case typing rule is non-modular and can lead to surprising
results.  For example, if we define an alias \textsf{app} for
\textsf{\$}, then \textsf{runST \$ (return True)} typechecks, but
$\mathsf{runST\ {}^\backprime app^\backprime\ (return\ True)}$ does not (unless the user is using the
\emph{ImpredicativeTypes} extension).

The reason for this special case rule is that GHC's type inference has built on a line of
work~\citep{oderskyPuttingTypeAnnotations1996,peytonjonesPracticalTypeInference2007,dunfieldCompleteEasyBidirectional2013}
which restricts the \textsc{$\forall$L} rule to be \emph{predicative}:
universally quantified types can only be instantiated at monotypes (i.e. types which do
not contain any quantifiers). This recovers decidability of
subsumption, but at the price of giving up any inference for uses of
impredicative types.

\paragraph{Type Syntax} Another line of work sites the difficulty of
type inference in the inexpressivity of the type language of System F.
The work on \mlf~\citep{botlanMLFRaisingML2003,Remy08} extends System F
with a form of bounded quantification. These bounds are rich enough
that the grammar of types can precisely record exactly which types a
polymorphic quantifier might be instantiated at, which is enough to
make the type inference problem tractable once more.

Of course, not just any language of bounds will work --- the \mlf system
was very carefully crafted to make inference both decidable and
expressive, and is startlingly successful at this goal, only
requiring type annotations on variables which are used at two different
polymorphic types. Unfortunately, the \mlf algorithm is somewhat
notorious for its implementation complexity, with multiple attempts to
integrate it into
GHC~\citep{leijenHMFSimpleType2008,vytiniotisFPHFirstclassPolymorphism2008}
failing due to the difficulty of software maintenance.

\paragraph{Heuristic Approaches} A third line of work observes that
type inference is actually easy in most cases, and that an algorithm which
only does the easy things and fails otherwise may well suffice to be
an algorithm which works in practice.

One of the oldest such approaches is that of
\citet{cardelliImplementation1993}, which was invented as part of the
implementation of F{$<:$}, a language with impredicative polymorphism
and recursive types. The heuristic algorithm Cardelli proposed was to
permit impredicative instantiation without tracking bounds --- his
algorithm simply took any bound it saw as an equation telling it what
to instantiate the quantifier to.
\citet{pierceLocalTypeInference2000}, seeing how well this algorithm
worked in practice, formalized it and proved it sound (naming it \emph{local type inference}), but were not able to give a clear
non-algorithmic specification.

\subsection{Contributions}

Historically, approaches to impredicative type inference have been
willing to pay a high cost in complexity in order to achieve high
levels of expressiveness and power of type inference. In this paper,
we rethink that historical approach by focusing upon the opposite
tradeoff. We prioritize simplicity --- of implementation,
specification, and connections to existing type theories --- over the
expressiveness of type inference.

By being willing to make this sacrifice, we discover a easy-to-implement
type inference algorithm for impredicative polymorphism, which has a
clear declarative specification and is even easier to implement than
traditional Damas-Milner type inference. Our algorithm sheds light on
a classical implementation technique of Cardelli, and works over a
foundational calculus, \cbpv.

\Cbpv was invented to study the interaction of
evaluation order and side-effects, and has been used extensively in
semantics. In this paper, we show that it is
also well-adapted to being a kernel calculus for type inference
by introducing \PSystemF, a polarized variant of System F.

Specifically, our contributions are:

\begin{itemize}
\item To model type inference declaratively, we first introduce a
  subtyping relation for \PSystemF that models the specialization
  order.  Our subtyping relation is extremely simple, and is
  \emph{almost} the ``off-the-shelf'' relation that one would first
  write down. The only unusual restriction is that subtyping for
  shifts (the modalities connecting value and computation types) is
  invariant rather than covariant.

  We then give a type system for our variant of \cbpv, using subtyping
  as a component. Surprisingly, features of \cbpv invented for
  semantic or type-theoretic reasons turn out to be perfectly suited
  for specifying type inference.

  For example, argument lists (sometimes called spines in the
  literature~\cite{cervesatoLinearSpineCalculus2003,spiwackDissection2014})
  are commonly used in practical type inference
  systems~\citep{leijenHMFSimpleType2008,serranoGuardedImpredicativePolymorphism2018,serranoQuickLookImpredicativity2020}. They
  also arise in \cbpv due to the shift structure mediating between
  functions and values. This lets us naturally scope inference to
  individual function applications. So we infer types for fully
  unambiguous function calls, and otherwise require type annotations.

  This illustrates our commitment to simplicity over powerful
  inference, and also results in very regular and predictable
  annotation behavior --- our language looks like ordinary (polarized)
  System F, except that many of the ``obvious'' type instantiations
  are implicit.

\item We give algorithmic subtyping rules corresponding to the
  specialization order, and algorithmic typing rules corresponding
  to declarative typing. 

  Both of these algorithms are startlingly simple, as well --- neither
  algorithm even needs unification, making them \emph{simpler} than
  traditional
  Hindley-Damas-Milner~\citep{milnerTheoryTypePolymorphism1978} type
  inference!  We prove that our algorithm is decidable, and that it is
  sound and complete with respect to the declarative specification.

\item In fact, the combination of spine inference plus specialization
  yields an algorithm that is very similar to Cardelli's inference
  algorithm for bounded System F (as well as
  \citeauthor{pierceLocalTypeInference2000}'s local type inference).
  It has been known for decades that Cardelli's algorithm works
  extremely well in practice. However, its type-theoretic origins have
  not been fully understood, nor has it been understood how to
  characterize the kinds of problems that Cardelli's algorithm
  succeeds on using a declarative specification.

  The restrictions we impose to recover decidability of specialization
  turn out to be natural both from the perspective of type theory and
  implementation, and thereby offer a theoretically motivated
  reconstruction of Cardelli's algorithm. Polarity also ends up
  shedding light on many of the design choices underpinning a number
  of other type inference algorithms for System
  F~\cite{leijenHMFSimpleType2008,serranoQuickLookImpredicativity2020,serranoGuardedImpredicativePolymorphism2018}.
\end{itemize}

An \emph{explicit non-contribution} of this work is to propose a
practical type inference algorithm for any existing language. For
example, we work with a call-by-push-value calculus, which is a type
structure no practical language currently has. Furthermore, we make no
attempt to infer the types of arguments to functions; instead we focus solely
on inferring type applications in a (polarized) variant of System
F. Restricting the scope so much lets us isolate the key design issues
underpinning a foundational problem in type inference.

\section{\PSystemF}

We first present \PSystemF, a language with type inference that combines \cbpv style polarization with System F style polymorphism\footnote{
  Adding polarization does not result in us losing any expressiveness compared to typed System F: we demonstrate in the appendix that typeable System F can be embedded within \PSystemF.
}.

Like \cbpv, \PSystemF partitions terms into \textit{values} and \textit{computations}, ascribing \textit{positive types} to values and \textit{negative types} to computations.
\citet{levyCallbypushvalueDecomposingCallbyvalue2006} described the difference between values and computations in terms of the operational semantics: \enquote{a value \textit{is}, whereas a computation \textit{does}}.
Another rule of thumb for differentiating between values and computations is to look at how we eliminate them: we tend to eliminate values by using pattern matching and computations by supplying an argument.
For example, datatypes are values because they are eliminated by pattern matching, whereas functions are computations because they are eliminated by supplying arguments.

\begin{figure}
  \input{shared-definitions-polarized-system-f.tex}

  \caption{\PSystemF}
  \Description{Definition of Polarized System F}
  \label{fig:Polarized System F}
\end{figure}

We present the term language for \PSystemF in \cref{fig:Polarized System F}.
Terms are split between values and computations as follows:

\begin{itemize}
  \item Variables $x$ are values, since this is one of the invariants of \cbpv.

  \item Lambda abstractions $\lamterm{x}{P}{t}$ are computations, not values, since they are eliminated by passing an argument.
  We can think of a lambda abstraction as a computation $t$ of type $N$ with holes in it representing the (positive) bound variable, so lambda abstractions have the negative type $\arrowtype{P}{N}$.

  \item Thunks $\thunk{t}$ are values that suspend arbitrary computations.
  Surrounding a computation $t$ of type $N$ with braces suspends that computation, giving us a thunk $\thunk{t}$ of type $\shiftd{N}$.

  To create a traditional function value, we make a lambda abstraction $\lamterm{x}{P}{t}$ into a thunk $\thunk{\lamterm{x}{P}{t}}$.
  Traditional thunks therefore have the type $\shiftd{(\arrowtype{P}{N})}$.

  \item System-F style type abstractions $\gen{\alpha}{t}$ are computations of type $\forall \alpha \ldotp N$. 
  Our choice of polarities for the types of polymorphic terms is motivated by the correspondence between function abstraction $\lamterm{x}{P}{t}$ and type abstraction $\gen{\alpha}{t}$.

  \item The let forms $\letplain{x}{v}{s}{t}$ and $\letanno{x}{P}{v}{s}{t}$ are sequencing computations that operate much like the bind operation in a monad.
  Each form takes the thunked computation $v$, passes it the arguments $s$ it needs, binds the result to the variable $x$ and continues the computation $t$.

  \item $\return{v}$ is a trivial computation that just returns a value, completing the sequencing monad.
  We can also use $\mathsf{return}$ to return values from a function --- in fact since all the terminal symbols for terms live in the grammar of values, the syntax mandates that functions eventually return a value.
  The type structure of returned values is symmetric to that of thunks: returned values have type $\shiftu P$.
\end{itemize}

Besides the addition of type annotations to guide inference and some differences to the syntax, our language is nothing more than \cbpv plus polymorphism.
Our use of argument lists serves only to reverse the order of argument lists from Levy's original work so that our syntax matches traditional languages with n-ary functions like JavaScript and C.
Just like those languages, all of a function's arguments must be passed at once in \PSystemF.

\subsection{Declarative typing}

\begin{figure}
  \input{shared-definitions-declarative-type-system.tex}
  \caption{Declarative type system}
  \Description{Definition of a declarative type system for \PSystemF}
  \label{fig:Declarative type system}
\end{figure}

We present in \cref{fig:Declarative type system} a declarative type system for \PSystemF.
Our system has simple, mostly syntax directed rules, with its main complexity lying in the unusual premise to \Dunambiguouslet and the rules for typing argument lists.

The system has five main judgments:

\begin{itemize}
  \item $\declsynjudg{\Theta; \Gamma}{v}{P}$: In the context $\Theta$ and typing environment $\Gamma$, the value $v$ synthesizes the positive type $P$.
  \item $\declsynjudg{\Theta; \Gamma}{t}{N}$: In the context $\Theta$ and typing environment $\Gamma$, the computation $t$ synthesizes the negative type $N$.
  \item $\declspinejudg{\Theta; \Gamma}{s}{\spine{N}{M}}$: In the context $\Theta$ and typing environment $\Gamma$, and when passed to a head of type $N$, the argument list $s$ synthesizes the type $M$.
  \item $\DPosSubtypeJudg{\Theta}{P}{Q}$: In the context $\Theta$, $P$ is a positive subtype of $Q$.
  \item $\DNegSubtypeJudg{\Theta}{N}{M}$: In the context $\Theta$, $N$ is a negative subtype of $M$.
  We reverse the alphabetical order of $M$ and $N$ in the negative judgment to better indicate the symmetry between positive and negative subtyping --- we will see what this symmetry is in \cref{section:groundness-invariant}.
\end{itemize}

When we want to be ambiguous between values and computations, we write $\declsynjudg{\Theta; \Gamma}{e}{A}$ and $\DPnSubtypeJudge{\Theta}{A}{B}$.

Our inference rules are the following:

\begin{itemize}
  \item \Dvar lets us reference variables within the type environment $\typeenv$.

  \item \Dthunk states that if we know that a computation has type $N$, then thunking it produces a value of type $\shiftd{N}$.

  \item \Dfunabs is the standard typing rule for lambda abstraction.
  The value hypothesis $x : P$ in this rule maintains the invariant that the type environment only contains bindings to values.

  \item \Dtypeabs is to \Dfunabs as type abstraction is to function abstraction.
  We add a type variable to the typing context, as opposed to adding a variable binding to the typing environment.

  \item \Dreturn complements \Dthunk: if a value has type $P$, then returning it produces a computation of type $\shiftu{P}$.

  \item \Dambiguouslet allows us to let-bind the results of function applications.

  \begin{itemize}
    \item In the first premise, $\declsynjudg{\dcontext; \typeenv}{v}{\shiftd M}$, we require $v$ to be a thunked computation.

    \item We then take the type $M$ of this computation and use the spine judgment $\declspinejudg{\Theta; \Gamma}{s}{\spine{M}{\shiftu{Q}}}$ to type the function application $v(s)$.
    The requirement that the argument list produces an upshifted type $\shiftu{Q}$ means that the type has to be maximally applied.
    This encodes the fact that partial application is forbidden.

    Typically $v$ will be a downshifted function and $s$ will be the argument list to pass to that function.
    For example, $v$ could be a function stored within the type environment, \eg $f$ when $f : \shiftd{(\arrowtype{P}{N})} \in \typeenv$, or one written inline $\thunk{\lamterm{x}{P}{t}}$.
    However we can also use an embedded value $\thunk{\return{v}}$ as a head along with an empty argument list to let-bind values.

    \item Now that we have the type $\shiftu{Q}$ that the argument list takes the head to, we use the subtyping judgment $\DNegSubtypeJudge{\Theta}{\shiftu{Q}}{\shiftu{P}}$ to check that the type synthesized by the function application is compatible with the annotation $P$.
    For our algorithm to work, we need to check the stronger compatibility constraint that $\shiftu{Q}$ is a subtype of $\shiftu{P}$, rather than just checking that $Q$ is a subtype of $P$.

    \item Finally, $\declsynjudg{\Theta; \Gamma, x : P}{t}{N}$ tells us that the type of the let term is given by the type of $t$ in the type environment $\typeenv$ extended with $x$ bound to $P$.
  \end{itemize}

  \item \Dunambiguouslet is a variant of \Dambiguouslet that lets us infer the return types of function applications when they are unambiguous.
  The first and second premises are identical to those of \Dambiguouslet, and the third premise is the same as the last premise of \Dambiguouslet except that it binds $x$ directly to the type synthesized by the function application.

  However the final premise is unusual, and one might wonder whether it is in fact well founded.
  This premise encodes the condition that in order to omit the annotation, the return type of the function application must be unambiguous.
  By quantifying over all possible inferred types, we check that there is only one type (modulo isomorphism) that can be inferred for the return type of the function application.
  If every inferred type is equivalent to $Q$, then we can arbitrarily choose $Q$ to bind $x$ to, since any other choice would synthesize the same type for the let-binding.

  The well-foundedness of our rule stems from the use of a syntactically smaller subterm in the premise of \Dunambiguouslet compared to the conclusion.
  In $\declspinejudg{\Theta;\Gamma}{s}{\spine{M}{\shiftu{P}}}$, $s$ is a smaller subterm than $\letplain{x}{v}{s}{t}$, and every subderivation of that premise acts on a smaller subterm too.

  A consequence of this unusual premise is that the soundness and completeness theorems end up being mutually recursive.
\end{itemize}

\begin{figure}
  \input{shared-definitions-well-formed-declarative-type.tex}
  \caption{Well-formedness of declarative types}
  \Description{Definition of when a declarative \PSystemF type is well-formed}
  \label{figure:well-formed-declarative}
\end{figure}

The last three rules teach us how to type function applications:

\begin{itemize}
  \item \Dspinenil states that an empty argument list does not change the type of the head of a function application.
  \item \Dspinecons tells us how to type a non-empty argument list.
  The argument list $v, s$ takes $\arrowtype{Q}{N}$ to $M$ if
  the value $v$ synthesizes a type $P$,
  this type $P$ is either $Q$ or a subtype of it, and
  the remaining argument list $s$ takes $N$ to $M$.
  \item \Dspinetypeabs lets us instantiate the head that an argument list can be passed to by replacing a type variable $\alpha$ with a well-formed (see \cref{figure:well-formed-declarative}) type $P$.
  Note that the instantiation is implicit — there is no annotation for $P$, so we are doing inference here.
\end{itemize}

\subsection{Subtyping}

\begin{figure}
  \input{shared-definitions-declarative-subtyping.tex}
  \caption{Declarative subtyping}
  \Description{Definition of the declarative subtyping judgment for \PSystemF}
  \label{fig:Declarative subtyping}
\end{figure}

To enable implicit type application, our type system uses a subtyping relation $\DPnSubtypeJudge{\dcontext}{A}{B}$ which reifies the specialization order, expressing when a term of type $A$ can be safely used where a term of type $B$ is expected.
This relation is defined in \cref{fig:Declarative subtyping} and consists of the following rules:

\begin{itemize}
  \item \drefl is the standard rule for type variables: if we are expecting a term of type $\alpha$, then the only terms we can safely use in its place are those that also have type $\alpha$.

  \item \darrow is the classic rule for function subtyping: we can replace a term of type $\arrowtype{Q}{M}$ with a function that takes inputs that are at least as general as $Q$ and produces outputs that are at least as specific as $M$.

  \item \dforalll and \dforallr correspond to the left and right rules governing $\forall$ quantifiers in the LK sequent calculus system.
  Surprisingly, the \dforalll rule does not have any restrictions beyond well-formedness: $P$ is an arbitrary well-formed type, so full impredicative instantiation is possible.

  \item \dshiftdown and \dshiftup are unusual since the standard denotational and operational semantics of shifts both give rise to covariant rules.
  However using these covariant rules would lead to an undecidable system~\citep{chrzaszcz-98,tiuryn-urzczyn-96}.
  To restore decidability, we use more restrictive invariant shift rules.

  This idea is motivated by a similar restriction that \citet{cardelliImplementation1993} introduced as a heuristic for type inference in an implementation of \systemfsub.
  Instead of using subtyping to infer the types assigned to type parameters, \citeauthor{cardelliImplementation1993} used first-order unification to \enquote{synthesize} these types.

  \begin{figure}
    \input{shared-definitions-isomorphic-types.tex}
    \caption{Isomorphic types}
    \Description{Definition of when two types are isomorphic}
    \label{figure:isomorphic-types}
  \end{figure}

  Since the goal of unification is to make expressions equal, an equivalent notion of \citeauthor{cardelliImplementation1993}'s restriction in our system is requiring the types to be \textit{isomorphic}.
  We say that types $A$ and $B$ are isomorphic, denoted as $\MutualSubtypePosNegJudge{\dcontext}{A}{B}$, if both $\DPnSubtypeJudge{\dcontext}{A}{B}$ and $\DPnSubtypeJudge{\dcontext}{B}{A}$ hold (see \cref{figure:isomorphic-types}).
  Indeed, this requirement corresponds exactly to the premises of the shift rules.
  To our knowledge, this is the first time that \citeauthor{cardelliImplementation1993}'s restriction has been presented as part of a declarative specification of typing.
\end{itemize}

\newcommand{\listtype}[1]{[#1]}

\section{Examples}

In this section we characterize the behavior of our system by giving some examples of terms and subtyping relationships that are permitted and some that are not.

\subsection{Subtyping}

\begin{itemize}
  \item Our subtyping relation allows us to swap quantifiers at identical depths.
    As indicated by the subtyping in both directions below, we can substitute a term of either of the types below for a term of the other type:

    \begin{align*}
      \forall \alpha, \beta \ldotp \shiftd{(\alpha \funarrow \shiftu{\beta})} \funarrow \listtype{\alpha} \funarrow \shiftu{\listtype{\beta}}
      \quad \genfrac{}{}{0pt}{0}{\negsubtype}{\rotatebox[origin=c]{180}{$\negsubtype$}} \quad
      \forall \beta, \alpha \ldotp \shiftd{(\alpha \funarrow \shiftu{\beta})} \funarrow \listtype{\alpha} \funarrow \shiftu{\listtype{\beta}}
    \end{align*}

  \item Subtypes can push in quantifiers as long as those quantifiers do not cross a shift boundary:

    \begin{align*}
      \forall \alpha \ldotp \alpha \funarrow (\forall \beta \ldotp \beta \funarrow \shiftu{(\alpha \times \beta)})
      \enspace\negsubtype\enspace
      \forall \alpha, \beta \ldotp \alpha \funarrow \beta \funarrow \shiftu{(\alpha \times \beta)}
    \end{align*}

  \item Unusually, our subtyping relation allows the instantiations of type variables to be impredicative.
    For example, here we can instantiate the $\alpha$ in $\listtype{\alpha}$ to be an identity function $\forall \beta \ldotp \beta \funarrow \shiftu{\beta}$, giving us a list of \textit{polymorphic} identity functions.
    Predicative systems would allow us to produce a list of identity functions that are all parameterized over the same type, i.e. $\forall \beta \ldotp \listtype{\beta \funarrow \shiftu{\beta}}$, but the list produced by our system $\listtype{\forall \beta \ldotp \beta \funarrow \shiftu{\beta}}$ is truly polymorphic with each element of the list having its own type variable.

    \begin{gather*}
      \forall \alpha \ldotp \shiftu \listtype{\alpha}
      \enspace\negsubtype\enspace
      \shiftu \listtype{\shiftd (\forall \beta \ldotp \beta \funarrow \shiftu{\beta})}
    \end{gather*}

  \item An important restriction of our system that is essential to decidability is that types underneath shifts must be isomorphic.
    For example, we must establish both $M \negsubtype N$ and $N \negsubtype M$ before we can infer that $\shiftd M \possubtype \shiftd N$.

    This means that terms of either of the types below may not be substituted for terms of the other type.
    While $\forall \alpha, \beta \ldotp \alpha \funarrow \beta \funarrow \shiftu{(\alpha \times \beta)}$ is indeed a subtype of $\mathsf{\inttype} \funarrow \mathsf{\stringtype} \funarrow \shiftu{(\mathsf{\inttype \times \stringtype})}$, the subtyping does not hold the other way around.
    Therefore in this example where we zip a list of integers and a list of strings together, the type of the pairing function $\mathsf{\shiftd (\inttype \funarrow \stringtype \funarrow \shiftu{(\inttype \times \stringtype)})}$ is neither a subtype nor a supertype of the type of the generic pairing function $\shiftd (\forall \alpha, \beta \ldotp \alpha \funarrow \beta \funarrow \shiftu{(\alpha \times \beta)})$.

    \begin{gather*}
      \mathsf{\shiftd{(\inttype \funarrow \stringtype \funarrow \shiftu{(\inttype \times \stringtype)})} \funarrow \listtype{\inttype} \funarrow \listtype{\stringtype} \funarrow \shiftu{\listtype{\inttype \times \stringtype}}} \\
      \rotatebox[origin=c]{-90}{$\not\negsubtype$} \quad \rotatebox[origin=c]{90}{$\not\negsubtype$} \\
      \mathsf{\shiftd{(\forall \alpha, \beta \ldotp \alpha \funarrow \beta \funarrow \shiftu{(\alpha \times \beta)})} \funarrow \listtype{\inttype} \funarrow \listtype{\stringtype} \funarrow \shiftu{\listtype{\inttype \times \stringtype}}}
    \end{gather*}
\end{itemize}

\subsection{Typing}

\newcommand{\booltype}{\datatype{Bool}}
\newcommand{\sttype}[2]{\datatype{ST}\ #1\ #2}

\begin{figure}
  \small{
    \begin{alignat*}{4}
      &\textsf{head} &&: \mathsf{\shiftd{(\foralltype{\alpha}{\arrowtype{\listtype{\alpha}}{\shiftu{\alpha}}})}}
      &&\qquad\textsf{inc} &&: \mathsf{\shiftd{(\arrowtype{\inttype}{\shiftu{\inttype}})}} \\
      &\textsf{tail} &&: \mathsf{\shiftd{(\foralltype{\alpha}{\arrowtype{\listtype{\alpha}}{\shiftu{\listtype{\alpha}}}})}}
      &&\qquad\textsf{choose} &&: \mathsf{\shiftd{(\foralltype{\alpha}{\arrowtype{\alpha}{\arrowtype{\alpha}{\shiftu{\alpha}}}})}} \\
      &\listterm{} &&: \mathsf{\shiftd{(\foralltype{\alpha}{\shiftu{\listtype{\alpha}}})}}
      &&\qquad\textsf{poly} &&: \mathsf{\shiftd{(\arrowtype{\shiftd{(\foralltype{\alpha}{\arrowtype{\alpha}{\shiftu{\alpha}}})}}{\shiftu{(\prodtype{\inttype}{\booltype})}})}} \\
      &(::) &&: \mathsf{\shiftd{(\foralltype{\alpha}{\arrowtype{\alpha}{\arrowtype{\listtype{\alpha}}{\shiftu{\listtype{\alpha}}}}})}}
      &&\qquad\textsf{auto} &&: \mathsf{\shiftd{(\arrowtype{\shiftd{(\foralltype{\alpha}{\arrowtype{\alpha}{\shiftu{\alpha}}})}}{(\foralltype{\alpha}{\arrowtype{\alpha}{\shiftu{\alpha}}})})}} \\
      &\textsf{single} &&: \mathsf{\shiftd{(\foralltype{\alpha}{\arrowtype{\alpha}{\shiftu{\listtype{\alpha}}}})}}
      &&\qquad\textsf{auto'} &&: \mathsf{\shiftd{(\foralltype{\alpha}{\arrowtype{\shiftd{(\foralltype{\beta}{\arrowtype{\beta}{\shiftu{\beta}}})}}{\arrowtype{\alpha}{\shiftu{\alpha}}}})}} \\
      &\textsf{append} &&: \mathsf{\shiftd{(\foralltype{\alpha}{\arrowtype{\listtype{\alpha}}{\arrowtype{\listtype{\alpha}}{\shiftu{\listtype{\alpha}}}}})}}
      &&\qquad\textsf{app} &&: \mathsf{\shiftd{(\foralltype{\alpha,\beta}{\arrowtype{\shiftd{(\arrowtype{\alpha}{\shiftu{\beta}})}}{\arrowtype{\alpha}{\shiftu{\beta}}}})}} \\
      &\textsf{length} &&: \mathsf{\shiftd{(\foralltype{\alpha}{\arrowtype{\listtype{\alpha}}{\shiftu{\inttype}}})}}
      &&\qquad\textsf{revapp} &&: \mathsf{\shiftd{(\foralltype{\alpha,\beta}{\arrowtype{\alpha}{\arrowtype{\shiftd{(\arrowtype{\alpha}{\shiftu{\beta}})}}{\shiftu{\beta}}}})}} \\
      &\textsf{map} &&: \mathsf{\shiftd{(\foralltype{\alpha,\beta}{\arrowtype{\shiftd{(\arrowtype{\alpha}{\beta})}}{\arrowtype{\listtype{\alpha}}{\shiftu{\listtype{\beta}}}}})}}
      &&\qquad\textsf{flip} &&: \mathsf{\shiftd{(\foralltype{\alpha,\beta,\gamma}{\arrowtype{
        \shiftd{(\arrowtype{\alpha}{\arrowtype{\beta}{\shiftu{\gamma}}})}
      }{
        \arrowtype{\beta}{\arrowtype{\alpha}{\shiftu{\gamma}}}
      }})}} \\
      &\textsf{id} &&: \mathsf{\shiftd{(\foralltype{\alpha}{\arrowtype{\alpha}{\shiftu{\alpha}}})}}
      &&\qquad\textsf{runST} &&: \mathsf{\shiftd{(\foralltype{\alpha}{\arrowtype{
        \shiftd{(\foralltype{\beta}{\sttype{\beta}{\alpha}})}
      }{
        \shiftu{\alpha}
      }})}} \\
      &\textsf{ids} &&: \mathsf{\listtype{\shiftd{(\foralltype{\alpha}{\arrowtype{\alpha}{\shiftu{\alpha}}})}}}
      &&\qquad\textsf{argST} &&: \mathsf{\shiftd{(\foralltype{\alpha}{\sttype{\alpha}{\inttype}})}} \\
      &\textsf{a9} &&: \mathsf{\shiftd{(\foralltype{\alpha}{\arrowtype{\shiftd{(\arrowtype{\alpha}{\shiftu{\alpha}})}}{\arrowtype{\listtype{\alpha}}{\shiftu{\alpha}}}})}}
      &&\qquad\textsf{c8} &&: \mathsf{\shiftd{(\foralltype{\alpha}{\arrowtype{\listtype{\alpha}}{\arrowtype{\listtype{\alpha}}{\shiftu{\alpha}}}})}} \\
      &\textsf{h} &&: \mathsf{\shiftd{(\arrowtype{\inttype}{(\foralltype{\alpha}{\arrowtype{\alpha}{\shiftu{\alpha}}})})}}
      &&\qquad\textsf{k} &&: \mathsf{\shiftd{(\foralltype{\alpha}{\arrowtype{\alpha}{\arrowtype{\listtype{\alpha}}{\shiftu{\alpha}}}})}} \\
      &\textsf{lst} &&: \mathsf{\listtype{\shiftd{(\foralltype{\alpha}{\arrowtype{\inttype}{\arrowtype{\alpha}{\shiftu{\alpha}}}})}}}
      &&\qquad\textsf{r} &&: \mathsf{\shiftd{(\arrowtype{\shiftd{(\foralltype{\alpha}{\arrowtype{\alpha}{\foralltype{\beta}{\arrowtype{\beta}{\shiftu{\beta}}}})}}}{\shiftu{\inttype}})}}
    \end{alignat*}
  }
  \caption{Environment for the examples from \citet{serranoGuardedImpredicativePolymorphism2018}, translated into \PSystemF}
  \Description{Environment for the examples from \citet{serranoGuardedImpredicativePolymorphism2018}, translated into \PSystemF}
  \label{figure:gi-examples-environment}
\end{figure}

\begin{figure}
  \small{
    \rowcolors{2}{white}{gray!20}
    \begin{tabularx}{\linewidth}{
        p{0.6cm}
        p{3.0cm}
        X
      | >{\centering\arraybackslash}m{0.6cm} }
      & \textbf{GI example} & \textbf{\PSystemF translation} & \\
      A1 & $\mathsf{const2 = \lambda x\ y\ldotp y}$ &
      $\mathsf{\letvar{const2}{\thunk{\return{\thunk{
        \gen{\alpha}{\gen{\beta}{\lamterm{x}{\alpha}{\lamterm{y}{\beta}{\return{y}}}}}}
      }}}{\ldots}}$ &
      Ann \\
      A2 & $\mathsf{choose\ id}$ &
      $\mathsf{\lamterm{x}{\shiftd{(\foralltype{\alpha}{\arrowtype{\alpha}{\shiftu{\alpha}}})}}
        {(\letplainnobrackets{t}{\textsf{choose}}{\textsf{id} \ x}{\return{t}})}}$ &
      Ann \\
      A3 & \textsf{choose [] ids} &
      $\mathsf{\letvar{n : \listtype{\shiftd{(\foralltype{\alpha}{\arrowtype{\alpha}{\shiftu{\alpha}}})}}}{[]}
      {\letplainnobrackets{t}{choose}{n\ ids}{\return{t}}}}$ &
      Ann \\
      A4 & \textsf{$\lambda$x : ($\forall \alpha \ldotp \alpha \funarrow \alpha$). x x} &
      $\mathsf{\lamterm{x}{(\foralltype{\alpha}{\arrowtype{\alpha}{\shiftu{\alpha}}})}{(\letplainnobrackets{y}{x}{x}{\return{y}})}}$ &
      \checkmark \\
      A5 & \textsf{id auto} &
      $\mathsf{\letplainnobrackets{t}{id}{auto}{\return{t}}}$ &
      \checkmark \\
      A6 & $\mathsf{id\ auto'}$ &
      $\mathsf{\letplainnobrackets{t}{id}{auto'}{\return{t}}}$ &
      \checkmark \\
      A7 & \textsf{choose id auto} &
      $\mathsf{\letplainnobrackets{t}{choose}{id\ auto}{\return{t}}}$ &
      $\times$ \\
      A8 & $\mathsf{choose\ id\ auto'}$ &
      $\mathsf{\letplainnobrackets{t}{choose}{id\ auto'}{\return{t}}}$ &
      $\times$ \\
      A9 & \textsf{a9 (choose id) ids} & %
      $\begin{aligned}[t]
        \letvar{\mathsf{f}}{\thunk{\return{
        \thunk{&\lamterm{\mathsf{x}}{\shiftd{(\foralltype{\alpha}{\arrowtype{\alpha}{\shiftu{\alpha}}})}}
          {\\ &\mathsf{\letplainnobrackets{y}{choose}{id\ x}{\return{y}}}}
        }
      }}}
      {\\ \noalign{$\mathsf{\letplainnobrackets{t}{a9}{f\ ids}{\return{t}}}$}}
      \end{aligned}$ &
      Ann \\
      A10 & $\mathsf{poly\ id}$ &
      $\mathsf{\letplainnobrackets{t}{poly}{id}{\return{t}}}$ &
      \checkmark \\
      A11 & $\mathsf{poly\ (\lambda x \ldotp x)}$ &
      $\mathsf{\letplainnobrackets{t}{poly}{\thunk{\gen{\alpha}\lamterm{x}{\alpha}{\return{x}}}}{\return{t}}}$ &
      Ann \\
      A12 & $\mathsf{id\ poly\ (\lambda x \ldotp x)}$ &
      $\mathsf{\letplainnobrackets{x}{id}{poly}
      {\letplainnobrackets{t}{x}{\thunk{\gen{\alpha}\lamterm{y}{\alpha}{\return{y}}}}{\return{t}}}}$ &
      Ann \\
      B1 & $\mathsf{\lambda f \ldotp (f\ 1, f\ True)}$ &
      $\mathsf{\lamterm{f}
        {\shiftd{(\foralltype{\alpha}{\arrowtype{\alpha}{\shiftu{\alpha}}})}}
        {(\letplainnobrackets{l}{f}{1}{\letplainnobrackets{r}{f}{\textsf{true}}{\return{(l, r)}}}})}$ &
      Ann \\
      B2 & $\mathsf{\lambda xs \ldotp poly\ (head\ xs)}$ &
      $\mathsf{\lamterm{xs}{\listtype{\shiftd{(\foralltype{\alpha}{\arrowtype{\alpha}{\shiftu{\alpha}}})}}}{(\textsf{let x = head xs; let y = poly x; return y})}}$ &
      Ann \\
      C1 & \textsf{length ids} &
      $\mathsf{\letplainnobrackets{t}{length}{ids}{\return{t}}}$ &
      \checkmark \\
      C2 & \textsf{tail ids} &
      $\mathsf{\letplainnobrackets{t}{tail}{ids}{\return{t}}}$ &
      \checkmark \\
      C3 & \textsf{head ids} &
      $\mathsf{\letplainnobrackets{t}{head}{ids}{\return{t}}}$ &
      \checkmark \\
      C4 & \textsf{single id} &
      $\mathsf{\letplainnobrackets{t}{single}{id}{\return{t}}}$ &
      \checkmark \\
      C5 & \textsf{cons id ids} &
      $\mathsf{\letplainnobrackets{t}{cons}{id\ ids}{\return{t}}}$ &
      \checkmark \\
      C6 & \textsf{cons ($\mathsf{\lambda x \ldotp x}$) ids} &
      \textsf{let t $=$ cons $\mathsf{\thunk{\gen{\alpha}\lamterm{x}{\alpha}{\return{x}}}}$ ids; return t} &
      Ann \\
      C7 & \textsf{append (single inc) \newline (single ids)} &
      \textsf{let x $=$ single inc; let y $=$ single id; let t $=$ append x y; return t} &
      $\times$ \\
      C8 & \textsf{c8 (single id) ids} &
      \textsf{let x $=$ single id; let t $=$ c8 x ids; return t} &
      \checkmark \\
      C9 & \textsf{map poly (single id)} &
      \textsf{let x $=$ single id; let t $=$ map poly x; return t} &
      \checkmark \\
      C10 & \textsf{map head (single ids)} &
      \textsf{let x $=$ single ids; let t $=$ map head x; return t} &
      $\times$ \\
      D1 & \textsf{app poly id} &
      \textsf{let t $=$ app poly id; return t} &
      \checkmark \\
      D2 & \textsf{revapp id poly} &
      \textsf{let t $=$ revapp id poly; return t} &
      \checkmark \\
      D3 & \textsf{runST argST} &
      \textsf{let t $=$ runST argST; return t} &
      \checkmark \\
      D4 & \textsf{app runST argST} &
      \textsf{let t $=$ app runST argST; return t} &
      $\times$ \\
      D5 & \textsf{revapp argST runST} &
      \textsf{let t $=$ revapp argST runST; return t} &
      $\times$ \\
      E1 & \textsf{k h lst} &
      \textsf{let t $=$ k h lst; return t} &
      $\times$ \\
      E2 & \textsf{k ($\lambda x \ldotp h\ x$) lst} &
      $\begin{aligned}[t]
        \letvar{\mathsf{\mathsf{f}}}{\thunk{\return{
          \thunk{&\gen{\alpha}{\lamterm{\mathsf{y}}{\mathsf{\inttype}}
          {\\ &\return{\thunk{\lamterm{\mathsf{x}}{\alpha}{
            \mathsf{\letplainnobrackets{z}{h}{y \ z}{\return{z}}}
          }}}}}}
        }}}
      {\\ \noalign{$\mathsf{\letplainnobrackets{t}{k}{f\ lst}{\return{t}}}$}}
      \end{aligned}$ &
      $\times$ \\
      E3 & \textsf{r ($\lambda x\ y \ldotp y$)} &
      \textsf{let t $=$ r $\mathsf{\thunk{\gen{\alpha}\gen{\beta}{\lamterm{x}{\alpha}{\lamterm{y}{\beta}{\return{y}}}}}}$; return t} &
      $\times$
    \end{tabularx}
  }
  \captionsetup{singlelinecheck=off}
  \caption[Examples used in the comparison of type systems in \citet{serranoGuardedImpredicativePolymorphism2018} along with their translations to \PSystemF]{
    Examples used in the comparison of type systems in \citet{serranoGuardedImpredicativePolymorphism2018} along with their translations to \PSystemF.
    In the right hand column:
    \begin{itemize}
      \item \checkmark \xspace indicates that an example typechecks without needing additional annotations beyond the GI example.
      \item Ann indicates that the example typechecks, but additional annotations are required beyond the GI example.
      \item $\times$ indicates that the example does not typecheck in \PSystemF.
    \end{itemize}
  }
  \Description{
    Examples used in the comparison of type systems in~\citet{serranoGuardedImpredicativePolymorphism2018} along with their translations to \PSystemF.
  }
  \label{fig:guarded-examples}
\end{figure}

To illustrate \PSystemF's type system, we translate the examples used in the comparison of type systems in~\citet{serranoGuardedImpredicativePolymorphism2018} (GI) to \PSystemF and investigate which of these translations (a) typecheck without needing additional annotations beyond the GI example, (b) typecheck but require additional annotations beyond the GI example, and (c) do not typecheck.
These examples make use of a few common functions whose types we give in \cref{figure:gi-examples-environment}.

Despite the extreme bias towards simplicity that is reflected in our straightforward local heuristic, surprisingly many of these examples typecheck.
We discuss a few characteristic ones below; the full set are in \cref{fig:guarded-examples}.

\paragraph{Shifts make types much more fine-grained}

The shift structure in \PSystemF types makes them much more fine-grained compared to systems like GI.
For example, in \PSystemF the number of arguments we can pass to a function is encoded in its type.
If we instantiate \textsf{id} at $\shiftd{(\foralltype{\alpha}{\arrowtype{\alpha}{\shiftu{\alpha}}})}$, it still takes only a single argument.
In contrast, in systems like GI, instantiating \textsf{id} at $\foralltype{\alpha}{\arrowtype{\alpha}{\alpha}}$ gives us a function that can take two arguments: first \textsf{id} and then an arbitrary value.

As a result, while \textsf{id} instantiated at $\foralltype{\alpha}{\arrowtype{\alpha}{\alpha}}$ and \textsf{auto} represent the same thing in systems like GI, \textsf{id} instantiated at $\shiftd{(\foralltype{\alpha}{\arrowtype{\alpha}{\shiftu{\alpha}}})}$ and \textsf{auto} represent fundamentally different things in \PSystemF.
Therefore in A7 we can not synthesize a consistent return type for \textsf{choose id auto} and so it does not typecheck.

Examples A8, C7, and E2 do not work for similar reasons.
For instance, in C7 our use of invariant shifts prevents us from using \textsf{single ids} at a less polymorphic type.

Another example of the shift structure in \PSystemF making types much more fine-grained is A3: a direct translation $\mathsf{\letplainnobrackets{t}{choose}{[]\ ids}{\return{t}}}$ will not typecheck since $[]$ has a shift at its top-level while \textsf{ids} does not.
As a result, we need to unwrap the top-level shift in the type of $[]$ using a let-binding like so: $\mathsf{let\ n} : \listtype{\shiftd{(\foralltype{\alpha}{\arrowtype{\alpha}{\shiftu{\alpha}}})}} = []$.

\paragraph{Hyperlocal inference}

Inference in \PSystemF refuses to incorporate far away information.
As encoded in the last premise of the \Dunambiguouslet rule, we must annotate a let-bound variable unless its type is completely determined by the application $v(s)$.

For example, in A3 there are many types we can infer for the empty list constructor \textsf{[]}, for instance $\listtype{\mathsf{\inttype}}$, $\listtype{\forall \alpha \ldotp \alpha}$, and $\forall \alpha \ldotp \listtype{\alpha}$.
We can see from the later application \textsf{choose n ids} that $\listtype{\shiftd{(\foralltype{\alpha}{\arrowtype{\alpha}{\shiftu{\alpha}}})}}$ is the right choice for this type.
However, \PSystemF only considers the immediate context of the function application.
In this context the type for \textsf{n} is ambiguous, so we need to annotate the let-binding.

While A3 needs an annotation, examples like C4 do not.
While the type $\listtype{\shiftd{(\foralltype{\alpha}{\arrowtype{\alpha}{\shiftu{\alpha}}})}}$ that is synthesized for the let-bound variable in C4 contains universal quantification, this quantification occurs underneath a shift.
Since shifts are invariant, we can not perform subtyping underneath them, so this type is the only one modulo isomorphism that we can infer for the let-bound variable.

\paragraph{We sometimes need to eta-expand terms to get them to type check}

An example of this is C10 \textsf{map head (single ids)}.
Here \textsf{map} expects an argument of type $\shiftd{(\arrowtype{\alpha}{\shiftu{\beta}})}$, but \textsf{head} has type $\shiftd{(\foralltype{\alpha}{\arrowtype{\listtype{\alpha}}{\shiftu{\alpha}}})}$.
Since shifts are invariant in \PSystemF, these types are incompatible.
We can remedy this by eta-expanding the term, for instance for C10 we can create a new term $\lamterm{x}{\shiftd{(\foralltype{\alpha}{\arrowtype{\alpha}{\shiftu{\alpha}}})}}{\letplainnobrackets{y}{\textsf{head}}{x}{\return{y}}}$ that specializes \textsf{head} to the type $\shiftd{(\arrowtype{\listtype{\shiftd{\foralltype{\alpha}{\arrowtype{\alpha}{\shiftu{\alpha}}}}}}{\shiftu{\shiftd{(\foralltype{\alpha}{\arrowtype{\alpha}{\shiftu{\alpha}}}}})})}$.
Examples D4, D5, E1, and E3 fail to typecheck for similar reasons.

Another situation where eta-expansion is necessary is when replicating partial application.
Since \PSystemF does not allow partial application, partial applications like \textsf{choose id} in GI must be represented as $\lamterm{x}{\shiftd{(\foralltype{\alpha}{\arrowtype{\alpha}{\shiftu{\alpha}}})}}
{\letplainnobrackets{y}{\textsf{choose}}{\textsf{id} \ x}{\return{y}}}$ in \PSystemF.
Since we syntactically require users to annotate lambda-bound variables, these examples all require annotations in \PSystemF.
This affects examples A2, A9, and E2.

\paragraph{\PSystemF tends to prefer inferring impredicative types to inferring type schemes}

Adding impredicativity to Hindley-Milner style systems means that, without changes to the type language like the introduction of bounded quantification, terms are no longer guaranteed to have a most general type.
For instance, when typing \textsf{single id} we find that $\listtype{\foralltype{\alpha}{\arrowtype{\alpha}{\alpha}}}$ and $\foralltype{\alpha}{\listtype{\arrowtype{\alpha}{\alpha}}}$ are equally general.
Due to its simple local heuristic, \PSystemF will instantiate the type variable in \textsf{single} impredicatively, thereby typing \textsf{single} as $\listtype{\foralltype{\alpha}{\arrowtype{\alpha}{\alpha}}}$.
In GI, the choice we make depends on whether the type variable $\alpha$ in the definition of $\textsf{single}$ appears under a type constructor in any of the arguments to it.
Since it does not, GI assigns $\alpha$ a \enquote{top-level monomorphic} type, thereby typing \textsf{single} as $\foralltype{\alpha}{\listtype{\arrowtype{\alpha}{\alpha}}}$.

\PSystemF's preference for inferring impredicative types is not an unquestionable pro or con: while it allows us to infer examples like C8 and C9 that systems that prefer type schemes can not, systems that prefer type schemes can deal better with eta-reduced terms.

\section{Algorithmic typing}

In this section, we discuss how to implement our declarative system algorithmically.

There are two rules that prevent us from directly implementing the declarative system.
These are \Dspinetypeabs and \dforalll, both of which have a type $P$ that appears in the premises but not in the conclusion of each rule.
As a result, a direct implementation of type inference would need to somehow invent a type $P$ out of thin air.

\begin{figure}
  \input{shared-definitions-algorithmic-system.tex}

  \caption{Additions to declarative types and contexts to form their algorithmic counterparts}
  \Description{Additions to declarative types and contexts to form their algorithmic counterparts}
  \label{fig:algorithmic-types-and-contexts}
\end{figure}

To get an algorithm, we follow~\citet{dunfieldCompleteEasyBidirectional2013} in replacing these types $P$ with \textit{existential type variables} $\guess{\alpha}$.
These represent unsolved types that the algorithm will solve at a later stage.
Since type variables and their solutions are positive, existential type variables must also be positive.

The introduction of existential variables has some knock-ons on our system: each judgment needs to be adapted to manage existential variables and their instantiation.
For instance, as shown in \cref{fig:algorithmic-types-and-contexts}, the contexts of algorithmic judgments will contain not only type variables, but also solved and unsolved existential type variables.

\subsection{Algorithmic subtyping}
\label{section:groundness-invariant}

We first describe the algorithmic judgments for subtyping.
These will be used to solve all relevant existential type variables.

Our algorithm for subtyping consists of two mutually recursive judgments $\APosSubtypeJudg{\acontext}{P}{Q}{\acontext'}$ and $\ANegSubtypeJudg{\acontext}{N}{M}{\acontext'}$.
Each judgment takes an input context $\acontext$, checks that the \lhs type is a subtype of the \rhs type, and produces an output context $\acontext'$.
The shape of this output context is identical to that of the corresponding input context: the only difference between them is that the output context contains the solutions to any newly solved existential variables.
Since existential type variables can only have a single solution, output contexts can only solve existential variables that are unsolved in the input context.

As a result of polarization, existential type variables only appear on the left-hand side of negative subtyping judgments and the right-hand side of positive subtyping judgments.
This \emph{groundness invariant} is critical to the workings of our algorithm.
Since we will frequently make use of this invariant, we briefly introduce some terminology related to it.
We call types that do not contain any existential variables \textit{ground}.
The side of a judgment that can not contain any existential type variables is the \textit{ground side}, and the opposite side is the \textit{non-ground side}.
We will observe symmetries between properties of the positive and negative judgments based on the groundness of the types (see \cref{lemma:well-formedness-pnsubtype} for instance).
This is why we swap the alphabetical order of $M$ and $N$ when writing negative judgments.

\begin{figure}
  \input{shared-definitions-algorithmic-subtyping.tex}

  \caption{Algorithmic subtyping}
  \Description{Definition of the algorithmic subtyping judgment for \PSystemF}
  \label{figure:algorithmic-subtyping}
\end{figure}

We present our algorithmic subtyping rules in \cref{figure:algorithmic-subtyping}.

\begin{itemize}
  \item \arefl is identical to our declarative rule for type variables, however we have moved the $\wfpostypeJudg{\acontext}{\alpha}$ judgment into an equivalent requirement that the input and output contexts have the form $\aconwithhole{\alpha}$.

  \item \ainst is a new rule that describes how to solve existential type variables.
    The idea is that if we reach a subtyping judgment like $P \possubtype \guess{\alpha}$, then there is no further work to be done, and we can just set $\guess{\alpha}$ to be equal to $P$.
    Since this rule is the only way we solve existential variables, all solutions to existential variables are ground.

    The premises of \ainst ensure that the rule preserves well-formedness: as shown in \cref{figure:context-wellformedness}, our contexts are ordered, and a solution $P$ to an existential variable $\guess{\alpha}$ must be well-formed with respect to the context items prior to $\guess{\alpha}$, \ie $\acontext_L$.
    This means that the solution for $\guess{\alpha}$ can only contain type variables that were in scope when $\guess{\alpha}$ was introduced: an important property for the soundness of the system.

  \item \aforalll implements the strategy we described to make the \dforalll rule algorithmic: rather than replacing the type variable $\alpha$ with a ground type $P$, we procrastinate deciding what $P$ should be and instead replace $\alpha$ with a fresh existential type variable $\guess{\alpha}$.
    The bracketed $\guess{\alpha}[ = P]$ in the output context indicates that the algorithm doesn't have to solve $\guess{\alpha}$: we will later see in \cref{section:subtyping-wellformedness} that it solves $\guess{\alpha}$ if and only if the type variable $\alpha$ appears in the term $N$.

    The second premise of \aforalll states that the right-hand side must not be a prenex polymorphic form.
    We introduce this premise to ensure that the system is strictly syntax-oriented: we have to eliminate all the quantifiers on the right with \aforallr before eliminating any quantifiers on the left with \aforalll.
    In general, introducing a premise like this could be problematic for completeness, as in completeness (see \cref{theorem:Completeness of algorithmic subtyping}) we want to be able to take apart the declarative derivation in the same order that the algorithmic system would derive the term.
    Thankfully, the invertibility of \dforallr ensures that we can transform declarative derivations concluding with \dforalll into equivalent ones that, like the algorithmic systems, use \dforallr as much as possible to eliminate all the prenex quantifiers on the right-hand side before using \dforalll.

  \item \aforallr is almost an exact copy of \dforalll.
    The forall rules must preserve the groundness invariant, so while \aforalll can introduce existential type variables when eliminating a quantifier on the \lhs, \aforallr can not introduce any existential variables while eliminating a quantifier on the \rhs.
    This ensures that $M$ remains ground from the conclusion of the rule to its premise.

\begin{figure}
  \input{shared-definitions-applying-context-to-term.tex}
  \caption{Applying a context to a type}
  \Description{Definition of applying a context to a \PSystemF type}
  \label{figure:context-application-to-terms}
\end{figure}

  \item \aarrow describes how to check function types.
    This rule is interesting since it has multiple premises.
    After checking the first premise $\APosSubtypeJudg{\acontext}{Q}{P}{\acontext'}$, we have an output context $\acontext'$ that might solve some existential variables that appear in $P$.
    Since some of these existential variables might also appear in $N$, we need to ensure that we propagate their solutions when we check $N \negsubtype M$.
    We do this by substituting all the existential variables that appear in the possibly non-ground $N$ by their solutions in $\acontext'$.
    We denote this operation as $\subcon{\acontext'} N$ and define it formally in \cref{figure:context-application-to-terms}.

  \item \ashiftdown and the symmetric \ashiftup are our shift rules.
    In these rules the groundness invariant defines the order in which the premises need to be be checked.
    For instance, in \ashiftdown only $N$ is ground in the conclusion, therefore we need to check $M \negsubtype N$ first.
    As we will see in \cref{section:subtyping-wellformedness}, checking this gives us an algorithmic context $\acontext'$ that solves all the existential variables that appear in $M$.
    We can use this context to \emph{complete} $M$ by substituting all the existential variables in $M$ by their solutions in $\acontext'$, giving us the ground type $\subcon{\acontext'} M$.
    Now that we have a ground type $\subcon{\acontext'} M$, we can check $N \negsubtype M$ by verifying $\ANegSubtypeJudg{\acontext'}{N}{\subcon{\acontext'} M}{\acontext''}$.
\end{itemize}

\subsection{Algorithmic type system}

\begin{figure}
  \input{shared-definitions-algorithmic-type-system.tex}
  \caption{Algorithmic type system}
  \Description{Definition of an algorithmic type system for \PSystemF}
  \label{figure:algorithmic-type-system}
\end{figure}

With a mechanism to solve existential variables in hand, we can now construct an algorithmic system to implement our declarative typing rules.
We present our algorithmic type system in \cref{figure:algorithmic-type-system}.

\begin{itemize}
  \item Most of the rules (\Avar, \Afunabs, \Atypeabs, \Athunk, \Areturn, \Aspinenil, and \Aspinecons) are identical to their declarative counterparts, modulo adding output contexts and applying output contexts to subsequent premises.

  \item We split the declarative \Dspinetypeabs rule into two rules depending on whether the new universal variable $\alpha$ appears in the type we are quantifying over $N$.
    If it does not, as in \Aspinetypeabsnotin, then we do not need to introduce a new existential variable.
    If it does, as in \Aspinetypeabsin, then as with \aforalll we introduce a fresh existential variable $\guess{\alpha}$ to replace $\alpha$.
    Note that unlike most rules, \Aspinetypeabsin adds a new existential variable to the output context of the conclusion \emph{that does not appear in its input context}.
    This will prove to be important in the \Aunambiguouslet and \Aambiguouslet rules.

\begin{figure}
  \input{shared-definitions-restricted-context.tex}
  \caption{Definition of context restriction}
  \Description{Definition of context restriction}
  \label{figure:context-restriction}
\end{figure}

  \item The first, second, and last premises of \Aambiguouslet are the same as their declarative counterparts.
  The third and fourth just inline the algorithmic \ashiftup rule corresponding to the declarative premise $\DNegSubtypeJudge{\acontext}{\shiftu{Q}}{\shiftu{P}}$ in \Dambiguouslet.

  We saw in the \Aspinetypeabsin rule that performing type inference on spines can introduce new existential variables.
    To simplify our proofs, we do not want these to leak.
    Therefore we introduce a notion of \emph{context restriction} in \cref{figure:context-restriction}.
    The fifth premise $\acontext^{(5)} = \restrictcontext{\acontext^{(4)}}{\acontext}$ creates a new context that restricts the context $\acontext^{(4)}$ to contain only the existential variables in $\acontext$.
    So if $\acontext^{(4)}$ contains new existential variables compared to $\acontext$, then these will not be present in $\acontext^{(5)}$.
    However any new solutions in $\acontext^{(4)}$ to existential variables already in $\acontext$ will be present in $\acontext^{(5)}$, so the algorithm will not solve any existential variable as two incompatible solutions.

  \item The first, second, and last premises of \Aunambiguouslet are lifted from the declarative rule \Dunambiguouslet.
    We replace the quantification in the last premise of \Dunambiguouslet with a statement that there are no existential variables left in the type $Q$, \ie $\FreeEV(Q) = \emptyset$.
    Since no instantiations are possible, there are not any other types left that $Q$ could have.
    Like \Aambiguouslet, we restrict the output context of the spine judgment to remove any existential variables newly introduced by the spine.
\end{itemize}

Our algorithm is really easy to implement --- we wrote a bare-bones implementation in 250 lines of OCaml.

\section{Properties of declarative typing}

\PSystemF should support the full complement of metatheoretic
properties, but in this paper we focus on the properties needed to
establish our theorems about type inference. For example, we will only
prove substitution at the type level, and totally ignore term-level
substitution.

\subsection{Subtyping}

Because we use subtyping to model type instantiation, we need to know
quite a few properties about how subtyping works. For example, we are
able to show that the subtyping relation admits both reflexivity and
transitivity:

\begin{restatable}[Declarative subtyping is reflexive]{lemma}{RestateReflexivityPnSubtype}
  \label{lemma:Reflexivity of declarative pnsubtype}
  If $\wfposnegtypeJudg{\dcontext} {A}$ then $\DPnSubtypeJudge{\dcontext} {A} {A}$.
\end{restatable}

\begin{restatable}[Declarative subtyping is transitive]{lemma}{RestateTransitivityPnSubtype}
  \label{lemma:Transitivity of declarative pnsubtype}
  If $\wfposnegtypeJudg{\dcontext}{A}$, $\wfposnegtypeJudg{\dcontext}{B}$, $\wfposnegtypeJudg{\dcontext}{C}$, $\DPnSubtypeJudge{\dcontext} {A} {B}$, and $\DPnSubtypeJudge{\dcontext} {B} {C}$, then $\DPnSubtypeJudge{\dcontext} {A} {C}$.
\end{restatable}

Because of impredicativity, transitivity is surprisingly subtle to get
right. We discuss the needed metric in \cref{section:decidability} as this metric is also used to show the decidability of algorithmic subtyping.

We also show that subtyping is stable under substitution, which
is useful for proving properties of type instantiation:

\begin{restatable}[Declarative subtyping is stable under substitution]{lemma}{RestateDeclarativeSubtypingSubstitutionLemma}
  \label{lemma:Declarative subtyping substitution lemma}
  If $\wfpostypeJudg{\dcontext_L, \dcontext_R}{P}$, then:

  \begin{itemize}
    \item If $\wfpostypeJudg{\dconwithhole{\alpha}}{Q}$, $\wfpostypeJudg{\dconwithhole{\alpha}}{R}$, and $\DPosSubtypeJudge{\dconwithhole{\alpha}}{Q}{R}$, then $\DPosSubtypeJudge{\dcontext_L, \dcontext_R}{\subterm{P / \alpha} Q} {\subterm{P / \alpha} R}$.

    \item If $\wfnegtypeJudg{\dconwithhole{\alpha}}{N}$, $\wfnegtypeJudg{\dconwithhole{\alpha}}{M}$, and $\DNegSubtypeJudge{\dconwithhole{\alpha}}{N}{M}$, then $\DNegSubtypeJudge{\dcontext_L, \dcontext_R} {\subterm{P / \alpha} N} {\subterm{P / \alpha} M}$.
  \end{itemize}
\end{restatable}

\subsection{Typing}

\begin{figure}
  \input{shared-definitions-isomorphic-environments.tex}
  \caption{Isomorphic environments}
  \Description{Definition of when two typing environments are isomorphic}
  \label{figure:isomorphic-environments}
\end{figure}

We also show an important property about our typing judgment.
Suppose we have two types $A$ and $B$ that are isomorphic, \ie $\MutualSubtypePosNegJudge{\dcontext}{A}{B}$.
Then, by the intuitive definition of subtyping, it should be possible to safely use any term of type A where a term of type B is expected, and vice versa.
Concretely, 
the $\mathsf{map}$ function can be given both the type 
$P = \shiftd{(\foralltype{\alpha,\beta}{\arrowtype{\shiftd{(\arrowtype{\alpha}{\beta})}}{\arrowtype{\listtype{\alpha}}{\shiftu{\listtype{\beta}}}}})}$
and also the type $Q = \shiftd{(\foralltype{\beta, \alpha}{\arrowtype{\shiftd{(\arrowtype{\alpha}{\beta})}}{\arrowtype{\listtype{\alpha}}{\shiftu{\listtype{\beta}}}}})}$, and there is no reason to prefer one to the other.
Therefore if we have an environment $\typeenv$ where $\mathsf{map} : P$ and an environment $\typeenv'$ where $\mathsf{map} : Q$, the inference we get within each environment should be the same.

We formalize this idea with the following lemma, which makes use of the isomorphic environment judgment we define in \cref{figure:isomorphic-environments}:

\begin{restatable}[Isomorphic environments type the same terms]{lemma}{IsomorphicTypeTypeSameExpressions}
  \label{lemma:isomorphic types check expressions}
  If $\declisovarctxjudg{\acontext}{\typeenv}{\typeenv'}$, then:
  \begin{itemize}
    \item If $\declsynjudg{\acontext; \typeenv}{v}{P}$ then $\exists P'$ such that
      $\declnisotypejudg{\acontext}{P}{P'}$ and
      $\declsynjudg{\acontext; \typeenv'}{v}{P'}$.

    \item If $\declsynjudg{\acontext; \typeenv}{t}{N}$ then $\exists N'$ such that
      $\declnisotypejudg{\acontext}{N}{N'}$ and
      $\declsynjudg{\acontext; \typeenv'}{t}{N'}$.

    \item If
      $\declspinejudg{\acontext; \typeenv}{s}{\spine{N}{M}}$ and
      $\declnisotypejudg{\acontext}{N}{N'}$, then $\exists M'$ such that
      $\declnisotypejudg{\acontext}{M}{M'}$ and
      $\declspinejudg{\acontext; \typeenv}{s}{\spine{N'}{M'}}$.
  \end{itemize}
\end{restatable}

This lemma tells us that regardless of which type we give the
variable $\mathsf{map}$, any term which uses it will have the
same type (up to isomorphism).

\section{Properties of algorithmic typing}

\subsection{Well-formedness}
\label{section:weak-context-extension}

We first establish some of the invariants our type system maintains. 

\subsubsection{Subtyping}
\label{section:subtyping-wellformedness}

\paragraph{Context well-formedness}

\begin{figure}
  \input{shared-definitions-well-formed-context.tex}
  \caption{Well-formedness of contexts}
  \Description{Definition of the well-formedness of declarative and algorithmic \PSystemF contexts}
  \label{figure:context-wellformedness}
\end{figure}

\begin{figure}
  \judgbox{\wfposnegtypeJudg{\acontext}{A}}{In the context $\acontext$, $A$ is a well-formed positive/negative type}

  \begin{mathpar}
    \Infer{\twfguess}
      {\guess{\alpha} \in \EV(\acontext)}
      {\wfpostypeJudg{\acontext} {\guess{\alpha}}}
  \end{mathpar}
  \caption{
    Additional well-formedness rules for algorithmic types.
    $\EV(\acontext)$ contains all the existential type variables in $\acontext$, independently of whether they are solved or unsolved.
  }
  \Description{
    Additional well-formedness rules for algorithmic types.
    $\EV(\acontext)$ contains all the existential type variables in $\acontext$, independently of whether they are solved or unsolved.
  }
  \label{figure:algorithmic-wellformedness-additions}
\end{figure}

A simple property that our algorithm maintains is that any solutions the algorithm chooses are well-formed: they are ground and only contain universal variables in scope at the time of the corresponding existential variable's creation.
We formalize this notion of the well-formedness of contexts in \cref{figure:context-wellformedness,figure:algorithmic-wellformedness-additions}.

\paragraph{Output context solves all existential variables}

Another property of our algorithm is that given well-formed inputs, the algorithm will solve all the necessary solutions: in other words, the output context contains solutions to all existential variables appearing in the non-ground type.

\paragraph{Context extension}

\begin{figure}
  \input{shared-definitions-context-extension.tex}
  \caption{Context extension}
  \Description{Definition of the context extension judgment}
  \label{figure:context-extension}
\end{figure}

Finally, we want to show that the algorithm produces an output context that only adds solutions to the input context.
For instance, our algorithm should not change an existing solution to an incompatible one.
We formalize this idea with a context extension judgment based on the earlier work of \citet{dunfieldCompleteEasyBidirectional2013}.
This judgment $\congoesJudg{\acontext}{\acontext'}$, as defined in \cref{figure:context-extension}, indicates information gain: $\acontext'$ must contain at least as much information as $\acontext$.

Most of the rules defining context extension are homomorphic.
The exceptions are \Csolveguess, which allows the algorithm to solve an existential variable, and \Csolvedguess, which allows the algorithm to change the solution $P$ for an existential variable $\guess{\alpha}$ to an isomorphic type $Q$.
This ability is not crucial to the algorithm; in fact the algorithm will maintain syntactically identical solutions.
However since the declarative system can invent any solutions it likes when instantiating quantified types, we need this little bit of flex to prove completeness.

\begin{figure}
  \input{shared-definitions-producing-declarative-context.tex}

  \caption{Producing a declarative context from an algorithmic context}
  \Description{Definition of how a declarative context can be produced from an algorithmic context}
  \label{figure:producing-declarative-context}
\end{figure}

Solutions to existential types are ground, so we do not need to introduce a new notion of isomorphism over algorithmic types.
However since the context $\acontext$ is algorithmic rather than declarative, we do need to introduce a simple operation $\makedec{\acontext}$ that converts an algorithmic context into a declarative one by dropping all the context items involving existential variables.
This operation is defined in \cref{figure:producing-declarative-context}.

We formalize all these properties in the well-formedness statement about the algorithmic subtyping relation below.

\begin{restatable}[Algorithmic subtyping is \wellformed]{lemma}{RestateWellFormednessPNSubtype} \
  \label{lemma:well-formedness-pnsubtype}
  \begin{itemize}
      \item If $\acontext \entails P \possubtype Q \prodcon \acontext'$, $\acontext \conwf$, $P$ ground, and $\subcon{\acontext} Q = Q$, then $\acontext' \conwf$, $\congoesJudg{\acontext}{\acontext'}$, and $\subcon{\acontext'} Q$ ground.

      \item If $\acontext \entails N \negsubtype M \prodcon \acontext'$, $\acontext \conwf$, $M$ ground, and $\subcon{\acontext} N = N$, then $\acontext' \conwf$, $\congoesJudg{\acontext}{\acontext'}$, and $\subcon{\acontext'} N$ ground.
  \end{itemize}
\end{restatable}

The premises of each case encode our expectations about the inputs to the subtyping algorithm: the input context should be well-formed, the type on the ground side should be ground, and the type on the non-ground side should not contain any existential variables that have already been solved.
We prove that when given these inputs, the subtyping algorithm produces an output context that is well-formed, is compatible with the input context, and that solves all the existential variables that appear in the input types.

\subsubsection{Typing}

\begin{figure}
  \input{shared-definitions-weak-context-extension.tex}
  \caption{Weak context extension. We highlight the rules that are \enquote{new} compared with context extension.}
  \Description{Definition of the weak context extension judgment}
  \label{figure:weak-context-extension}
\end{figure}

\begin{figure}
  \input{shared-definitions-well-formed-environment.tex}
  \caption{Well-formedness of typing environments}
  \Description{Definition of the well-formedness of typing environments}
  \label{figure:wf-environments}
\end{figure}

The well-formedness statement for typing is very similar to the one for subtyping.
However stating well-formedness for the spine judgment is more complex because this judgment (specifically the \Aspinetypeabsin rule) can introduce new existential variables that do not appear within its input context.

To tackle this, we introduce a new notion of context extension, \emph{weak context extension}, in \cref{figure:weak-context-extension}.
This is identical to normal context extension, except it has two additional rules to permit adding new existential variables: \Wcunsolvedextend lets it add an unsolved variable, and \Wcsolvedextend lets it add a solved variable.
We also extend in \cref{figure:wf-environments} our notion of the well-formedness of algorithmic contexts to typing environments in the obvious way, with \Ewfvar paralleling \cwfsolvedguess.

With this weaker notion of context extension we can state well-formedness of typing as follows:

\begin{restatable}[Algorithmic typing is w.f.]{lemma}{AlgorithmicTypingWellFormed} \
  \label{lemma:algorithmic-typing-well-formed}
  Given a typing context $\acontext$ and typing environment $\typeenv$ such that $\conwfJudg{\acontext}$ and $\envwfJudg{\acontext}{\typeenv}$:
  \begin{itemize}
      \item If
      $\algosynjudg{\acontext; \typeenv}{v}{P}{\acontext'}$,
      then
      $\conwfJudg{\acontext'}$,
      $\congoesJudg{\acontext}{\acontext'}$,
      $\wfpostypeJudg{\acontext'}{P}$, and
      $\groundJudge{P}$.

      \item If
      $\algosynjudg{\acontext; \typeenv}{t}{N}{\acontext'}$,
      then
      $\conwfJudg{\acontext'}$,
      $\congoesJudg{\acontext}{\acontext'}$,
      $\wfnegtypeJudg{\acontext'}{N}$, and
      $\groundJudge{N}$.

      \item If $\algospinejudg{\acontext; \typeenv}{s}{\spine{N}{M}}{\acontext'}$,
      $\wfnegtypeJudg{\acontext}{N}$, and
      $\subcon{\acontext} N = N$,
      then
      $\conwfJudg{\acontext'}$,
      $\congoesWeakJudg{\acontext}{\acontext'}$,
      $\wfnegtypeJudg{\acontext'}{M}$,
      $\subcon{\acontext'} M = M$, and
      $\FreeEV(M) \subseteq \FreeEV(N) \cup (\FreeEV(\acontext') \setminus \FreeEV(\acontext))$.
  \end{itemize}
\end{restatable}

In addition to the standard postconditions, we also prove in the spine judgment case that the free existential variables in the output type $M$ either come from the input type $N$ or were added to the context by $\Aspinetypeabsin$ while instantiating universal quantifiers.

\subsection{Determinism and decidability}
\label{section:decidability}

Since our system is syntax-directed, determinism of algorithmic typing follows from straightforward rule inductions on the typing and subtyping rules.
However, decidability is more intricate.

\subsubsection{Subtyping}

Our goal in proving the decidability of subtyping is finding a metric that decreases from the conclusion to each premise of the algorithmic rules.
However the obvious metrics do not work.
For instance, a metric based on the size of the type will not work because types can get bigger when we instantiate type variables.
Predicative systems might use the lexicographic ordering of the number of prenex quantifiers followed by the size of the type as a metric.
But while instantiations will not increase the number of quantifiers in predicative systems, they can do in impredicative systems.

\begin{figure}
  \input{shared-definitions-type-size-ignoring-quantification.tex}
  \input{shared-definitions-num-prenex-quantifiers.tex}
  \caption{Definitions used in the decidability metric for \PSystemF}
  \Description{Definitions used in the decidability metric for \PSystemF}
  \label{figure:definitions-for-decidability}
\end{figure}

In order to find a metric that works, we go back to the declarative system. Our declarative subtyping relation has the
property that (using a size metric $\termsize{\_}$ for terms that ignores quantifiers), a well-formed%
\footnote{One for which the well-formedness assumptions in \lemmaref{lemma:well-formedness-pnsubtype} hold.}
$\DPosSubtypeJudge{\dcontext}{P}{Q}$ judgment implies that $\termsize{Q} \leq \termsize{P}$, and similarly a well-formed $\DNegSubtypeJudge{\dcontext}{N}{M}$ judgment implies that $\termsize{N} \leq \termsize{M}$.
The intuition behind this property is that subtyping reflects the specialization order, and when a polymorphic type is instantiated, occurrences of type variables $\alpha$ get replaced with instantiations $P$. As a result, the size of a subtype has to be smaller than its supertypes.
In ordinary System F, this intuition is actually false due to the contravariance of the function type. However, in \PSystemF, universal quantification is a \emph{negative type} of the form $\foralltype{\alpha}{N}$, and
ranges over \emph{positive types}. Since positive type variables $\alpha$ are occurring within a negative type $N$, each occurrence of $\alpha$ must be underneath a shift, which is invariant in our system.
As a result, the intuitive property is true. 

Now, note that our algorithm always solves all of the existential problems in a well-formed subtyping problem. In particular,
given an algorithmic derivation $\APosSubtypeJudg{\acontext} {P} {Q} {\acontext'}$, we know that $\subcon{\acontext'} Q$ will be ground. So we
therefore expect $\subcon{\acontext'} Q$  to be the same size as it would be in a declarative derivation.
This gives us a hint about what to try for the metric.

Since the size of the ground side (\eg $P$) bounds the size of the completed non-ground side (\eg $\subcon{\acontext'} Q$), we will incorporate the size of the ground side into our metric.
This size decreases between the conclusions and the premises of all of the rules apart from \aforalll and \aforallr, where it stays the same.
\aforalll and \aforallr both however pick off a prenex quantifier, so we can take the lexicographic ordering of this size followed by the total number of prenex quantifiers in the subtyping judgment.
This is $(\termsize{P}, \numprenex{P} + \numprenex{Q})$ for positive judgments $\APosSubtypeJudg{\acontext}{P}{Q}{\acontext'}$ and $(\termsize{M}, \numprenex{M} + \numprenex{N})$ for negative judgments $\ANegSubtypeJudg{\acontext}{N}{M}{\acontext'}$.
We give formal definitions of $\termsize{\_}$ and $\numprenex{\_}$ in \cref{figure:definitions-for-decidability}.

Note that this is somewhat backwards from a typical metric for predicative systems where we count quantifiers first, then the size.

We prove that this metric assigns a total ordering to derivations in the algorithmic subtyping system.
In each rule with multiple hypotheses, we invoke the bounding property described above, which we formalize in the following lemma:

\begin{restatable}[Completed non-ground size bounded by ground size]{lemma}{RestateCompletedNonGroundSizeBounded} \ \
  \label{lemma:Completed non-ground size bounded}
  \begin{itemize}
    \item If
      $\APosSubtypeJudg{\acontext}{P}{Q}{\acontext'}$,
      $\acontext \conwf$,
      $P$ ground, and
      $\subcon{\acontext} Q = Q$,
      then $\termsize{\subcon{\acontext'} Q} \leq \termsize{P}$.
  
    \item If
      $\ANegSubtypeJudg{\acontext}{N}{M}{\acontext'}$,
      $\acontext \conwf$,
      $M$ ground, and
      $\subcon{\acontext} N = N$,
      then $\termsize{\subcon{\acontext'} N} \leq \termsize{M}$.
  \end{itemize}
\end{restatable}

As a result of our key tactic for deriving this metric coming from the declarative system, it turns out that this metric assigns total orderings to derivations both in the algorithmic and the declarative subtyping systems.
This allows us to reuse it in both the proof of completeness and the proof of transitivity, where in both cases using the height of the derivation as an induction metric is too weak.

\subsubsection{Typing}

\newcommand{\standardsize}[1]{|#1|}

The decidability of algorithmic typing ends up being relatively straightforward.
Almost every rule decreases in the standard structural notions of size $\standardsize{\_}$ on terms and spines.
The \Dspinetypeabs rule however preserves the size of the spine involved, so taking this size alone is insufficient.
Therefore when comparing two spine judgments, we take the lexicographic ordering of $(\standardsize{s}, \numprenex{N})$, where $s$ is the spine and $N$ is the input type.
We prove that this metric assigns a total ordering to algorithmic typing derivations, thus demonstrating that our typing algorithm is decidable.
Alongside our proof, we include an exact statement of this metric in the appendix.

Note that, as with subtyping, this metric assigns a total ordering not only to algorithmic typing derivations but also to declarative typing derivations.
Therefore we reuse this metric in the proofs of soundness, completeness, and the behavior of isomorphic types.
In each of these cases, using the height of the derivation as an induction metric is too weak.

\section{Soundness and completeness}

We have now set out declarative subtyping and typing systems for \PSystemF as well as algorithms to implement them.
In this section, we will demonstrate that the algorithms are sound and complete with respect to their declarative counterparts.
Proofs of all of these results are in the appendix.

\subsection{Subtyping}

\subsubsection{Soundness}

Consider the algorithmic subtyping judgment $\APosSubtypeJudg{\acontext} {P} {Q} {\acontext'}$.
The non-ground input $Q$ might contain some unsolved existential variables, which are solved in $\acontext'$. For soundness, we want to say that no matter what solutions are ever made for these existential variables, there will be a corresponding declarative derivation.
To state this, we introduce a notion of a \emph{complete context}.
These are algorithmic contexts which contain no unsolved existential variables:
\begin{align*}
  \text{Complete contexts} \quad
  \ccontext &::= \emptyccontext \alt \ccontext, \alpha \alt
  \ccontext, \guess{\alpha} = P
\end{align*}
We can now use the context extension judgment $\congoesJudg{\acontext'}{\ccontext}$ to state that $\ccontext$ is a complete context that contains solutions to all the unsolved existential variables in $\acontext'$.
Applying this complete context to $Q$ gives us our hoped-for declarative subtyping judgment $\DPosSubtypeJudge{\makedec{\acontext}}{P}{\subcon{\ccontext} Q}$. 
(The $\makedec{\acontext}$ judgment (defined in \cref{figure:producing-declarative-context}) drops the existential variables to create a declarative context.)

This gives us the following statement of soundness, which we prove in the appendix:

\begin{restatable}[Soundness of algorithmic subtyping]{theorem}{RestateSoundnessPNSubtype}
  \label{theorem:soundness-pnsubtype}
  Given a well-formed algorithmic context $\acontext$ and a well-formed complete context $\ccontext$:
  \begin{itemize}
      \item If
        $\acontext \entails P \possubtype Q \prodcon \acontext'$,
        $\congoesJudg{\acontext'}{\ccontext}$,
        $P$ ground,
        $\subcon{\acontext} Q = Q$,
        $\wfpostypeJudg{\acontext} {P}$, and
        $\wfpostypeJudg{\acontext} {Q}$,

        then $\CnewPosJudge{\acontext} {P} {Q}$.

      \item If
        $\acontext \entails N \negsubtype M \prodcon \acontext'$,
        $\congoesJudg{\acontext'}{\ccontext}$,
        $M$ ground,
        $\subcon{\acontext} N = N$,
        $\wfnegtypeJudg{\acontext} {N}$, and
        $\wfnegtypeJudg{\acontext} {M}$,

        then $\CnewNegJudge{\acontext} {N} {M}$.
  \end{itemize}
\end{restatable}

The side conditions for soundness are similar to those of well-formedness (\cref{lemma:well-formedness-pnsubtype}), except we also require the complete context and the types in the algorithmic judgment to be well-formed.

\subsubsection{Completeness}

Completeness is effectively the reverse of soundness.
For each positive declarative judgment $\CnewPosJudge{\acontext}{P}{Q}$ where $\congoesJudg{\acontext}{\ccontext}$ we want to find a corresponding algorithmic derivation $\APosSubtypeJudg{\acontext}{P}{Q}{\acontext'}$, and likewise for negative judgments.
No matter what declarative solutions we had for a solved type, the algorithmic system infers compatible solutions.

Since context extension $\congoesJudg{\acontext}{\ccontext}$ now appears in the premise, we need to strengthen our induction hypothesis to prove completeness for rules with multiple premises.
To do this, we prove that the algorithm's output context $\acontext'$  is also compatible with $\ccontext$, \ie $\congoesJudg{\acontext'}{\ccontext}$.
This gives us the following statement of completeness, in which the side conditions are identical to soundness.

\begin{restatable}[Completeness of algorithmic subtyping]{theorem}{RestateCompletenessPNSubtype}
  \label{theorem:Completeness of algorithmic subtyping}

  If $\acontext \conwf$, $\acontext \congoes \ccontext$, and $\ccontext \conwf$, then:

  \begin{itemize}
    \item
      If
      $\CnewPosJudge{\acontext} {P} {Q}$,
      $\acontext \entails P \postype$,
      $\acontext \entails Q \postype$,
      $P$ ground, and
      $\subcon{\acontext} Q = Q$,
      then $\exists \acontext'$ such that
      $\acontext \entails P \possubtype Q \prodcon \acontext'$ and
      $\acontext' \congoes \ccontext$.

    \item
      If
      $\CnewNegJudge{\acontext} {N} {M}$,
      $\acontext \entails M \negtype$,
      $\acontext \entails N \negtype$,
      $M \text{ ground}$, and
      $\subcon{\acontext} N = N$,
      then $\exists \acontext'$ such that
      $\acontext \entails N \negsubtype M \prodcon \acontext'$ and
      $\acontext' \congoes \ccontext$.

  \end{itemize}
\end{restatable}

\subsection{Typing}

Due to the unusual implication within \Dunambiguouslet, soundness and completeness of typing are mutually recursive in our system.
In soundness we use completeness while proving this implication, and in completeness we use soundness while unpacking the implication.
We justify each of these uses by only applying either soundness or completeness to a judgment involving a subterm.

\subsubsection{Soundness}

Soundness of typing is formulated in the same way as soundness of subtyping. We introducing a complete context $\ccontext$ which the output context extends to.
For the spine judgment we allow the completed output type $\subcon{\ccontext} M$ to be isomorphic to the declarative output type $M'$.

\begin{restatable}[Soundness of algorithmic typing]{theorem}{Soundness}
  \label{theorem:soundness}
  If $\conwfJudg{\acontext}$,
  $\envwfJudg{\acontext}{\typeenv}$,
  $\congoesJudg{\acontext'}{\ccontext}$, and
  $\conwfJudg{\ccontext}$, then:

  \begin{itemize}
    \item If
    $\algosynjudg{\acontext; \typeenv}{v}{P}{\acontext'}$,
    then
    $\declsynjudg{\makedec{\acontext}; \typeenv}{v}{[\ccontext]P}$.

    \item If
    $\algosynjudg{\acontext; \typeenv}{t}{N}{\acontext'}$,
    then
    $\declsynjudg{\makedec{\acontext}; \typeenv}{t}{[\ccontext]N}$.

    \item If
    $\algospinejudg{\acontext; \typeenv}{s}{\spine{N}{M}}{\acontext'}$,
    $\wfnegtypeJudg{\acontext}{N}$, and
    $[\acontext]N = N$,
    then
    $\exists M'$ such that $\MutualSubtypeNegJudge{\makedec{\acontext}}{[\ccontext]M}{M'}$ and $\declspinejudg{\makedec{\acontext}; \typeenv}{s}{\spine{[\ccontext]N}{M'}}$.
  \end{itemize}
\end{restatable}

\subsection{Completeness}

The main challenge in stating completeness of typing is taking into account the fact that spine judgments can introduce new existential variables.
Practically, the introduction of new existential variables means that the output context of the algorithmic spine judgment $\acontext'$ will not necessarily extend to the complete context $\ccontext$.
To deal with this, we make use of weak context extension and instead require there to exist some new complete context $\ccontext'$ containing the new existential variables.
The original complete context $\ccontext$ should then weakly extend to $\ccontext'$, and the output context of the spine judgment $\acontext'$ should strongly extend to $\ccontext'$.

This gives us the following statement of the completeness of typing:

\begin{restatable}[Completeness of algorithmic typing]{theorem}{Completeness}
  \label{theorem:typing-completeness}
  If $\conwfJudg{\acontext}$, $\envwfJudg{\acontext}{\typeenv}$, $\congoesJudg{\acontext}{\ccontext}$, and $\conwfJudg{\ccontext}$, then:

  \begin{itemize}
    \item If
      $\declsynjudg{\makedec{\acontext}; \typeenv}{v}{P}$
      then $\exists \acontext'$ such that
      $\algosynjudg{\acontext; \typeenv}{v}{P}{\acontext'}$ and
      $\congoesJudg{\acontext'}{\ccontext}$.

    \item If
      $\declsynjudg{\makedec{\acontext}; \typeenv}{t}{N}$
      then $\exists \acontext'$ such that
      $\algosynjudg{\acontext; \typeenv}{t}{N}{\acontext'}$ and
      $\congoesJudg{\acontext'}{\ccontext}$.

    \item If
      $\declspinejudg{\makedec{\acontext}; \typeenv}{s}{\spine{\subcon{\ccontext} N}{M}}$,
      $\wfnegtypeJudg{\acontext}{N}$, and
      $\subcon{\acontext} N = N$,
      then $\exists \acontext', \ccontext'$ and $M'$ such that
      $\algospinejudg{\acontext; \typeenv}{s}{\spine{N}{M'}}{\acontext'}$,
      $\congoesWeakJudg{\ccontext}{\ccontext'}$,
      $\congoesJudg{\acontext'}{\ccontext'}$,
      $\MutualSubtypeNegJudge{\makedec{\acontext}}{\subcon{\ccontext'} M'}{M}$,
      $\subcon{\acontext'} M' = M'$, and
      $\conwfJudg{\ccontext'}$.
  \end{itemize}
\end{restatable}

\section{Related work}

There has been considerable research into working around the
undecidability of type inference for System F. Broadly, it falls into
three main categories: enriching the language of types to make
inference possible, restricting the subtype relation, and using
heuristics to knock off the easy cases.

\paragraph{Enriching the Type Language}
The ``gold standard'' for System F type inference is the \mlf system
of \citet{botlanMLFRaisingML2003}. They observe that type inference for System F is
complicated by the lack of principal types. To create
principal types, they extend the type language of System F with
bounded type constraints, and show that this admits a type inference
algorithm with a simple and powerful specification: only arguments to
functions used at multiple parametric types need annotation.

Unfortunately, \mlf is notoriously difficult to implement (see
\citet{Remy08} for the state-of-the-art), and there have been
multiple attempts to find simpler subsystems.
FPH (\citet{vytiniotisFPHFirstclassPolymorphism2008}) attempted to
simplify \mlf by limiting the user-visible language of types to the
standard System F types and only using \mlf{}-style constrained types
internally within the type checker. Another attempt to simplify \mlf was
HML (\citet{leijenRobustTypeInference2008}). Unlike FPH, HML exposes
part of the \mlf machinery (flexible types) to the user, which
changes the language of types but also types more programs. Both FPH and HML require annotations only for polymorphic arguments, however
unfortunately, both approaches proved too intricate to adapt to GHC.

\emph{Restricting the subtype relation.} The most widely used approach
for System F type inference simply abandons impredicativity. This line
of work was originated by \citet{oderskyPuttingTypeAnnotations1996},
who proposed restricting type instantiation in the subtype relation to
monotypes.  This made subtyping decidable, and forms the basis for
type inference in
Haskell~\cite{peytonjonesPracticalTypeInference2007}.
\citet{dunfieldCompleteEasyBidirectional2013} give a simple variant of
this algorithm based on bidirectional typechecking.

This style of inference omits impredicativity from the subtype
relation altogether. \emph{Boxy types}~\cite{vytiniotisBoxyTypesInference2006} combine
predicative subtyping with a generalization of bidirectional
typechecking to support
impredicativity. \citeauthor{vytiniotisBoxyTypesInference2006}
introduce a new type system feature, \textit{boxes}, which merge the
synthesis and checking judgments from bidirectional typechecking into
marks on types (\ie the boxes) which
indicate whether part of a type came from inference or annotation.
(This is somewhat reminiscent of ``colored local type
inference''~\cite{odersky03}.)  Unfortunately, boxy types lack a clear
non-algorithmic specification of when type annotations are required.

HMF~\cite{leijenHMFSimpleType2008} introduced another approach to restricting the specialization relation. It restricted
subtyping for the function type constructor --- instead
of being contravariant, function types were
invariant. This meant that inference could be done with only a modest
modification to the Damas-Milner algorithm. Our work retains function
contravariance, and only imposes invariance at shifts. 

\emph{Heuristic Approaches.} Many of the type inference problems
which arise in practice are actually easy to
solve. \citet{cardelliImplementation1993} invented one of the oldest
such techniques while constructing a type inference algorithm for a
language with F-bounded impredicative polymorphism. Rather than doing
anything difficult, Cardelli's algorithm simply instantiated each
quantifier with the first type constraint it ran
into. \citet{pierceLocalTypeInference2000} formalized Cardelli's approach, and
noticed that it did not need unification to implement. To make their
algorithm work, they needed to use a seemingly ad-hoc representation
of types, with a single type constructor that was simultaneously an
$n$-ary small and big lambda. Furthermore, while they proved that
inference was sound, they did not offer a declarative characterization
of inference.

\citet{serranoGuardedImpredicativePolymorphism2018} recently revisited
the idea of controlling inference with heuristics with their system
GI. They restrict impredicative instantiation to \textit{guarded
  instantiations}, which are (roughly speaking) the cases when the
type variable being instantiated is underneath a type
constructor. This restriction is automatically achieved in
our setting via the presence of shifts, which suggests that this
syntactic restriction actually arises for deeper type-theoretic
reasons. In follow up work, \citet{serranoQuickLookImpredicativity2020}
further simplify their approach, making some dramatic simplifications
to the type theory (e.g., giving up function type contravariance) in
order to achieve a simpler implementation. 

\emph{Conclusions.} Many researchers have noticed that type inference
algorithms benefit from being able to look at the entire argument list
to a function. From the perspective of plain lambda calculus, this
looks ad-hoc and non-compositional. For example, HMF was originally
described in two variants, one using argument lists and one without,
and only the weaker algorithm without argument lists has a correctness
proof. However, from the perspective of polarized type theory,
argument lists are entirely type-theoretically natural: they mark the
change of a polarity boundary!

This explains why \citeauthor{pierceLocalTypeInference2000}'s use of
``jumbo'' function types which combine multiple quantifiers and
arguments makes sense: the merged connectives are all negative, with
no interposed shifts. Making shifts explicit means that small
connectives can have the same effect, which makes it possible to give
a clear specification for the system. 

We have also seen that many algorithms omit function contravariance
from the specialization order to support impredicative inference. Our
work clarifies that contravariance \emph{per se} is not problematic,
but rather that the benefits for inference arise from controlling the
crossing of polarity boundaries. This again permits a simpler and more
regular specification of subtyping.

\begin{acks}
  This research was supported in part by a European Research Council (ERC) Consolidator Grant
  for the project \emph{TypeFoundry}, funded under the European Union's Horizon 2020 Framework Programme
  (grant agreement ID: 101002277).
\end{acks}

\bibliographystyle{ACM-Reference-Format}
\bibliography{paper-refs}


\begin{thebibliography}{25}


\ifx \showCODEN    \undefined \def \showCODEN     #1{\unskip}     \fi
\ifx \showDOI      \undefined \def \showDOI       #1{#1}\fi
\ifx \showISBNx    \undefined \def \showISBNx     #1{\unskip}     \fi
\ifx \showISBNxiii \undefined \def \showISBNxiii  #1{\unskip}     \fi
\ifx \showISSN     \undefined \def \showISSN      #1{\unskip}     \fi
\ifx \showLCCN     \undefined \def \showLCCN      #1{\unskip}     \fi
\ifx \shownote     \undefined \def \shownote      #1{#1}          \fi
\ifx \showarticletitle \undefined \def \showarticletitle #1{#1}   \fi
\ifx \showURL      \undefined \def \showURL       {\relax}        \fi
\providecommand\bibfield[2]{#2}
\providecommand\bibinfo[2]{#2}
\providecommand\natexlab[1]{#1}
\providecommand\showeprint[2][]{arXiv:#2}

\bibitem[Botlan and R{\'e}my(2003)]%
        {botlanMLFRaisingML2003}
\bibfield{author}{\bibinfo{person}{Didier~Le Botlan} {and}
  \bibinfo{person}{Didier R{\'e}my}.} \bibinfo{year}{2003}\natexlab{}.
\newblock \showarticletitle{{{MLF Raising ML}} to the {{Power}} of {{System
  F}}}. In \bibinfo{booktitle}{\emph{{{ICFP}} '03}}. \bibinfo{publisher}{{ACM
  Press}}, \bibinfo{address}{{Uppsala, Sweden}}, \bibinfo{pages}{52--63}.
\newblock


\bibitem[Cardelli(1993)]%
        {cardelliImplementation1993}
\bibfield{author}{\bibinfo{person}{Luca Cardelli}.}
  \bibinfo{year}{1993}\natexlab{}.
\newblock \bibinfo{booktitle}{\emph{An Implementation of {{F}}{$<$}:}}.
\newblock \bibinfo{type}{{T}echnical {R}eport}. \bibinfo{institution}{{Systems
  Research Center, Digital Equipment Corporation}}.
\newblock


\bibitem[Cervesato(2003)]%
        {cervesatoLinearSpineCalculus2003}
\bibfield{author}{\bibinfo{person}{I. Cervesato}.}
  \bibinfo{year}{2003}\natexlab{}.
\newblock \showarticletitle{A {{Linear Spine Calculus}}}.
\newblock \bibinfo{journal}{\emph{Journal of Logic and Computation}}
  \bibinfo{volume}{13}, \bibinfo{number}{5} (\bibinfo{date}{Oct.}
  \bibinfo{year}{2003}), \bibinfo{pages}{639--688}.
\newblock
\showISSN{0955-792X, 1465-363X}
\urldef\tempurl%
\url{https://doi.org/10.1093/logcom/13.5.639}
\showDOI{\tempurl}


\bibitem[Chrzaszcz(1998)]%
        {chrzaszcz-98}
\bibfield{author}{\bibinfo{person}{Jacek Chrzaszcz}.}
  \bibinfo{year}{1998}\natexlab{}.
\newblock \showarticletitle{Polymorphic Subtyping without Distributivity}. In
  \bibinfo{booktitle}{\emph{Proceedings of the 23rd International Symposium on
  Mathematical Foundations of Computer Science}}
  \emph{(\bibinfo{series}{{{MFCS}} '98})}.
  \bibinfo{publisher}{{Springer-Verlag}}, \bibinfo{address}{{Berlin,
  Heidelberg}}, \bibinfo{pages}{346--355}.
\newblock
\showISBNx{3-540-64827-5}


\bibitem[Dunfield and Krishnaswami(2013)]%
        {dunfieldCompleteEasyBidirectional2013}
\bibfield{author}{\bibinfo{person}{Jana Dunfield} {and} \bibinfo{person}{Neel
  Krishnaswami}.} \bibinfo{year}{2013}\natexlab{}.
\newblock \showarticletitle{Complete and Easy Bidirectional Typechecking for
  Higher-Rank Polymorphism}. In \bibinfo{booktitle}{\emph{Proceedings of the
  18th {{ACM SIGPLAN}} International Conference on {{Functional}} Programming -
  {{ICFP}} '13}}. \bibinfo{publisher}{{ACM Press}}, \bibinfo{address}{{Boston,
  Massachusetts, USA}}, \bibinfo{pages}{429}.
\newblock
\showISBNx{978-1-4503-2326-0}
\urldef\tempurl%
\url{https://doi.org/10.1145/2500365.2500582}
\showDOI{\tempurl}


\bibitem[Girard(1971)]%
        {girard-system-f}
\bibfield{author}{\bibinfo{person}{Jean-Yves Girard}.}
  \bibinfo{year}{1971}\natexlab{}.
\newblock \showarticletitle{Une Extension de {{\v{L}Interpretation}} de
  {{G\"odel}} \`a {{\v{L}Analyse}}, et Son Application \`a {{\v{L}Elimination}}
  Des Coupures Dans {{\v{L}Analyse}} et La Theorie Des Types}.
\newblock In \bibinfo{booktitle}{\emph{Proceedings of the Second Scandinavian
  Logic Symposium}}, \bibfield{editor}{\bibinfo{person}{J.E. Fenstad}} (Ed.).
  \bibinfo{series}{Studies in Logic and the Foundations of Mathematics},
  Vol.~\bibinfo{volume}{63}. \bibinfo{publisher}{{Elsevier}},
  \bibinfo{address}{{North Holland}}, \bibinfo{pages}{63--92}.
\newblock
\showISSN{0049-237X}
\urldef\tempurl%
\url{https://doi.org/10.1016/S0049-237X(08)70843-7}
\showDOI{\tempurl}


\bibitem[Leijen(2008a)]%
        {leijenHMFSimpleType2008}
\bibfield{author}{\bibinfo{person}{Daan Leijen}.}
  \bibinfo{year}{2008}\natexlab{a}.
\newblock \bibinfo{booktitle}{\emph{{{HMF}}: {{Simple}} Type Inference for
  First-Class Polymorphism}}.
\newblock \bibinfo{type}{{T}echnical {R}eport} MSR-TR-2008-65.
  \bibinfo{institution}{{Microsoft Research}}. \bibinfo{pages}{15} pages.
\newblock


\bibitem[Leijen(2008b)]%
        {leijenRobustTypeInference2008}
\bibfield{author}{\bibinfo{person}{Daan Leijen}.}
  \bibinfo{year}{2008}\natexlab{b}.
\newblock \bibinfo{booktitle}{\emph{Robust Type Inference for First-Class
  Polymorphism}}.
\newblock \bibinfo{type}{{T}echnical {R}eport} MSR-TR-2008-55.
  \bibinfo{institution}{{Microsoft Research}}. \bibinfo{pages}{10} pages.
\newblock


\bibitem[Levy(2006)]%
        {levyCallbypushvalueDecomposingCallbyvalue2006}
\bibfield{author}{\bibinfo{person}{Paul~Blain Levy}.}
  \bibinfo{year}{2006}\natexlab{}.
\newblock \showarticletitle{Call-by-Push-Value: {{Decomposing}} Call-by-Value
  and Call-by-Name}.
\newblock \bibinfo{journal}{\emph{Higher-Order and Symbolic Computation}}
  \bibinfo{volume}{19}, \bibinfo{number}{4} (\bibinfo{date}{Dec.}
  \bibinfo{year}{2006}), \bibinfo{pages}{377--414}.
\newblock
\showISSN{1388-3690, 1573-0557}
\urldef\tempurl%
\url{https://doi.org/10.1007/s10990-006-0480-6}
\showDOI{\tempurl}


\bibitem[Milner(1978)]%
        {milnerTheoryTypePolymorphism1978}
\bibfield{author}{\bibinfo{person}{Robin Milner}.}
  \bibinfo{year}{1978}\natexlab{}.
\newblock \showarticletitle{A Theory of Type Polymorphism in Programming}.
\newblock \bibinfo{journal}{\emph{J. Comput. System Sci.}}
  \bibinfo{volume}{17}, \bibinfo{number}{3} (\bibinfo{date}{Dec.}
  \bibinfo{year}{1978}), \bibinfo{pages}{348--375}.
\newblock
\showISSN{00220000}
\urldef\tempurl%
\url{https://doi.org/10.1016/0022-0000(78)90014-4}
\showDOI{\tempurl}


\bibitem[Odersky and L{\"a}ufer(1996)]%
        {oderskyPuttingTypeAnnotations1996}
\bibfield{author}{\bibinfo{person}{Martin Odersky} {and}
  \bibinfo{person}{Konstantin L{\"a}ufer}.} \bibinfo{year}{1996}\natexlab{}.
\newblock \showarticletitle{Putting Type Annotations to Work}. In
  \bibinfo{booktitle}{\emph{Proceedings of the 23rd {{ACM SIGPLAN-SIGACT}}
  Symposium on {{Principles}} of Programming Languages - {{POPL}} '96}}.
  \bibinfo{publisher}{{ACM Press}}, \bibinfo{address}{{St. Petersburg Beach,
  Florida, United States}}, \bibinfo{pages}{54--67}.
\newblock
\showISBNx{978-0-89791-769-8}
\urldef\tempurl%
\url{https://doi.org/10.1145/237721.237729}
\showDOI{\tempurl}


\bibitem[Odersky et~al\mbox{.}(2001)]%
        {odersky03}
\bibfield{author}{\bibinfo{person}{Martin Odersky}, \bibinfo{person}{Christoph
  Zenger}, {and} \bibinfo{person}{Matthias Zenger}.}
  \bibinfo{year}{2001}\natexlab{}.
\newblock \showarticletitle{Colored Local Type Inference}. In
  \bibinfo{booktitle}{\emph{Proceedings of the 28th {{ACM SIGPLAN-SIGACT}}
  Symposium on {{Principles}} of Programming Languages - {{POPL}} '01}}.
  \bibinfo{publisher}{{ACM Press}}, \bibinfo{address}{{London, United
  Kingdom}}, \bibinfo{pages}{41--53}.
\newblock
\showISBNx{978-1-58113-336-3}
\urldef\tempurl%
\url{https://doi.org/10.1145/360204.360207}
\showDOI{\tempurl}


\bibitem[Peyton~Jones et~al\mbox{.}(2007)]%
        {peytonjonesPracticalTypeInference2007}
\bibfield{author}{\bibinfo{person}{Simon Peyton~Jones},
  \bibinfo{person}{Dimitrios Vytiniotis}, \bibinfo{person}{Stephanie Weirich},
  {and} \bibinfo{person}{Mark Shields}.} \bibinfo{year}{2007}\natexlab{}.
\newblock \showarticletitle{Practical Type Inference for Arbitrary-Rank Types}.
\newblock \bibinfo{journal}{\emph{Journal of Functional Programming}}
  \bibinfo{volume}{17}, \bibinfo{number}{01} (\bibinfo{date}{Jan.}
  \bibinfo{year}{2007}), \bibinfo{pages}{1}.
\newblock
\showISSN{0956-7968, 1469-7653}
\urldef\tempurl%
\url{https://doi.org/10.1017/S0956796806006034}
\showDOI{\tempurl}


\bibitem[Pierce and Turner(2000)]%
        {pierceLocalTypeInference2000}
\bibfield{author}{\bibinfo{person}{Benjamin Pierce} {and}
  \bibinfo{person}{David Turner}.} \bibinfo{year}{2000}\natexlab{}.
\newblock \showarticletitle{Local {{Type Inference}}}.
\newblock \bibinfo{journal}{\emph{ACM Transactions on Programming Languages and
  Systems}} \bibinfo{volume}{22}, \bibinfo{number}{1} (\bibinfo{date}{Jan.}
  \bibinfo{year}{2000}), \bibinfo{pages}{1--44}.
\newblock
\showISSN{0164-0925}
\urldef\tempurl%
\url{https://doi.org/10.1145/345099.345100}
\showDOI{\tempurl}


\bibitem[R{\'e}my and Yakobowski(2008)]%
        {Remy08}
\bibfield{author}{\bibinfo{person}{Didier R{\'e}my} {and}
  \bibinfo{person}{Boris Yakobowski}.} \bibinfo{year}{2008}\natexlab{}.
\newblock \showarticletitle{From {{ML}} to {{MLF}}: Graphic Type Constraints
  with Efficient Type Inference}. In \bibinfo{booktitle}{\emph{{{ICFP}}}}.
  \bibinfo{publisher}{{ACM}}, \bibinfo{address}{{Victoria, BC, Canada}},
  \bibinfo{pages}{63--74}.
\newblock


\bibitem[Reynolds(1974)]%
        {jcr-system-f}
\bibfield{author}{\bibinfo{person}{John~C Reynolds}.}
  \bibinfo{year}{1974}\natexlab{}.
\newblock \showarticletitle{Towards a Theory of Type Structure}. In
  \bibinfo{booktitle}{\emph{Programming Symposium}}. {Springer},
  \bibinfo{publisher}{{Springer}}, \bibinfo{address}{{Paris, France}},
  \bibinfo{pages}{408--425}.
\newblock


\bibitem[Reynolds(1983)]%
        {reynolds-parametricity}
\bibfield{author}{\bibinfo{person}{John~C. Reynolds}.}
  \bibinfo{year}{1983}\natexlab{}.
\newblock \showarticletitle{Types, Abstraction and Parametric Polymorphism}. In
  \bibinfo{booktitle}{\emph{{{IFIP}} Congress}}.
  \bibinfo{publisher}{{North-Holland/IFIP}}, \bibinfo{address}{{Paris,
  France}}, \bibinfo{pages}{513--523}.
\newblock


\bibitem[Serrano et~al\mbox{.}(2020)]%
        {serranoQuickLookImpredicativity2020}
\bibfield{author}{\bibinfo{person}{Alejandro Serrano},
  \bibinfo{person}{Jurriaan Hage}, \bibinfo{person}{Simon Peyton~Jones}, {and}
  \bibinfo{person}{Dimitrios Vytiniotis}.} \bibinfo{year}{2020}\natexlab{}.
\newblock \showarticletitle{A Quick Look at Impredicativity}.
\newblock \bibinfo{journal}{\emph{Proceedings of the ACM on Programming
  Languages}} \bibinfo{volume}{4}, \bibinfo{number}{ICFP} (\bibinfo{date}{Aug.}
  \bibinfo{year}{2020}), \bibinfo{pages}{1--29}.
\newblock
\showISSN{2475-1421, 2475-1421}
\urldef\tempurl%
\url{https://doi.org/10.1145/3408971}
\showDOI{\tempurl}


\bibitem[Serrano et~al\mbox{.}(2018)]%
        {serranoGuardedImpredicativePolymorphism2018}
\bibfield{author}{\bibinfo{person}{Alejandro Serrano},
  \bibinfo{person}{Jurriaan Hage}, \bibinfo{person}{Dimitrios Vytiniotis},
  {and} \bibinfo{person}{Simon Peyton~Jones}.} \bibinfo{year}{2018}\natexlab{}.
\newblock \showarticletitle{Guarded Impredicative Polymorphism}. In
  \bibinfo{booktitle}{\emph{Proceedings of the 39th {{ACM SIGPLAN Conference}}
  on {{Programming Language Design}} and {{Implementation}} - {{PLDI}} 2018}}.
  \bibinfo{publisher}{{ACM Press}}, \bibinfo{address}{{Philadelphia, PA, USA}},
  \bibinfo{pages}{783--796}.
\newblock
\showISBNx{978-1-4503-5698-5}
\urldef\tempurl%
\url{https://doi.org/10.1145/3192366.3192389}
\showDOI{\tempurl}


\bibitem[Spiwack(2014)]%
        {spiwackDissection2014}
\bibfield{author}{\bibinfo{person}{Arnaud Spiwack}.}
  \bibinfo{year}{2014}\natexlab{}.
\newblock \bibinfo{title}{A Dissection of {{L}}}.  (\bibinfo{year}{2014}).
\newblock


\bibitem[Sulzmann et~al\mbox{.}(2007)]%
        {haskell-fc}
\bibfield{author}{\bibinfo{person}{Martin Sulzmann}, \bibinfo{person}{Manuel
  M.~T. Chakravarty}, \bibinfo{person}{Simon~Peyton Jones}, {and}
  \bibinfo{person}{Kevin Donnelly}.} \bibinfo{year}{2007}\natexlab{}.
\newblock \showarticletitle{System {{F}} with Type Equality Coercions}. In
  \bibinfo{booktitle}{\emph{Proceedings of the 2007 {{ACM SIGPLAN}}
  International Workshop on Types in Languages Design and Implementation}}
  \emph{(\bibinfo{series}{{{TLDI}} '07})}. \bibinfo{publisher}{{Association for
  Computing Machinery}}, \bibinfo{address}{{New York, NY, USA}},
  \bibinfo{pages}{53--66}.
\newblock
\showISBNx{1-59593-393-X}
\urldef\tempurl%
\url{https://doi.org/10.1145/1190315.1190324}
\showDOI{\tempurl}


\bibitem[Tiuryn and Urzyczyn(1996)]%
        {tiuryn-urzczyn-96}
\bibfield{author}{\bibinfo{person}{Jerzy Tiuryn} {and} \bibinfo{person}{Pawel
  Urzyczyn}.} \bibinfo{year}{1996}\natexlab{}.
\newblock \showarticletitle{The Subtyping Problem for Second-Order Types Is
  Undecidable}. In \bibinfo{booktitle}{\emph{Proceedings of the 11th Annual
  {{IEEE}} Symposium on Logic in Computer Science}}
  \emph{(\bibinfo{series}{{{LICS}} '96})}. \bibinfo{publisher}{{IEEE Computer
  Society}}, \bibinfo{address}{{USA}}, \bibinfo{pages}{74}.
\newblock
\showISBNx{0-8186-7463-6}


\bibitem[Vytiniotis et~al\mbox{.}(2006)]%
        {vytiniotisBoxyTypesInference2006}
\bibfield{author}{\bibinfo{person}{Dimitrios Vytiniotis},
  \bibinfo{person}{Stephanie Weirich}, {and} \bibinfo{person}{Simon
  Peyton~Jones}.} \bibinfo{year}{2006}\natexlab{}.
\newblock \showarticletitle{Boxy {{Types}}: {{Inference}} for {{Higher-Rank
  Types}} and {{Impredicativity}}}. In \bibinfo{booktitle}{\emph{{{ICFP}}
  '06}}. \bibinfo{publisher}{{ACM Press}}, \bibinfo{address}{{Portland, Oregon,
  USA}}, \bibinfo{pages}{251--262}.
\newblock


\bibitem[Vytiniotis et~al\mbox{.}(2008)]%
        {vytiniotisFPHFirstclassPolymorphism2008}
\bibfield{author}{\bibinfo{person}{Dimitrios Vytiniotis},
  \bibinfo{person}{Stephanie Weirich}, {and} \bibinfo{person}{Simon
  Peyton~Jones}.} \bibinfo{year}{2008}\natexlab{}.
\newblock \showarticletitle{{{FPH}}: {{First-class Polymorphism}} for
  {{Haskell}}}. In \bibinfo{booktitle}{\emph{{{ICFP}} '08}}.
  \bibinfo{publisher}{{ACM Press}}, \bibinfo{address}{{Victoria, BC, Canada}},
  \bibinfo{pages}{295--306}.
\newblock


\bibitem[Wadler(1989)]%
        {wadler-free-theorems}
\bibfield{author}{\bibinfo{person}{Philip Wadler}.}
  \bibinfo{year}{1989}\natexlab{}.
\newblock \showarticletitle{Theorems for Free!}. In
  \bibinfo{booktitle}{\emph{Proceedings of the Fourth International Conference
  on {{Functional}} Programming Languages and Computer Architecture - {{FPCA}}
  '89}}. \bibinfo{publisher}{{ACM Press}}, \bibinfo{address}{{Imperial College,
  London, United Kingdom}}, \bibinfo{pages}{347--359}.
\newblock
\showISBNx{978-0-89791-328-7}
\urldef\tempurl%
\url{https://doi.org/10.1145/99370.99404}
\showDOI{\tempurl}


\end{thebibliography}

\end{document}


\maketitle

\tableofcontents
\newpage

\section*{Definitions}
\addcontentsline{toc}{section}{Definitions}

\begin{figure}[ht!]
  \input{shared-definitions-polarized-system-f.tex}
  
  \caption{\PSystemF}
  \label{fig:Polarized System F}
\end{figure}

\begin{figure}[ht!]
  \input{shared-definitions-well-formed-declarative-type.tex}
  \caption{Well-formedness of declarative types}
\end{figure}

\begin{figure}[ht!]
  \input{shared-definitions-declarative-type-system.tex}
  \caption{Declarative type system}
  \label{fig:Declarative type system}
\end{figure}

\begin{figure}[ht!]
  \input{shared-definitions-declarative-subtyping.tex}
  \caption{Declarative subtyping}
  \label{fig:Declarative subtyping}
\end{figure}

\begin{figure}[ht!]
  \input{shared-definitions-isomorphic-types.tex}
  \caption{Isomorphic types}
\end{figure}

\begin{figure}[ht!]
  \input{shared-definitions-isomorphic-environments.tex}
  \caption{Isomorphic environments}
\end{figure}

\begin{figure}[ht!]
  \input{shared-definitions-applying-context-to-term.tex}
  \caption{Applying a context to a type}
  \label{figure:context-application-to-terms}
\end{figure}

\begin{figure}[ht!]
  \input{shared-definitions-algorithmic-system.tex}
  \caption{Additions to declarative types and contexts to form their algorithmic counterparts}
  \label{fig:algorithmic-types-and-contexts}
\end{figure}

\begin{figure}[ht!]
  \input{shared-definitions-algorithmic-subtyping.tex}

  \caption{Algorithmic subtyping}
  \label{figure:algorithmic-subtyping}
\end{figure}

\begin{figure}[ht!]
  \input{shared-definitions-restricted-context.tex}
  \caption{Definition of context restriction}
\end{figure}

\begin{figure}[ht!]
  \input{shared-definitions-algorithmic-type-system.tex}
  \caption{Algorithmic type system}
  \label{figure:algorithmic-type-system}
\end{figure}

\begin{figure}[ht!]
  \input{shared-definitions-well-formed-context.tex}
  \caption{Well-formedness of contexts}
\end{figure}

\begin{figure}[ht!]
  \judgbox{\wfposnegtypeJudg{\acontext}{A}}{In the context $\acontext$, $A$ is a well-formed positive/negative type}

  \begin{mathpar}
    \Infer{\twfguess}
      {\guess{\alpha} \in \EV(\acontext)}
      {\wfpostypeJudg{\acontext} {\guess{\alpha}}}
  \end{mathpar}
  \caption{
    Additional well-formedness rules for algorithmic types.
    $\EV(\acontext)$ contains all the existential type variables in $\acontext$, independently of whether they are solved or unsolved.
  }
\end{figure}

\begin{figure}[ht!]
  \input{shared-definitions-context-extension.tex}
  \caption{Context extension}
  \label{fig:Context extension}
\end{figure}

\begin{figure}[ht!]
  \input{shared-definitions-weak-context-extension.tex}
  \caption{Weak context extension. We highlight the rules that are \enquote{new} compared with context extension.}
  \label{figure:weak-context-extension}
\end{figure}

\begin{figure}[ht!]
  \input{shared-definitions-well-formed-environment.tex}
  \caption{Well-formedness of typing environments}
\end{figure}

\begin{figure}[ht!]
  \input{shared-definitions-type-size-ignoring-quantification.tex}
  \caption{The size of a type, ignoring universal quantification}
  \label{figure:type-size-ignoring-quantification}
\end{figure}

\begin{figure}[ht!]
  \input{shared-definitions-num-prenex-quantifiers.tex}
  \caption{The number of prenex quantifiers in a type $A$}
  \label{figure:num-prenex-quantifiers}
\end{figure}

\begin{figure}[ht!]
  \input{shared-definitions-producing-declarative-context.tex}
  \caption{Producing a declarative context from an algorithmic context}
  \label{figure:producing-declarative-context}
\end{figure}

\begin{figure}
  \input{shared-definitions-system-f-cbv-embedding.tex}
  \caption{
    An embedding of typeable terms in System F under a call-by-value evaluation order in \PSystemF.
  }
  \label{fig:Embedding of System F}
\end{figure}

\clearpage

\part*{Lemmas}
\addcontentsline{toc}{part}{Lemmas}

\setcounter{section}{0}
\renewcommand\thesection{\Alph{section}}

\section{Weakening}

\begin{restatable}[Pushing uvars right preserves \termwellformedness]{lemma}{RestatePushingUvarsRightPreservesTwf}
  \label{lemma:Pushing uvars right preserves term well-formedness}
  Let $\aconwithholeshort{\acontext_M}$ abbreviate $\aconwithhole{\acontext_M}$.
  Then if $\wfposnegtypeJudg{\pushstartcontext}{A}$, $\wfposnegtypeJudg{\pushresultcontext}{A}$.
\end{restatable}

\begin{restatable}[Term well-formedness weakening]{lemma}{RestateTwfWeakening}
  \label{lemma:Term well-formedness weakening}
  If $\wfposnegtypeJudg{\acontext}{A}$ then $\wfposnegtypeJudg{\acontext, \acontext'}{A}$.
\end{restatable}

\begin{restatable}[Pushing uvars right in declarative judgment]{lemma}{RestatePushingUvarsRightInDJudge}
  \label{lemma:Pushing uvars right in declarative judgment}
  Let $\aconwithholeshort{\acontext_M}$ abbreviate $\aconwithhole{\acontext_M}$.
  Then if $\DPosNegSubtypeJudg{\pushstartcontext} {A} {B}$, $\DPosNegSubtypeJudg{\pushresultcontext} {A} {B}$.
\end{restatable}

\begin{restatable}[Declarative subtyping weakening]{lemma}{RestateDJudgeWeakening}
  \label{lemma:Declarative subtyping weakening}
  If $\DPosNegSubtypeJudg{\acontext} {A} {B}$ then $\DPosNegSubtypeJudg{\acontext, \acontext'} {A} {B}$.
\end{restatable}

\section{Declarative subtyping}

\begin{restatable}[Declarative subtyping is reflexive]{lemma}{RestateReflexivityPnSubtype}
  \label{lemma:Reflexivity of declarative pnsubtype}
  If $\wfposnegtypeJudg{\dcontext} {A}$ then $\DPnSubtypeJudge{\dcontext} {A} {A}$.
\end{restatable}

\begin{restatable}[Declarative substitution \wellformedness]{lemma}{RestateSubstitutionWf}
  \label{lemma:Declarative substitution well-formedness}
  If $\wfpostypeJudg{\dcontext_L, \dcontext_R} {P}$ and $\wfposnegtypeJudg{\dconwithhole{\alpha}}{A}$, then $\wfposnegtypeJudg{\dcontext_L, \dcontext_R}{\subterm{P / \alpha} A}$.
\end{restatable}

\begin{restatable}[Declarative subtyping is stable under substitution]{lemma}{RestateDeclarativeSubtypingSubstitutionLemma}
  \label{lemma:Declarative subtyping substitution lemma}
  If $\wfpostypeJudg{\dcontext_L, \dcontext_R}{P}$, then:

  \begin{itemize}
    \item If $\wfpostypeJudg{\dconwithhole{\alpha}}{Q}$, $\wfpostypeJudg{\dconwithhole{\alpha}}{R}$, and $\DPosSubtypeJudge{\dconwithhole{\alpha}}{Q}{R}$, then $\DPosSubtypeJudge{\dcontext_L, \dcontext_R}{\subterm{P / \alpha} Q} {\subterm{P / \alpha} R}$.

    \item If $\wfnegtypeJudg{\dconwithhole{\alpha}}{N}$, $\wfnegtypeJudg{\dconwithhole{\alpha}}{M}$, and $\DNegSubtypeJudge{\dconwithhole{\alpha}}{N}{M}$, then $\DNegSubtypeJudge{\dcontext_L, \dcontext_R} {\subterm{P / \alpha} N} {\subterm{P / \alpha} M}$.
  \end{itemize}
\end{restatable}

\begin{restatable}[Symmetry of positive declarative subtyping]{lemma}{RestateSymmetryPositiveDeclarativeSubtyping}
  \label{lemma:Symmetry of positive declarative subtyping}
  If $\DPosSubtypeJudge{\dcontext} {P} {Q}$ then $\DPosSubtypeJudge{\dcontext} {Q} {P}$ by a derivation of equal height.
\end{restatable}

\subsection{Isomorphic types}

\begin{restatable}[Mutual subtyping substitution]{lemma}{RestateDeclarativeMutualSubtypingLemma} \ \
  \label{lemma:Declarative mutual subtyping lemma}
  Given $\wfpostypeJudg{\dcontext, \vv{\alpha}} {\vv{P}}$ and $\wfpostypeJudg{\dcontext, \vv{\beta}} {\vv{Q}}$:
  \begin{itemize}
    \begin{minipage}[t]{0.5\linewidth}
      \item If:
    
        \begin{enumerate}[noitemsep]
          \item $\wfpostypeJudg{\dcontext, \vv{\alpha}} {R}$
          \item $\wfpostypeJudg{\dcontext, \vv{\beta}} {S}$
          \item $\DPosSubtypeJudge{\dcontext, \vv{\alpha}} {R} {\subcon{\vv{P / \beta}} S}$
          \item $\DPosSubtypeJudge{\dcontext, \vv{\beta}} {S} {\subcon{\vv{Q / \alpha}} R}$
        \end{enumerate}

        then:
        
        \begin{enumerate}[noitemsep]
          \item $\forall \beta_i \in \vv{\beta} \ldotp \beta_i \in \FreeUV(S) \implies \\ \exists \gamma \ldotp P_i = \gamma$
          \item $\forall \alpha_i \in \vv{\alpha} \ldotp \alpha_i \in \FreeUV(R) \implies \\ \exists \gamma \ldotp Q_i = \gamma$
        \end{enumerate}
    \end{minipage}
    \begin{minipage}[t]{0.5\linewidth}
      \item If:
        
        \begin{enumerate}[noitemsep]
          \item $\wfnegtypeJudg{\dcontext, \vv{\alpha}} {M}$
          \item $\wfnegtypeJudg{\dcontext, \vv{\beta}} {N}$
          \item $\DNegSubtypeJudge{\dcontext, \vv{\alpha}} {\subcon{\vv{P / \beta}} N} {M}$
          \item $\DNegSubtypeJudge{\dcontext, \vv{\beta}} {\subcon{\vv{Q / \alpha}} M} {N}$
        \end{enumerate}

        then:
        
        \begin{enumerate}[noitemsep]
          \item $\forall \beta_i \in \vv{\beta} \ldotp \beta_i \in \FreeUV(N) \implies \\ \exists \gamma \ldotp P_i = \gamma$
          \item $\forall \alpha_i \in \vv{\alpha} \ldotp \alpha_i \in \FreeUV(M) \implies \\ \exists \gamma \ldotp Q_i = \gamma$
        \end{enumerate}
    \end{minipage}
  \end{itemize}
\end{restatable}

\begin{restatable}[Isomorphic types are the same size]{lemma}{RestateDeclarativeMutualSubtypingSizeLemma}
  \label{lemma:Isomorphic types are the same size}
  If:
  
  \begin{enumerate}
    \item $\wfpostypeJudg{\dcontext} {A}$
    \item $\wfpostypeJudg{\dcontext} {B}$
    \item $\MutualSubtypePosNegJudge{\dcontext} {A} {B}$
  \end{enumerate}
  
  then $\termsize{A} = \termsize{B}$.
\end{restatable}

\subsection{Transitivity}

\begin{restatable}[Declarative subtyping is transitive]{lemma}{RestateTransitivityPnSubtype}
  \label{lemma:Transitivity of declarative pnsubtype}
  If $\wfposnegtypeJudg{\dcontext}{A}$, $\wfposnegtypeJudg{\dcontext}{B}$, $\wfposnegtypeJudg{\dcontext}{C}$, $\DPnSubtypeJudge{\dcontext} {A} {B}$, and $\DPnSubtypeJudge{\dcontext} {B} {C}$, then $\DPnSubtypeJudge{\dcontext} {A} {C}$.
\end{restatable}

\section{Weak context extension}

\begin{restatable}[$\congoesweak$ subsumes $\congoes$]{lemma}{WeakCtxExtSubsumes}
  \label{lemma:weak context extension subsumes normal}
  If $\congoesJudg{\acontext}{\acontext'}$, then $\congoesWeakJudg{\acontext}{\acontext'}$.
\end{restatable}

\begin{restatable}[Weak context extension is reflexive]{lemma}{WeakCtxExtRefl}
  \label{lemma:weak context extension reflexive}
  For all contexts $\Theta$, $\congoesWeakJudg{\Theta}{\Theta}$.
\end{restatable}

\begin{restatable}[Equality of declarative contexts (weak)]{lemma}{WeakContextDeclCxt}
  \label{lemma:weak context extension equal declarative contexts}
  If $\congoesWeakJudg{\acontext}{\acontext'}$,
  then $\makedec{\acontext} = \makedec{\acontext'}$.
\end{restatable}

\begin{restatable}[Weak context extension is transitive]{lemma}{WeakCtxExtTransitive}
  \label{lemma:weak context extension transitive}
  If $\congoesWeakJudg{\Theta}{\Theta'}$ and $\congoesWeakJudg{\Theta'}{\Theta''}$, then $\congoesWeakJudg{\Theta}{\Theta''}$.
\end{restatable}

\begin{restatable}[Weak context extension preserves well-formedness]{lemma}{WeakContextExtensionPreservesWF}
  \label{lemma:weak context extension preserves well-formedness}
  If $\wfposnegtypeJudg{\Theta}{A}$ and $\congoesWeakJudg{\Theta}{\Theta'}$ then $\wfposnegtypeJudg{\Theta'}{A}$.
\end{restatable}

\begin{restatable}[Weak context extension preserves w.f. envs]{lemma}{ContextExtWFEnvs}
  \label{lemma:context extension preserves wf envs}
  If $\congoesWeakJudg{\Theta}{\Theta'}$ and $\envwfJudg{\Theta}{\Gamma}$, then $\envwfJudg{\Theta'}{\Gamma}$.
\end{restatable}

\begin{restatable}[The extended context makes the type ground (weak)]{lemma}{WeakExtendedContextMakesGround}
  \label{lemma:The extended context makes the type ground (weak)}
  If $\acontext \conwf$, $\acontext' \conwf$, $\congoesWeakJudg{\acontext}{\acontext'}$, and $\subcon{\acontext'} \subcon{\acontext} A$ ground, then $\subcon{\acontext'} A$ ground.
\end{restatable}

\begin{restatable}[Extending context preserves groundness (weak)]{lemma}{WeakExtendingContextPreservesGroundness}
  \label{lemma:extending-context-preserves-groundness-weak}
  If $\acontext \conwf$, $\acontext' \conwf$, $\congoesWeakJudg{\acontext}{\acontext'}$, and $\subcon{\acontext} A$ ground, then $\subcon{\acontext'} A$ ground.
\end{restatable}

\section{Context extension}

\begin{restatable}[Context extension is reflexive]{lemma}{ContextExtensionReflexive}
  \label{lemma:context extension reflexive}
  For all contexts $\Theta$, $\congoesJudg{\Theta}{\Theta}$.
\end{restatable}

\begin{restatable}[Equality of declarative contexts]{lemma}{EqualityOfDeclarativeContexts}
  \label{lemma:equal declarative contexts}
  If $\congoesJudg{\acontext}{\acontext'}$,
  then $\makedec{\acontext} = \makedec{\acontext'}$.
\end{restatable}

\begin{restatable}[Context extension is transitive]{lemma}{congoesTransitive}
  \label{lemma:context extension transitive}
  If $\congoesJudg{\acontext}{\acontext'}$ and $\congoesJudg{\acontext'}{\acontext''}$, then $\congoesJudg{\acontext}{\acontext''}$.
\end{restatable}

\begin{restatable}[Context extension preserves \wellformedness]{lemma}{RestateContextExtensionTermWf}
  \label{lemma:Context extension preserves term well-formedness}
  If $\wfposnegtypeJudg{\acontext}{A}$ and $\congoesJudg{\acontext}{\acontext'}$, then $\wfposnegtypeJudg{\acontext'}{A}$.
\end{restatable}

\begin{restatable}[Applying a context to a ground type]{lemma}{RestateApplyingContextToGroundTypes}
  \label{lemma:Context substitution on ground terms}
  If $A$ ground, then $\subcon{\acontext} A = A$.
\end{restatable}

\begin{restatable}[Context application is idempotent]{lemma}{RestateContextSubstitutionIdempotent}
  \label{lemma:Context substitution idempotence}
  If $\acontext \conwf$, then $\subcon{\acontext}\subcon{\acontext} A = \subcon{\acontext} A$.
\end{restatable}

\begin{restatable}[The extended context makes the type ground]{lemma}{ExtendedContextMakesGround}
  \label{lemma:The extended context makes the type ground}
  If $\acontext \conwf$, $\acontext' \conwf$, $\congoesJudg{\acontext}{\acontext'}$, and $\subcon{\acontext'} \subcon{\acontext} A$ ground, then $\subcon{\acontext'} A$ ground.
\end{restatable}

\begin{restatable}[Extending context preserves groundness]{lemma}{ExtendingContextPreservesGroundness}
  \label{lemma:extending-context-preserves-groundness}
  If $\acontext \conwf$, $\acontext' \conwf$, $\congoesJudg{\acontext}{\acontext'}$, and $\subcon{\acontext} A$ ground, then $\subcon{\acontext'} A$ ground.
\end{restatable}

\section{Well-formedness of subtyping}

\begin{restatable}[Applying context to the type preserves \wellformedness]{lemma}{RestateCwfSubTwf}
  \label{lemma:Well-formed context substitution preserves term well-formedness}
  If $\acontext \conwf$ and $\wfposnegtypeJudg{\acontext}{A}$,
  then $\wfposnegtypeJudg{\acontext}{\subcon{\acontext} A}$.
\end{restatable}

\begin{restatable}[Algorithmic subtyping is \wellformed]{lemma}{RestateWellFormednessPNSubtype} \
  \label{lemma:well-formedness-pnsubtype}
  \begin{itemize}
      \item If $\acontext \entails P \possubtype Q \prodcon \acontext'$, $\acontext \conwf$, $P$ ground, and $\subcon{\acontext} Q = Q$, then $\acontext' \conwf$, $\congoesJudg{\acontext}{\acontext'}$, and $\subcon{\acontext'} Q$ ground.

      \item If $\acontext \entails N \negsubtype M \prodcon \acontext'$, $\acontext \conwf$, $M$ ground, and $\subcon{\acontext} N = N$, then $\acontext' \conwf$, $\congoesJudg{\acontext}{\acontext'}$, and $\subcon{\acontext'} N$ ground.
  \end{itemize}
\end{restatable}

\section{Soundness of subtyping}

\subsection{Lemmas for soundness}

\begin{restatable}[Completing context preserves \wellformedness]{lemma}{RestateCompletingContextPreservesWf}
  \label{lemma:Completing a context preserves well-formedness}
  If $\wfposnegtypeJudg{\acontext} {A}$ and $\groundJudge{A}$ then $\wfposnegtypeJudg{\makedec{\acontext}} {A}$.
\end{restatable}

\begin{restatable}[$\congoesweak$ leads to isomorphic types]{lemma}{RestateWeakExtendedContextMutualSubtyping}
  \label{lemma:Weak context extension leads to isomorphic types}
  If:
  
  \begin{enumerate}
    \item $\wfposnegtypeJudg{\acontext} {A}$
    \item $\congoesWeakJudg{\acontext}{\acontext'}$
    \item $\groundJudge{\subcon{\acontext'} A}$
    \item $\acontext \conwf$
    \item $\acontext' \conwf$
  \end{enumerate}
  
  then $\MutualSubtypePosNegJudge{\makedec{\acontext}} {\subcon{\acontext'} \subcon{\acontext} A} {\subcon{\acontext'} A}$.
\end{restatable}

\begin{restatable}[$\congoesweak$ leads to isomorphic types (ground)]{lemma}{RestateWeakExtendedContextGroundMutualSubtyping}
  \label{lemma:Weak context extension leads to isomorphic types (ground)}
  If:
  
  \begin{enumerate}
    \item $\wfposnegtypeJudg{\acontext} {A}$
    \item $\subcon{\acontext} A$ ground
    \item $\acontext \congoesweak \acontext'$
    \item $\acontext \conwf$
    \item $\acontext' \conwf$
  \end{enumerate}
  
  then $\MutualSubtypePosNegJudge{\makedec{\acontext}} {\subcon{\acontext} A} {\subcon{\acontext'} A}$.
\end{restatable}

\begin{restatable}[$\congoes$ leads to isomorphic types]{lemma}{RestateExtendedContextMutualSubtyping}
  \label{lemma:Context extension leads to isomorphic types}
  If:
  
  \begin{enumerate}
    \item $\wfposnegtypeJudg{\acontext} {A}$
    \item $\congoesJudg{\acontext}{\acontext'}$
    \item $\groundJudge{\subcon{\acontext'} A}$
    \item $\acontext \conwf$
    \item $\acontext' \conwf$
  \end{enumerate}
  
  then $\MutualSubtypePosNegJudge{\makedec{\acontext}} {\subcon{\acontext'} \subcon{\acontext} A} {\subcon{\acontext'} A}$.
\end{restatable}

\begin{restatable}[$\congoes$ leads to isomorphic types (ground)]{lemma}{RestateExtendedContextGroundMutualSubtyping}
  \label{lemma:Context extension leads to isomorphic types (ground)}
  If:
  
  \begin{enumerate}
    \item $\wfposnegtypeJudg{\acontext} {A}$
    \item $\subcon{\acontext} A$ ground
    \item $\acontext \congoes \acontext'$
    \item $\acontext \conwf$
    \item $\acontext' \conwf$
  \end{enumerate}
  
  then $\MutualSubtypePosNegJudge{\makedec{\acontext}} {\subcon{\acontext} A} {\subcon{\acontext'} A}$.
\end{restatable}

\subsection{Statement}

\begin{restatable}[Soundness of algorithmic subtyping]{theorem}{RestateSoundnessPNSubtype}
  \label{theorem:soundness-pnsubtype}
  Given a well-formed algorithmic context $\acontext$ and a well-formed complete context $\ccontext$:
  \begin{itemize}
      \item If
        $\acontext \entails P \possubtype Q \prodcon \acontext'$,
        $\congoesJudg{\acontext'}{\ccontext}$,
        $P$ ground,
        $\subcon{\acontext} Q = Q$,
        $\wfpostypeJudg{\acontext} {P}$, and
        $\wfpostypeJudg{\acontext} {Q}$,

        then $\CnewPosJudge{\acontext} {P} {Q}$.

      \item If
        $\acontext \entails N \negsubtype M \prodcon \acontext'$,
        $\congoesJudg{\acontext'}{\ccontext}$,
        $M$ ground,
        $\subcon{\acontext} N = N$,
        $\wfnegtypeJudg{\acontext} {N}$, and
        $\wfnegtypeJudg{\acontext} {M}$,

        then $\CnewNegJudge{\acontext} {N} {M}$.
  \end{itemize}
\end{restatable}

\section{Completeness of subtyping}

\subsection{Lemmas for completeness}

\begin{restatable}[Completion preserves \wellformedness]{lemma}{RestateCompletingContextAndTypePreservesWf}
  \label{lemma:Completing a context and type preserves well-formedness}
  If
    $\conwfJudg{\acontext}$,
    $\wfposnegtypeJudg{\acontext} {A}$, and
    $\acontext \congoes \ccontext$, then
    $\wfposnegtypeJudg{\makedec{\acontext}} {\subcon{\ccontext} A}$.
\end{restatable}

\begin{restatable}[Extension solving guess]{lemma}{ExtensionSolvingGuess}
  \label{lemma:extension solving guess}
  If $\aconwithhole{\guess{\alpha}} \congoes \cconwithhole{\guess{\alpha} = Q}$ and $\MutualSubtypePosJudge{\subcon{\ccontext_L} \acontext_L} {P} {Q}$, then $\aconwithhole{\guess{\alpha} = P} \congoes \cconwithhole{\guess{\alpha} = Q}$.
\end{restatable}

\begin{restatable}[Context extension substitution size]{lemma}{ContextExtensionSubstitutionSizeLemma}
  \label{lemma:context extension substitution size lemma}
    If:

    \begin{enumerate}
      \item $\conwfJudg{\acontext}$
      \item $\wfposnegtypeJudg{\acontext}{A}$
      \item $\congoesJudg{\acontext}{\ccontext}$
      \item $\conwfJudg{\ccontext}$
    \end{enumerate}

    then $\termsize{\subcon{\ccontext} \subcon{\acontext} A} = \termsize{\subcon{\ccontext} A}$.
\end{restatable}

\begin{restatable}[Context extension ground substitution size]{lemma}{ContextExtensionGroundSubstitutionSizeLemma}
  \label{lemma:context extension ground substitution size lemma}
    If:

    \begin{enumerate}
      \item $\conwfJudg{\acontext}$
      \item $\wfposnegtypeJudg{\acontext}{A}$
      \item $\groundJudge{\subcon{\acontext} A}$
      \item $\congoesJudg{\acontext}{\ccontext}$
      \item $\conwfJudg{\ccontext}$
    \end{enumerate}

    then $\termsize{\subcon{\acontext} A} = \termsize{\subcon{\ccontext} A}$.
\end{restatable}

\subsection{Statement}

\begin{restatable}[Completeness of algorithmic subtyping]{theorem}{RestateCompletenessPNSubtype}
  \label{theorem:Completeness of algorithmic subtyping}

  If $\acontext \conwf$, $\acontext \congoes \ccontext$, and $\ccontext \conwf$, then:

  \begin{itemize}
    \item
      If
      $\CnewPosJudge{\acontext} {P} {Q}$,
      $\acontext \entails P \postype$,
      $\acontext \entails Q \postype$,
      $P$ ground, and
      $\subcon{\acontext} Q = Q$,
      then $\exists \acontext'$ such that
      $\acontext \entails P \possubtype Q \prodcon \acontext'$ and
      $\acontext' \congoes \ccontext$.

    \item
      If
      $\CnewNegJudge{\acontext} {N} {M}$,
      $\acontext \entails M \negtype$,
      $\acontext \entails N \negtype$,
      $M \text{ ground}$, and
      $\subcon{\acontext} N = N$,
      then $\exists \acontext'$ such that
      $\acontext \entails N \negsubtype M \prodcon \acontext'$ and
      $\acontext' \congoes \ccontext$.

  \end{itemize}
\end{restatable}

\section{Determinism of subtyping}

\begin{restatable}[Algorithmic subtyping is deterministic]{lemma}{RestateSubtypingDeterministic} \
  \label{lemma:subtyping determinacy}
  \begin{itemize}
    \item 
      If $\APosSubtypeJudg{\Theta}{P}{Q}{\Theta_1'}$ and $\APosSubtypeJudg{\Theta}{P}{Q}{\Theta_2'}$, then $\Theta_1' = \Theta_2'$.
    \item 
      If $\ANegSubtypeJudg{\Theta}{N}{M}{\Theta_1'}$ and $\ANegSubtypeJudg{\Theta}{N}{M}{\Theta_2'}$, then $\Theta_1' = \Theta_2'$.
  \end{itemize}
\end{restatable}

\section{Decidability of subtyping}

\subsection{Lemmas for decidability}

\begin{restatable}[Completed non-ground size bounded by ground size]{lemma}{RestateCompletedNonGroundSizeBounded} \ \
  \label{lemma:Completed non-ground size bounded}
  \begin{itemize}
    \item If
      $\APosSubtypeJudg{\acontext}{P}{Q}{\acontext'}$,
      $\acontext \conwf$,
      $P$ ground, and
      $\subcon{\acontext} Q = Q$,
      then $\termsize{\subcon{\acontext'} Q} \leq \termsize{P}$.
  
    \item If
      $\ANegSubtypeJudg{\acontext}{N}{M}{\acontext'}$,
      $\acontext \conwf$,
      $M$ ground, and
      $\subcon{\acontext} N = N$,
      then $\termsize{\subcon{\acontext'} N} \leq \termsize{M}$.
  \end{itemize}
\end{restatable}

\subsection{Statement}

\begin{restatable}[Decidability of algorithmic subtyping]{lemma}{RestateDecidabilityAlgorithmicSubtyping}
  \label{lemma:Decidability of algorithmic subtyping}
  There exists a total order $\ajudgelt$ on well-formed algorithmic subtyping judgments such that for each derivation with subtyping judgment premises $A_i$ and conclusion $B$, each $A_i$ compares less than $B$, i.e. $\forall i \ldotp A_i \ajudgelt B$.
\end{restatable}

\section{Isomorphic types}

\begin{restatable}[Isomorphic environments type the same terms]{lemma}{IsomorphicTypeTypeSameExpressions}
  \label{lemma:isomorphic types check expressions}
  If $\declisovarctxjudg{\acontext}{\typeenv}{\typeenv'}$, then:
  \begin{itemize}
    \item If $\declsynjudg{\acontext; \typeenv}{v}{P}$ then $\exists P'$ such that
      $\declnisotypejudg{\acontext}{P}{P'}$ and
      $\declsynjudg{\acontext; \typeenv'}{v}{P'}$.

    \item If $\declsynjudg{\acontext; \typeenv}{t}{N}$ then $\exists N'$ such that
      $\declnisotypejudg{\acontext}{N}{N'}$ and
      $\declsynjudg{\acontext; \typeenv'}{t}{N'}$.

    \item If
      $\declspinejudg{\acontext; \typeenv}{s}{\spine{N}{M}}$ and
      $\declnisotypejudg{\acontext}{N}{N'}$, then $\exists M'$ such that
      $\declnisotypejudg{\acontext}{M}{M'}$ and
      $\declspinejudg{\acontext; \typeenv}{s}{\spine{N'}{M'}}$.
  \end{itemize}
\end{restatable}

\section{Well-formedness of typing}

\begin{restatable}[Well-formedness of restricted contexts]{lemma}{RestrictedContextWf} \
  \label{lemma:restricted context wf}
  If $\conwfJudg{\acontext}$,
  $\conwfJudg{\acontext'}$,
  $\congoesWeakJudg{\acontext}{\acontext'}$, then
  $\conwfJudg{\restrictcontext{\acontext'}{\acontext}}$,
  $\congoesJudg{\acontext}{\restrictcontext{\acontext'}{\acontext}}$, and
  $\congoesWeakJudg{\restrictcontext{\acontext'}{\acontext}}{\acontext'}$.
\end{restatable}

\begin{restatable}[Type well-formed with type variable removed]{lemma}{AlphaNotInTypeWellFormed}
  \label{lemma:type well-formed with alpha removed}
  If $\wfposnegtypeJudg{\dcontext_L, \alpha, \dcontext_R}{T}$ and $\alpha \notin \FreeUV(T)$,
  then $\wfposnegtypeJudg{\dcontext_L, \dcontext_R}{T}$.
\end{restatable}

\begin{restatable}[Substitution preserves well-formedness of types]{lemma}{SubstitutionPreservesWFTypes}
  \label{lemma:substituion preserves well-formedness of types}
  If
  $\wfposnegtypeJudg{\acontext_L, \alpha, \acontext_R}{T}$,
  then
  $\wfposnegtypeJudg{\acontext_L, \guess{\alpha}, \acontext_R}{\subterm{\guess{\alpha}/\alpha} T}$.
\end{restatable}

\begin{restatable}[Context extension maintains variables]{lemma}{ContextExtensionVarSame}
  \label{lemma:Context extension maintains variables}
  If $\congoesJudg{\acontext}{\ccontext}$, then $\FreeUV(\acontext) = \FreeUV(\ccontext)$ and $\FreeEV(\acontext) = \FreeEV(\ccontext)$.
\end{restatable}

\begin{restatable}[Algorithmic typing is w.f.]{lemma}{AlgorithmicTypingWellFormed} \
  \label{lemma:algorithmic-typing-well-formed}
  Given a typing context $\acontext$ and typing environment $\typeenv$ such that $\conwfJudg{\acontext}$ and $\envwfJudg{\acontext}{\typeenv}$:
  \begin{itemize}
      \item If
      $\algosynjudg{\acontext; \typeenv}{v}{P}{\acontext'}$,
      then
      $\conwfJudg{\acontext'}$,
      $\congoesJudg{\acontext}{\acontext'}$,
      $\wfpostypeJudg{\acontext'}{P}$, and
      $\groundJudge{P}$.

      \item If
      $\algosynjudg{\acontext; \typeenv}{t}{N}{\acontext'}$,
      then
      $\conwfJudg{\acontext'}$,
      $\congoesJudg{\acontext}{\acontext'}$,
      $\wfnegtypeJudg{\acontext'}{N}$, and
      $\groundJudge{N}$.

      \item If $\algospinejudg{\acontext; \typeenv}{s}{\spine{N}{M}}{\acontext'}$,
      $\wfnegtypeJudg{\acontext}{N}$, and
      $\subcon{\acontext} N = N$,
      then
      $\conwfJudg{\acontext'}$,
      $\congoesWeakJudg{\acontext}{\acontext'}$,
      $\wfnegtypeJudg{\acontext'}{M}$,
      $\subcon{\acontext'} M = M$, and
      $\FreeEV(M) \subseteq \FreeEV(N) \cup (\FreeEV(\acontext') \setminus \FreeEV(\acontext))$.
  \end{itemize}
\end{restatable}

\section{Determinism of typing}

\begin{restatable}[Algorithmic typing is deterministic]{lemma}{TypingDeterminacy} \
  \label{lemma:typing determinacy}
  \begin{itemize}
    \item
      If $\algosynjudg{\acontext; \typeenv}{e}{A_1}{\acontext'_1}$ and
      $\algosynjudg{\acontext; \typeenv}{e}{A_2}{\acontext'_2}$, then
      $A_1 = A_2$ and
      $\acontext'_1 = \acontext'_2$.

    \item
      If $\algospinejudg{\acontext; \typeenv}{t}{\spine{N}{M_1}}{\acontext'_1}$ and
      $\algospinejudg{\acontext; \typeenv}{t}{\spine{N}{M_2}}{\acontext'_2}$, then
      $M_1 = M_2$ and
      $\acontext'_1 = \acontext'_2$.
  \end{itemize}
\end{restatable}

\section{Decidability of typing}

\begin{restatable}[Decidability of algorithmic typing]{lemma}{RestateDecidabilityAlgorithmicTyping}
  \label{lemma:Decidability of algorithmic typing}
  There exists a total order $\ajudgelt$ on well-formed algorithmic typing judgments such that for each derivation with typing judgment premises $A_i$ and conclusion $B$, each $A_i$ compares less than $B$, i.e. $\forall i \ldotp A_i \ajudgelt B$.
\end{restatable}

\section{Soundness of typing}

\subsection{Lemmas}

\begin{restatable}[Extended complete context]{lemma}{ExtendedCompleteContext}
  \label{lemma:Extended complete context}
  If
  $\conwfJudg{\acontext'}$,
  $\conwfJudg{\ccontext}$,
  $\congoesJudg{\acontext}{\ccontext}$,
  $\congoesWeakJudg{\acontext}{\acontext'}$, and
  $\congoesJudg{\restrictcontext{\acontext'}{\acontext}}{\ccontext}$,
  then $\exists \ccontext'$ such that
  $\conwfJudg{\ccontext'}$,
  $\congoesJudg{\acontext'}{\ccontext'}$, and
  $\congoesWeakJudg{\ccontext}{\ccontext'}$.
\end{restatable}

\begin{restatable}[Identical restricted contexts]{lemma}{IdenticalRestrictedContexts}
  \label{lemma:Identical restricted contexts}
  If
  $\conwfJudg{\acontext'}$ and
  $\congoesJudg{\acontext}{\acontext'}$,
  then $\restrictcontext{\acontext''}{\acontext} = \restrictcontext{\acontext''}{\acontext'}$.
\end{restatable}

\subsection{Statement}

\begin{restatable}[Soundness of algorithmic typing]{theorem}{Soundness}
  \label{theorem:soundness}
  If $\conwfJudg{\acontext}$,
  $\envwfJudg{\acontext}{\typeenv}$,
  $\congoesJudg{\acontext'}{\ccontext}$, and
  $\conwfJudg{\ccontext}$, then:

  \begin{itemize}
    \item If
    $\algosynjudg{\acontext; \typeenv}{v}{P}{\acontext'}$,
    then
    $\declsynjudg{\makedec{\acontext}; \typeenv}{v}{[\ccontext]P}$.

    \item If
    $\algosynjudg{\acontext; \typeenv}{t}{N}{\acontext'}$,
    then
    $\declsynjudg{\makedec{\acontext}; \typeenv}{t}{[\ccontext]N}$.

    \item If
    $\algospinejudg{\acontext; \typeenv}{s}{\spine{N}{M}}{\acontext'}$,
    $\wfnegtypeJudg{\acontext}{N}$, and
    $[\acontext]N = N$,
    then
    $\exists M'$ such that $\MutualSubtypeNegJudge{\makedec{\acontext}}{[\ccontext]M}{M'}$ and $\declspinejudg{\makedec{\acontext}; \typeenv}{s}{\spine{[\ccontext]N}{M'}}$.
  \end{itemize}
\end{restatable}

\section{Completeness of typing}

\subsection{Lemmas}

\begin{restatable}[Weak context extension maintains variables]{lemma}{WeakContextExtensionSameVars} \
  \label{lemma:weak context extension maintains variables}
  If $\congoesWeakJudg{\Theta}{\Theta'}$ then $\FreeEV(\Theta) \subseteq \FreeEV(\Theta')$ and $\FreeUV(\Theta) = \FreeUV(\Theta')$.
\end{restatable}

\begin{restatable}[Reversing context extension from a complete context]{lemma}{ReversingExtensionFromComplete} \
  \label{lemma:Reversing context extension from a complete context}
  If $\congoesJudg{\ccontext}{\acontext}$ then $\congoesJudg{\acontext}{\ccontext}$.
\end{restatable}

\begin{restatable}[Pulling back restricted contexts]{lemma}{PullingBackRestrictedContexts} \
  \label{lemma:Pulling back restricted contexts}
  If $\congoesJudg{\acontext}{\acontext'}$ and
  $\congoesJudg{\restrictcontext{\acontext'}{\acontext''}}{\acontext'''}$, then
  $\congoesJudg{\restrictcontext{\acontext}{\acontext''}}{\acontext'''}$.
\end{restatable}

\subsection{Statement}

\begin{restatable}[Completeness of algorithmic typing]{theorem}{Completeness}
  \label{theorem:typing-completeness}
  If $\conwfJudg{\acontext}$, $\envwfJudg{\acontext}{\typeenv}$, $\congoesJudg{\acontext}{\ccontext}$, and $\conwfJudg{\ccontext}$, then:

  \begin{itemize}
    \item If
      $\declsynjudg{\makedec{\acontext}; \typeenv}{v}{P}$
      then $\exists \acontext'$ such that
      $\algosynjudg{\acontext; \typeenv}{v}{P}{\acontext'}$ and
      $\congoesJudg{\acontext'}{\ccontext}$.

    \item If
      $\declsynjudg{\makedec{\acontext}; \typeenv}{t}{N}$
      then $\exists \acontext'$ such that
      $\algosynjudg{\acontext; \typeenv}{t}{N}{\acontext'}$ and
      $\congoesJudg{\acontext'}{\ccontext}$.

    \item If
      $\declspinejudg{\makedec{\acontext}; \typeenv}{s}{\spine{\subcon{\ccontext} N}{M}}$,
      $\wfnegtypeJudg{\acontext}{N}$, and
      $\subcon{\acontext} N = N$,
      then $\exists \acontext', \ccontext'$ and $M'$ such that
      $\algospinejudg{\acontext; \typeenv}{s}{\spine{N}{M'}}{\acontext'}$,
      $\congoesWeakJudg{\ccontext}{\ccontext'}$,
      $\congoesJudg{\acontext'}{\ccontext'}$,
      $\MutualSubtypeNegJudge{\makedec{\acontext}}{\subcon{\ccontext'} M'}{M}$,
      $\subcon{\acontext'} M' = M'$, and
      $\conwfJudg{\ccontext'}$.
  \end{itemize}
\end{restatable}

\part*{Proofs}
\addcontentsline{toc}{part}{Proofs}

\setcounter{section}{0}
\renewcommand\thesection{\Alph{section}'}
\renewcommand\theHsection{proofs.\thesection}

\newcommand{\proofcomment}[1]{\proofsep \noalign{#1} \proofsep}
\newcommand{\caseitem}[1]{\item \textbf{Case} #1:}

\newcommand{\Label}[1]{\text{({#1})~~~~}}

\newcommand{\NoSolvedVarsJudg}[2]{\subcon{#1} #2 = #2}

\newcommand{\congoesPf}[3]{\Pf{#1}{\congoes}{#2} {#3}}
\newcommand{\conwfPf}[2]{\Pf{#1}{}{\conwf} {#2}}
\newcommand{\groundPf}[2]{\Pf{#1}{}{\text{ground}} {#2}}
\newcommand{\arbsubtypePf}[4]{\Pf{#1}{\begingroup \color{dDigPurple} #2 \endgroup}{#3} {#4}}
\newcommand{\negsubtypePf}[3]{\arbsubtypePf{#1}{\negsubtype}{#2} {#3}}
\newcommand{\possubtypePf}[3]{\arbsubtypePf{#1}{\possubtype}{#2} {#3}}
\newcommand{\continuenegsubtypePf}[2]{\negsubtypePf{}{#1}{#2}}
\newcommand{\continuepossubtypePf}[2]{\possubtypePf{}{#1}{#2}}

\newcommand{\wfposnegtypePf}[3]{\Pf{#1 \entails #2}{}{\posnegtype} {#3}}
\newcommand{\wfnegtypePf}[3]{\Pf{#1 \entails #2}{}{\negtype} {#3}}
\newcommand{\wfpostypePf}[3]{\Pf{#1 \entails #2}{}{\postype} {#3}}
\newcommand{\NoSolvedVarsPf}[3]{\eqPf{\subcon{#1} #2}{#2}{#3}}

\newcommand{\DJudgePosPf}[4]{\possubtypePf{#1 \entails #2}{#3}{#4}}
\newcommand{\AJudgePosPf}[5]{\possubtypePf{#1 \entails #2}{#3 \prodcon #4}{#5}}
\newcommand{\MutualJudgePosPf}[4]{\Pf{#1 \entails #2}{\mutualSubtypePos}{#3}{#4}}
\newcommand{\NotMutualJudgePosPf}[4]{\Pf{#1 \entails #2}{\not\mutualSubtypePos}{#3}{#4}}
\newcommand{\CCustomJudgePosPf}[5]{\DJudgePosPf{\makedec{#2}}
                              {\subcon{#1} #3}
                              {\subcon{#1} #4}
                              {#5}}
\newcommand{\CnewCustomJudgePosPf}[5]{\DJudgePosPf{\makedec{#2}}
                                      {#3}
                                      {\subcon{#1} #4}
                                      {#5}}
\newcommand{\CJudgePosPf}[4]{\CCustomJudgePosPf{\ccontext}{#1}{#2}{#3}{#4}}
\newcommand{\CnewJudgePosPf}[4]{\CnewCustomJudgePosPf{\ccontext}{#1}{#2}{#3}{#4}}

\newcommand{\DJudgeNegPf}[4]{\negsubtypePf{#1 \entails #2}{#3}{#4}}
\newcommand{\AJudgeNegPf}[5]{\negsubtypePf{#1 \entails #2}{#3 \prodcon #4}{#5}}
\newcommand{\MutualJudgeNegPf}[4]{\Pf{#1 \entails #2}{\mutualSubtypeNeg}{#3}{#4}}
\newcommand{\CCustomJudgeNegPf}[5]{\DJudgeNegPf{\makedec{#2}}
                              {\subcon{#1} #3}
                              {\subcon{#1} #4}
                              {#5}}
\newcommand{\CJudgeNegPf}[4]{\CCustomJudgeNegPf{\ccontext}{#1}{#2}{#3}{#4}}
\newcommand{\CnewCustomJudgeNegPf}[5]{\DJudgeNegPf{\makedec{#2}}
                              {\subcon{#1} #3}
                              {#4}
                              {#5}}
\newcommand{\CnewJudgeNegPf}[4]{\CnewCustomJudgeNegPf{\ccontext}{#1}{#2}{#3}{#4}}

\newcommand{\subconunderscores}{\ensuremath{\subcon{-} -}}

\newcommand{\baseacontext}{\acontext_b}
\newcommand{\bydefsubcon}{By definition of \subconunderscores\xspace}
\newcommand{\defsubcon}{definition of \subconunderscores\xspace}

\newcommand{\termsizeEqPf}[3]{\eqPf{\termsize{#1}}{\termsize{#2}}{#3}}
\newcommand{\MutualJudgePosNegPf}[4]{\Pf{#1 \entails #2}{\mutualSubtypePosNeg}{#3}{#4}}
\newcommand{\bydeftermsize}{By definition of $\termsize{-}$\xspace}
\newcommand{\bydefground}{By definition of ground\xspace}
\newcommand{\bydeffev}{By definition of $\FreeEV$\xspace}
\newcommand{\bydefmakedec}{By definition of $\makedec{-}$}

\newcommand{\termsizeLeqPf}[3]{\Pf{\termsize{#1}}{\leq}{\termsize{#2}}{#3}}
\newcommand{\continueLeqPf}[2]{\Pf{}{\leq}{#1}{#2}}
\newcommand{\continueTermsizeLeqPf}[2]{\continueLeqPf{\termsize{#1}}{#2}}
\newcommand{\byih}{By i.h.\xspace}

\newcommand{\ltPf}[3]{\Pf{#1}{<}{#2}{#3}}
\newcommand{\continueLtPf}[2]{\ltPf{}{#1}{#2}}
\newcommand{\gtPf}[3]{\Pf{#1}{>}{#2}{#3}}
\newcommand{\continueGtPf}[2]{\gtPf{}{#1}{#2}}

\newcommand{\supersetPf}[3]{\Pf{#1}{\supseteq}{#2}{#3}}

\newcommand{\hugePf}[3]{
    &\multicolumn{2}{c}{$#1$} & \multirow{3}{*}{#3} \\
    &\multicolumn{2}{c}{$\sqcap$}\\
    &\multicolumn{2}{c}{$#2$}\\
}

\newcommand{\numPrenexGtPf}[3]{\gtPf{\numprenex{#1}}{\numprenex{#2}}{#3}}

\section{Weakening}

\RestatePushingUvarsRightPreservesTwf*

\begin{proof}
  By rule induction on $\wfposnegtypeJudg{\pushstartcontext}{A}$.

  \begin{itemize}
    \DerivationProofCase{\twfuvar}
      {\beta \in \ConUV(\pushstartcontext)}
      {\wfpostypeJudg{\pushstartcontext} {\beta}}
  
      \begin{llproof}
        \inPf{\beta}
          {\ConUV(\pushstartcontext)}
          {Subderivation}
        \inPf{\beta}
          {\ConUV(\pushresultcontext)}
          {Since $\ConUV$ ignores order}
  \Hand \wfpostypePf{\pushresultcontext}
          {\beta}
          {By \twfuvar}
      \end{llproof}
  
    \DerivationProofCase{\twfguess}
        {\guess{\alpha} \in \EV(\pushstartcontext)}
        {\wfpostypeJudg{\pushstartcontext} {\guess{\alpha}}}
  
      \begin{llproof}
        \inPf{\guess{\alpha}}
          {\EV(\pushstartcontext)}
          {Subderivation}
        \inPf{\guess{\alpha}}
          {\EV(\pushresultcontext)}
          {Since $\EV$ ignores order}
  \Hand \wfpostypePf{\pushresultcontext}
          {\guess{\alpha}}
          {By \twfguess}
      \end{llproof}
  
    \DerivationProofCase{\twfshiftdown}
        {\wfnegtypeJudg{\pushstartcontext} {N}}
        {\wfpostypeJudg{\pushstartcontext} {\shiftdown N}}
  
      \begin{llproof}
        \wfnegtypePf{\pushstartcontext}
          {N}
          {Subderivation}
        \wfnegtypePf{\pushresultcontext}
          {N}
          {By i.h.}
  \Hand \wfpostypePf{\pushresultcontext}
          {\shiftdown N}
          {By \twfshiftdown}
      \end{llproof}
  
    \DerivationProofCase{\twfforall}
        {\wfnegtypeJudg{\pushstartcontextlong, \beta} {N}}
        {\wfnegtypeJudg{\pushstartcontextlong} {\forall \beta \ldotp N}}
  
      \begin{llproof}
        \wfnegtypePf{\pushstartcontextlong, \beta}
          {N}
          {Subderivation}
        \wfnegtypePf{\pushresultcontextlong, \beta}
          {N}
          {By i.h.}
  \Hand \wfnegtypePf{\pushresultcontextlong}
          {\forall \beta \ldotp N}
          {By \twfforall}
      \end{llproof}
  
    \DerivationProofCase{\twfarrow}
        {\wfpostypeJudg{\pushstartcontext} {P} \\ \wfnegtypeJudg{\pushstartcontext} {N}}
        {\wfnegtypeJudg{\pushstartcontext} {P \funarrow N}}
      
      \begin{llproof}
        \wfpostypePf{\pushstartcontext}
          {P}
          {Subderivation}
        \wfpostypePf{\pushresultcontext}
          {P}
          {By i.h.}
        \wfnegtypePf{\pushstartcontext}
          {N}
          {Subderivation}
        \wfnegtypePf{\pushresultcontext}
          {N}
          {By i.h.}
  \Hand \wfnegtypePf{\pushresultcontext}
          {P \funarrow N}
          {By \twfarrow}
      \end{llproof}
  
    \DerivationProofCase{\twfshiftup}
      {\wfpostypeJudg{\pushstartcontext} {P}}
      {\wfnegtypeJudg{\pushstartcontext} {\shiftup P}}
  
      \begin{llproof}
        \wfpostypePf{\pushstartcontext}
          {P}
          {Subderivation}
        \wfpostypePf{\pushresultcontext}
          {P}
          {By i.h.}
  \Hand \wfnegtypePf{\pushresultcontext}
          {\shiftup P}
          {By \twfshiftup}
      \end{llproof}
  \end{itemize}
\end{proof}

\RestateTwfWeakening*

\begin{proof}
  By rule induction on $\wfposnegtypeJudg{\acontext}{A}$.
  \begin{itemize}
    \DerivationProofCase{\twfuvar}
      {\alpha \in \ConUV(\acontext)}
      {\wfpostypeJudg{\acontext} {\alpha}}
  
      \begin{llproof}
        \inPf{\alpha}
          {\ConUV(\acontext)}
          {Subderivation}
        \inPf{\alpha}
          {\ConUV(\acontext, \acontext')}
          {Since $\ConUV(\acontext) \subseteq \ConUV(\acontext, \acontext')$}
  \Hand \wfpostypePf{\acontext, \acontext'}
          {\alpha}
          {By \twfuvar}
      \end{llproof}
  
    \DerivationProofCase{\twfguess}
        {\guess{\alpha} \in \EV(\acontext)}
        {\wfpostypeJudg{\acontext} {\guess{\alpha}}}
  
      \begin{llproof}
        \inPf{\guess{\alpha}}
          {\EV(\acontext)}
          {Subderivation}
        \inPf{\guess{\alpha}}
          {\EV(\acontext, \acontext')}
          {Since $\EV(\acontext) \subseteq \EV(\acontext, \acontext')$}
  \Hand \wfpostypePf{\acontext, \acontext'}
          {\guess{\alpha}}
          {By \twfguess}
      \end{llproof}
  
    \DerivationProofCase{\twfshiftdown}
        {\wfnegtypeJudg{\acontext} {N}}
        {\wfpostypeJudg{\acontext} {\shiftdown N}}
  
      \begin{llproof}
        \wfnegtypePf{\acontext}
          {N}
          {Subderivation}
        \wfnegtypePf{\acontext, \acontext'}
          {N}
          {By i.h.}
  \Hand \wfpostypePf{\acontext, \acontext'}
          {\shiftdown N}
          {By \twfshiftdown}
      \end{llproof}
  
    \DerivationProofCase{\twfforall}
        {\wfnegtypeJudg{\acontext, \alpha} {N}}
        {\wfnegtypeJudg{\acontext} {\forall \alpha \ldotp N}}
  
      \begin{llproof}
        \wfnegtypePf{\acontext, \alpha}
          {N}
          {Subderivation}
        \wfnegtypePf{\acontext, \alpha, \acontext'}
          {N}
          {By i.h.}
        \wfnegtypePf{\acontext, \acontext', \alpha}
          {N}
          {By \lemmaref{lemma:Pushing uvars right preserves term well-formedness}}
  \Hand \wfnegtypePf{\acontext, \acontext'}
          {\forall \alpha \ldotp N}
          {By \twfforall}
      \end{llproof}
  
    \DerivationProofCase{\twfarrow}
        {\wfpostypeJudg{\acontext} {P} \\ \wfnegtypeJudg{\acontext} {N}}
        {\wfnegtypeJudg{\acontext} {P \funarrow N}}
      
      \begin{llproof}
        \wfpostypePf{\acontext}
          {P}
          {Subderivation}
        \wfpostypePf{\acontext, \acontext'}
          {P}
          {By i.h.}
        \wfnegtypePf{\acontext}
          {N}
          {Subderivation}
        \wfnegtypePf{\acontext, \acontext'}
          {N}
          {By i.h.}
  \Hand \wfnegtypePf{\acontext, \acontext'}
          {P \funarrow N}
          {By \twfarrow}
      \end{llproof}
  
    \DerivationProofCase{\twfshiftup}
      {\wfpostypeJudg{\acontext} {P}}
      {\wfnegtypeJudg{\acontext} {\shiftup P}}
  
      \begin{llproof}
        \wfpostypePf{\acontext}
          {P}
          {Subderivation}
        \wfpostypePf{\acontext, \acontext'}
          {P}
          {By i.h.}
  \Hand \wfnegtypePf{\acontext, \acontext'}
          {\shiftup P}
          {By \twfshiftup}
      \end{llproof}
  \end{itemize}
\end{proof}

\RestatePushingUvarsRightInDJudge*

\begin{proof}
  By rule induction on $\DPnSubtypeJudge{\pushstartcontext}{A}{B}$.
  
  \begin{itemize}
    \DerivationProofCase{\drefl}
        {\wfpostypeJudg{\pushstartcontext}{\beta}}
        {\DPosSubtypeJudge{\pushstartcontext} {\beta} {\beta}}
  
      \begin{llproof}
        \wfpostypePf{\pushstartcontext}
          {\beta}
          {Subderivation}
        \wfpostypePf{\pushresultcontext}
          {\beta}
          {By \lemmaref{lemma:Pushing uvars right preserves term well-formedness}}
  \Hand \DJudgePosPf{\pushresultcontext}
          {\beta}
          {\beta}
          {By \drefl}
      \end{llproof}
  
    \DerivationProofCase{\dshiftdown}
        {\DNegSubtypeJudge{\pushstartcontext} {M} {N} \\ \DNegSubtypeJudge{\pushstartcontext} {N} {M}}
        {\DPosSubtypeJudge{\pushstartcontext} {\shiftdown N} {\shiftdown M}}
  
      \begin{llproof}
        \DJudgeNegPf{\pushstartcontext}
          {M}
          {N}
          {Subderivation}
        \DJudgeNegPf{\pushresultcontext}
          {M}
          {N}
          {By i.h.}
        \DJudgeNegPf{\pushstartcontext}
          {N}
          {M}
          {Subderivation}
        \DJudgeNegPf{\pushresultcontext}
          {N}
          {M}
          {By i.h.}
  \Hand \DJudgePosPf{\pushresultcontext}
          {\shiftdown N}
          {\shiftdown M}
          {By \dshiftdown}
      \end{llproof}
  
    \DerivationProofCase{\dforallr}
        {\DNegSubtypeJudge{\pushstartcontextlong, \beta} {N} {M}}
        {\DNegSubtypeJudge{\pushstartcontextlong} {N} {\forall \beta \ldotp M}}
  
      \begin{llproof}
        \DJudgeNegPf{\pushstartcontextlong, \beta}
          {N}
          {M}
          {Subderivation}
        \DJudgeNegPf{\pushresultcontextlong, \beta}
          {N}
          {M}
          {By i.h.}
  \Hand \DJudgeNegPf{\pushresultcontextlong}
          {N}
          {\forall \beta \ldotp M}
          {By \dforallr}
      \end{llproof}
  
    \DerivationProofCase{\dforalll}
        {\wfpostypeJudg{\pushstartcontext} {P} \\ \DNegSubtypeJudge{\pushstartcontext} {\subterm{P / \beta} N} {M}}
        {\DNegSubtypeJudge{\pushstartcontext} {\forall \beta \ldotp N} {M}}
  
      \begin{llproof}
        \wfpostypePf{\pushstartcontext}
          {P}
          {Subderivation}
        \wfpostypePf{\pushresultcontext}
          {P}
          {By \lemmaref{lemma:Pushing uvars right preserves term well-formedness}}
        \DJudgeNegPf{\pushstartcontext}
          {\subterm{P / \beta} N}
          {M}
          {Subderivation}
        \DJudgeNegPf{\pushresultcontext}
          {\subterm{P / \beta} N}
          {M}
          {By i.h.}
  \Hand \DJudgeNegPf{\pushresultcontext}
          {\forall \beta \ldotp N}
          {M}
          {By \dforalll}
      \end{llproof}
    
    \DerivationProofCase{\darrow}
        {\DPosSubtypeJudge{\dcontext} {Q} {P} \\ \DNegSubtypeJudge{\dcontext} {N} {M}}
        {\DNegSubtypeJudge{\dcontext} {P \funarrow N} {Q \funarrow M}}
  
      \begin{llproof}
        \DJudgePosPf{\pushstartcontext}
          {Q}
          {P}
          {Subderivation}
        \DJudgePosPf{\pushresultcontext}
          {Q}
          {P}
          {By i.h.}
        \DJudgeNegPf{\pushstartcontext}
          {N}
          {M}
          {Subderivation}
        \DJudgeNegPf{\pushresultcontext}
          {N}
          {M}
          {By i.h.}
  \Hand \DJudgeNegPf{\pushresultcontext}
          {P \funarrow N}
          {Q \funarrow M}
          {By \darrow}
      \end{llproof}
  
    \DerivationProofCase{\dshiftup}
        {\DPosSubtypeJudge{\dcontext} {Q} {P} \\ \DPosSubtypeJudge{\dcontext} {P} {Q}}
        {\DNegSubtypeJudge{\dcontext} {\shiftup P} {\shiftup Q}}
  
      \begin{llproof}
        \DJudgePosPf{\pushstartcontext}
          {Q}
          {P}
          {Subderivation}
        \DJudgePosPf{\pushresultcontext}
          {Q}
          {P}
          {By i.h.}
        \DJudgePosPf{\pushstartcontext}
          {P}
          {Q}
          {Subderivation}
        \DJudgePosPf{\pushresultcontext}
          {P}
          {Q}
          {By i.h.}
  \Hand \DJudgeNegPf{\pushresultcontext}
          {\shiftup P}
          {\shiftup Q}
          {By \dshiftup}
      \end{llproof}
  \end{itemize}
\end{proof}

\RestateDJudgeWeakening*

\begin{proof}
  By rule induction on $\DPnSubtypeJudge{\dcontext} {A} {B}$.

  \begin{itemize}
    \DerivationProofCase{\drefl}
        {\wfpostypeJudg{\dcontext}{\alpha}}
        {\DPosSubtypeJudge{\dcontext} {\alpha} {\alpha}}
  
      \begin{llproof}
        \wfpostypePf{\dcontext}
          {\alpha}
          {Subderivation}
        \wfpostypePf{\dcontext, \dcontext'}
          {\alpha}
          {By \lemmaref{lemma:Term well-formedness weakening}}
  \Hand \DJudgePosPf{\dcontext, \dcontext'}
          {\alpha}
          {\alpha}
          {By \drefl}
      \end{llproof}
  
    \DerivationProofCase{\dshiftdown}
        {\DNegSubtypeJudge{\dcontext} {M} {N} \\ \DNegSubtypeJudge{\dcontext} {N} {M}}
        {\DPosSubtypeJudge{\dcontext} {\shiftdown N} {\shiftdown M}}
  
      \begin{llproof}
        \DJudgeNegPf{\dcontext}
          {M}
          {N}
          {Subderivation}
        \DJudgeNegPf{\dcontext, \dcontext'}
          {M}
          {N}
          {By i.h.}
        \DJudgeNegPf{\dcontext}
          {N}
          {M}
          {Subderivation}
        \DJudgeNegPf{\dcontext, \dcontext'}
          {N}
          {M}
          {By i.h.}
  \Hand \DJudgePosPf{\dcontext, \dcontext'}
          {\shiftdown N}
          {\shiftdown M}
          {By \dshiftdown}
      \end{llproof}
  
    \DerivationProofCase{\dforallr}
        {\DNegSubtypeJudge{\dcontext, \alpha} {N} {M}}
        {\DNegSubtypeJudge{\dcontext} {N} {\forall \alpha \ldotp M}}
  
      \begin{llproof}
        \DJudgeNegPf{\dcontext, \alpha}
          {N}
          {M}
          {Subderivation}
        \DJudgeNegPf{\dcontext, \alpha, \dcontext'}
          {N}
          {M}
          {By i.h.}
        \DJudgeNegPf{\dcontext, \dcontext', \alpha}
          {N}
          {M}
          {By \lemmaref{lemma:Pushing uvars right in declarative judgment}}
  \Hand \DJudgeNegPf{\dcontext, \dcontext'}
          {N}
          {\forall \alpha \ldotp M}
          {By i.h.}
      \end{llproof}
  
    \DerivationProofCase{\dforalll}
        {\wfpostypeJudg{\dcontext} {P} \\ \DNegSubtypeJudge{\dcontext} {\subterm{P / \alpha} N} {M}}
        {\DNegSubtypeJudge{\dcontext} {\forall \alpha \ldotp N} {M}}
  
      \begin{llproof}
        \wfpostypePf{\dcontext}
          {P}
          {Subderivation}
        \wfpostypePf{\dcontext, \dcontext'}
          {P}
          {By \lemmaref{lemma:Term well-formedness weakening}}
        \DJudgeNegPf{\dcontext}
          {\subterm{P / \alpha} N}
          {M}
          {Subderivation}
        \DJudgeNegPf{\dcontext, \dcontext'}
          {\subterm{P / \alpha} N}
          {M}
          {By i.h.}
  \Hand \DJudgeNegPf{\dcontext, \dcontext'}
          {\forall \alpha \ldotp N}
          {M}
          {By \dforalll}
      \end{llproof}
    
    \DerivationProofCase{\darrow}
        {\DPosSubtypeJudge{\dcontext} {Q} {P} \\ \DNegSubtypeJudge{\dcontext} {N} {M}}
        {\DNegSubtypeJudge{\dcontext} {P \funarrow N} {Q \funarrow M}}
  
      \begin{llproof}
        \DJudgePosPf{\dcontext}
          {Q}
          {P}
          {Subderivation}
        \DJudgePosPf{\dcontext, \dcontext'}
          {Q}
          {P}
          {By i.h.}
        \DJudgeNegPf{\dcontext}
          {N}
          {M}
          {Subderivation}
        \DJudgeNegPf{\dcontext, \dcontext'}
          {N}
          {M}
          {By i.h.}
  \Hand \DJudgeNegPf{\dcontext, \dcontext'}
          {P \funarrow N}
          {Q \funarrow M}
          {By \darrow}
      \end{llproof}
  
    \DerivationProofCase{\dshiftup}
        {\DPosSubtypeJudge{\dcontext} {Q} {P} \\ \DPosSubtypeJudge{\dcontext} {P} {Q}}
        {\DNegSubtypeJudge{\dcontext} {\shiftup P} {\shiftup Q}}
  
      \begin{llproof}
        \DJudgePosPf{\dcontext}
          {Q}
          {P}
          {Subderivation}
        \DJudgePosPf{\dcontext, \dcontext'}
          {Q}
          {P}
          {By i.h.}
        \DJudgePosPf{\dcontext}
          {P}
          {Q}
          {Subderivation}
        \DJudgePosPf{\dcontext, \dcontext'}
          {P}
          {Q}
          {By i.h.}
  \Hand \DJudgeNegPf{\dcontext, \dcontext'}
          {\shiftup P}
          {\shiftup Q}
          {By \dshiftup}
      \end{llproof}
  \end{itemize}
\end{proof}

\section{Declarative subtyping}

\RestateReflexivityPnSubtype*

\begin{proof}
  By rule induction on $\wfposnegtypeJudg{\dcontext} {A}$.

  \begin{itemize}
    \DerivationProofCase{\twfuvar}
      {\alpha \in \ConUV(\dcontext)}
      {\wfpostypeJudg{\dcontext} {\alpha}}
  
      \begin{llproof}
        \wfpostypePf{\dcontext}
          {\alpha}
          {Assumption}
  \Hand \DJudgePosPf{\dcontext}
          {\alpha}
          {\alpha}
          {By \drefl}
      \end{llproof}
  
    \DerivationProofCase{\twfshiftdown}
      {\wfnegtypeJudg{\dcontext} {N}}
      {\wfpostypeJudg{\dcontext} {\shiftdown N}}
  
      \begin{llproof}
        \wfnegtypePf{\dcontext}
          {N}
          {Subderivation}
        \DJudgeNegPf{\dcontext}
          {N}
          {N}
          {By i.h.}
  \Hand \DJudgePosPf{\dcontext}
          {\shiftdown N}
          {\shiftdown N}
          {By \dshiftdown}
      \end{llproof}
  
    \DerivationProofCase{\twfforall}
      {\wfnegtypeJudg{\dcontext, \alpha} {N}}
      {\wfnegtypeJudg{\dcontext} {\forall \alpha \ldotp N}}
  
      \begin{llproof}
        \wfnegtypePf{\dcontext, \alpha}
          {N}
          {Subderivation}
        \DJudgeNegPf{\dcontext, \alpha}
          {N}
          {N}
          {By i.h.}
        \inPf{\alpha}
          {\ConUV(\dcontext, \alpha)}
          {By definition of $\ConUV$}
        \wfpostypePf{\dcontext, \alpha}
          {\alpha}
          {By \twfuvar}
        \DJudgeNegPf{\dcontext, \alpha}
          {\forall \alpha \ldotp N}
          {N}
          {By \dforalll}
  \Hand \DJudgeNegPf{\dcontext}
          {\forall \alpha \ldotp N}
          {\forall \alpha \ldotp N}
          {By \dforallr}
      \end{llproof}
  
    \DerivationProofCase{\twfarrow}
      {\wfpostypeJudg{\dcontext} {P} \\ \wfnegtypeJudg{\dcontext} {N}}
      {\wfnegtypeJudg{\dcontext} {P \funarrow N}}
      
      \begin{llproof}
        \wfpostypePf{\dcontext}
          {P}
          {Subderivation}
        \DJudgePosPf{\dcontext}
          {P}
          {P}
          {By i.h.}
        \wfnegtypePf{\dcontext}
          {N}
          {Subderivation}
        \DJudgeNegPf{\dcontext}
          {N}
          {N}
          {By i.h.}
  \Hand \DJudgeNegPf{\dcontext}
          {P \funarrow N}
          {P \funarrow N}
          {By \darrow}
      \end{llproof}
  
    \DerivationProofCase{\twfshiftup}
      {\wfpostypeJudg{\dcontext} {P}}
      {\wfnegtypeJudg{\dcontext} {\shiftup P}}
  
      \begin{llproof}
        \wfpostypePf{\dcontext}
          {P}
          {Subderivation}
        \DJudgePosPf{\dcontext}
          {P}
          {P}
          {By i.h.}
  \Hand \DJudgeNegPf{\dcontext}
          {\shiftup P}
          {\shiftup P}
          {By \dshiftup}
      \end{llproof}
  \end{itemize}
\end{proof}

\RestateSubstitutionWf*

\begin{proof}
  By rule induction on $\wfposnegtypeJudg{\dconwithhole{\alpha}}{A}$.

  \begin{itemize}
      \DerivationProofCase{\twfuvar}
        {\beta \in \ConUV(\dconwithhole{\alpha})}
        {\wfpostypeJudg{\dconwithhole{\alpha}} {\beta}}
      
      \begin{llproof}
        \proofcomment{Case $\beta = \alpha$:}
        \eqPf{\subterm{P / \alpha} \beta}
          {P}
          {\bydefsubcon}
  \Hand \wfpostypePf{\dcontext_L, \dcontext_R}
          {P}
          {Assumption}
  
        \proofcomment{Case $\beta \neq \alpha$:}
        \eqPf{\subterm{P / \alpha} \beta}
          {\beta}
          {\bydefsubcon}
        \inPf{\beta}
          {\ConUV(\dcontext_L, \alpha, \dcontext_R)}
          {Subderivation}
        \inPf{\beta}
          {\ConUV(\dcontext_L, \dcontext_R)}
          {Since $\beta \neq \alpha$}
  \Hand \wfpostypePf{\dcontext_L, \dcontext_R}
          {\beta}
          {By \twfuvar}
      \end{llproof}
  
      \DerivationProofCase{\twfshiftdown}
        {\wfnegtypeJudg{\dconwithhole{\alpha}} {N}}
        {\wfpostypeJudg{\dconwithhole{\alpha}} {\shiftdown N}}
  
      \begin{llproof}
        \wfpostypePf{\dcontext_L, \dcontext_R}
          {P}
          {Assumption}
        \wfnegtypePf{\dconwithhole{\alpha}}
          {N}
          {Subderivation}
        \wfnegtypePf{\dcontext_L, \dcontext_R}
          {\subterm{P / \alpha} N}
          {By i.h.}
        \wfpostypePf{\dcontext_L, \dcontext_R}
          {\shiftdown \subterm{P / \alpha} N}
          {By \twfshiftdown}
  \Hand \wfpostypePf{\dcontext_L, \dcontext_R}
          {\subterm{P / \alpha} \shiftdown N}
          {\bydefsubcon}
      \end{llproof}
  
      \DerivationProofCase{\twfforall}
        {\wfnegtypeJudg{\dconwithhole{\alpha}, \beta} {N}}
        {\wfnegtypeJudg{\dconwithhole{\alpha}} {\forall \beta \ldotp N}}
      
      \begin{llproof}
        \wfpostypePf{\dcontext_L, \dcontext_R}
          {P}
          {Assumption}
        \wfnegtypePf{\dconwithhole{\alpha}, \beta}
          {N}
          {Subderivation}
        \wfpostypePf{\dcontext_L, \dcontext_R, \beta}
          {P}
          {By \lemmaref{lemma:Term well-formedness weakening}}
        \wfnegtypePf{\dcontext_L, \dcontext_R, \beta}
          {\subterm{P / \alpha} N}
          {By i.h.}
        \wfnegtypePf{\dcontext_L, \dcontext_R}
          {\forall \beta \ldotp \subterm{P / \alpha} N}
          {By \twfforall}
  \Hand \wfnegtypePf{\dcontext_L, \dcontext_R}
          {\subterm{P / \alpha} \forall \beta \ldotp N}
          {\bydefsubcon}
      \end{llproof}
    
      \DerivationProofCase{\twfarrow}
        {\wfpostypeJudg{\dconwithhole{\alpha}} {Q} \\ \wfnegtypeJudg{\dconwithhole{\alpha}} {N}}
        {\wfnegtypeJudg{\dconwithhole{\alpha}} {Q \funarrow N}}
  
        \begin{llproof}
          \wfpostypePf{\dcontext_L, \dcontext_R}
            {P}
            {Assumption}
          \wfpostypePf{\dconwithhole{\alpha}}
            {Q}
            {Subderivation}
          \wfpostypePf{\dcontext_L, \dcontext_R}
            {\subterm{P / \alpha} Q}
            {By i.h.}
          \wfnegtypePf{\dconwithhole{\alpha}}
            {N}
            {Subderivation}
          \wfnegtypePf{\dcontext_L, \dcontext_R}
            {\subterm{P / \alpha} N}
            {By i.h.}
          \wfnegtypePf{\dcontext_L, \dcontext_R}
            {\subterm{P / \alpha} Q \funarrow \subterm{P / \alpha} N}
            {By \twfarrow}
    \Hand \wfnegtypePf{\dcontext_L, \dcontext_R}
            {\subterm{P / \alpha} (Q \funarrow N)}
            {\bydefsubcon}
        \end{llproof}
  
      \DerivationProofCase{\twfshiftup}
        {\wfpostypeJudg{\dconwithhole{\alpha}} {Q}}
        {\wfnegtypeJudg{\dconwithhole{\alpha}} {\shiftup Q}}
  
        \begin{llproof}
          \wfpostypePf{\dcontext_L, \dcontext_R}
            {P}
            {Assumption}
          \wfpostypePf{\dconwithhole{\alpha}}
            {Q}
            {Subderivation}
          \wfpostypePf{\dcontext_L, \dcontext_R}
            {\subterm{P / \alpha} Q}
            {By i.h.}
          \wfnegtypePf{\dcontext_L, \dcontext_R}
            {\shiftup \subterm{P / \alpha} Q}
            {By \twfshiftup}
    \Hand \wfnegtypePf{\dcontext_L, \dcontext_R}
            {\subterm{P / \alpha} \shiftup Q}
            {\bydefsubcon}
        \end{llproof}
  \end{itemize}
\end{proof}

\RestateDeclarativeSubtypingSubstitutionLemma*

\begin{proof}
  By mutual rule induction on $\DPosSubtypeJudge{\dconwithhole{\alpha}}{Q}{R}$ and $\DNegSubtypeJudge{\dconwithhole{\alpha}}{N}{M}$.

  \begin{itemize}
      \DerivationProofCase{\drefl}
        {\wfpostypeJudg{\dconwithhole{\alpha}}{\beta}}
        {\DPosSubtypeJudge{\dconwithhole{\alpha}} {\beta} {\beta}}
  
      \begin{llproof}
        \proofcomment{Case $\beta \neq \alpha$:}
        \eqPf{\subterm{P / \alpha} \beta}
          {\beta}
          {\bydefsubcon}
        \wfpostypePf{\dconwithhole{\alpha}}
          {\beta}
          {Subderivation}
        \inPf{\beta}
          {\ConUV(\dcontext_L, \alpha, \dcontext_R)}
          {Inversion (\twfuvar)}
        \inPf{\beta}
          {\ConUV(\dcontext_L, \dcontext_R)}
          {Since $\beta \neq \alpha$}
        \wfpostypePf{\dcontext_L, \dcontext_R}
          {\beta}
          {By \twfuvar}
  \Hand \DJudgePosPf{\dcontext_L, \dcontext_R}
          {\beta}
          {\beta}
          {By \drefl}
  
        \proofcomment{Case $\beta = \alpha$:}
        \eqPf{\subterm{P / \alpha} \beta}
          {P}
          {\bydefsubcon}
        \wfpostypePf{\dcontext_L, \dcontext_R}
          {P}
          {Assumption}
  \Hand \DJudgePosPf{\dcontext_L, \dcontext_R}
          {P}
          {P}
          {By \lemmaref{lemma:Reflexivity of declarative pnsubtype}}
      \end{llproof}
  
      \DerivationProofCase{\dshiftdown}
        {\dconwithhole{\alpha} \entails M \negsubtype N \\ \dconwithhole{\alpha} \entails N \negsubtype M}
        {\dconwithhole{\alpha} \entails \shiftdown N \possubtype \shiftdown M}
  
      \begin{llproof}
        \wfpostypePf{\dcontext_L, \dcontext_R}
          {P}
          {Assumption}
        \wfpostypePf{\dconwithhole{\alpha}}
          {\shiftdown N}
          {\ditto}
        \wfnegtypePf{\dconwithhole{\alpha}}
          {N}
          {Inversion (\twfshiftdown)}
        \wfpostypePf{\dconwithhole{\alpha}}
          {\shiftdown M}
          {Assumption}
        \wfnegtypePf{\dconwithhole{\alpha}}
          {M}
          {Inversion (\twfshiftdown)}
        
        \proofsep
        \DJudgeNegPf{\dconwithhole{\alpha}}
          {M}
          {N}
          {Subderivation}
        \DJudgeNegPf{\dcontext_L, \dcontext_R}
          {\subterm{P / \alpha} M}
          {\subterm{P / \alpha} N}
          {By i.h.}
        \DJudgeNegPf{\dconwithhole{\alpha}}
          {M}
          {N}
          {Subderivation}
        \DJudgeNegPf{\dcontext_L, \dcontext_R}
          {\subterm{P / \alpha} N}
          {\subterm{P / \alpha} M}
          {\ditto}
  \Hand \DJudgePosPf{\dcontext_L, \dcontext_R}
          {\shiftdown \subterm{P / \alpha} N}
          {\shiftdown \subterm{P / \alpha} M}
          {By \dshiftdown}
  \Hand \DJudgePosPf{\dcontext_L, \dcontext_R}
          {\subterm{P / \alpha} \shiftdown N}
          {\subterm{P / \alpha} \shiftdown M}
          {\bydefsubcon}
      \end{llproof}
  
      \DerivationProofCase{\dforallr}
        {\dconwithhole{\alpha}, \beta \entails N \negsubtype M}
        {\dconwithhole{\alpha} \entails N \negsubtype \forall \beta \ldotp M}
  
      \begin{llproof}
        \wfpostypePf{\dcontext_L, \dcontext_R}
          {P}
          {Assumption}
        \wfnegtypePf{\dconwithhole{\alpha}}
          {N}
          {\ditto}
        \wfnegtypePf{\dconwithhole{\alpha}}
          {\forall \beta \ldotp M}
          {\ditto}
  
        \proofsep
        \wfpostypePf{\dcontext_L, \dcontext_R, \beta}
          {P}
          {By \lemmaref{lemma:Term well-formedness weakening}}
        \wfnegtypePf{\dconwithhole{\alpha}, \beta}
          {N}
          {By \lemmaref{lemma:Term well-formedness weakening}}
        \wfnegtypePf{\dconwithhole{\alpha}, \beta}
          {M}
          {Inversion (\twfforall)}
        \DJudgeNegPf{\dconwithhole{\alpha}, \beta}
          {N}
          {M}
          {Subderivation}
        \DJudgeNegPf{\dcontext_L, \dcontext_R, \beta}
          {\subterm{P / \alpha} N}
          {\subterm{P / \alpha} M}
          {By i.h.}
  \Hand \DJudgeNegPf{\dcontext_L, \dcontext_R}
          {\subterm{P / \alpha} N}
          {\forall \beta \ldotp \subterm{P / \alpha} M}
          {By \dforallr}
  \Hand \DJudgeNegPf{\dcontext_L, \dcontext_R}
          {\subterm{P / \alpha} N}
          {\subterm{P / \alpha} \forall \beta \ldotp M}
          {\bydefsubcon}
      \end{llproof}
  
      \DerivationProofCase{\dforalll}
        {\wfpostypeJudg{\dconwithhole{\alpha}} {Q} \\ \DNegSubtypeJudge{\dconwithhole{\alpha}} {\subterm{Q / \beta} N} {M}}
        {\DNegSubtypeJudge{\dconwithhole{\alpha}} {\forall \beta \ldotp N} {M}}
  
      \begin{llproof}
        \wfpostypePf{\dcontext_L, \dcontext_R}
          {P}
          {Assumption}
        \wfnegtypePf{\dconwithhole{\alpha}}
          {\forall \beta \ldotp N}
          {\ditto}
        \wfnegtypePf{\dconwithhole{\alpha}, \beta}
          {N}
          {Inversion (\twfforall)}
        \wfpostypePf{\dconwithhole{\alpha}}
          {Q}
          {Subderivation}
        \wfnegtypePf{\dconwithhole{\alpha}}
          {\subterm{Q / \beta} N}
          {By \lemmaref{lemma:Declarative substitution well-formedness}}
        \wfnegtypePf{\dconwithhole{\alpha}}
          {M}
          {Assumption}
        \DJudgeNegPf{\dconwithhole{\alpha}}
          {\subterm{Q / \beta} N}
          {M}
          {Subderivation}
        \DJudgeNegPf{\dcontext_L, \dcontext_R}
          {\subterm{P / \alpha} \subterm{Q / \beta} N}
          {\subterm{P / \alpha} M}
          {By i.h.}
        \DJudgeNegPf{\dcontext_L, \dcontext_R}
          {\subterm{(\subterm{P / \alpha} Q) / \beta} \subterm{P / \alpha} N}
          {\subterm{P / \alpha} M}
          {Reordering substitutions}
        \wfpostypePf{\dcontext_L, \dcontext_R}
          {\subterm{P / \alpha} Q}
          {By \lemmaref{lemma:Declarative substitution well-formedness}}
  \Hand \DJudgeNegPf{\dcontext_L, \dcontext_R}
          {\forall \beta \ldotp \subterm{P / \alpha} N}
          {\subterm{P / \alpha} M}
          {By \dforalll (using $\subterm{P / \alpha} Q$ as the ground term)}
  \Hand \DJudgeNegPf{\dcontext_L, \dcontext_R}
          {\subterm{P / \alpha} \forall \beta \ldotp N}
          {\subterm{P / \alpha} M}
          {\bydefsubcon}
      \end{llproof}
  
      \DerivationProofCase{\darrow}
        {\DPosSubtypeJudge{\dconwithhole{\alpha}} {R} {Q} \\ \DNegSubtypeJudge{\dconwithhole{\alpha}} {N} {M}}
        {\DNegSubtypeJudge{\dconwithhole{\alpha}} {Q \funarrow N} {R \funarrow M}}
  
      \begin{llproof}
        \wfnegtypePf{\dcontext_L, \dcontext_R}
          {P}
          {Assumption}
        \wfnegtypePf{\dconwithhole{\alpha}}
          {Q \funarrow N}
          {\ditto}
        \wfnegtypePf{\dconwithhole{\alpha}}
          {R \funarrow M}
          {\ditto}
        
        \proofsep
        \wfpostypePf{\dconwithhole{\alpha}}
          {R}
          {Inversion (\twfarrow)}
        \wfpostypePf{\dconwithhole{\alpha}}
          {Q}
          {\ditto}
        \DJudgePosPf{\dconwithhole{\alpha}}
          {R}
          {Q}
          {Subderivation}
        \DJudgePosPf{\dcontext_L, \dcontext_R}
          {\subterm{P / \alpha} R}
          {\subterm{P / \alpha} Q}
          {By i.h.}
        
        \proofsep
        \wfnegtypePf{\dconwithhole{\alpha}}
          {N}
          {Inversion (\twfarrow)}
        \wfnegtypePf{\dconwithhole{\alpha}}
          {M}
          {\ditto}
        \DJudgeNegPf{\dconwithhole{\alpha}}
          {N}
          {M}
          {Subderivation}
        \DJudgeNegPf{\dcontext_L, \dcontext_R}
          {\subterm{P / \alpha} N}
          {\subterm{P / \alpha} M}
          {By i.h.}
        
        \proofsep
  \Hand \DJudgeNegPf{\dcontext_L, \dcontext_R}
          {\subterm{P / \alpha} Q \funarrow \subterm{P / \alpha} N}
          {\subterm{P / \alpha} R \funarrow \subterm{P / \alpha} M}
          {By \darrow}
  \Hand \DJudgeNegPf{\dcontext_L, \dcontext_R}
          {\subterm{P / \alpha} (Q \funarrow N)}
          {\subterm{P / \alpha} (R \funarrow M)}
          {\bydefsubcon}
      \end{llproof}
  
      \DerivationProofCase{\dshiftup}
        {\DPosSubtypeJudge{\dconwithhole{\alpha}} {R} {Q} \\ \DPosSubtypeJudge{\dconwithhole{\alpha}} {Q} {R}}
        {\DNegSubtypeJudge{\dconwithhole{\alpha}} {\shiftup Q} {\shiftup R}}
  
      \begin{llproof}
        \wfpostypePf{\dcontext_L, \dcontext_R}
          {P}
          {Assumption}
        \wfnegtypePf{\dconwithhole{\alpha}}
          {\shiftup Q}
          {\ditto}
        \wfpostypePf{\dconwithhole{\alpha}}
          {Q}
          {Inversion (\twfshiftup)}
        \wfnegtypePf{\dconwithhole{\alpha}}
          {\shiftup R}
          {Assumption}
        \wfpostypePf{\dconwithhole{\alpha}}
          {R}
          {Inversion (\twfshiftup)}
        \DJudgePosPf{\dcontext_L, \dcontext_R}
          {\subterm{P / \alpha} R}
          {\subterm{P / \alpha} Q}
          {By i.h.}
        \DJudgePosPf{\dcontext_L, \dcontext_R}
          {\subterm{P / \alpha} Q}
          {\subterm{P / \alpha} R}
          {\ditto}
  \Hand \DJudgeNegPf{\dcontext_L, \dcontext_R}
          {\shiftup \subterm{P / \alpha} Q}
          {\shiftup \subterm{P / \alpha} R}
          {By \dshiftup}
          \Hand \DJudgeNegPf{\dcontext_L, \dcontext_R}
          {\subterm{P / \alpha} \shiftup Q}
          {\subterm{P / \alpha} \shiftup R}
          {\bydefsubcon}
      \end{llproof}
  \end{itemize}
\end{proof}

\RestateSymmetryPositiveDeclarativeSubtyping*

\begin{proof}
  By rule induction on $\DPosSubtypeJudge{\dcontext} {P} {Q}$.

  \begin{itemize}
      \DerivationProofCase{\drefl}
        {\wfpostypeJudg{\dcontext}{\alpha}}
        {\DPosSubtypeJudge{\dcontext} {\alpha} {\alpha}}
  
      \begin{llproof}
  \Hand \DJudgePosPf{\dcontext}
          {\alpha}
          {\alpha}
          {Assumption}
      \end{llproof}
  
      \DerivationProofCase{\dshiftdown}
        {\DNegSubtypeJudge{\dcontext} {M} {N} \\ \DNegSubtypeJudge{\dcontext} {N} {M}}
        {\DPosSubtypeJudge{\dcontext} {\shiftdown N} {\shiftdown M}}
  
      \begin{llproof}
        \DJudgeNegPf{\dcontext}
          {N}
          {M}
          {Subderivation}
        \DJudgeNegPf{\dcontext}
          {M}
          {N}
          {\ditto}
  \Hand \DJudgePosPf{\dcontext}
          {\shiftdown M}
          {\shiftdown N}
          {By \dshiftdown}
      \end{llproof}
  \end{itemize}
\end{proof}

\subsection{Isomorphic types}

\RestateDeclarativeMutualSubtypingLemma*

\begin{proof}
  By strong mutual rule induction on the pair of $\DPosSubtypeJudge{\dcontext, \vv{\alpha}} {R} {\subcon{\vv{P / \beta}} S}$ and $\DPosSubtypeJudge{\dcontext, \vv{\beta}} {S} {\subcon{\vv{Q / \alpha}} R}$, and the pair of $\DNegSubtypeJudge{\dcontext, \vv{\alpha}} {\subcon{\vv{P / \beta}} N} {M}$ and $\DNegSubtypeJudge{\dcontext, \vv{\beta}} {\subcon{\vv{Q / \alpha}} M} {N}$.

  \begin{itemize}
    \DoubleDerivationProofCase
      {\drefl}
      {\wfpostypeJudg{\dcontext, \vv{\alpha}} {\gamma}}
      {\DPosSubtypeJudge{\dcontext, \vv{\alpha}} {\gamma} {\gamma}}
      {\drefl}
      {\wfpostypeJudg{\dcontext, \vv{\beta}} {\gamma}}
      {\DPosSubtypeJudge{\dcontext, \vv{\beta}} {\gamma} {\gamma}}
  
      By the \drefl rule, we must have the same universal variable on both sides of both judgments.
  
      \begin{llproof}
        \eqPf{\subcon{\vv{P / \beta}} S}
          {\gamma}
          {Since we have an instance of \drefl}
  \Hand \eqPf{P_i}
          {\gamma}
          {For all $P_i$ such that $\beta_i \in \vv{\beta}$ and $\beta_i \in \FreeUV(S)$}
        
        \proofsep
        \eqPf{\subcon{\vv{Q / \alpha}} R}
          {\gamma}
          {Since we have an instance of \drefl}
  \Hand \eqPf{Q_i}
          {\gamma}
          {For all $Q_i$ such that $\alpha_i \in \vv{\alpha}$ and $\alpha_i \in \FreeUV(R)$}
      \end{llproof}
    
    \DerivationProofCase
      {\dshiftdown}
      {\DNegSubtypeJudge{\dcontext, \vv{\alpha}} {\subcon{\vv{P / \beta}} N} {M} \\
        \DNegSubtypeJudge{\dcontext, \vv{\alpha}} {M} {\subcon{\vv{P / \beta}} N}}
      {\DPosSubtypeJudge{\dcontext, \vv{\alpha}} {\shiftdown M} {\subcon{\vv{P / \beta}} \shiftdown N}}
  
      If we have an instance of \dshiftdown, then the types on both sides of the other judgment must also start with $\shiftdown$, so we must have another instance of \dshiftdown:

      \begin{mathpar}
        \Infer{\dshiftdown}
          {\DNegSubtypeJudge{\dcontext, \vv{\beta}} {\subcon{\vv{Q / \alpha}} M} {N} \\
            \DNegSubtypeJudge{\dcontext, \vv{\beta}} {N} {\subcon{\vv{Q / \alpha}} M}}
          {\DPosSubtypeJudge{\dcontext, \vv{\beta}} {\shiftdown N} {\subcon{\vv{Q / \alpha}} \shiftdown M}}
      \end{mathpar}

      \begin{llproof}
        \wfnegtypePf{\dcontext, \vv{\alpha}}
          {M}
          {Inversion (\twfshiftdown)}
        \wfnegtypePf{\dcontext, \vv{\beta}}
          {N}
          {Inversion (\twfshiftdown)}
        \DJudgeNegPf{\dcontext, \vv{\alpha}}
          {\subcon{\vv{P / \beta}} N}
          {M}
          {Subderivation}
        \DJudgeNegPf{\dcontext, \vv{\beta}}
          {\subcon{\vv{Q / \alpha}} M}
          {N}
          {\ditto}
        \eqPf{P_i}
          {\gamma}
          {For all $P_i$ such that $\beta_i \in \vv{\beta}$ and $\beta_i \in \FreeUV(N)$ (by i.h.)}
        \eqPf{Q_i}
          {\gamma}
          {For all $Q_i$ such that $\alpha_i \in \vv{\alpha}$ and $\alpha_i \in \FreeUV(M)$ (by i.h.)}
  \Hand \eqPf{P_i}
          {\gamma}
          {For all $P_i$ such that $\beta_i \in \vv{\beta}$ and $\beta_i \in \FreeUV(\shiftdown N)$ (by definition of $\FreeUV$)}
  \Hand \eqPf{Q_i}
          {\gamma}
          {For all $Q_i$ such that $\alpha_i \in \vv{\alpha}$ and $\alpha_i \in \FreeUV(\shiftdown M)$ (by definition of $\FreeUV$)}
      \end{llproof}
  
    \DerivationProofCase{\dforallr}
      {\DNegSubtypeJudge{\dcontext, \vv{\alpha}, \gamma} {\subcon{\vv{P / \beta}} N} {M}}
      {\DNegSubtypeJudge{\dcontext, \vv{\alpha}} {\subcon{\vv{P / \beta}} N} {\forall \gamma \ldotp M}}

      By induction on the number of consecutive instances of \dforallr in the derivation of the second judgment.

      \begin{itemize}
        \DerivationProofCase{\dforalll}
          {\wfpostypeJudg{\dcontext, \vv{\beta}} {R} \\
            \DNegSubtypeJudge{\dcontext, \vv{\beta}} {\subcon{\vv{Q / \alpha}, R / \gamma} M} {N}}
          {\DNegSubtypeJudge{\dcontext, \vv{\beta}} {\forall \gamma \ldotp \subcon{\vv{Q / \alpha}} M} {N}}

          This is the base case of the inner induction.
          Our use of the outer induction hypothesis in this case is why we needed to perform a strong rule induction.

          \begin{llproof}
            \wfnegtypePf{\dcontext, \vv{\alpha}, \gamma}
              {M}
              {Inversion (\twfforall)}
            \wfnegtypePf{\dcontext, \vv{\beta}}
              {N}
              {Assumption}
            \DJudgeNegPf{\dcontext, \vv{\alpha}, \gamma}
              {\subcon{\vv{P / \beta}} N}
              {M}
              {Subderivation}
            \DJudgeNegPf{\dcontext, \vv{\beta}}
              {\subcon{\vv{Q / \alpha}, R / \gamma} M}
              {N}
              {\ditto}
      \Hand \eqPf{P_i}
              {\delta}
              {For all $P_i$ such that $\beta_i \in \vv{\beta}$ and $\beta_i \in \FreeUV(N)$ (by outer i.h.)}
            \eqPf{Q_i}
              {\delta}
              {For all $Q_i$ such that $\alpha_i \in \vv{\alpha}, \gamma$ and $\alpha_i \in \FreeUV(M)$ (by outer i.h.)}
      \Hand \eqPf{Q_i}
              {\delta}
              {For all $Q_i$ such that $\alpha_i \in \vv{\alpha}$ and $\alpha_i \in \FreeUV(\forall \gamma \ldotp M)$}
              \trailingjust{(by definition of $\FreeUV$)}
          \end{llproof}

        \DerivationProofCase{\dforallr}
          {\DNegSubtypeJudge{\dcontext, \vv{\beta}, \delta} {\subcon{\vv{Q / \alpha}} \forall \gamma \ldotp M} {N'}}
          {\DNegSubtypeJudge{\dcontext, \vv{\beta}} {\subcon{\vv{Q / \alpha}} \forall \gamma \ldotp M} {\forall \delta \ldotp N'}}

          This is the inductive step of the inner induction.
          Here we have $n = k + 1$ consecutive instances of \dforallr in the derivation of the second judgment.

          \begin{llproof}
            \wfnegtypePf{\dcontext, \vv{\beta}, \delta}
              {N'}
              {Inversion (\twfforall)}
            \DJudgeNegPf{\dcontext, \vv{\beta}, \delta}
              {\subcon{\vv{Q / \alpha}} \forall \gamma \ldotp M}
              {N'}
              {Subderivation}
            
            \proofsep
            \wfnegtypePf{\dcontext, \delta, \vv{\beta}}
              {N'}
              {By \lemmaref{lemma:Pushing uvars right preserves term well-formedness}}
            \DJudgeNegPf{\dcontext, \delta, \vv{\beta}}
              {\subcon{\vv{Q / \alpha}} \forall \gamma \ldotp M}
              {N'}
              {By \lemmaref{lemma:Pushing uvars right in declarative judgment}}
            \eqPf{P_i}
              {\eta}
              {For all $P_i$ such that $\beta_i \in \vv{\beta}$ and $\beta_i \in \FreeUV(N')$ (by inner i.h.)}
            \eqPf{Q_i}
              {\eta}
              {For all $Q_i$ such that $\alpha_i \in \vv{\alpha}, \gamma$ and $\alpha_i \in \FreeUV(M)$ (by inner i.h.)}
      \Hand \eqPf{P_i}
              {\eta}
              {For all $P_i$ such that $\beta_i \in \vv{\beta}$ and $\beta_i \in \FreeUV(\forall \delta \ldotp N')$}
              \trailingjust{(by definition of $\FreeUV$)}
      \Hand \eqPf{Q_i}
              {\eta}
              {For all $Q_i$ such that $\alpha_i \in \vv{\alpha}$ and $\alpha_i \in \FreeUV(\forall \gamma \ldotp M)$}
              \trailingjust{(by definition of $\FreeUV$)}
          \end{llproof}
      \end{itemize}

    \DerivationProofCase
      {\dforalll}
      {\wfpostypeJudg{\dcontext, \vv{\alpha}} {R} \\
        \DNegSubtypeJudge{\dcontext, \vv{\alpha}} {\subcon{\vv{P / \beta}, R / \gamma} N} {M}}
      {\DNegSubtypeJudge{\dcontext, \vv{\alpha}} {\forall \gamma \ldotp \subcon{\vv{P / \beta}} N} {M}}

      We perform a case split over the derivation of the second judgment.

      \begin{itemize}
        \DerivationProofCase{\dforalll}
          {\wfpostypeJudg{\dcontext, \vv{\beta}} {S} \\
          \DNegSubtypeJudge{\dcontext, \vv{\beta}} {\subcon{\vv{Q / \alpha}, S / \delta} M'} {\forall \gamma \ldotp N}}
          {\DNegSubtypeJudge{\dcontext, \vv{\beta}} {\forall \delta \ldotp \subcon{\vv{Q / \alpha}} M'} {\forall \gamma \ldotp N}}

        \begin{llproof}
          \wfnegtypePf{\dcontext, \vv{\alpha}, \delta}
            {M'}
            {Inversion (\twfforall)}
          \wfnegtypePf{\dcontext, \vv{\beta}, \gamma}
            {N}
            {\ditto}
          \DJudgeNegPf{\dcontext, \vv{\alpha}}
            {\subcon{\vv{P / \beta}, R / \gamma} N}
            {\forall \delta \ldotp M'}
            {Subderivation}
          \DJudgeNegPf{\dcontext, \vv{\beta}}
            {\subcon{\vv{Q / \alpha}, S / \delta} M'}
            {\forall \gamma \ldotp N}
            {\ditto}
          \DJudgeNegPf{\dcontext, \vv{\alpha}, \delta}
            {\subcon{\vv{P / \beta}, R / \gamma} N}
            {M'}
            {Inversion (\dforallr)}
          \DJudgeNegPf{\dcontext, \vv{\beta}, \gamma}
            {\subcon{\vv{Q / \alpha}, S / \delta} M'}
            {N}
            {\ditto}
          \eqPf{P_i}
            {\eta}
            {For all $P_i$ such that $\beta_i \in \vv{\beta}, \gamma$ and $\beta_i \in \FreeUV(N)$ (by i.h.)}
          \eqPf{Q_i}
            {\eta}
            {For all $Q_i$ such that $\alpha_i \in \vv{\alpha}, \delta$ and $\alpha_i \in \FreeUV(M')$ (by i.h.)}
    \Hand \eqPf{P_i}
            {\eta}
            {For all $P_i$ such that $\beta_i \in \vv{\beta}$ and $\beta_i \in \FreeUV(\forall \gamma \ldotp N)$}
            \trailingjust{(by definition of $\FreeUV$)}
    \Hand \eqPf{Q_i}
            {\eta}
            {For all $Q_i$ such that $\alpha_i \in \vv{\alpha}$ and $\alpha_i \in \FreeUV(\forall \delta \ldotp M')$}
            \trailingjust{(by definition of $\FreeUV$)}
        \end{llproof}

        Here the application of the inductive hypothesis states that every universal variable in the arrays $\vv{\beta}, \gamma$ and $\vv{\alpha}, \delta$ that appears in the corresponding type is substituted by a universal variable (including $\gamma$ and $\delta$).
        As a result, the conclusion holds for just the universal variables in $\vv{\beta}$ and $\vv{\alpha}$.

        \DerivationProofCase{\dforallr}
          {\DNegSubtypeJudge{\dcontext, \vv{\beta}, \gamma} {\subcon{\vv{Q / \alpha}} M} {N}}
          {\DNegSubtypeJudge{\dcontext, \vv{\beta}} {\subcon{\vv{Q / \alpha}} M} {\forall \gamma \ldotp N}}

        \begin{llproof}
          \wfnegtypePf{\dcontext, \vv{\alpha}}
            {M}
            {Assumption}
          \wfnegtypePf{\dcontext, \vv{\beta}, \gamma}
            {N}
            {Inversion (\twfforall)}
          \DJudgeNegPf{\dcontext, \vv{\alpha}}
            {\subcon{\vv{P / \beta}, R / \gamma} N}
            {M}
            {Subderivation}
          \DJudgeNegPf{\dcontext, \vv{\beta}, \gamma}
            {\subcon{\vv{Q / \alpha}} M}
            {N}
            {\ditto}
          \eqPf{P_i}
            {\delta}
            {For all $P_i$ such that $\beta_i \in \vv{\beta}, \gamma$ and $\beta_i \in \FreeUV(N)$ (by outer i.h.)}
          \Hand \eqPf{P_i}
            {\delta}
            {For all $P_i$ such that $\beta_i \in \vv{\beta}$ and $\beta_i \in \FreeUV(\forall \gamma \ldotp N)$}
            \trailingjust{(by definition of $\FreeUV$)}
    \Hand \eqPf{Q_i}
            {\delta}
            {For all $Q_i$ such that $\alpha_i \in \vv{\alpha}$ and $\alpha_i \in \FreeUV(M)$ (by outer i.h.)}
        \end{llproof}
    \end{itemize}
  
    \DerivationProofCase
      {\darrow}
      {\DPosSubtypeJudge{\dcontext, \vv{\alpha}} {R} {\subcon{\vv{P / \beta}} S} \\
        \DNegSubtypeJudge{\dcontext, \vv{\alpha}} {\subcon{\vv{P / \beta}} N} {M}}
      {\DNegSubtypeJudge{\dcontext, \vv{\alpha}} {\subcon{\vv{P / \beta}} (S \funarrow N)} {R \funarrow M}}
  
      If we have an instance of \darrow, then the types on both sides of the other judgment must be function types, so we must have another instance of \darrow:

      \begin{mathpar}
        \Infer{\darrow}
          {\DPosSubtypeJudge{\dcontext, \vv{\beta}} {S} {\subcon{\vv{Q / \alpha}} R} \\
            \DNegSubtypeJudge{\dcontext, \vv{\beta}} {\subcon{\vv{Q / \alpha}} M} {N}}
          {\DNegSubtypeJudge{\dcontext, \vv{\beta}} {\subcon{\vv{Q / \alpha}} (R \funarrow M)} {S \funarrow N}}
      \end{mathpar}

      \begin{llproof}
        \wfpostypePf{\dcontext, \vv{\alpha}}
          {R}
          {Inversion (\twfarrow)}
        \wfpostypePf{\dcontext, \vv{\beta}}
          {S}
          {Inversion (\twfarrow)}
        \DJudgePosPf{\dcontext, \vv{\alpha}}
          {R}
          {\subcon{\vv{P / \beta}} S}
          {Subderivation}
        \DJudgePosPf{\dcontext, \vv{\beta}}
          {S}
          {\subcon{\vv{Q / \alpha}} R}
          {\ditto}
        \eqPf{P_i}
          {\gamma}
          {For all $P_i$ such that $\beta_i \in \vv{\beta}$ and $\beta_i \in \FreeUV(S)$ (by i.h.)}
        \eqPf{Q_i}
          {\gamma}
          {For all $Q_i$ such that $\alpha_i \in \vv{\alpha}$ and $\alpha_i \in \FreeUV(R)$ (by i.h.)}

        \proofsep
        \wfnegtypePf{\dcontext, \vv{\alpha}}
          {M}
          {Inversion (\twfarrow)}
        \wfnegtypePf{\dcontext, \vv{\beta}}
          {N}
          {Inversion (\twfarrow)}
        \DJudgeNegPf{\dcontext, \vv{\alpha}}
          {\subcon{\vv{P / \beta}} N}
          {M}
          {Subderivation}
        \DJudgeNegPf{\dcontext, \vv{\beta}}
          {\subcon{\vv{Q / \alpha}} M}
          {N}
          {\ditto}
        \eqPf{P_i}
          {\gamma}
          {For all $P_i$ such that $\beta_i \in \vv{\beta}$ and $\beta_i \in \FreeUV(N)$ (by i.h.)}
        \eqPf{Q_i}
          {\gamma}
          {For all $Q_i$ such that $\alpha_i \in \vv{\alpha}$ and $\alpha_i \in \FreeUV(M)$ (by i.h.)}

        \proofsep
  \Hand \eqPf{P_i}
          {\gamma}
          {For all $P_i$ such that $\beta_i \in \vv{\beta}$ and $\beta_i \in \FreeUV(S \funarrow N)$}
          \trailingjust{(by definition of $\FreeUV$)}
  \Hand \eqPf{Q_i}
          {\gamma}
          {For all $Q_i$ such that $\alpha_i \in \vv{\alpha}$ and $\alpha_i \in \FreeUV(R \funarrow M)$}
          \trailingjust{(by definition of $\FreeUV$)}
      \end{llproof}
  
    \DerivationProofCase
      {\dshiftup}
      {\DPosSubtypeJudge{\dcontext, \vv{\alpha}} {\subcon{\vv{P / \beta}} S} {R} \\
        \DPosSubtypeJudge{\dcontext, \vv{\alpha}} {R} {\subcon{\vv{P / \beta}} S}}
      {\DNegSubtypeJudge{\dcontext, \vv{\alpha}} {\shiftup R} {\subcon{\vv{P / \beta}} \shiftup S}}

    If we have an instance of \dshiftup, then the types on both sides of the other judgment must also start with $\shiftup$, so we must have another instance of \dshiftup:

    \begin{mathpar}
      \Infer{\dshiftup}
        {\DPosSubtypeJudge{\dcontext, \vv{\beta}} {\subcon{\vv{Q / \alpha}} R} {S} \\
          \DPosSubtypeJudge{\dcontext, \vv{\beta}} {S} {\subcon{\vv{Q / \alpha}} R}}
        {\DNegSubtypeJudge{\dcontext, \vv{\beta}} {\shiftup S} {\subcon{\vv{Q / \alpha}} \shiftup R}}
    \end{mathpar}

    \begin{llproof}
      \wfpostypePf{\dcontext, \vv{\alpha}}
        {R}
        {Inversion (\twfshiftup)}
      \wfpostypePf{\dcontext, \vv{\beta}}
        {S}
        {Inversion (\twfshiftup)}
      \DJudgePosPf{\dcontext, \vv{\alpha}}
        {R}
        {\subcon{\vv{P / \beta}} S}
        {Subderivation}
      \DJudgePosPf{\dcontext, \vv{\beta}}
        {S}
        {\subcon{\vv{Q / \alpha}} R}
        {\ditto}
      \eqPf{P_i}
        {\gamma}
        {For all $P_i$ such that $\beta_i \in \vv{\beta}$ and $\beta_i \in \FreeUV(S)$ (by i.h.)}
      \eqPf{Q_i}
        {\gamma}
        {For all $Q_i$ such that $\alpha_i \in \vv{\alpha}$ and $\alpha_i \in \FreeUV(R)$ (by i.h.)}
\Hand \eqPf{P_i}
        {\gamma}
        {For all $P_i$ such that $\beta_i \in \vv{\beta}$ and $\beta_i \in \FreeUV(\shiftup S)$}
        \trailingjust{(by definition of $\FreeUV$)}
\Hand \eqPf{Q_i}
        {\gamma}
        {For all $Q_i$ such that $\alpha_i \in \vv{\alpha}$ and $\alpha_i \in \FreeUV(\shiftup R)$}
        \trailingjust{(by definition of $\FreeUV$)}
    \end{llproof}
  \end{itemize}  
\end{proof}

\RestateDeclarativeMutualSubtypingSizeLemma*

\begin{proof}
  By rule induction on $\DPnSubtypeJudge{\dcontext} {A} {B}$ and $\DPnSubtypeJudge{\dcontext} {B} {A}$.

\begin{itemize}
  \DoubleDerivationProofCase
    {\drefl}
    {\wfpostypeJudg{\dcontext} {\gamma}}
    {\DPosSubtypeJudge{\dcontext} {\gamma} {\gamma}}
    {\drefl}
    {\wfpostypeJudg{\dcontext} {\gamma}}
    {\DPosSubtypeJudge{\dcontext} {\gamma} {\gamma}}

    \begin{llproof}
\Hand \eqPf{\termsize{\gamma}}
        {\termsize{\gamma}}
        {Identical LHS and RHS}
    \end{llproof}

  \DoubleDerivationProofCase
    {\dshiftdown}
    {\DNegSubtypeJudge{\dcontext} {N} {M} \\ \DNegSubtypeJudge{\dcontext} {M} {N}}
    {\DPosSubtypeJudge{\dcontext} {\shiftdown M} {\shiftdown N}}
    {\dshiftdown}
    {\DNegSubtypeJudge{\dcontext} {M} {N} \\ \DNegSubtypeJudge{\dcontext} {N} {M}}
    {\DPosSubtypeJudge{\dcontext} {\shiftdown N} {\shiftdown M}}

    \begin{llproof}
      \wfpostypePf{\dcontext}
        {\shiftdown M}
        {Assumption}
      \wfnegtypePf{\dcontext}
        {M}
        {Inversion (\twfshiftdown)}
      \wfpostypePf{\dcontext}
        {\shiftdown N}
        {Assumption}
      \wfnegtypePf{\dcontext}
        {N}
        {Inversion (\twfshiftdown)}
      
      \proofsep
      \DJudgeNegPf{\dcontext}
        {M}
        {N}
        {Subderivation}
      \DJudgeNegPf{\dcontext}
        {N}
        {M}
        {Subderivation}
      \termsizeEqPf{M}
        {N}
        {By i.h.}
\Hand \termsizeEqPf{\shiftdown M}
        {\shiftdown N}
        {\bydeftermsize}
    \end{llproof}

  \DoubleDerivationProofCase
    {\dforalll}
    {\wfpostypeJudg{\dcontext} {P} \\ \DNegSubtypeJudge{\dcontext} {\subcon{P / \alpha} M} {N}}
    {\DNegSubtypeJudge{\dcontext} {\forall \alpha \ldotp M} {N}}
    {\dforallr}
    {\DNegSubtypeJudge{\dcontext, \alpha} {N} {M}}
    {\DNegSubtypeJudge{\dcontext} {N} {\forall \alpha \ldotp M}}

    \begin{llproof}
      \wfpostypePf{\dcontext}
        {P}
        {Subderivation}
      \wfnegtypePf{\dcontext}
        {N}
        {Assumption}
      \wfnegtypePf{\dcontext, \alpha}
        {N}
        {By \lemmaref{lemma:Term well-formedness weakening}}
      \wfnegtypePf{\dcontext}
        {\forall \alpha \ldotp M}
        {Assumption}
      \wfnegtypePf{\dcontext, \alpha}
        {M}
        {Inversion (\twfforall)}
      \DJudgeNegPf{\dcontext}
        {\subcon{P / \alpha} M}
        {N}
        {Subderivation}
      \DJudgeNegPf{\dcontext, \alpha}
        {N}
        {M}
        {Subderivation}
      \inPf{P}
        {\text{Uvar}}
        {By \lemmaref{lemma:Declarative mutual subtyping lemma}}

      \proofsep
      \eqPf{\termsize{\subcon{P / \alpha} M}}
        {\termsize{M}}
        {Since $\termsize{P} = 1 = \termsize{\alpha}$}
      \wfnegtypePf{\dcontext}
        {\subterm{P / \alpha} M}
        {By \lemmaref{lemma:Declarative substitution well-formedness}}
      \DJudgeNegPf{\dcontext}
        {N}
        {\subcon{P / \alpha} M}
        {By \lemmaref{lemma:Declarative subtyping substitution lemma}}
      \eqPf{\termsize{\subcon{P / \alpha} M}}
        {\termsize{N}}
        {By i.h. (since $P \in \text{Uvar}$, the derivation of $\DNegSubtypeJudge{\dcontext} {N} {\subcon{P / \alpha} M}$ is}
        \trailingjust{the same size as the derivation of $\DNegSubtypeJudge{\dcontext, \alpha} {N} {M}$)}
      \eqPf{\termsize{M}}
        {\termsize{N}}
        {Using $\termsize{\subterm{P / \alpha} M} = \termsize{M}$}
\Hand \eqPf{\termsize{\forall \alpha \ldotp M}}
        {\termsize{N}}
        {By definition of type size}
    \end{llproof}

  \DoubleDerivationProofCase
    {\dforallr}
    {\DNegSubtypeJudge{\dcontext, \alpha} {M} {N}}
    {\DNegSubtypeJudge{\dcontext} {M} {\forall \alpha \ldotp N}}
    {\dforalll}
    {\wfpostypeJudg{\dcontext} {P} \DNegSubtypeJudge{\dcontext} {\subcon{P / \alpha} N} {M}}
    {\DNegSubtypeJudge{\dcontext} {\forall \alpha \ldotp N} {M}}

    Symmetrical to the previous case ($M$ and $N$ are swapped).

  \DoubleDerivationProofCase
    {\darrow}
    {\DPosSubtypeJudge{\dcontext} {Q} {P} \\ \DNegSubtypeJudge{\dcontext} {M} {N}}
    {\DNegSubtypeJudge{\dcontext} {P \funarrow M} {Q \funarrow N}}
    {\darrow}
    {\DPosSubtypeJudge{\dcontext} {P} {Q} \\ \DNegSubtypeJudge{\dcontext} {N} {M}}
    {\DNegSubtypeJudge{\dcontext} {Q \funarrow N} {P \funarrow M}}

    \begin{llproof}
      \wfnegtypePf{\dcontext}
        {P \funarrow N}
        {Assumption}
      \wfpostypePf{\dcontext}
        {P}
        {Inversion (\twfarrow)}
      \wfnegtypePf{\dcontext}
        {N}
        {\ditto}
      \wfnegtypePf{\dcontext}
        {Q \funarrow M}
        {Assumption}
      \wfpostypePf{\dcontext}
        {Q}
        {Inversion (\twfarrow)}
      \wfnegtypePf{\dcontext}
        {M}
        {\ditto}

      \proofsep
      \DJudgePosPf{\dcontext}
        {P}
        {Q}
        {Subderivation}
      \DJudgePosPf{\dcontext}
        {Q}
        {P}
        {\ditto}
      \eqPf{\termsize{P}}
        {\termsize{Q}}
        {By i.h.}
      \DJudgeNegPf{\dcontext}
        {M}
        {N}
        {Subderivation}
      \DJudgeNegPf{\dcontext}
        {N}
        {M}
        {\ditto}
      \eqPf{\termsize{M}}
        {\termsize{N}}
        {By i.h.}
\Hand \eqPf{\termsize{P \funarrow M}}
        {\termsize{Q \funarrow N}}
        {By definition of type size}
    \end{llproof}

  \DoubleDerivationProofCase
    {\dshiftup}
    {\DPosSubtypeJudge{\dcontext} {Q} {P} \\ \DPosSubtypeJudge{\dcontext} {P} {Q}}
    {\DNegSubtypeJudge{\dcontext} {\shiftup P} {\shiftup Q}}
    {\dshiftup}
    {\DPosSubtypeJudge{\dcontext} {Q} {P} \\ \DPosSubtypeJudge{\dcontext} {Q} {P}}
    {\DNegSubtypeJudge{\dcontext} {\shiftup Q} {\shiftup P}}

    \begin{llproof}
      \wfnegtypePf{\dcontext}
        {\shiftup P}
        {Assumption}
      \wfpostypePf{\dcontext}
        {P}
        {Inversion (\twfshiftup)}
      \wfnegtypePf{\dcontext}
        {\shiftup Q}
        {Assumption}
      \wfpostypePf{\dcontext}
        {Q}
        {Inversion (\twfshiftup)}

      \proofsep
      \DJudgePosPf{\dcontext}
        {P}
        {Q}
        {Subderivation}
      \DJudgePosPf{\dcontext}
        {Q}
        {P}
        {\ditto}
      \eqPf{\termsize{P}}
        {\termsize{Q}}
        {By i.h.}
      \eqPf{\termsize{\shiftup P}}
        {\termsize{\shiftup Q}}
        {\bydeftermsize}
    \end{llproof}
\end{itemize}
\end{proof}

\subsection{Transitivity}

\RestateTransitivityPnSubtype*

\begin{proof}
  By rule induction on $\DPnSubtypeJudge{\dcontext} {B} {C}$ weighted by the lexicographic ordering of ($\termsize{B}$, $\numprenex{B} + \numprenex{C}$) in the positive case and ($\termsize{C}$, $\numprenex{B} + \numprenex{C}$) in the negative case.

  \begin{itemize}
    \DoubleDerivationProofCase{\drefl}
      {\wfpostypeJudg{\dcontext}{\alpha}}
      {\DPosSubtypeJudge{\dcontext} {\alpha} {\alpha}}
      {\drefl}
      {\wfpostypeJudg{\dcontext}{\alpha}}
      {\DPosSubtypeJudge{\dcontext} {\alpha} {\alpha}}
  
      \begin{llproof}
        \wfpostypePf{\dcontext}
          {\alpha}
          {Subderivation}
  \Hand \DJudgePosPf{\dcontext}
          {\alpha}
          {\alpha}
          {By \drefl}
      \end{llproof}

    \DoubleDerivationProofCase{\dshiftdown}
      {\DNegSubtypeJudge{\dcontext} {M} {N} \\ \DNegSubtypeJudge{\dcontext} {N} {M}}
      {\DPosSubtypeJudge{\dcontext} {\shiftdown N} {\shiftdown M}}
      {\dshiftdown}
      {\DNegSubtypeJudge{\dcontext} {N'} {M} \\ \DNegSubtypeJudge{\dcontext} {M} {N'}}
      {\DPosSubtypeJudge{\dcontext} {\shiftdown M} {\shiftdown N'}}
  
      The second judgment must be an instance of \dshiftdown due to the structure of $\shiftdown M$.
  
      \begin{llproof}
        \wfpostypePf{\dcontext}
          {\shiftdown N}
          {Assumption}
        \wfpostypePf{\dcontext}
          {\shiftdown M}
          {\ditto}
        \wfpostypePf{\dcontext}
          {\shiftdown N'}
          {\ditto}
        \wfnegtypePf{\dcontext}
          {N}
          {Inversion (\twfshiftdown)}
        \wfnegtypePf{\dcontext}
          {M}
          {\ditto}
        \wfnegtypePf{\dcontext}
          {N'}
          {\ditto}
  
        \proofsep
        \DJudgeNegPf{\dcontext}
          {M}
          {N}
          {Subderivation}
        \DJudgeNegPf{\dcontext}
          {N}
          {M}
          {\ditto}
        \eqPf{\termsize{M}}
          {\termsize{N}}
          {By \lemmaref{lemma:Isomorphic types are the same size}}
        \DJudgeNegPf{\dcontext}
          {M}
          {N'}
          {Subderivation}
        \DJudgeNegPf{\dcontext}
          {N'}
          {M}
          {\ditto}
        \eqPf{\termsize{M}}
          {\termsize{N'}}
          {By \lemmaref{lemma:Isomorphic types are the same size}}
  
        \proofsep
        \DJudgeNegPf{\dcontext}
          {N'}
          {M}
          {Above}
        \DJudgeNegPf{\dcontext}
          {M}
          {N}
          {\ditto}
        \DJudgeNegPf{\dcontext}
          {N'}
          {N}
          {By i.h. ($\termsize{N} = \termsize{M} < \termsize{\shiftdown M}$)}
        \DJudgeNegPf{\dcontext}
          {N}
          {M}
          {Above}
        \DJudgeNegPf{\dcontext}
          {M}
          {N'}
          {\ditto}
        \DJudgeNegPf{\dcontext}
          {N}
          {N'}
          {By i.h. ($\termsize{N'} = \termsize{M} < \termsize{\shiftdown M}$)}
  \Hand \DJudgePosPf{\dcontext}
          {\shiftdown N}
          {\shiftdown N'}
          {By \dshiftdown}
      \end{llproof}

    \DerivationProofCase{\dforallr}
      {\DNegSubtypeJudge{\dcontext, \alpha} {M} {N'}}
      {\DNegSubtypeJudge{\dcontext} {M} {\forall \alpha \ldotp N'}}
  
      Here we only need to decompose the second declarative judgment.
  
      \begin{llproof}
        \wfnegtypePf{\dcontext}
          {N}
          {Assumption}
        \wfnegtypePf{\dcontext, \alpha}
          {N}
          {By \lemmaref{lemma:Term well-formedness weakening}}
        \wfnegtypePf{\dcontext}
          {M}
          {Assumption}
        \wfnegtypePf{\dcontext, \alpha}
          {M}
          {By \lemmaref{lemma:Term well-formedness weakening}}
        \wfnegtypePf{\dcontext, \alpha}
          {\forall \alpha \ldotp N'}
          {Assumption}
        \wfnegtypePf{\dcontext, \alpha}
          {N'}
          {Inversion (\twfforall)}
        
        \proofsep
        \DJudgeNegPf{\dcontext}
          {N}
          {M}
          {Assumption}
        \DJudgeNegPf{\dcontext, \alpha}
          {N}
          {M}
          {By \lemmaref{lemma:Declarative subtyping weakening}}
        \DJudgeNegPf{\dcontext, \alpha}
          {M}
          {N'}
          {Subderivation}
        \DJudgeNegPf{\dcontext, \alpha}
          {N}
          {N'}
          {By i.h. ($\termsize{N'} = \termsize{\forall \alpha \ldotp N'}$ and the number of prenex quantifiers in}
          \trailingjust{the second judgment has reduced by 1)}
  \Hand \DJudgeNegPf{\dcontext}
          {N}
          {\forall \alpha \ldotp N'}
          {By \dforallr ($\alpha \notin \FreeUV(N)$ since $\alpha \notin \ConUV(\dcontext)$ (because $\alpha$ fresh)}
          \trailingjust{and also $\wfnegtypeJudg{\dcontext} {N}$)}
      \end{llproof}

    \DoubleDerivationProofCase{\dforallr}
      {\DNegSubtypeJudge{\dcontext, \alpha} {N} {M}}
      {\DNegSubtypeJudge{\dcontext} {N} {\forall \alpha \ldotp M}}
      {\dforalll}
      {\wfpostypeJudg{\dcontext} {P} \\ \DNegSubtypeJudge{\dcontext} {\subterm{P / \alpha} M} {N'}}
      {\DNegSubtypeJudge{\dcontext} {\forall \alpha \ldotp M} {N'}}
  
      \begin{llproof}
        \wfnegtypePf{\dcontext}
          {N}
          {Assumption}
        \wfnegtypePf{\dcontext}
          {\forall \alpha \ldotp M}
          {Assumption}
        \wfnegtypePf{\dcontext, \alpha}
          {M}
          {Inversion (\twfforall)}
        \wfnegtypePf{\dcontext}
          {N'}
          {Assumption}
  
        \proofsep
        \DJudgeNegPf{\dcontext, \alpha}
          {N}
          {M}
          {Subderivation}
        \wfpostypePf{\dcontext}
          {P}
          {Subderivation}
        \DJudgeNegPf{\dcontext}
          {N}
          {\subterm{P / \alpha} M}
          {By \lemmaref{lemma:Declarative subtyping substitution lemma}}
          \trailingjust{($\alpha \notin \FreeUV(N)$ by side condition of \dforallr)}
        \DJudgeNegPf{\dcontext}
          {\subterm{P / \alpha} M}
          {N'}
          {Subderivation}
  \Hand \DJudgeNegPf{\dcontext}
          {N}
          {N'}
          {By i.h. ($\termsize{N'} = \termsize{N'}$ and the number of prenex quantifiers in}
          \trailingjust{the second judgment has reduced by 1.  Substitution}
          \trailingjust{can only replace positive types, so it cannot change the}
          \trailingjust{number of prenex quantifiers in a negative type)}
      \end{llproof}

    \DoubleDerivationProofCase{\darrow}
      {\DPosSubtypeJudge{\dcontext} {Q} {P} \\ \DNegSubtypeJudge{\dcontext} {N} {M}}
      {\DNegSubtypeJudge{\dcontext} {P \funarrow N} {Q \funarrow M}}
      {\darrow}
      {\DPosSubtypeJudge{\dcontext} {P'} {Q} \\ \DNegSubtypeJudge{\dcontext} {M} {N'}}
      {\DNegSubtypeJudge{\dcontext} {Q \funarrow M} {P' \funarrow N'}}
  
      \begin{llproof}
        \wfnegtypePf{\dcontext}
          {P \funarrow N}
          {Assumption}
        \wfpostypePf{\dcontext}
          {P}
          {Inversion (\twfarrow)}
        \wfnegtypePf{\dcontext}
          {N}
          {\ditto}
        \wfnegtypePf{\dcontext}
          {Q \funarrow M}
          {Assumption}
        \wfpostypePf{\dcontext}
          {Q}
          {Inversion (\twfarrow)}
        \wfnegtypePf{\dcontext}
          {M}
          {\ditto}
        \wfnegtypePf{\dcontext}
          {P' \funarrow N'}
          {Assumption}
        \wfpostypePf{\dcontext}
          {P'}
          {Inversion (\twfarrow)}
        \wfnegtypePf{\dcontext}
          {N'}
          {\ditto}
  
        \proofsep
        \DJudgePosPf{\dcontext}
          {P'}
          {Q}
          {Subderivation}
        \DJudgePosPf{\dcontext}
          {Q}
          {P'}
          {By \lemmaref{lemma:Symmetry of positive declarative subtyping}}
        \termsizeEqPf{P'}
          {Q}
          {By \lemmaref{lemma:Isomorphic types are the same size}}
  
        \proofsep
        \DJudgePosPf{\dcontext}
          {P'}
          {Q}
          {Subderivation}
        \DJudgePosPf{\dcontext}
          {Q}
          {P}
          {\ditto}
        \DJudgePosPf{\dcontext}
          {P'}
          {P}
          {By i.h. ($\termsize{Q} = \termsize{P'} < \termsize{P' \funarrow N'}$)}
        \DJudgeNegPf{\dcontext}
          {N}
          {M}
          {Subderivation}
        \DJudgeNegPf{\dcontext}
          {M}
          {N'}
          {\ditto}
        \DJudgeNegPf{\dcontext}
          {N}
          {N'}
          {By i.h. ($\termsize{N'} < \termsize{P' \funarrow N'}$)}
  \Hand \DJudgeNegPf{\dcontext}
          {P \funarrow N}
          {P' \funarrow N'}
          {By \darrow}
      \end{llproof}

    \DoubleDerivationProofCase{\dshiftup}
      {\DPosSubtypeJudge{\dcontext} {Q} {P} \\ \DPosSubtypeJudge{\dcontext} {P} {Q}}
      {\DNegSubtypeJudge{\dcontext} {\shiftup P} {\shiftup Q}}
      {\dshiftup}
      {\DPosSubtypeJudge{\dcontext} {P'} {Q} \\ \DPosSubtypeJudge{\dcontext} {Q} {P'}}
      {\DNegSubtypeJudge{\dcontext} {\shiftup Q} {\shiftup P'}}

      Symmetrical to \dshiftdown case.
  \end{itemize}
\end{proof}

\section{Weak context extension}

\WeakCtxExtSubsumes*

\begin{proof}
  By rule induction over the $\congoesJudg{\acontext}{\acontext'}$ judgment.

  \begin{itemize}
    \DerivationProofCase{\Cempty}
      {}
      {\congoesJudg{\emptyacontext}{\emptyacontext}}

    \begin{llproof}
\Hand \congoesWeakPf{\emptyacontext}{\emptyacontext}{By {\Wcempty}}
    \end{llproof}

    \DerivationProofCase{\Cuvar}
      {\congoesJudg{\acontext}{\acontext'}}
      {\congoesJudg{\acontext, \alpha}{\acontext', \alpha}}

    \begin{llproof}
      \congoesWeakPf{\acontext}{\acontext'}{\byih}
\Hand \congoesWeakPf{\acontext, \alpha}{\acontext', \alpha}{By {\Wcuvar}}
    \end{llproof}

    \DerivationProofCase{\Cunsolvedguess}
      {\congoesJudg{\acontext}{\acontext'}}
      {\congoesJudg{\acontext, \guess{\alpha}}{\acontext', \guess{\alpha}}}

    \begin{llproof}
      \congoesWeakPf{\acontext}{\acontext'}{\byih}
\Hand \congoesWeakPf{\acontext, \guess{\alpha}}{\acontext', \guess{\alpha}}{By {\Wcunsolvedguess}}
    \end{llproof}

    \DerivationProofCase{\Csolveguess}
      {\congoesJudg{\acontext}{\acontext'}}
      {\congoesJudg{\acontext, \guess{\alpha}}{\acontext', \guess{\alpha} = P}}

    \begin{llproof}
      \congoesWeakPf{\acontext}{\acontext'}{\byih}
\Hand \congoesWeakPf{\acontext, \guess{\alpha}}{\acontext', \guess{\alpha} = P}{By {\Wcsolveguess}}
    \end{llproof}

    \DerivationProofCase{\Csolvedguess}
      {\congoesJudg{\acontext}{\acontext'} \\ \MutualSubtypePosJudge{\makedec{\acontext}} {P} {Q}}
      {\congoesJudg{\acontext, \guess{\alpha} = P}{\acontext', \guess{\alpha} = Q}}

    \begin{llproof}
      \congoesWeakPf{\acontext}{\acontext'}{\byih}
      \MutualJudgePosPf{\makedec{\acontext}} {P} {Q} {Premise}
\Hand \congoesWeakPf{\acontext, \guess{\alpha} = P}{\acontext', \guess{\alpha} = P}{By {\Wcsolvedguess}}
    \end{llproof}
  \end{itemize}
\end{proof}

\WeakCtxExtRefl*

\begin{proof}
  Corollary of \lemmaref{lemma:context extension reflexive}.

  \begin{llproof}
    \congoesPf{\Theta}{\Theta}{By \lemmaref{lemma:context extension reflexive}}
\Hand \congoesWeakPf{\Theta}{\Theta}{By \lemmaref{lemma:weak context extension subsumes normal}}
  \end{llproof}
\end{proof}

\WeakContextDeclCxt*

\begin{proof}
  By rule induction over the $\congoesWeakJudg{\acontext}{\acontext'}$ judgment.

  \begin{itemize}
    \DerivationProofCase{\Wcempty}
      {}
      {\congoesWeakJudg{\emptyacontext}{\emptyacontext}}

    \begin{llproof}
\Hand \eqPf{\makedec{\emptyacontext}}{\makedec{\emptyacontext}}{}
    \end{llproof}

    \DerivationProofCase{\Wcuvar}
      {\congoesWeakJudg{\Theta}{\Theta'}}
      {\congoesWeakJudg{\Theta, \alpha}{\Theta', \alpha}}

    \begin{llproof}
      \eqPf{\makedec{\Theta, \alpha}}{\makedec{\Theta}, \alpha}{\bydefmakedec}
      \continueeqPf{\makedec{\Theta'}, \alpha}{\byih}
      \continueeqPf{\makedec{\Theta', \alpha}}{\bydefmakedec}
    \end{llproof}

    \DerivationProofCase{\Wcunsolvedguess}
      {\congoesWeakJudg{\Theta}{\Theta'}}
      {\congoesWeakJudg{\Theta, \guess{\alpha}}{\Theta', \guess{\alpha}}}

    \begin{llproof}
      \eqPf{\makedec{\Theta, \guess{\alpha}}}{\makedec{\Theta}}{\bydefmakedec}
      \continueeqPf{\makedec{\Theta'}}{\byih}
      \continueeqPf{\makedec{\Theta', \guess{\alpha}}}{\bydefmakedec}
    \end{llproof}

    \DerivationProofCase{\Wcsolveguess}
      {\congoesWeakJudg{\Theta}{\Theta'}}
      {\congoesWeakJudg{\Theta, \guess{\alpha}}{\Theta', \guess{\alpha} = P}}

    \begin{llproof}
      \eqPf{\makedec{\Theta, \guess{\alpha}}}{\makedec{\Theta}}{\bydefmakedec}
      \continueeqPf{\makedec{\Theta'}}{\byih}
      \continueeqPf{\makedec{\Theta', \guess{\alpha} = P}}{\bydefmakedec}
    \end{llproof}

    \DerivationProofCase{\Wcsolvedguess}
      {
        \congoesWeakJudg{\acontext}{\acontext'} \\
        \MutualSubtypePosJudge{\makedec{\acontext}} {P} {Q}
      }
      {\congoesWeakJudg{\acontext, \guess{\alpha} = P}{\acontext', \guess{\alpha} = Q}}

    \begin{llproof}
      \eqPf{\makedec{\Theta, \guess{\alpha} = P}}{\makedec{\Theta}}{\bydefmakedec}
      \continueeqPf{\makedec{\Theta'}}{\byih}
      \continueeqPf{\makedec{\Theta', \guess{\alpha} = Q}}{\bydefmakedec}
    \end{llproof}

    \DerivationProofCase{\Wcunsolvedextend}
    {\congoesWeakJudg{\Theta}{\Theta'}}
    {\congoesWeakJudg{\Theta}{\Theta', \guess{\alpha}}}

    \begin{llproof}
      \eqPf{\makedec{\Theta}}{\makedec{\Theta'}}{\byih}
      \continueeqPf{\makedec{\Theta', \guess{\alpha}}}{\bydefmakedec}
    \end{llproof}

    \DerivationProofCase{\Wcsolvedextend}
    {\congoesWeakJudg{\Theta}{\Theta'}}
    {\congoesWeakJudg{\Theta}{\Theta', \guess{\alpha}} = P}

    \begin{llproof}
      \eqPf{\makedec{\Theta}}{\makedec{\Theta'}}{\byih}
      \continueeqPf{\makedec{\Theta', \guess{\alpha} = P}}{\bydefmakedec}
    \end{llproof}
  \end{itemize}
\end{proof}

\WeakCtxExtTransitive*

\begin{proof}
  By rule induction over the $\congoesWeakJudg{\acontext'}{\acontext''}$ judgment.

  \begin{itemize}
    \item Neither \Wcunsolvedextend nor \Wcunsolvedextend:
    
    By rule induction over the $\congoesWeakJudg{\acontext}{\acontext'}$ judgment.

    \begin{itemize}
      \DerivationProofCase{\Wcempty}
        {}
        {\congoesWeakJudg{\emptyacontext}{\emptyacontext}}

      \begin{llproof}
  \Hand \congoesWeakPf{\emptyacontext}{\acontext''}{Assumption}
      \end{llproof}

      \DerivationProofCase{\Wcuvar}
        {\congoesWeakJudg{\Theta}{\Theta'}}
        {\congoesWeakJudg{\Theta, \alpha}{\Theta', \alpha}}

        By inversion on the second assumption (\Wcuvar), the last context must be $\acontext'', \alpha$.
    
        \begin{llproof}
          \congoesWeakPf{\acontext'}{\acontext''}{Inversion (\Wcuvar)}
          \congoesWeakPf{\acontext}{\acontext''}{\byih}
  \Hand   \congoesWeakPf{\acontext, \alpha}{\acontext'', \alpha}{By {\Wcuvar}}
        \end{llproof}

      \DerivationProofCase{\Wcunsolvedguess}
        {\congoesWeakJudg{\Theta}{\Theta'}}
        {\congoesWeakJudg{\Theta, \guess{\alpha}}{\Theta', \guess{\alpha}}}

      By inversion on the second assumption, we must have either $\congoesWeakJudg{\Theta', \guess{\alpha}}{\Theta'', \guess{\alpha}}$ (\Wcunsolvedguess) or
      $\congoesWeakJudg{\Theta', \guess{\alpha}}{\Theta'', \guess{\alpha} = P}$ (\Wcsolveguess):

      \begin{itemize}
        \caseitem{$\congoesWeakJudg{\Theta', \guess{\alpha}}{\Theta'', \guess{\alpha}}$}

        \begin{llproof}
          \congoesWeakPf{\Theta'}{\Theta''}{Inversion (\Wcunsolvedguess)}
          \proofsep
          \congoesWeakPf{\Theta}{\Theta''}{\byih}
  \Hand   \congoesWeakPf{\Theta, \guess{\alpha}}{\Theta'', \guess{\alpha}}{By {\Wcunsolvedguess}}
        \end{llproof}

        \caseitem{$\congoesWeakJudg{\Theta', \guess{\alpha}}{\Theta'', \guess{\alpha} = P}$}

        \begin{llproof}
          \congoesWeakPf{\Theta'}{\Theta''}{Inversion (\Wcsolveguess)}
          \proofsep
          \congoesWeakPf{\Theta}{\Theta''}{\byih}
  \Hand   \congoesWeakPf{\Theta, \guess{\alpha}}{\Theta'', \guess{\alpha} = P}{By {\Wcsolveguess}}
        \end{llproof}
    \end{itemize}

      \DerivationProofCase{\Wcsolvedguess}
        {
          \congoesWeakJudg{\acontext}{\acontext'} \\
          \MutualSubtypePosJudge{\makedec{\acontext}} {P} {Q}
        }
        {\congoesWeakJudg{\acontext, \guess{\alpha} = P}{\acontext', \guess{\alpha} = Q}}

      By inversion on the second assumption (\Wcsolvedguess), the last context must be of the form $\acontext'', \guess{\alpha} = R$.

      \begin{llproof}
        \congoesWeakPf{\Theta', \guess{\alpha} = Q}{\Theta'', \guess{\alpha} = R}{Assumption}
        \congoesWeakPf{\Theta'}{\Theta''}{Inversion (\Wcsolvedguess)}
        \MutualJudgePosPf{\makedec{\acontext'}}{Q}{R}{\ditto}
        \proofsep
        \congoesWeakPf{\Theta}{\Theta''}{\byih}
        \MutualJudgePosPf{\makedec{\acontext}}{P}{Q}{Premise}
        \MutualJudgePosPf{\makedec{\acontext}}{Q}{R}{By \lemmaref{lemma:weak context extension equal declarative contexts}}
        \MutualJudgePosPf{\makedec{\acontext}}{P}{R}{By \lemmaref{lemma:Transitivity of declarative pnsubtype}}
  \Hand \congoesWeakPf{\Theta, \guess{\alpha} = P}{\Theta'', \guess{\alpha} = R}{By {\Wcsolvedguess}}
      \end{llproof}

      \DerivationProofCase{\Wcsolveguess}
        {\congoesWeakJudg{\Theta}{\Theta'}}
        {\congoesWeakJudg{\Theta, \guess{\alpha}}{\Theta', \guess{\alpha} = P}}

      By inversion on the second assumption (\Wcsolvedguess), the last context must be of the form $\acontext'', \guess{\alpha} = Q$.

      \begin{llproof}
        \congoesWeakPf{\Theta', \guess{\alpha} = P}{\Theta'', \guess{\alpha} = Q}{Assumption}
        \congoesWeakPf{\Theta'}{\Theta''}{Inversion (\Wcsolvedguess)}
        \proofsep
        \congoesWeakPf{\Theta}{\Theta''}{\byih}
  \Hand \congoesWeakPf{\Theta, \guess{\alpha}}{\Theta'', \guess{\alpha} = Q}{By {\Wcsolveguess}}
      \end{llproof}

      \DerivationProofCase{\Wcunsolvedextend}
        {\congoesWeakJudg{\Theta}{\Theta'}}
        {\congoesWeakJudg{\Theta}{\Theta', \guess{\alpha}}}

      By inversion on the second assumption, we must have either $\congoesWeakJudg{\Theta', \guess{\alpha}}{\Theta'', \guess{\alpha}}$ (\Wcunsolvedguess) or $\congoesWeakJudg{\Theta', \guess{\alpha}}{\Theta'', \guess{\alpha} = P}$ (\Wcsolvedguess):

      \begin{itemize}
        \caseitem{$\congoesWeakJudg{\Theta', \guess{\alpha}}{\Theta'', \guess{\alpha}}$}

        \begin{llproof}
          \congoesWeakPf{\Theta', \guess{\alpha}}{\Theta'', \guess{\alpha}}{Assumption}
          \congoesWeakPf{\Theta'}{\Theta''}{Inversion (\Wcunsolvedguess)}
          \proofsep
          \congoesWeakPf{\Theta}{\Theta''}{\byih}
          \congoesWeakPf{\Theta}{\Theta'', \guess{\alpha}}{By {\Wcunsolvedextend}}
        \end{llproof}

        \caseitem{$\congoesWeakJudg{\Theta', \guess{\alpha}}{\Theta'', \guess{\alpha} = P}$}

        \begin{llproof}
          \congoesWeakPf{\Theta', \guess{\alpha}}{\Theta'', \guess{\alpha} = P}{Assumption}
          \congoesWeakPf{\Theta'}{\Theta''}{Inversion (\Wcsolvedguess)}
          \proofsep
          \congoesWeakPf{\Theta}{\Theta''}{\byih}
          \congoesWeakPf{\Theta}{\Theta'', \guess{\alpha} = P}{By {\Wcsolvedextend}}
        \end{llproof}
      \end{itemize}

      \DerivationProofCase{\Wcsolvedextend}
        {\congoesWeakJudg{\Theta}{\Theta'}}
        {\congoesWeakJudg{\Theta}{\Theta', \guess{\alpha}} = P}

      By inversion on the second assumption (\Wcsolvedguess), the last context must be of the form $\acontext'', \guess{\alpha} = Q$.

      \begin{llproof}
        \congoesWeakPf{\Theta', \guess{\alpha} = P}{\Theta'', \guess{\alpha} = Q}{Assumption}
        \congoesWeakPf{\Theta'}{\Theta''}{Inversion (\Wcsolvedguess)}
        \proofsep
        \congoesWeakPf{\Theta}{\Theta''}{\byih}
        \congoesWeakPf{\Theta}{\Theta'', \guess{\alpha} = Q}{By {\Wcsolvedextend}}
      \end{llproof}
    \end{itemize}

    \DerivationProofCase{\Wcunsolvedextend}
      {\congoesWeakJudg{\Theta'}{\Theta''}}
      {\congoesWeakJudg{\Theta'}{\Theta'', \guess{\alpha}}}

      \begin{llproof}
        \congoesWeakPf{\Theta}{\Theta''}{\byih}
\Hand   \congoesWeakPf{\Theta}{\Theta'', \guess{\alpha}}{By \Wcunsolvedextend}
      \end{llproof}

    \DerivationProofCase{\Wcunsolvedextend}
      {\congoesWeakJudg{\Theta'}{\Theta''}}
      {\congoesWeakJudg{\Theta'}{\Theta'', \guess{\alpha} = P}}

    \begin{llproof}
      \congoesWeakPf{\Theta}{\Theta''}{\byih}
\Hand   \congoesWeakPf{\Theta}{\Theta'', \guess{\alpha} = P}{By \Wcsolvedextend}
    \end{llproof}
  \end{itemize}
\end{proof}

\begin{center}
  \WeakContextExtensionPreservesWF*
\end{center}

\begin{proof}
  By rule induction over the $\wfposnegtypeJudg{\acontext}{A}$ judgment.
  
  \begin{itemize}
    \DerivationProofCase{\twfuvar}
      {\alpha \in \FreeUV(\acontext)}
      {\wfpostypeJudg{\acontext}{\alpha}}

      \begin{llproof}
        \inPf{\alpha}{\FreeUV(\acontext)}{Premise}
        \inPf{\alpha}{\FreeUV(\acontext')}{Inversion (must have instance of \Wcuvar)}
  \Hand \wfpostypePf{\acontext'}{\alpha}{By {\twfuvar}}
      \end{llproof}

    \DerivationProofCase{\twfguess}
      {\guess{\alpha} \in \FreeEV(\acontext)}
      {\wfpostypeJudg{\acontext}{\guess{\alpha}}}

      \begin{llproof}
        \inPf{\guess{\alpha}}{\FreeEV(\acontext)}{Premise}
        \inPf{\guess{\alpha}}{\FreeEV(\acontext')}{Inversion (must have instance of \Wcunsolvedguess, \Wcsolveguess, or \Wcsolvedguess)}
  \Hand \wfpostypePf{\acontext'}{\guess{\alpha}}{By {\twfguess}}
      \end{llproof}

    \DerivationProofCase{\twfshiftdown}
      {\wfnegtypeJudg{\acontext}{N}}
      {\wfpostypeJudg{\acontext}{\shiftd{N}}}

      \begin{llproof}
        \wfnegtypePf{\acontext'}{N}{\byih}
  \Hand \wfnegtypePf{\acontext'}{\shiftd{N}}{By {\twfshiftdown}}
      \end{llproof}

    \DerivationProofCase{\twfforall}
      {\wfnegtypeJudg{\acontext, \alpha}{N}}
      {\wfnegtypeJudg{\acontext}{\forall\alpha\ldotp N}}
      
      \begin{llproof}
        \congoesWeakPf{\acontext}{\acontext'}{Assumption}
        \congoesWeakPf{\acontext, \alpha}{\acontext', \alpha}{By {\Wcuvar}}
        \wfnegtypePf{\acontext',\alpha}{N}{\byih}
  \Hand \wfnegtypePf{\acontext'}{\forall\alpha\ldotp N}{By {\twfforall}}
      \end{llproof}

    \DerivationProofCase{\twfarrow}
      {\wfpostypeJudg{\acontext}{P} \\ \wfnegtypeJudg{\acontext}{N}}
      {\wfnegtypeJudg{\acontext}{P \funarrow N}}

      \begin{llproof}
        \wfpostypePf{\acontext'}{P}{\byih}
        \wfnegtypePf{\acontext'}{N}{\byih}
  \Hand \wfnegtypePf{\acontext'}{P \funarrow N}{By {\twfarrow}}
      \end{llproof}

    \DerivationProofCase{\twfshiftup}
      {\wfpostypeJudg{\acontext}{P}}
      {\wfnegtypeJudg{\acontext}{\shiftu{P}}}

      \begin{llproof}
        \wfpostypePf{\acontext'}{P}{\byih}
        \wfnegtypePf{\acontext'}{\shiftu{P}}{By {\twfshiftup}}
      \end{llproof}
  \end{itemize}
\end{proof}

\begin{center}
  \ContextExtWFEnvs*
\end{center}

\begin{proof}
  By rule induction over the definition of well-formed typing environments.

  \begin{itemize}
    \DerivationProofCase{\Ewfempty}
    {}
    {\envwfJudg{\Theta}{\cdot}}

    \begin{llproof}
\Hand   \envwfPf{\Theta'}{\cdot}{By {\Ewfempty}}
    \end{llproof}

    \DerivationProofCase{\Ewfvar}
    {\envwfJudg{\Theta}{\Gamma} \\ \wfpostypeJudg{\Theta}{P} \\ \groundJudge{P}}
    {\envwfJudg{\Theta}{\Gamma, x : P}}

    \begin{llproof}
      \envwfPf{\Theta}{\Gamma, x : P}{Assumption}
      \proofsep
      \envwfPf{\Theta}{\Gamma}{By premise}
      \envwfPf{\Theta'}{\Gamma}{\byih}
      \proofsep
      \wfposnegtypePf{\Theta}{P}{By premise}
      \wfposnegtypePf{\Theta'}{P}{By \cref{lemma:weak context extension preserves well-formedness}}
      \proofsep
      \groundPf{P}{By premise}
      \proofsep
\Hand \envwfPf{\Theta'}{\Gamma, x : P}{By {\Ewfvar}}
    \end{llproof}
  \end{itemize}
\end{proof}

\WeakExtendedContextMakesGround*

\begin{proof}
  Consider an arbitrary existential variable $\guess{\alpha}$ in $A$.
  Then for $\subcon{\acontext'} \subcon{\acontext} A$ to be ground, we must have at least one of $\guess{\alpha} = P \in \acontext$, or $\guess{\alpha} = Q \in \acontext'$.
  We know that applying the contexts to the type will never introduce a non-ground type since $\conwfJudg{\acontext}$ and $\conwfJudg{\acontext'}$.

  By inversion on $\congoesWeakJudg{\acontext}{\acontext'}$, we can also see that if an existential variable is solved in the \lhs context, then it must also be solved in the \rhs context.
  Therefore we must have that $\guess{\alpha} = Q \in \acontext'$, and by $\conwfJudg{\acontext'}$ we know that $Q$ is ground.

  We now know that every existential variable in $A$ is solved as a ground type by $\acontext'$, hence $\subcon{\acontext'} A$ must be ground.
\end{proof}

\WeakExtendingContextPreservesGroundness*

\begin{proof}
  Corollary of \lemmaref{lemma:The extended context makes the type ground (weak)}.

  \begin{llproof}
    \conwfPf{\acontext}
      {Assumption}
    \conwfPf{\acontext'}
      {Assumption}
    \congoesWeakPf{\acontext}
      {\acontext'}
      {Assumption}
    \groundPf{\subcon{\acontext} A}
      {Assumption}
    \groundPf{\subcon{\acontext'} \subcon{\acontext} A}
      {By \lemmaref{lemma:Context substitution on ground terms}}
\Hand \groundPf{\subcon{\acontext'} A}
      {By \lemmaref{lemma:The extended context makes the type ground (weak)}}
  \end{llproof}
\end{proof}

\section{Context extension}

\ContextExtensionReflexive*

\newcommand{\congoesReflPf}[2]{\congoesPf{#1}{#1}{#2}}

\begin{proof}
  By structural induction on $\acontext$.

  \begin{itemize}
    \caseitem{$\emptyacontext$}

    \begin{llproof}
\Hand \congoesReflPf{\emptyacontext}{By \Cempty}
    \end{llproof}
  
    \caseitem{$\acontext, \alpha$}

    \begin{llproof}
      \congoesReflPf{\acontext}{\byih}
\Hand \congoesReflPf{\acontext, \alpha}{By \Cuvar}
    \end{llproof}
  
    \caseitem{$\acontext, \guess{\alpha}$}
    
    \begin{llproof}
      \congoesReflPf{\acontext}{\byih}
\Hand \congoesReflPf{\acontext, \guess{\alpha}}{By \Cunsolvedguess}
    \end{llproof}
  
    \caseitem{$\acontext, \guess{\alpha} = P$}

    \begin{llproof}
      \congoesPf{\acontext}{\acontext}{\byih}
      \MutualJudgePosPf{\acontext}{P}{P}{By \lemmaref{lemma:Reflexivity of declarative pnsubtype}}
\Hand \congoesPf{\acontext, \guess{\alpha} = P}{\acontext, \guess{\alpha} = P}{By \Csolvedguess}
    \end{llproof}
  \end{itemize}
\end{proof}

\EqualityOfDeclarativeContexts*

\begin{proof}
  Corollary of \lemmaref{lemma:weak context extension equal declarative contexts}.

  \begin{llproof}
    \congoesPf{\acontext}{\acontext'}{Assumption}
    \congoesWeakPf{\acontext}{\acontext'}{By \lemmaref{lemma:weak context extension subsumes normal}}
\Hand \eqPf{\makedec{\acontext}}{\makedec{\acontext'}}{By \lemmaref{lemma:weak context extension equal declarative contexts}}
  \end{llproof}
\end{proof}

\congoesTransitive*

\begin{proof}
  By rule induction over the $\congoesJudg{\acontext}{\acontext'}$ judgment.

  \begin{itemize}
    \DerivationProofCase{\Cempty}
      {}
      {\congoesJudg{\emptyacontext}{\emptyacontext}}

    \begin{llproof}
\Hand \congoesPf{\emptyacontext}{\acontext''}{Assumption}
    \end{llproof}

    \DerivationProofCase{\Cuvar}
      {\congoesJudg{\Theta}{\Theta'}}
      {\congoesJudg{\Theta, \alpha}{\Theta', \alpha}}

      By inversion on the second assumption (\Cuvar), the last context must be $\acontext'', \alpha$.
  
      \begin{llproof}
        \congoesPf{\acontext'}{\acontext''}{Inversion (\Cuvar)}
        \congoesPf{\acontext}{\acontext''}{\byih}
\Hand   \congoesPf{\acontext, \alpha}{\acontext'', \alpha}{By {\Cuvar}}
      \end{llproof}

    \DerivationProofCase{\Cunsolvedguess}
      {\congoesJudg{\Theta}{\Theta'}}
      {\congoesJudg{\Theta, \guess{\alpha}}{\Theta', \guess{\alpha}}}

    By inversion on the second assumption, we must have either $\congoesJudg{\Theta', \guess{\alpha}}{\Theta'', \guess{\alpha}}$ (\Cunsolvedguess) or
    $\congoesJudg{\Theta', \guess{\alpha}}{\Theta'', \guess{\alpha} = P}$ (\Csolveguess):

    \begin{itemize}
      \caseitem{$\congoesJudg{\Theta', \guess{\alpha}}{\Theta'', \guess{\alpha}}$}

      \begin{llproof}
        \congoesPf{\Theta'}{\Theta''}{Inversion (\Cunsolvedguess)}
        \proofsep
        \congoesPf{\Theta}{\Theta''}{\byih}
\Hand   \congoesPf{\Theta, \guess{\alpha}}{\Theta'', \guess{\alpha}}{By {\Cunsolvedguess}}
      \end{llproof}

      \caseitem{$\congoesJudg{\Theta', \guess{\alpha}}{\Theta'', \guess{\alpha} = P}$}

      \begin{llproof}
        \congoesPf{\Theta'}{\Theta''}{Inversion (\Csolveguess)}
        \proofsep
        \congoesPf{\Theta}{\Theta''}{\byih}
\Hand   \congoesPf{\Theta, \guess{\alpha}}{\Theta'', \guess{\alpha} = P}{By {\Csolveguess}}
      \end{llproof}
  \end{itemize}

    \DerivationProofCase{\Csolvedguess}
      {
        \congoesJudg{\acontext}{\acontext'} \\
        \MutualSubtypePosJudge{\makedec{\acontext}} {P} {Q}
      }
      {\congoesJudg{\acontext, \guess{\alpha} = P}{\acontext', \guess{\alpha} = Q}}

    By inversion on the second assumption (\Csolvedguess), the last context must be of the form $\acontext'', \guess{\alpha} = R$.

    \begin{llproof}
      \congoesPf{\Theta', \guess{\alpha} = Q}{\Theta'', \guess{\alpha} = R}{Assumption}
      \congoesPf{\Theta'}{\Theta''}{Inversion (\Csolvedguess)}
      \MutualJudgePosPf{\makedec{\acontext'}}{Q}{R}{\ditto}
      \proofsep
      \congoesPf{\Theta}{\Theta''}{\byih}
      \MutualJudgePosPf{\makedec{\acontext}}{P}{Q}{Premise}
      \MutualJudgePosPf{\makedec{\acontext}}{Q}{R}{By \lemmaref{lemma:equal declarative contexts}}
      \MutualJudgePosPf{\makedec{\acontext}}{P}{R}{By \lemmaref{lemma:Transitivity of declarative pnsubtype}}
\Hand \congoesPf{\Theta, \guess{\alpha} = P}{\Theta'', \guess{\alpha} = R}{By {\Csolvedguess}}
    \end{llproof}

    \DerivationProofCase{\Csolveguess}
      {\congoesJudg{\Theta}{\Theta'}}
      {\congoesJudg{\Theta, \guess{\alpha}}{\Theta', \guess{\alpha} = P}}

    By inversion on the second assumption (\Csolvedguess), the last context must be of the form $\acontext'', \guess{\alpha} = Q$.

    \begin{llproof}
      \congoesPf{\Theta', \guess{\alpha} = P}{\Theta'', \guess{\alpha} = Q}{Assumption}
      \congoesPf{\Theta'}{\Theta''}{Inversion (\Csolvedguess)}
      \proofsep
      \congoesPf{\Theta}{\Theta''}{\byih}
\Hand \congoesPf{\Theta, \guess{\alpha}}{\Theta'', \guess{\alpha} = Q}{By {\Csolveguess}}
    \end{llproof}
  \end{itemize}
\end{proof}

\RestateContextExtensionTermWf*

\begin{proof}
  By rule induction on $\wfposnegtypeJudg{\acontext} {A}$.

  \begin{itemize}
    \DerivationProofCase{\twfuvar}
      {\alpha \in \ConUV(\acontext)}
      {\wfpostypeJudg{\acontext} {\alpha}}
    
    \begin{llproof}
      \inPf{\alpha}
        {\ConUV(\acontext)}
        {Subderivation}
      \congoesPf{\acontext}
        {\acontext'}
        {Assumption}
      \inPf{\alpha}
        {\ConUV(\acontext')}
        {Inversion (\Cuvar)}
\Hand \wfpostypePf{\acontext'}
        {\alpha}
        {By \twfuvar}
    \end{llproof}

    \DerivationProofCase{\twfguess}
      {\guess{\alpha} \in \EV(\acontext)}
      {\wfpostypeJudg{\acontext} {\guess{\alpha}}}
    
      \begin{llproof}
        \congoesPf{\acontext}
          {\acontext'}
          {Assumption}
        \inPf{\guess{\alpha}}
          {\EV(\acontext)}
          {Subderivation}
        \inPf{\guess{\alpha}}
          {\EV(\acontext')}
          {Must have an instance of \Cunsolvedguess, \Csolveguess, or \Csolvedguess}
  \Hand \wfpostypePf{\acontext'}
          {\guess{\alpha}}
          {By \twfuvar}
      \end{llproof}

    \DerivationProofCase{\twfshiftdown}
      {\wfnegtypeJudg{\acontext} {N}}
      {\wfpostypeJudg{\acontext} {\shiftdown N}}

    \begin{llproof}
      \wfnegtypePf{\acontext}
        {N}
        {Subderivation}
      \congoesPf{\acontext}
        {\acontext'}
        {Assumption}
      \wfnegtypePf{\acontext'}
        {N}
        {By i.h.}
\Hand \wfpostypePf{\acontext'}
        {\shiftdown N}
        {By \twfshiftdown}
    \end{llproof}

    \DerivationProofCase{\twfforall}
      {\wfnegtypeJudg{\acontext, \alpha} {N}}
      {\wfnegtypeJudg{\acontext} {\forall \alpha \ldotp N}}
    
    \begin{llproof}
      \wfnegtypePf{\acontext, \alpha}
        {N}
        {Subderivation}
      \congoesPf{\acontext}
        {\acontext'}
        {Assumption}
      \congoesPf{\acontext, \alpha}
        {\acontext', \alpha}
        {By \Cuvar}
      \wfnegtypePf{\acontext', \alpha}
        {N}
        {By i.h.}
\Hand \wfnegtypePf{\acontext'}
        {\forall \alpha \ldotp N}
        {By \twfforall}
    \end{llproof}
  
    \DerivationProofCase{\twfarrow}
      {\wfpostypeJudg{\acontext} {P} \\ \wfnegtypeJudg{\acontext} {N}}
      {\wfnegtypeJudg{\acontext} {P \funarrow N}}

    \begin{llproof}
      \wfpostypePf{\acontext}
        {P}
        {Subderivation}
      \congoesPf{\acontext}
        {\acontext'}
        {Assumption}
      \wfpostypePf{\acontext'}
        {P}
        {By i.h.}
      \wfnegtypePf{\acontext}
        {N}
        {Subderivation}
      \wfnegtypePf{\acontext'}
        {N}
        {By i.h.}
\Hand \wfnegtypePf{\acontext'}
        {P \funarrow N}
        {By \twfarrow}
    \end{llproof}

    \DerivationProofCase{\twfshiftup}
      {\wfpostypeJudg{\acontext} {P}}
      {\wfnegtypeJudg{\acontext} {\shiftup P}}

    \begin{llproof}
      \wfpostypePf{\acontext}
        {P}
        {Subderivation}
      \congoesPf{\acontext}
        {\acontext'}
        {\ditto}
      \wfpostypePf{\acontext'}
        {P}
        {By i.h.}
\Hand \wfnegtypePf{\acontext'}
        {\shiftup P}
        {By \twfshiftup}
    \end{llproof}
  \end{itemize}
\end{proof}

\RestateApplyingContextToGroundTypes*

\begin{proof}
  By structural induction on $\acontext$.

  \begin{itemize}
    \caseitem{$\emptyacontext$}

    \begin{llproof}
      \eqPf{\subcon{\emptyacontext} A}
        {A}
        {\bydefsubcon}
    \end{llproof}

    \caseitem{$\acontext, \alpha$}

    \begin{llproof}
      \eqPf{\subcon{\acontext, \alpha} A}
        {\subcon{\acontext} A}
        {\bydefsubcon}
      \continueeqPf{A}
        {\byih}
    \end{llproof}

    \caseitem{$\acontext, \guess{\alpha}$}

    \begin{llproof}
      \eqPf{\subcon{\acontext, \guess{\alpha}} A}
        {\subcon{\acontext} A}
        {\bydefsubcon}
      \continueeqPf{A}
        {\byih}
    \end{llproof}

    \caseitem{$\acontext, \guess{\alpha} = P$}

    \begin{llproof}
      \eqPf{\subcon{\acontext, \guess{\alpha} = P} A}
        {\subcon{\acontext} (\subcon{P / \guess{\alpha}} A)}
        {\bydefsubcon}
      \continueeqPf{\subcon{\acontext} A}
        {$\groundJudge{A}$, so no $\guess{\alpha}$s to substitute}
      \continueeqPf{A}
        {\byih}
    \end{llproof}
  \end{itemize}
\end{proof}

\RestateContextSubstitutionIdempotent*

\begin{proof}
  By structural induction on $\acontext$:.

  \begin{itemize}
    \caseitem{$\emptyacontext$}

    \begin{llproof}
      \eqPf{\subcon{\emptyacontext} \subcon{\emptyacontext} A}
        {A}
        {\bydefsubcon}
    \end{llproof}

    \caseitem{$\acontext, \alpha$}

    \begin{llproof}
      \eqPf{\subcon{\acontext, \alpha} \subcon{\acontext, \alpha} A}
        {\subcon{\acontext} \subcon{\acontext} A}
        {\bydefsubcon}
      \continueeqPf{A}
        {\byih}
    \end{llproof}

    \caseitem{$\acontext, \guess{\alpha}$}

    \begin{llproof}
      \eqPf{\subcon{\acontext, \guess{\alpha}} \subcon{\acontext, \guess{\alpha}} A}
        {\subcon{\acontext} \subcon{\acontext} A}
        {\bydefsubcon}
      \continueeqPf{A}
        {\byih}
    \end{llproof}

    \caseitem{$\acontext, \guess{\alpha} = P$}

    \begin{llproof}
      \eqPf{\subcon{\acontext, \guess{\alpha} = P} \subcon{\acontext, \guess{\alpha} = P} A}
        {\subcon{\acontext} [P / \guess{\alpha}] \subcon{\acontext} [P / \guess{\alpha}] A}
        {\bydefsubcon}
      \continueeqPf{[P / \guess{\alpha}] \subcon{\acontext} \subcon{\acontext} A}
        {$\groundJudge{P}$ and $\conwfJudg{\acontext, \guess{\alpha} = P}$, so $\guess{\alpha}$ does not reappear}
      \continueeqPf{[P / \guess{\alpha}] \subcon{\acontext} A}
        {\byih}
      \continueeqPf{\subcon{\acontext} [P / \guess{\alpha}] A}
        {$\groundJudge{P}$ and $\conwfJudg{\acontext, \guess{\alpha} = P}$, so $\guess{\alpha}$ does not reappear}
      \continueeqPf{\subcon{\acontext, \guess{\alpha} = P} A}
        {\bydefsubcon}
    \end{llproof}
  \end{itemize}
\end{proof}

\ExtendedContextMakesGround*

\begin{proof}
  Corollary of \lemmaref{lemma:The extended context makes the type ground (weak)}.

  \begin{llproof}
    \conwfPf{\acontext}{Assumption}
    \conwfPf{\acontext'}{Assumption}
    \congoesPf{\acontext}{\acontext'}{Assumption}
    \congoesWeakPf{\acontext}{\acontext'}{By \lemmaref{lemma:weak context extension subsumes normal}}
    \groundPf{\subcon{\acontext'} \subcon{\acontext} A}{Assumption}
    \groundPf{\subcon{\acontext'} A}{By \lemmaref{lemma:The extended context makes the type ground (weak)}}
  \end{llproof}
\end{proof}

\ExtendingContextPreservesGroundness*

\begin{proof}
  Corollary of \lemmaref{lemma:extending-context-preserves-groundness-weak}.

  \begin{llproof}
    \conwfPf{\acontext}{Assumption}
    \conwfPf{\acontext'}{Assumption}
    \congoesPf{\acontext}{\acontext'}{Assumption}
    \congoesWeakPf{\acontext}{\acontext'}{By \lemmaref{lemma:weak context extension subsumes normal}}
    \groundPf{\subcon{\acontext} A}{Assumption}
    \groundPf{\subcon{\acontext'} A}{By \lemmaref{lemma:extending-context-preserves-groundness-weak}}
  \end{llproof}
\end{proof}

\section{Well-formedness of subtyping}

\RestateCwfSubTwf*

\begin{proof}
  By rule induction on $\wfposnegtypeJudg{\acontext} {A}$.

  \begin{itemize}
    \DerivationProofCase{\twfuvar}
      {\alpha \in \ConUV(\acontext)}
      {\wfpostypeJudg{\acontext} {\alpha}}
    
    \begin{llproof}
      \wfpostypePf{\acontext}
        {\alpha}
        {Assumption}
      \eqPf{\subcon{\acontext} \alpha}
        {\alpha}
        {\bydefsubcon}
\Hand \wfpostypePf{\acontext}
        {\subcon{\acontext} \alpha}
        {By above two statements}
    \end{llproof}

    \DerivationProofCase{\twfguess}
      {\guess{\alpha} \in \EV(\acontext)}
      {\wfpostypeJudg{\acontext} {\guess{\alpha}}}
    
    \begin{llproof}
      \proofcomment{\textbf{Case} $(\guess{\alpha} = P) \in \acontext$:}
      \conwfPf{\acontext}
        {Assumption}
      \wfpostypePf{\acontext}
        {P}
        {Must have an instance of \cwfsolvedguess}
      \eqPf{\subcon{\acontext} \guess{\alpha}}
        {P}
        {\bydefsubcon}
\Hand \wfpostypePf{\acontext}
        {\subcon{\acontext} \guess{\alpha}}
        {By above two statements}

      \proofcomment{\textbf{Case} $(\guess{\alpha} = P) \notin \acontext$:}
      \wfpostypePf{\acontext}
        {\guess{\alpha}}
        {Assumption}
      \NoSolvedVarsPf{\acontext}
        {\guess{\alpha}}
        {Since $(\guess{\alpha} = P) \notin \acontext$}
\Hand \wfpostypePf{\acontext}
        {\subcon{\acontext} \guess{\alpha}}
        {By above two statements}
    \end{llproof}

    \DerivationProofCase{\twfshiftdown}
      {\wfnegtypeJudg{\acontext} {N}}
      {\wfpostypeJudg{\acontext} {\shiftdown N}}

    \begin{llproof}
      \conwfPf{\acontext}
        {Assumption}
      \wfnegtypePf{\acontext}
        {N}
        {Subderivation}
      \wfnegtypePf{\acontext}
        {\subcon{\acontext} N}
        {By i.h.}
      \wfpostypePf{\acontext}
        {\shiftdown \subcon{\acontext} N}
        {By \twfshiftdown}
\Hand \wfpostypePf{\acontext}
        {\subcon{\acontext} \shiftdown N}
        {\bydefsubcon}
    \end{llproof}

    \DerivationProofCase{\twfforall}
      {\wfnegtypeJudg{\acontext, \alpha} {N}}
      {\wfnegtypeJudg{\acontext} {\forall \alpha \ldotp N}}
    
    \begin{llproof}
      \conwfPf{\acontext}
        {Assumption}
      \conwfPf{\acontext, \alpha}
        {By \cwfuvar}
      \wfnegtypePf{\acontext, \alpha}
        {N}
        {Subderivation}
      \wfnegtypePf{\acontext, \alpha}
        {\subcon{\acontext, \alpha} N}
        {By i.h.}
      \wfnegtypePf{\acontext, \alpha}
        {\subcon{\acontext} N}
        {\bydefsubcon}
      \wfnegtypePf{\acontext}
        {\forall \alpha \ldotp \subcon{\acontext} N}
        {By \twfforall}
\Hand \wfnegtypePf{\acontext}
        {\subcon{\acontext} \forall \alpha \ldotp N}
        {\bydefsubcon}
    \end{llproof}
  
    \DerivationProofCase{\twfarrow}
      {\wfpostypeJudg{\acontext} {P} \\ \wfnegtypeJudg{\acontext} {N}}
      {\wfnegtypeJudg{\acontext} {P \funarrow N}}

      \begin{llproof}
        \conwfPf{\acontext}
          {Assumption}
        \wfpostypePf{\acontext}
          {P}
          {Subderivation}
        \wfpostypePf{\acontext}
          {\subcon{\acontext} P}
          {By i.h.}
        \wfnegtypePf{\acontext}
          {N}
          {Subderivation}
        \wfnegtypePf{\acontext}
          {\subcon{\acontext} N}
          {By i.h.}
        \wfnegtypePf{\acontext}
          {\subcon{\acontext} P \funarrow \subcon{\acontext} N}
          {By \twfarrow}
  \Hand \wfnegtypePf{\acontext}
          {\subcon{\acontext} (P \funarrow N)}
          {\bydefsubcon}
      \end{llproof}

    \DerivationProofCase{\twfshiftup}
      {\wfpostypeJudg{\acontext} {P}}
      {\wfnegtypeJudg{\acontext} {\shiftup P}}

      \begin{llproof}
        \conwfPf{\acontext}
          {Assumption}
        \wfpostypePf{\acontext}
          {P}
          {Subderivation}
        \wfpostypePf{\acontext}
          {\subcon{\acontext} P}
          {By i.h.}
        \wfnegtypePf{\acontext}
          {\shiftup \subcon{\acontext} P}
          {By \twfshiftup}
  \Hand \wfnegtypePf{\acontext}
          {\subcon{\acontext} \shiftup P}
          {\bydefsubcon}
      \end{llproof}
  \end{itemize}
\end{proof}

\RestateWellFormednessPNSubtype*

\begin{proof}
  By mutual induction on the derivation of $\acontext \entails A \pnsubtype B \prodcon \acontext'$.

  \begin{itemize}
    \DerivationProofCase{\arefl}
      {}
      {\aconwithhole{\alpha} \entails \alpha \possubtype \alpha \prodcon \aconwithhole{\alpha}}

      \begin{llproof}
  \Hand \conwfPf{\aconwithhole{\alpha}}
          {Assumption}
  \Hand \congoesPf{\aconwithhole{\alpha}}
          {\aconwithhole{\alpha}}
          {By \lemmaref{lemma:context extension reflexive}}

  \proofsep
        \eqPf{\subcon{\aconwithhole{\alpha}} \alpha}
          {\alpha}
          {Assumption}
        \groundPf{\alpha}
          {Assumption}
  \Hand \groundPf{\subcon{\aconwithhole{\alpha}} \alpha}
          {By the previous two statements}
      \end{llproof}

    \DerivationProofCase{\ainst}
      {\acontext_L \entails P \postype \\ \groundJudge{P}}
      {\aconwithhole{\guess{\alpha}} \entails P \possubtype \guess{\alpha} \prodcon \aconwithhole{\guess{\alpha} = P}}

      \begin{llproof}
        \conwfPf{\aconwithhole{\guess{\alpha}}}
          {Assumption}
        \wfpostypePf{\acontext_L}
          {P}
          {Subderivation}
        \groundPf{P}
          {Assumption}
  \Hand \conwfPf{\aconwithhole{\guess{\alpha} = P}}
          {Replacing the instance of \cwfunsolvedguess corresponding}
          \trailingjust{to $\guess{\alpha}$ with an instance of \cwfsolvedguess}

  \proofsep
        \congoesPf{\acontext_L}
          {\acontext_L}
          {By \lemmaref{lemma:context extension reflexive}}
        \congoesPf{\acontext_L, \guess{\alpha}}
          {\acontext_L, \guess{\alpha} = P}
          {By \Csolveguess}
        \congoesPf{\acontext_R}
          {\acontext_R}
          {By \lemmaref{lemma:context extension reflexive}}
        \congoesPf{\aconwithhole{\guess{\alpha}}}
          {\aconwithhole{\guess{\alpha} = P}}
          {Reapplying rules from $\congoesJudg{\acontext_R}{\acontext_R}$}

  \proofsep
        \eqPf{\subcon{\aconwithhole{\guess{\alpha} = P}} \guess{\alpha}}
          {P}
          {\bydefsubcon}
  \Hand \groundPf{\subcon{\aconwithhole{\guess{\alpha} = P}} \guess{\alpha}}
          {By the previous two statements}
      \end{llproof}

    \DerivationProofCase{\ashiftdown}
      {\acontext \entails M \negsubtype N \prodcon \acontext' \\
       \acontext' \entails N \negsubtype \subcon{\acontext'} M \prodcon \acontext''}
      {\acontext \entails \shiftdown N \possubtype \shiftdown M \prodcon \acontext''}

      \begin{llproof}
        \groundPf{\shiftdown N}
          {Assumption}
        \eqPf{\subcon{\acontext} \shiftdown M}
          {\shiftdown M}
          {Assumption}

      \proofcomment{We have:}
        \negsubtypePf{\acontext \entails M}
          {N \prodcon \acontext'}
          {Subderivation}
        \conwfPf{\acontext}
          {Assumption}
        \groundPf{N}
          {By definition of ground}
        \eqPf{\subcon{\acontext} M}
          {M}
          {\bydefsubcon}

      \proofcomment{Therefore:}
        \conwfPf{\acontext'}
          {By i.h.}
        \congoesPf{\acontext}
          {\acontext'}
          {\ditto}
        \groundPf{\subcon{\acontext'} M}
          {\ditto}

    \proofsep
      \proofcomment{Now, looking at the second premise, we have:}
        \negsubtypePf{\acontext' \entails N}
          {\subcon{\acontext'} M \prodcon \acontext''}
          {Subderivation}
        \conwfPf{\acontext'}
          {Above}
        \groundPf{\subcon{\acontext'} M}
          {Above}
        \eqPf{\subcon{\acontext'} N}
          {N}
          {By \lemmaref{lemma:Context substitution on ground terms}}

      \proofcomment{Therefore:}
        \congoesPf{\acontext'}
          {\acontext''}
          {By i.h.}
  \Hand \conwfPf{\acontext''}
          {\ditto}

      \proofsep
  \Hand \congoesPf{\acontext}
          {\acontext''}
          {By \lemmaref{lemma:context extension transitive}}
        \groundPf{\subcon{\acontext''} M}
          {By \lemmaref{lemma:extending-context-preserves-groundness}}
  \Hand \groundPf{\subcon{\acontext''} \shiftdown M}
          {By definition of ground}
      \end{llproof}

    \DerivationProofCase{\aforallr}
      {\acontext, \alpha \entails N \negsubtype M \prodcon \acontext', \alpha}
      {\acontext \entails N \negsubtype \forall \alpha \ldotp M \prodcon \acontext'}

      \begin{llproof}
        \conwfPf{\acontext}
          {Assumption}
        \groundPf{\forall \alpha \ldotp M}
          {Assumption}
        \eqPf{\subcon{\acontext} N}
          {N}
          {Assumption}

        \proofcomment{We have:}
        \negsubtypePf{\acontext, \alpha \entails N}
          {M \prodcon \acontext', \alpha}
          {Subderivation}
        \conwfPf{\acontext, \alpha}
          {By \cwfuvar}
        \groundPf{M}
          {By definition of ground}
        \eqPf{\subcon{\acontext, \alpha} N}
          {N}
          {Since $\subcon{\acontext, \alpha} N = \subcon{\acontext} N$ by \defsubcon}

        \proofcomment{Therefore:}
        \conwfPf{\acontext', \alpha}
          {By i.h.}
        \congoesPf{\acontext, \alpha}
          {\acontext', \alpha}
          {\ditto}
        \groundPf{\subcon{\acontext', \alpha} N}
          {\ditto}

        \proofsep
  \Hand \conwfPf{\acontext'}
          {Inversion (\cwfuvar)}
  \Hand \congoesPf{\acontext}
          {\acontext'}
          {Inversion (\Cuvar)}
  \Hand \groundPf{\subcon{\acontext'} N}
          {Since $\subcon{\acontext'} N = \subcon{\acontext', \alpha} N$ by \defsubcon}
      \end{llproof}

    \DerivationProofCase{\aforalll}
      {\acontext, \guess{\alpha} \entails \subcon{\guess{\alpha} / \alpha} N \negsubtype M \prodcon \acontext', \guess{\alpha} \,[= P] \\ M \neq \forall \alpha \ldotp M'}
      {\acontext \entails \forall \alpha \ldotp N \negsubtype M \prodcon \acontext'}

      \begin{llproof}
        \conwfPf{\acontext}
          {Assumption}
        \eqPf{\subcon{\acontext} \forall \alpha \ldotp N}
          {\forall \alpha \ldotp N}
          {Assumption}
        \eqPf{\subcon{\acontext} N}
          {N}
          {\bydefsubcon}

        \proofcomment{We have:}
        \negsubtypePf{\acontext, \guess{\alpha} \entails \subcon{\guess{\alpha} / \alpha} N}
          {M \prodcon \acontext', \guess{\alpha} \,[= P]}
          {Subderivation}
        \conwfPf{\acontext, \guess{\alpha}}
          {By \cwfunsolvedguess}
        \groundPf{M}
          {Assumption}
        \NoSolvedVarsPf{\acontext}
          {\guess{\alpha}}
          {Since $\acontext, \guess{\alpha} \conwf$}
        \eqPf{\subcon{\acontext, \guess{\alpha}} \subcon{\guess{\alpha} / \alpha} N}
          {\subcon{\guess{\alpha} / \alpha} N}
          {Since $\subcon{\acontext} \guess{\alpha} = \guess{\alpha}$ and $\subcon{\acontext} N = N$}

        \proofcomment{Therefore:}
        \conwfPf{\acontext', \guess{\alpha} \,[= P]}
          {By i.h.}
        \congoesPf{\acontext, \guess{\alpha}}
          {\acontext', \guess{\alpha} \,[= P]}
          {\ditto}
        \groundPf{\subcon{\acontext', \guess{\alpha} \,[= P]} \subcon{\guess{\alpha} / \alpha} N}
          {\ditto}

        \proofsep
  \Hand \conwfPf{\acontext'}
          {Inversion (\cwfuvar)}
  \Hand \congoesPf{\acontext}
          {\acontext'}
          {Inversion (\Cuvar)}
        \groundPf{\subcon{\acontext'} N}
          {Using above, $\alpha$ ground, and $\guess{\alpha} \notin \FreeEV(N)$}
  \Hand \groundPf{\subcon{\acontext'} \forall \alpha \ldotp N}
          {By definition of ground and $\subconunderscores$}
      \end{llproof}

    \DerivationProofCase{\aarrow}
      {\acontext \entails Q \possubtype P \prodcon \acontext' \\
        \acontext' \entails \subcon{\acontext'} N \negsubtype M \prodcon \acontext''}
      {\acontext \entails P \funarrow N \negsubtype Q \funarrow M \prodcon \acontext''}

      \begin{llproof}
        \groundPf{Q \funarrow M}
          {Assumption}
        \eqPf{\subcon{\acontext} (P \funarrow N)}
          {P \funarrow N}
          {Assumption}

        \proofcomment{We have:}
        \possubtypePf{\acontext \entails Q}{P \prodcon \acontext'} {Subderivation}
        \conwfPf{\acontext}
          {Assumption}
        \groundPf{Q}
          {Since $Q \funarrow M$ ground}
        \eqPf{\subcon{\acontext} P}
          {P}
          {\bydefsubcon}

        \proofcomment{Therefore:}
        \conwfPf{\acontext'}
          {By i.h.}
        \congoesPf{\acontext}
          {\acontext'}
          {\ditto}
        \groundPf{\subcon{\acontext'} P}
          {\ditto}

        \proofcomment{Looking at the second premise, we have:}
        \negsubtypePf{\acontext' \entails \subcon{\acontext'} N}
          {M \prodcon \acontext''}
          {Subderivation}
        \conwfPf{\acontext'}
          {Above}
        \groundPf{M}
          {Since $Q \funarrow M$ ground}
        \eqPf{\subcon{\acontext'} \subcon{\acontext'} N}
          {\subcon{\acontext'} N}
          {By \lemmaref{lemma:Context substitution idempotence}}

        \proofcomment{Therefore:}
  \Hand \conwfPf{\acontext''}
          {By i.h.}
        \congoesPf{\acontext'}
          {\acontext''}
          {\ditto}
        \groundPf{\subcon{\acontext''} \subcon{\acontext'} N}
          {\ditto}

        \proofsep
  \Hand \congoesPf{\acontext}
          {\acontext''}
          {By \lemmaref{lemma:context extension transitive}}
        \groundPf{\subcon{\acontext''} N}
          {By \lemmaref{lemma:The extended context makes the type ground}}
        \groundPf{\subcon{\acontext''} P}
          {Applying \lemmaref{lemma:extending-context-preserves-groundness}}
          \trailingjust{with $\subcon{\acontext'} P$ ground}
  \Hand \groundPf{\subcon{\acontext''} P \funarrow N}
          {From equations above}
      \end{llproof}

    \DerivationProofCase{\ashiftup}
      {\acontext \entails Q \possubtype P \prodcon \acontext' \\
        \acontext' \entails \subcon{\acontext'} P \possubtype Q \prodcon \acontext''}
      {\acontext \entails \shiftup P \negsubtype \shiftup Q \prodcon \acontext''}

      \begin{llproof}
        \groundPf{\shiftup Q}
          {Assumption}
        \eqPf{\subcon{\acontext} \shiftup P}
          {\shiftup P}
          {Assumption}

        \proofcomment{We have:}
        \possubtypePf{\acontext \entails Q}
          {P \prodcon \acontext'}
          {Subderivation}
        \conwfPf{\acontext}
          {Assumption}
        \groundPf{Q}
          {Since $\shiftup Q$ ground}
        \eqPf{\subcon{\acontext} P}
          {P}
          {\bydefsubcon}

        \proofcomment{Therefore:}
        \conwfPf{\acontext'}
          {By i.h.}
        \congoesPf{\acontext}
          {\acontext'}
          {\ditto}
        \groundPf{\subcon{\acontext'} P}
          {\ditto}

        \proofcomment{Looking at the second premise, we have:}
        \possubtypePf{\acontext' \entails \subcon{\acontext'} P}
          {Q \prodcon \acontext''}
          {Subderivation}
        \conwfPf{\acontext'}
          {Above}
        \groundPf{\subcon{\acontext'} P}
          {Above}
        \eqPf{\subcon{\acontext'} Q}
          {Q}
          {By \lemmaref{lemma:Context substitution on ground terms}}

        \proofcomment{Therefore:}
  \Hand \conwfPf{\acontext''}
          {By i.h.}
        \congoesPf{\acontext'}
          {\acontext''}
          {\ditto}

        \proofsep
  \Hand \congoesPf{\acontext}
        {\acontext''}
        {By \lemmaref{lemma:context extension transitive}}
        \groundPf{\subcon{\acontext''} P}
          {Applying \lemmaref{lemma:extending-context-preserves-groundness}}
          \trailingjust{with $\subcon{\acontext'} P$ ground}
  \Hand \groundPf{\subcon{\acontext''} \shiftup P}
          {By definition of groundness and above}
      \end{llproof}
  \end{itemize}
\end{proof}

\section{Soundness of subtyping}

\subsection{Lemmas for soundness}

\RestateCompletingContextPreservesWf*

\begin{proof}
  By rule induction on $\wfposnegtypeJudg{\acontext} {A}$.

  \begin{itemize}
    \DerivationProofCase{\twfuvar}
      {\alpha \in \ConUV(\acontext)}
      {\wfpostypeJudg{\acontext} {\alpha}}
    
    \begin{llproof}
      \inPf{\alpha}
        {\ConUV(\acontext)}
        {Subderivation}
      \inPf{\alpha}
        {\ConUV(\subcon{\ccontext} \acontext)}
        {\bydefsubcon}
\Hand \wfpostypePf{\makedec{\acontext}}
        {\alpha}
        {By \twfuvar}
    \end{llproof}

    \DerivationProofCase{\twfguess}
      {\guess{\alpha} \in \EV(\acontext)}
      {\wfpostypeJudg{\acontext} {\guess{\alpha}}}
    
      Not possible, since $A$ is ground.

    \DerivationProofCase{\twfshiftdown}
      {\wfnegtypeJudg{\acontext} {N}}
      {\wfpostypeJudg{\acontext} {\shiftdown N}}

    \begin{llproof}
      \groundPf{\shiftdown N}
        {Assumption}
      \wfnegtypePf{\acontext}
        {N}
        {Subderivation}
      \groundPf{N}
        {\bydefground}
      \wfnegtypePf{\makedec{\acontext}}
        {N}
        {By i.h.}
\Hand \wfpostypePf{\makedec{\acontext}}
        {\shiftdown N}
        {By \twfshiftdown}
    \end{llproof}

    \DerivationProofCase{\twfforall}
      {\wfnegtypeJudg{\acontext, \alpha} {N}}
      {\wfnegtypeJudg{\acontext} {\forall \alpha \ldotp N}}
    
    \begin{llproof}
      \groundPf{\forall \alpha \ldotp N}
        {Assumption}
      \wfnegtypePf{\acontext, \alpha}
        {N}
        {Subderivation}
      \groundPf{N}
        {\bydefground}
      \wfnegtypePf{\subcon{\ccontext} (\acontext, \alpha)}
        {N}
        {By i.h.}
      \wfnegtypePf{\subcon{\ccontext} \acontext, \alpha}
        {N}
        {\bydefsubcon}
\Hand \wfnegtypePf{\makedec{\acontext}}
        {\forall \alpha \ldotp N}
        {By \twfforall}
    \end{llproof}
  
    \DerivationProofCase{\twfarrow}
      {\wfpostypeJudg{\acontext} {P} \\ \wfnegtypeJudg{\acontext} {N}}
      {\wfnegtypeJudg{\acontext} {P \funarrow N}}

    \begin{llproof}
      \groundPf{P \funarrow N}
        {Assumption}
      \wfpostypePf{\acontext}
        {P}
        {Subderivation}
      \groundPf{P}
        {\bydefground}
      \wfpostypePf{\makedec{\acontext}}
        {P}
        {By i.h.}
      \wfnegtypePf{\acontext}
        {N}
        {Subderivation}
      \groundPf{N}
        {\bydefground}
      \wfnegtypePf{\makedec{\acontext}}
        {N}
        {By i.h.}
\Hand \wfnegtypePf{\makedec{\acontext}}
        {P \funarrow N}
        {By \twfarrow}
    \end{llproof}

    \DerivationProofCase{\twfshiftup}
      {\wfpostypeJudg{\acontext} {P}}
      {\wfnegtypeJudg{\acontext} {\shiftup P}}

    \begin{llproof}
      \groundPf{\shiftup P}
        {Assumption}
      \wfpostypePf{\acontext}
        {P}
        {Subderivation}
      \groundPf{P}
        {\bydefground}
      \wfpostypePf{\makedec{\acontext}}
        {P}
        {By i.h.}
\Hand \wfnegtypePf{\makedec{\acontext}}
        {\shiftup P}
        {By \twfshiftup}
    \end{llproof}
  \end{itemize}
\end{proof}

\RestateWeakExtendedContextMutualSubtyping*

\begin{proof}
  By rule induction on $\wfposnegtypeJudg{\acontext} {A}$.

  \begin{itemize}
    \DerivationProofCase{\twfuvar}
      {\alpha \in \ConUV(\acontext)}
      {\wfpostypeJudg{\acontext} {\alpha}}

      \begin{llproof}
        \wfpostypePf{\acontext}
          {\alpha}
          {Subderivation}
        \wfpostypePf{\makedec{\acontext}}
          {\alpha}
          {By \lemmaref{lemma:Completing a context preserves well-formedness}}
        \DJudgePosPf{\makedec{\acontext}}
          {\alpha}
          {\alpha}
          {By \drefl}
  \Hand \MutualJudgePosPf{\makedec{\acontext}}
          {\subcon{\acontext'} \subcon{\acontext} \alpha}
          {\subcon{\acontext'} \alpha}
          {By \lemmaref{lemma:Context substitution on ground terms}}
      \end{llproof}

    \DerivationProofCase{\twfguess}
      {\guess{\alpha} \in \EV(\acontext)}
      {\wfpostypeJudg{\acontext} {\guess{\alpha}}}

      \begin{llproof}
        \proofcomment{Case $\subcon{\acontext} \guess{\alpha} = \guess{\alpha}$:}
        \inPf{(\guess{\alpha} = P)}
          {\acontext'}
          {Where $P = \subcon{\acontext'} \guess{\alpha}$, since $\groundJudge{\subcon{\acontext'} \guess{\alpha}}$}
        \wfpostypePf{\acontext'}
          {P}
          {Inversion on $\acontext' \conwf$ (must have instance of \cwfsolvedguess)}
        \wfpostypePf{\makedec{\acontext'}}
          {P}
          {By \lemmaref{lemma:Completing a context preserves well-formedness}}
        \wfpostypePf{\makedec{\acontext}}
          {P}
          {By \lemmaref{lemma:weak context extension equal declarative contexts}}
        \wfpostypePf{\makedec{\acontext}}
          {\subcon{\acontext'} \guess{\alpha}}
          {Substituting for $P$}
        \MutualJudgePosPf{\makedec{\acontext}}
          {\subcon{\acontext'} \guess{\alpha}}
          {\subcon{\acontext'} \guess{\alpha}}
          {By \lemmaref{lemma:Reflexivity of declarative pnsubtype}}
  \Hand \MutualJudgePosPf{\makedec{\acontext}}
          {\subcon{\acontext'} \subcon{\acontext} \guess{\alpha}}
          {\subcon{\acontext'} \guess{\alpha}}
          {As we are in the case that $\subcon{\acontext} \guess{\alpha} = \guess{\alpha}$}

        \proofcomment{Case $\subcon{\acontext} \guess{\alpha} \neq \guess{\alpha}$:}

        \MutualJudgePosPf{\makedec{\acontext_L}}
          {\subcon{\acontext_L} \guess{\alpha}}
          {\subcon{\acontext'_L} \guess{\alpha}}
          {Inversion on $\acontext \congoesweak \acontext'$ (must have instance of \Wcsolvedguess,}
        \trailingjust{\Wcunsolvedextend, or \Wcsolvedextend)}
        \wfpostypePf{\acontext_L}
          {\subcon{\acontext_L} \guess{\alpha}}
          {Inversion on $\acontext \conwf$ (must have instance of \cwfsolvedguess)}
        \MutualJudgePosPf{\makedec{\acontext}}
          {\subcon{\acontext} \guess{\alpha}}
          {\subcon{\acontext'} \guess{\alpha}}
          {By \lemmaref{lemma:Declarative subtyping weakening} and}
          \trailingjust{\lemmaref{lemma:Context substitution on ground terms}}
  \Hand \MutualJudgePosPf{\makedec{\acontext}}
          {\subcon{\acontext'} \subcon{\acontext} \guess{\alpha}}
          {\subcon{\acontext'} \guess{\alpha}}
          {By \lemmaref{lemma:Context substitution on ground terms}}
      \end{llproof}

    \DerivationProofCase{\twfshiftdown}
      {\wfnegtypeJudg{\acontext} {N}}
      {\wfpostypeJudg{\acontext} {\shiftdown N}}

      \begin{llproof}
        \wfnegtypePf{\acontext}
          {N}
          {Subderivation}
        \congoesWeakPf{\acontext}
          {\acontext'}
          {Assumption}
        \groundPf{\subcon{\acontext'} \shiftdown N}
          {Assumption}
        \groundPf{\subcon{\acontext'} N}
          {\bydefground}
        \conwfPf{\acontext}
          {Assumption}
        \conwfPf{\acontext'}
          {Assumption}
        \MutualJudgeNegPf{\makedec{\acontext}}
          {\subcon{\acontext'} \subcon{\acontext} N}
          {\subcon{\acontext'} N}
          {By i.h.}
  \Hand \MutualJudgePosPf{\makedec{\acontext}}
          {\subcon{\acontext'} \subcon{\acontext} \shiftdown N}
          {\subcon{\acontext'} \shiftdown N}
          {By \dshiftdown and \defsubcon}
      \end{llproof}

    \DerivationProofCase{\twfforall}
      {\wfnegtypeJudg{\acontext, \alpha} {N}}
      {\wfnegtypeJudg{\acontext} {\forall \alpha \ldotp N}}

      \begin{llproof}
        \wfnegtypePf{\acontext, \alpha}
          {N}
          {Subderivation}
        \congoesWeakPf{\acontext}
          {\acontext'}
          {Assumption}
        \congoesWeakPf{\acontext, \alpha}
          {\acontext', \alpha}
          {By \Wcuvar}
        \groundPf{\subcon{\acontext'} \forall \alpha \ldotp N}
          {Assumption}
        \groundPf{\subcon{\acontext', \alpha} N}
          {\bydefground}
        \conwfPf{\acontext}
          {Assumption}
        \conwfPf{\acontext, \alpha}
          {By \cwfuvar}
        \conwfPf{\acontext'}
          {Assumption}
        \conwfPf{\acontext', \alpha}
          {By \cwfuvar}
        \MutualJudgeNegPf{\makedec{\acontext, \alpha}}
          {\subcon{\acontext', \alpha} \subcon{\acontext, \alpha} N}
          {\subcon{\acontext', \alpha} N}
          {By i.h.}
        \MutualJudgeNegPf{\makedec{\acontext, \alpha}}
          {\subcon{\acontext'} \subcon{\acontext} N}
          {\subcon{\acontext'} N}
          {\bydefsubcon}
        \MutualJudgeNegPf{\makedec{\acontext}, \alpha}
          {\subcon{\acontext'} \subcon{\acontext} N}
          {\subcon{\acontext'} N}
          {\bydefmakedec}

        \proofsep
        \wfpostypePf{\makedec{\acontext}, \alpha}
          {\alpha}
          {By \twfuvar}
        \DJudgeNegPf{\makedec{\acontext}, \alpha}
          {\subcon{\acontext'} \subcon{\acontext} \forall \alpha \ldotp N}
          {\subcon{\acontext'} N}
          {By \dforalll and \defsubcon}
        \DJudgeNegPf{\makedec{\acontext}}
          {\subcon{\acontext'} \subcon{\acontext} \forall \alpha \ldotp N}
          {\subcon{\acontext'} \forall \alpha \ldotp N}
          {By \dforallr ($\alpha \notin \FreeUV(N)$) and \defsubcon}

        \proofsep
        \wfpostypePf{\makedec{\acontext}, \alpha}
          {\alpha}
          {Above}
        \DJudgeNegPf{\makedec{\acontext}, \alpha}
          {\subcon{\acontext'} \forall \alpha \ldotp N}
          {\subcon{\acontext'} \subcon{\acontext} N}
          {By \dforalll and \defsubcon}
        \DJudgeNegPf{\makedec{\acontext}}
          {\subcon{\acontext'} \forall \alpha \ldotp N}
          {\subcon{\acontext'} \subcon{\acontext} \forall \alpha \ldotp N}
          {By \dforallr ($\alpha \notin \FreeUV(N)$) and \defsubcon}

        \proofsep
  \Hand \MutualJudgeNegPf{\makedec{\acontext}}
          {\subcon{\acontext'} \subcon{\acontext} \forall \alpha \ldotp N}
          {\subcon{\acontext'} \forall \alpha \ldotp N}
          {Since we have the two component judgments}
      \end{llproof}

    \DerivationProofCase{\twfarrow}
      {\wfpostypeJudg{\acontext} {P} \\ \wfnegtypeJudg{\acontext} {N}}
      {\wfnegtypeJudg{\acontext} {P \funarrow N}}

      \begin{llproof}
        \wfpostypePf{\acontext}
          {P}
          {Subderivation}
        \congoesWeakPf{\acontext}
          {\acontext'}
          {Assumption}
        \groundPf{\subcon{\acontext'} (P \funarrow N)}
          {Assumption}
        \groundPf{\subcon{\acontext'} P}
          {\bydefground}
        \conwfPf{\acontext}
          {Assumption}
        \conwfPf{\acontext'}
          {Assumption}
        \MutualJudgePosPf{\makedec{\acontext}}
          {\subcon{\acontext'} \subcon{\acontext} P}
          {\subcon{\acontext'} P}
          {By i.h.}

        \proofsep
        \wfnegtypePf{\acontext}
          {N}
          {Subderivation}
        \congoesWeakPf{\acontext}
          {\acontext'}
          {Above}
        \groundPf{\subcon{\acontext'} N}
          {\bydefground}
        \conwfPf{\acontext}
          {Above}
        \conwfPf{\acontext'}
          {Above}
        \MutualJudgeNegPf{\makedec{\acontext}}
          {\subcon{\acontext'} \subcon{\acontext} N}
          {\subcon{\acontext'} N}
          {By i.h.}

        \proofsep
  \Hand \MutualJudgeNegPf{\makedec{\acontext}}
          {\subcon{\acontext'} \subcon{\acontext} (P \funarrow N)}
          {\subcon{\acontext'} (P \funarrow N)}
          {By \darrow and \defsubcon}
      \end{llproof}

    \DerivationProofCase{\twfshiftup}
      {\wfpostypeJudg{\acontext} {P}}
      {\wfnegtypeJudg{\acontext} {\shiftup P}}

      Symmetric to \twfshiftdown case.
  \end{itemize}
\end{proof}

\RestateWeakExtendedContextGroundMutualSubtyping*

\begin{proof}
  Corollary of \cref{lemma:Weak context extension leads to isomorphic types}.

  \begin{llproof}
    \wfposnegtypePf{\acontext}
      {A}
      {Assumption}
    \congoesWeakPf{\acontext}
      {\acontext'}
      {\ditto}
    \groundPf{\subcon{\acontext} A}
      {Assumption}
    \groundPf{\subcon{\acontext'} A}
      {By \lemmaref{lemma:extending-context-preserves-groundness-weak}}
    \conwfPf{\acontext}
      {\ditto}
    \conwfPf{\acontext'}
      {\ditto}
    \MutualJudgePosNegPf{\makedec{\acontext}}
      {\subcon{\acontext'} \subcon{\acontext} A}
      {\subcon{\acontext'} A}
      {By \lemmaref{lemma:Weak context extension leads to isomorphic types}}
\Hand \MutualJudgePosNegPf{\makedec{\acontext}}
      {\subcon{\acontext} A}
      {\subcon{\acontext'} A}
      {By \lemmaref{lemma:Context substitution on ground terms}}
  \end{llproof}
\end{proof}

\RestateExtendedContextMutualSubtyping*

\begin{proof}
  Corollary of \lemmaref{lemma:Weak context extension leads to isomorphic types}.

  \begin{llproof}
    \wfposnegtypePf{\acontext}{A}{Assumption}
    \congoesPf{\acontext}{\acontext'}{Assumption}
    \congoesWeakPf{\acontext}{\acontext'}{By \lemmaref{lemma:weak context extension subsumes normal}}
    \groundPf{\subcon{\acontext'} A}{Assumption}
    \conwfPf{\acontext}{Assumption}
    \conwfPf{\acontext'}{Assumption}
\Hand \MutualJudgePosNegPf{\makedec{\acontext}}
      {\subcon{\acontext'} \subcon{\acontext} A}
      {\subcon{\acontext'} A}
      {By \lemmaref{lemma:Weak context extension leads to isomorphic types}}
  \end{llproof}
\end{proof}

\RestateExtendedContextGroundMutualSubtyping*

\begin{proof}
  Corollary of \lemmaref{lemma:Weak context extension leads to isomorphic types (ground)}.

  \begin{llproof}
    \wfposnegtypePf{\acontext}{A}{Assumption}
    \groundPf{\subcon{\acontext} A}{Assumption}
    \congoesPf{\acontext}{\acontext'}{Assumption}
    \congoesWeakPf{\acontext}{\acontext'}{By \lemmaref{lemma:weak context extension subsumes normal}}
    \conwfPf{\acontext}{Assumption}
    \conwfPf{\acontext'}{Assumption}
\Hand \MutualJudgePosNegPf{\makedec{\acontext}}
      {\subcon{\acontext} A}
      {\subcon{\acontext'} A}
      {By \lemmaref{lemma:Weak context extension leads to isomorphic types (ground)}}
  \end{llproof}
\end{proof}

\subsection{Statement}

\RestateSoundnessPNSubtype*

\begin{proof}
  By mutual induction on the derivation of $\acontext \entails P \pnsubtype Q \prodcon \acontext'$.

  \begin{itemize}
    \DerivationProofCase{\arefl}
      {}
      {\aconwithhole{\alpha} \entails \alpha \possubtype \alpha \prodcon \aconwithhole{\alpha}}

      \begin{llproof}
        \inPf{\alpha}
          {\ConUV(\makedec{\aconwithhole{\alpha}})}
          {\bydefsubcon}
        \wfpostypePf{\makedec{\aconwithhole{\alpha}}}
          {\alpha}
          {By \twfuvar}
        \DJudgePosPf{\makedec{\aconwithhole{\alpha}}}
          {\alpha}
          {\alpha}
          {By \drefl}
  \Hand \CnewJudgePosPf{\aconwithhole{\alpha}}
          {\alpha}
          {\alpha}
          {By \lemmaref{lemma:Context substitution on ground terms}}
      \end{llproof}

    \DerivationProofCase{\ainst}
      {\acontext_L \entails P \postype \\ \groundJudge{P}}
      {\aconwithhole{\guess{\alpha}} \entails P \possubtype \guess{\alpha} \prodcon \aconwithhole{\guess{\alpha} = P}}

      \begin{llproof}
        \congoesPf{\aconwithhole{\guess{\alpha} = P}}
          {\ccontext}
          {Assumption}
        \eqPf{\ccontext}
          {\cconwithhole{\guess{\alpha} = Q}}
          {Inversion (must have instance of \Csolvedguess)}
        \MutualJudgePosPf{\makedec{\acontext_L}}{P}{Q}{\ditto}
        \eqPf{\subcon{\ccontext} \guess{\alpha}}
          {\subcon{\cconwithhole{\guess{\alpha} = Q}} \guess{\alpha}}
          {Substituting for $\ccontext$}
        \continueeqPf{Q}
          {\bydefsubcon}
        \DJudgePosPf{\makedec{\aconwithhole{\guess{\alpha}}}}
          {P}
          {Q}
          {By \lemmaref{lemma:Declarative subtyping weakening}}
        \CnewJudgePosPf{\ccontext}{P}{\guess{\alpha}}
          {Substituting using above equations}
      \end{llproof}

    \DerivationProofCase{\ashiftdown}
      {\acontext \entails M \negsubtype N \prodcon \acontext' \\
       \acontext' \entails N \negsubtype \subcon{\acontext'} M \prodcon \acontext''}
      {\acontext \entails \shiftdown N \possubtype \shiftdown M \prodcon \acontext''}

      \begin{llproof}
        \groundPf{\shiftdown N}
          {Assumption}
        \NoSolvedVarsPf{\acontext}
          {\shiftdown M}
          {\ditto}

        \proofcomment{We have:}
        \AJudgeNegPf{\acontext}
          {M}
          {N}
          {\acontext'}
          {Subderivation}
        \conwfPf{\acontext}
          {Assumption}
        \groundPf{N}
          {By definition of ground}
        \NoSolvedVarsPf{\acontext}
          {M}
          {\bydefsubcon}

        \proofcomment{Therefore:}
        \conwfPf{\acontext'}
          {By \lemmaref{lemma:well-formedness-pnsubtype}}
        \congoesPf{\acontext}
          {\acontext'}
          {\ditto}
        \groundPf{\subcon{\acontext'} M}
          {\ditto}

        \proofcomment{We have:}
        \AJudgeNegPf{\acontext'}
          {N}
          {\subcon{\acontext'} M}
          {\acontext''}
          {Subderivation}
        \conwfPf{\acontext'}
          {Above}
        \groundPf{\subcon{\acontext'} M}
          {Above}
        \NoSolvedVarsPf{\acontext'}
          {N}
          {By \lemmaref{lemma:Context substitution on ground terms}}

        \proofcomment{Therefore:}
        \congoesPf{\acontext'}
          {\acontext''}
          {By \lemmaref{lemma:well-formedness-pnsubtype}}

        \proofcomment{Now show the antecedents of the induction hypothesis applied to the first premise of the algorithmic rule:}
\Label{1} \conwfPf{\acontext}
          {Above}
\Label{2} \conwfPf{\ccontext}
          {Assumption}
\Label{3} \AJudgeNegPf{\acontext}{M}{N}{\acontext'}{Subderivation}
        \congoesPf{\acontext''}
          {\ccontext}
          {Assumption}
\Label{4} \congoesPf{\acontext'}
          {\ccontext}
          {By \lemmaref{lemma:context extension transitive}}
\Label{5} \groundPf{N}{Above}
\Label{6} \NoSolvedVarsPf{\acontext}{M}{Above}
\Label{7} \wfnegtypePf{\acontext}
          {N}
          {Inversion on assumption (\twfshiftdown)}
\Label{8} \wfnegtypePf{\acontext}
          {M}
          {\ditto}

        \proofcomment{We conclude:}
        \CnewJudgeNegPf{\acontext}
          {M}
          {N}
          {\byih applied to first premise, using (1--8)}

        \proofcomment{Show the antecedents of the induction hypothesis applied this time to the second premise of the algorithmic rule:}
\Label{9} \conwfPf{\acontext'}
          {Above}
\Label{10} \conwfPf{\ccontext}
          {Above}
\Label{11} \AJudgeNegPf{\acontext'}{N}{\subcon{\acontext'} M}{\acontext''}{Subderivation}
\Label{12} \congoesPf{\acontext''}
          {\ccontext}
          {Assumption}
\Label{13} \groundPf{\subcon{\acontext'} M}{Above}
\Label{14} \NoSolvedVarsPf{\acontext'}{N}{Above}
\Label{15} \wfnegtypePf{\acontext'}
          {N}
          {By \lemmaref{lemma:Context extension preserves term well-formedness}}
\Label{16} \wfnegtypePf{\acontext'}
          {M}
          {By \lemmaref{lemma:Context extension preserves term well-formedness}}

        \proofcomment{We conclude:}
        \CnewJudgeNegPf{\acontext'}
          {N}
          {\subcon{\acontext'} M}
          {\byih applied to second premise, using (9--16)}
        \DJudgeNegPf{\makedec{\acontext'}}
          {N}
          {\subcon{\acontext'} M}
          {By \lemmaref{lemma:Context substitution on ground terms}}
        \continuenegsubtypePf{\subcon{\ccontext} M}
          {By \lemmaref{lemma:Context extension leads to isomorphic types (ground)}}
        \DJudgeNegPf{\makedec{\acontext}}
          {N}
          {\subcon{\ccontext} M}
          {By \lemmaref{lemma:equal declarative contexts}}

        \proofcomment{Applying the declarative rule:}
        \DJudgePosPf{\makedec{\acontext}}
          {\shiftdown N}
          {\shiftdown \subcon{\ccontext} M}
          {By \dshiftdown}
  \Hand \CnewJudgePosPf{\acontext}
          {\shiftdown N}
          {\shiftdown M}
          {\bydefsubcon}
      \end{llproof}

    \DerivationProofCase{\aforallr}
      {\acontext, \alpha \entails N \negsubtype M \prodcon \acontext', \alpha}
      {\acontext \entails N \negsubtype \forall \alpha \ldotp M \prodcon \acontext'}

      \begin{llproof}
        \conwfPf{\acontext}
          {Assumption}
\Label{1} \conwfPf{\acontext, \alpha}
          {By \cwfuvar}
        \conwfPf{\ccontext}
          {Assumption}
\Label{2} \conwfPf{\ccontext, \alpha}
          {By \cwfuvar}
\Label{3} \AJudgeNegPf{\acontext, \alpha}
          {N}
          {M}
          {\acontext', \alpha}
          {Subderivation}
        \congoesPf{\acontext'}
          {\ccontext}
          {Assumption}
\Label{4} \congoesPf{\acontext', \alpha}
          {\ccontext, \alpha}
          {By \Cuvar}
        \groundPf{\forall \alpha \ldotp M}
          {Assumption}
\Label{5} \groundPf{M}
          {Definition of ground}
\Label{6} \NoSolvedVarsPf{\acontext}
          {N}
          {Assumption}
        \wfnegtypePf{\acontext}
          {N}
          {Assumption}
\Label{7} \wfnegtypePf{\acontext, \alpha}
          {N}
          {By \lemmaref{lemma:Term well-formedness weakening}}
        \wfnegtypePf{\acontext}
          {\forall \alpha \ldotp M}
          {Assumption}
\Label{8} \wfnegtypePf{\acontext, \alpha}
          {M}
          {Inversion (\twfforall)}

        \proofsep
        \CnewCustomJudgeNegPf{\ccontext, \alpha}
          {\acontext, \alpha}
          {N}
          {M}
          {\byih, using (1--8)}
        \CnewJudgeNegPf{\acontext, \alpha}
          {N}
          {M}
          {\bydefsubcon}
  \Hand \CnewJudgeNegPf{\acontext}
          {N}
          {\forall \alpha \ldotp M}
          {By \dforallr}
      \end{llproof}

    \DerivationProofCase{\aforalll}
      { \ANegSubtypeJudg{\acontext, \guess{\alpha}}
          {\subcon{\guess{\alpha} / \alpha} N}
          {M}
          {\acontext', \guess{\alpha} \,[= P]} \\
        M \neq \forall \alpha \ldotp M'
      }
      {\acontext \entails \forall \alpha \ldotp N \negsubtype M \prodcon \acontext'}

      \begin{llproof}
        \proofcomment{Apply well-formedness to the premise:}
\Label{1} \AJudgeNegPf{\acontext, \guess{\alpha}}
          {\subcon{\guess{\alpha} / \alpha} N}
          {M}
          {\acontext', \guess{\alpha} \, [= P]}
          {Subderivation}
        \conwfPf{\acontext}
          {Assumption}
\Label{2} \conwfPf{\acontext, \guess{\alpha}}
          {By \cwfunsolvedguess}
\Label{3} \groundPf{M}
          {Assumption}
        \NoSolvedVarsPf{\acontext}
          {\forall \alpha \ldotp N}
          {Assumption}
\Label{4} \NoSolvedVarsPf{\acontext}
          {N}
          {\bydefsubcon}

        \proofsep
        \conwfPf{\acontext', \guess{\alpha} \, [= P]}
          {By \lemmaref{lemma:well-formedness-pnsubtype}, using (1--4)}
        \congoesPf{\acontext, \guess{\alpha}}
          {\acontext', \guess{\alpha} \, [= P]}
          {\ditto}

        \proofcomment{Now apply the inductive hypothesis:}

\Label{5} \conwfPf{\acontext, \guess{\alpha}}
          {Above}
        \conwfPf{\ccontext}
          {Assumption}
        \wfpostypePf{\acontext'}
          {P}
          {Inversion \cwfsolvedguess)}
        \groundPf{P}
          {\ditto}
        \congoesPf{\acontext'}{\ccontext}{Assumption}
        \wfpostypePf{\ccontext}
          {P}
          {By \lemmaref{lemma:Context extension preserves term well-formedness}}
\Label{6} \conwfPf{\ccontext, \guess{\alpha} = P}
          {By \cwfsolvedguess}
\Label{7} \AJudgeNegPf{\acontext, \guess{\alpha}}
          {\subcon{\guess{\alpha} / \alpha} N}
          {M}
          {\acontext', \guess{\alpha} \, [= P]}
          {Subderivation}
        \congoesPf{\acontext'}
          {\ccontext}
          {Assumption}
\Label{8} \congoesPf{\acontext', \guess{\alpha} \, [= P]}
          {\ccontext, \guess{\alpha} = P}
          {By \Csolveguess / \Csolvedguess}
\Label{9} \groundPf{M}
          {Assumption}
        \NoSolvedVarsPf{\acontext}
          {\forall \alpha \ldotp N}
          {Assumption}
        \NoSolvedVarsPf{\acontext}
          {N}
          {\bydefsubcon}
\Label{10} \NoSolvedVarsPf{\acontext, \guess{\alpha}}
          {\subcon{\guess{\alpha} / \alpha} N}
          {$\acontext, \guess{\alpha} \conwf$, so $\guess{\alpha} \notin \EV(\acontext)$}
\Label{11} \wfnegtypePf{\acontext}
          {\forall \alpha \ldotp N}
          {Assumption}
\Label{12} \wfnegtypePf{\acontext, \alpha}
          {N}
          {Inversion (\twfforall)}

        \proofsep
        \DJudgeNegPf{\makedec{\acontext, \guess{\alpha}}}
          {\subcon{\ccontext, \guess{\alpha} = P} \subterm{\guess{\alpha} / \alpha} N}
          {M}
          {\byih, using (5--12)}
        \DJudgeNegPf{\makedec{\acontext}}
          {\subcon{\ccontext, \guess{\alpha} = P} \subcon{\guess{\alpha} / \alpha} N}
          {M}
          {\bydefmakedec}

        \proofsep
        \notinPf{\alpha}
          {\ConUV(\acontext')}
          {Since $\alpha \notin \ConUV(\acontext)$ as $\alpha$ fresh}
        \eqPf{\subcon{\ccontext, \guess{\alpha} = P} \subcon{\guess{\alpha} / \alpha} N}
          {\subcon{\ccontext} \subcon{P / \guess{\alpha}} \subcon{\guess{\alpha} / \alpha} N}
          {\bydefsubcon}
        \continueeqPf
          {\subcon{\ccontext} \subcon{P / \guess{\alpha}} \subcon{P / \alpha} N}
          {By composition of substitutions}
        \continueeqPf{\subcon{\ccontext} \subcon{P / \alpha} \subcon{P / \guess{\alpha}} N}
          {$\wfpostypeJudg{\acontext'} {P}$ and $\alpha \notin \ConUV(\acontext')$, so $\alpha \notin \FreeUV(P)$.}
          \trailingjust{Also $\groundJudge{P}$, so $\guess{\alpha} \notin \FreeEV(P)$.}
        \continueeqPf{\subcon{\ccontext} \subcon{P / \alpha} N}
          {Since $\guess{\alpha}$ fresh, $\guess{\alpha} \notin \FreeEV(N)$}
        \continueeqPf{\subcon{(\subcon{\ccontext} P) / \alpha} \subcon{\ccontext} N}
          {Since context application does not replace}
          \trailingjust{universal variables}

        \proofsep
        \DJudgeNegPf{\makedec{\acontext}}
          {\subcon{(\subcon{\ccontext} P) / \alpha} \subcon{\ccontext} N}
          {M}
          {Substituting for $\subcon{\ccontext, \guess{\alpha} = P} \subcon{\guess{\alpha} / \alpha} N$}
        \wfpostypePf{\acontext'}
          {P}
          {Above}
        \groundPf{P}
          {Above}
        \wfpostypePf{\makedec{\acontext'}}
          {P}
          {By \lemmaref{lemma:Completing a context preserves well-formedness}}
        \wfpostypePf{\makedec{\acontext}}
          {P}
          {By \lemmaref{lemma:equal declarative contexts}}
        \wfpostypePf{\makedec{\acontext}}
          {\subcon{\ccontext} P}
          {By \lemmaref{lemma:Context substitution on ground terms}}
        \DJudgeNegPf{\makedec{\acontext}}
          {\forall \alpha \ldotp \subcon{\ccontext} N}
          {M}
          {By \dforalll}
  \Hand \CnewJudgeNegPf{\acontext}
          {\forall \alpha \ldotp N}
          {M}
          {\bydefsubcon}
      \end{llproof}

    \DerivationProofCase{\aarrow}
      {\acontext \entails Q \possubtype P \prodcon \acontext' \\
        \acontext' \entails \subcon{\acontext'} N \negsubtype M \prodcon \acontext''}
      {\acontext \entails P \funarrow N \negsubtype Q \funarrow M \prodcon \acontext''}

      \begin{llproof}
        \groundPf{Q \funarrow M}
          {Assumption}
        \eqPf{\subcon{\acontext} (P \funarrow N)}
          {P \funarrow N}
          {Assumption}
        \wfnegtypePf{\acontext}
          {P \funarrow N}
          {Assumption}
        \wfnegtypePf{\acontext}
          {Q \funarrow M}
          {Assumption}

        \proofcomment{We have:}
        \AJudgePosPf{\acontext}
          {Q}
          {P}
          {\acontext'}
          {Subderivation}
        \conwfPf{\acontext}
          {Assumption}
        \groundPf{Q}
          {Since $Q \funarrow M$ ground}
        \NoSolvedVarsPf{\acontext}
          {P}
          {\bydefsubcon}

        \proofcomment{Therefore by well-formedness:}
        \conwfPf{\acontext'}
          {By \lemmaref{lemma:well-formedness-pnsubtype}}
        \congoesPf{\acontext}
          {\acontext'}
          {\ditto}

        \proofcomment{We have:}
        \AJudgeNegPf{\acontext'}
          {\subcon{\acontext'} N}
          {M}
          {\acontext''}
          {Subderivation}
        \conwfPf{\acontext'}
          {Above}
        \groundPf{M}
          {Since $Q \funarrow M$ ground}
        \NoSolvedVarsPf{\acontext'}
          {\subcon{\acontext'} N}
          {By \lemmaref{lemma:Context substitution idempotence}}

        \proofcomment{Therefore by well-formedness:}
        \congoesPf{\acontext'}
          {\acontext''}
          {By \lemmaref{lemma:well-formedness-pnsubtype}}

        \proofcomment{Applying the induction hypothesis to the first premise:}
        \congoesPf{\acontext''}
          {\ccontext}
          {Assumption}
        \congoesPf{\acontext'}
          {\ccontext}
          {By \lemmaref{lemma:context extension transitive}}
        \wfpostypePf{\acontext}
          {Q}
          {Inversion (\twfarrow)}
        \wfpostypePf{\acontext}
          {P}
          {\ditto}
        \conwfPf{\ccontext}
          {Assumption}
        \CnewJudgePosPf{\acontext}
          {Q}
          {P}
          {\byih applied to first premise}

        \proofcomment{Applying the induction hypothesis to the second premise:}
        \congoesPf{\acontext''}
          {\ccontext}
          {Assumption}
        \wfnegtypePf{\acontext}
          {N}
          {Inversion (\twfarrow)}
        \wfnegtypePf{\acontext}
          {M}
          {\ditto}
        \conwfPf{\ccontext}
          {Assumption}
        \CnewJudgeNegPf{\acontext'}
          {\subcon{\acontext'} N}
          {M}
          {\byih applied to second premise}

        \proofcomment{Rework the second declarative judgment:}
        \congoesPf{\acontext'}
          {\ccontext}
          {Above}
        \DJudgeNegPf{\makedec{\acontext'}}
          {\subcon{\ccontext} N}
          {\subcon{\ccontext} \subcon{\acontext'} N}
          {By \lemmaref{lemma:Context extension leads to isomorphic types}}
        \continuenegsubtypePf{M}
          {By \lemmaref{lemma:Transitivity of declarative pnsubtype}}
        \CnewJudgeNegPf{\acontext}
          {N}
          {M}
          {By \lemmaref{lemma:equal declarative contexts}}

        \proofcomment{Finally, apply the declarative rule:}
        \DJudgeNegPf{\makedec{\acontext}}
          {\subcon{\ccontext} P \funarrow \subcon{\ccontext} N}
          {Q \funarrow M}
          {By \darrow}
  \Hand \CnewJudgeNegPf{\acontext}
          {(P \funarrow N)}
          {Q \funarrow M}
          {By definition of \subconunderscores}
      \end{llproof}

    \DerivationProofCase{\ashiftup}
      {\acontext \entails Q \possubtype P \prodcon \acontext' \\
        \acontext' \entails \subcon{\acontext'} P \possubtype Q \prodcon \acontext''}
      {\acontext \entails \shiftup P \negsubtype \shiftup Q \prodcon \acontext''}

      Symmetric to \ashiftdown case.
  \end{itemize}
\end{proof}

\section{Completeness of subtyping}

\subsection{Lemmas for completeness}

\RestateCompletingContextAndTypePreservesWf*

\begin{proof}
  Corollary of \lemmaref{lemma:Completing a context preserves well-formedness}.

  By $\wfposnegtypeJudg{\acontext}{A}$, all existential variables in $A$ will appear in $\acontext$.
  By $\congoesJudg{\acontext}{\ccontext}$, these will also all appear in $\ccontext$ as ground types.
  Therefore $\subcon{\ccontext} A$ must be ground.
  Then:

  \begin{llproof}
    \wfposnegtypePf{\acontext}{A}{Assumption}
    \wfposnegtypePf{\acontext}{\subcon{\ccontext} A}{By \lemmaref{lemma:Well-formed context substitution preserves term well-formedness}}
    \groundPf{\subcon{\ccontext} A}{Above}
    \wfposnegtypePf{\makedec{\acontext}}{\subcon{\ccontext} A}{By \lemmaref{lemma:Completing a context preserves well-formedness}}
  \end{llproof}
\end{proof}

\ExtensionSolvingGuess*

\begin{proof}
  By structural induction on $\acontext_R$.

  \begin{itemize}
    \caseitem{$\acontext_R = \emptyacontext$}

    \begin{llproof}
      \congoesPf{\acontext_L, \guess{\alpha}}
        {\ccontext_L, \guess{\alpha} = Q}
        {Assumption}
      \congoesPf{\acontext_L}
        {\ccontext_L}
        {Inversion (\Csolveguess)}
      \MutualJudgePosPf{\subcon{\ccontext_L} \acontext_L}
        {P}
        {Q}
        {Assumption}
\Hand \congoesPf{\acontext_L, \guess{\alpha} = P}
        {\ccontext_L, \guess{\alpha} = Q}
        {By \Csolvedguess}
    \end{llproof}

    \caseitem{$\acontext_R = \acontext_R', \alpha$}

    \begin{llproof}
      \congoesPf{\acontext_L, \guess{\alpha}, \acontext_R', \alpha}
        {\ccontext_L, \guess{\alpha} = Q, \ccontext_R', \alpha}
        {By structure of $\acontext_R$, must have instance of}
        \trailingjust{\Cuvar}
      \MutualJudgePosPf{\subcon{\ccontext_L} \acontext_L}
        {P}
        {Q}
        {Assumption}
      \congoesPf{\acontext_L, \guess{\alpha}, \acontext_R'}
        {\ccontext_L, \guess{\alpha} = Q, \ccontext_R'}
        {Inversion (\Cuvar)}
      \congoesPf{\acontext_L, \guess{\alpha} = P, \acontext_R'}
        {\ccontext_L, \guess{\alpha} = Q, \ccontext_R'}
        {By i.h.}
\Hand \congoesPf{\acontext_L, \guess{\alpha} = P, \acontext_R', \alpha}
        {\ccontext_L, \guess{\alpha} = Q, \ccontext_R', \alpha}
        {By \Cuvar}
    \end{llproof}

    \caseitem{$\acontext_R = \acontext_R', \guess{\beta}$}

    \begin{llproof}
      \congoesPf{\acontext_L, \guess{\alpha}, \acontext_R', \guess{\beta}}
        {\ccontext_L, \guess{\alpha} = Q, \ccontext_R', \guess{\beta} = R}
        {By structure of $\acontext_R$, must have instance of}
        \trailingjust{\Csolveguess}
      \MutualJudgePosPf{\subcon{\ccontext_L} \acontext_L}
        {P}
        {Q}
        {Assumption}
      \congoesPf{\acontext_L, \guess{\alpha}, \acontext_R'}
        {\ccontext_L, \guess{\alpha} = Q, \ccontext_R'}
        {Inversion (\Csolveguess)}
      \congoesPf{\acontext_L, \guess{\alpha} = P, \acontext_R'}
        {\ccontext_L, \guess{\alpha} = Q, \ccontext_R'}
        {By i.h.}
\Hand \congoesPf{\acontext_L, \guess{\alpha} = P, \acontext_R', \guess{\beta}}
        {\ccontext_L, \guess{\alpha} = Q, \ccontext_R', \guess{\beta} = R}
        {By \Csolveguess}
    \end{llproof}

    \caseitem{$\acontext_R = \acontext_R', \guess{\beta} = R$}

    \begin{llproof}
      \congoesPf{\acontext_L, \guess{\alpha}, \acontext_R', \guess{\beta} = R}
        {\ccontext_L, \guess{\alpha} = Q, \ccontext_R', \guess{\beta} = S}
        {By structure of $\acontext_R$, must}
        \trailingjust{have instance of}
        \trailingjust{\Csolvedguess}
      \MutualJudgePosPf{\subcon{\ccontext_L, \guess{\alpha} = Q, \ccontext_R'} (\acontext_L, \guess{\alpha}, \acontext_R')}
        {R}
        {S}
        {\ditto}
      \MutualJudgePosPf{\subcon{\ccontext_L} \acontext_L}
        {P}
        {Q}
        {Assumption}
      \congoesPf{\acontext_L, \guess{\alpha}, \acontext_R'}
        {\ccontext_L, \guess{\alpha} = Q, \ccontext_R'}
        {Inversion}
        \trailingjust{(\Csolveguess)}
      \congoesPf{\acontext_L, \guess{\alpha} = P, \acontext_R'}
        {\ccontext_L, \guess{\alpha} = Q, \ccontext_R'}
        {By i.h.}
      \MutualJudgePosPf{\subcon{\ccontext_L, \guess{\alpha} = Q, \ccontext_R'} (\acontext_L, \guess{\alpha} = P, \acontext_R')}
        {R}
        {S}
        {Since $\subconunderscores$ ignores}
        \trailingjust{existential variables}
\Hand \congoesPf{\acontext_L, \guess{\alpha} = P, \acontext_R', \guess{\beta} = R}
        {\ccontext_L, \guess{\alpha} = Q, \ccontext_R', \guess{\beta} = S}
        {By \Csolvedguess}
    \end{llproof}
  \end{itemize}
\end{proof}

\ContextExtensionSubstitutionSizeLemma*

\begin{proof}
  Corollary of \lemmaref{lemma:Context extension leads to isomorphic types} and \lemmaref{lemma:Isomorphic types are the same size}.

  \begin{llproof}
    \wfposnegtypePf{\acontext}
      {A}
      {Assumption}
    \groundPf{\subcon{\ccontext} A}
      {$\ccontext$ completes all free existential variables in $A$}
    \congoesPf{\acontext}
      {\ccontext}
      {\ditto}
    \MutualJudgePosNegPf{\makedec{\acontext}}
      {\subcon{\ccontext} \subcon{\acontext} A}
      {\subcon{\ccontext} A}
      {By \lemmaref{lemma:Context extension leads to isomorphic types}}
    \conwfPf{\acontext}
      {Assumption}
    \wfposnegtypePf{\acontext}
      {\subcon{\acontext} A}
      {By \lemmaref{lemma:Well-formed context substitution preserves term well-formedness}}
    \wfposnegtypePf{\makedec{\acontext}}
      {\subcon{\ccontext} \subcon{\acontext} A}
      {By \lemmaref{lemma:Completing a context and type preserves well-formedness}}
    \conwfPf{\ccontext}
      {Assumption}
    \wfposnegtypePf{\makedec{\acontext}}
      {\subcon{\ccontext} A}
      {By \lemmaref{lemma:Completing a context and type preserves well-formedness}}
  \Hand \termsizeEqPf{\subcon{\ccontext} \subcon{\acontext} A}
      {\subcon{\ccontext} A}
      {By \lemmaref{lemma:Isomorphic types are the same size}}
  \end{llproof}
\end{proof}

\ContextExtensionGroundSubstitutionSizeLemma*

\begin{proof}
  Corollary of \lemmaref{lemma:context extension ground substitution size lemma}.

  \begin{llproof}
    \termsizeEqPf{\subcon{\ccontext} \subcon{\acontext} A}
      {\subcon{\ccontext} A}
      {By \lemmaref{lemma:Isomorphic types are the same size}}
\Hand \termsizeEqPf{\subcon{\acontext} A}
      {\subcon{\ccontext} A}
      {By \lemmaref{lemma:Context substitution on ground terms}}
  \end{llproof}
\end{proof}

\subsection{Statement}

\RestateCompletenessPNSubtype*

\begin{proof}
  By mutual rule induction on the declarative judgment weighted by the lexicographic ordering of ($\termsize{P}$, $\numprenex{P} + \numprenex{Q}$) in the positive case where we have $\CnewPosJudge{\acontext} {P} {Q}$, and ($\termsize{M}$, $\numprenex{N} + \numprenex{M}$) in the negative case where we have $\CnewNegJudge{\acontext} {N} {M}$.

  Firstly consider the case where $B = \guess{\alpha}$.  Suppose $\subcon{\ccontext} \guess{\alpha} = Q'$:

  \begin{llproof}
    \eqPf{\acontext}
      {\acontext_L, \guess{\alpha}, \acontext_R}
      {Since $\wfpostypeJudg{\acontext}{\guess{\alpha}}$ and $\NoSolvedVarsJudg{\acontext}{\guess{\alpha}}$ by assumption}
    \eqPf{\ccontext}
      {\ccontext_L, \guess{\alpha} = Q', \ccontext_R}
      {Since $\subcon{\ccontext} \guess{\alpha} = Q'$}
    \wfpostypePf{\ccontext_L}
      {Q'}
      {Inversion on $\ccontext \conwf$ (\cwfsolvedguess)}
    \groundPf{Q'}
      {\ditto}
    \congoesPf{\aconwithhole{\guess{\alpha}}}
      {\cconwithhole{\guess{\alpha} = Q'}}
      {Assumption}
    \congoesPf{\acontext_L}
      {\ccontext_L}
      {Inversion (must have instance of \Csolveguess)}
    \wfpostypePf{\acontext_L}
      {Q'}
      {By the rules defining $\congoes$, each uvar in $\ccontext_L$ must appear}
      \trailingjust{in $\acontext_L$}
    \DJudgePosPf{\makedec{\acontext}}
      {P}
      {Q'}
      {Assumption}
    \groundPf{P}
      {\ditto}
    \subseteqPf{\FreeUV(P)}
      {\ConUV(\acontext_L)}
      {Since $\DPosSubtypeJudg{\makedec{\acontext}}{P}{Q'}$ and $\wfpostypeJudg{\acontext_L}{Q'}$}
    \subseteqPf{\FreeEV(P)}
      {\EV(\acontext_L)}
      {$\FreeEV(P) = \varnothing$ since $P$ ground}
    \wfpostypePf{\acontext}
      {P}
      {Assumption}
    \wfpostypePf{\acontext_L}
      {P}
      {By above three equations}
\Hand \AJudgePosPf{\aconwithhole{\guess{\alpha}}}
      {P}
      {\guess{\alpha}}
      {\aconwithhole{\guess{\alpha} = P}}
      {By \ainst}

    \proofsep
    \DJudgePosPf{\makedec{\acontext}}
      {Q'}
      {P}
      {By \lemmaref{lemma:Symmetry of positive declarative subtyping}}
    \MutualJudgePosPf{\makedec{\acontext}}
      {P}
      {Q'}
      {Since we have both the component judgments}
    \MutualJudgePosPf{\subcon{\ccontext} \acontext_L}
      {P}
      {Q'}
      {Since $\MutualSubtypePosJudge{\makedec{\acontext}}{P}{Q'}$, $\wfpostypeJudg{\acontext_L}{P}$, and $\wfpostypeJudg{\acontext_L}{Q'}$}
    \MutualJudgePosPf{\subcon{\ccontext_L} \acontext_L}
      {P}
      {Q'}
      {Since $\acontext_L \congoes \ccontext_L$, $\wfpostypeJudg{\acontext_L}{P}$, and $\wfpostypeJudg{\acontext_L}{Q'}$}
    \congoesPf{\aconwithhole{\guess{\alpha}}}
      {\cconwithhole{\guess{\alpha} = Q'}}
      {Above}
\Hand \congoesPf{\aconwithhole{\guess{\alpha} = P}}
      {\cconwithhole{\guess{\alpha} = Q'}}
      {By \lemmaref{lemma:extension solving guess}}
  \end{llproof}

  Now consider the cases where $B \neq \guess{\alpha}$:

  \begin{itemize}
    \DerivationProofCase{\drefl}
      {\wfpostypeJudg{\makedec{\acontext}} {\alpha}}
      {\CnewPosJudge{\acontext}{\alpha}{\alpha}}

    \begin{llproof}
      \wfpostypePf{\makedec{\acontext}}
        {\alpha}
        {Subderivation}
      \inPf{\alpha}
        {\ConUV(\subcon{\ccontext} \acontext)}
        {Inversion (\twfuvar)}
      \inPf{\alpha}
        {\ConUV(\acontext)}
        {\bydefsubcon}
      \eqPf{\acontext}
        {\aconwithhole{\alpha}}
        {Since $\alpha \in \ConUV(\acontext)$}
\Hand \AJudgePosPf{\aconwithhole{\alpha}}
        {\alpha}
        {\alpha}
        {\aconwithhole{\alpha}}
        {By \arefl}
\Hand \congoesPf{\aconwithhole{\alpha}}
        {\ccontext}
        {Assumption}
    \end{llproof}

    \DerivationProofCase{\dshiftdown}
      {\CnewNegJudge{\acontext}{M}{N} \\ \DNegSubtypeJudge{\makedec{\acontext}}{N}{\subcon{\ccontext} M}}
      {\CnewPosJudge{\acontext}{\shiftdown N}{\shiftdown M}}

    \begin{llproof}
      \wfpostypePf{\acontext}
        {\shiftdown N}
        {Assumption}
      \wfpostypePf{\acontext}
        {\shiftdown M}
        {Assumption}
      \groundPf{\shiftdown N}
        {Assumption}
      \NoSolvedVarsPf{\acontext}
        {\shiftdown M}
        {Assumption}
      
      \proofsep
      \CnewJudgeNegPf{\acontext}
        {M}
        {N}
        {Subderivation}
      \conwfPf{\acontext}
        {Assumption}
      \wfnegtypePf{\acontext}
        {M}
        {Inversion (\twfshiftdown)}
      \wfnegtypePf{\acontext}
        {N}
        {\ditto}
      \congoesPf{\acontext}
        {\ccontext}
        {Assumption}
      \conwfPf{\ccontext}
        {\ditto}
      \groundPf{N}
        {By definition of ground}
      \NoSolvedVarsPf{\acontext}
        {M}
        {By definition of $\subconunderscores$}
      \AJudgeNegPf{\acontext}
        {M}
        {N}
        {\acontext'}
        {By i.h. (the type size of the ground side type in the declarative}
        \trailingjust{judgment has decreased)}
      \congoesPf{\acontext'}
        {\ccontext}
        {\ditto}
      \conwfPf{\acontext'}
        {By \lemmaref{lemma:well-formedness-pnsubtype}}
      \congoesPf{\acontext}
        {\acontext'}
        {\ditto}
      \groundPf{\subcon{\acontext'} M}
        {\ditto}

      \proofsep
      \DJudgeNegPf{\makedec{\acontext}}
        {N}
        {\subcon{\ccontext} M}
        {Subderivation}
      \DJudgeNegPf{\makedec{\acontext'}}
        {N}
        {\subcon{\ccontext} M}
        {By \lemmaref{lemma:equal declarative contexts}}
      \MutualJudgeNegPf{\makedec{\acontext'}}
        {\subcon{\acontext'} M}
        {\subcon{\ccontext} M}
        {By \lemmaref{lemma:Context extension leads to isomorphic types (ground)}}
      
      \proofsep
      \wfnegtypePf{\acontext'}
        {N}
        {By \lemmaref{lemma:Context extension preserves term well-formedness}}
      \wfnegtypePf{\makedec{\acontext'}}
        {N}
        {By \lemmaref{lemma:Completing a context preserves well-formedness}}
      \wfnegtypePf{\acontext'}
        {M}
        {By \lemmaref{lemma:Context extension preserves term well-formedness}}
      \wfnegtypePf{\makedec{\acontext'}}
        {\subcon{\ccontext} M}
        {By \lemmaref{lemma:Completing a context and type preserves well-formedness}}
      \wfnegtypePf{\acontext'}
        {\subcon{\acontext'} M}
        {By \lemmaref{lemma:Well-formed context substitution preserves term well-formedness}}
      \wfnegtypePf{\makedec{\acontext'}}
        {\subcon{\acontext'} M}
        {By \lemmaref{lemma:Completing a context preserves well-formedness}}
      \DJudgeNegPf{\makedec{\acontext'}}
        {N}
        {\subcon{\acontext'} M}
        {By \lemmaref{lemma:Transitivity of declarative pnsubtype}}

      \proofsep
      \CnewJudgeNegPf{\acontext'}
        {N}
        {\subcon{\acontext'} M}
        {By \lemmaref{lemma:Context substitution on ground terms}}
      \conwfPf{\acontext'}
        {Above}
      \wfnegtypePf{\acontext'}
        {N}
        {Above}
      \wfnegtypePf{\acontext'}
        {\subcon{\acontext'} M}
        {Above}
      \congoesPf{\acontext'}
        {\ccontext}
        {Above}
      \conwfPf{\ccontext}
        {Above}
      \groundPf{\subcon{\acontext'} M}
        {Above}
      \NoSolvedVarsPf{\acontext'}
        {N}
        {By \lemmaref{lemma:Context substitution on ground terms}}

      \proofsep
      \termsizeEqPf{\subcon{\acontext'} M}
        {\subcon{\ccontext} M}
        {By \lemmaref{lemma:context extension ground substitution size lemma}}
      \continueeqPf{\termsize{N}}
        {By \lemmaref{lemma:Isomorphic types are the same size}}
      \continueLtPf{\termsize{\shiftdown N}}
        {\bydeftermsize}
      \AJudgeNegPf{\acontext'}
        {N}
        {\subcon{\acontext'} M}
        {\acontext''}
        {By i.h. (the type size of the ground side type in the declarative}
        \trailingjust{judgment has decreased)}
\Hand \congoesPf{\acontext''}
        {\ccontext}
        {\ditto}
\Hand \AJudgePosPf{\acontext}
        {\shiftdown N}
        {\shiftdown M}
        {\acontext''}
        {By \ashiftdown}
    \end{llproof}

    \DerivationProofCase{\dforallr}
      {\CnewNegJudge{\acontext, \alpha} {N} {M}}
      {\CnewNegJudge{\acontext} {N} {\forall \alpha \ldotp M}}

    \begin{llproof}
      \conwfPf{\acontext}
        {Assumption}
      \wfnegtypePf{\acontext}
        {N}
        {Assumption}
      \wfnegtypePf{\acontext}
        {\forall \alpha \ldotp M}
        {Assumption}
      \congoesPf{\acontext}
        {\ccontext}
        {Assumption}
      \conwfPf{\ccontext}
        {Assumption}
      \groundPf{\forall \alpha \ldotp M}
        {Assumption}
      \NoSolvedVarsPf{\acontext}
        {N}
        {Assumption}

      \proofsep
      \CnewJudgeNegPf{\acontext, \alpha}
        {N}
        {M}
        {Subderivation}
      \CnewCustomJudgeNegPf{\ccontext, \alpha}
        {\acontext, \alpha}
        {N}
        {M}
        {\bydefsubcon}
      \conwfPf{\acontext, \alpha}
        {By \cwfuvar}
      \wfnegtypePf{\acontext, \alpha}
        {N}
        {By \lemmaref{lemma:Term well-formedness weakening}}
      \wfnegtypePf{\acontext, \alpha}
        {M}
        {Inversion (\twfforall)}
      \congoesPf{\acontext, \alpha}
        {\ccontext, \alpha}
        {By \Cuvar}
      \conwfPf{\ccontext, \alpha}
        {By \cwfuvar}
      \groundPf{M}
        {By definition of ground}
      \NoSolvedVarsPf{\acontext, \alpha}
        {N}
        {By definition of $\subconunderscores$}
      \AJudgeNegPf{\acontext, \alpha}
        {N}
        {M}
        {\acontext''}
        {By i.h. (the type size of the ground side type in the declarative}
        \trailingjust{judgment is the same and the total number of prenex}
        \trailingjust{quantifiers has decreased by 1)}
      \congoesPf{\acontext''}
        {\ccontext, \alpha}
        {\ditto}

      \proofsep
      \eqPf{\acontext''}
        {\acontext', \alpha}
        {Inversion (\Cuvar)}
\Hand \congoesPf{\acontext'}
        {\ccontext}
        {\ditto}
      \AJudgeNegPf{\acontext, \alpha}
        {N}
        {M}
        {\acontext', \alpha}
        {Substituting for $\acontext''$}
\Hand \AJudgeNegPf{\acontext}
        {N}
        {\forall \alpha \ldotp M}
        {\acontext'}
        {By \aforallr}
    \end{llproof}

  \DerivationProofCase{\dforalll}
    {\wfpostypeJudg{\makedec{\acontext}} {P} \\
      \CnewNegJudge{\acontext} {\subcon{P / \alpha} N} {M'}}
    {\CnewNegJudge{\acontext} {\forall \alpha \ldotp N} {M'}}

  Proof by induction on the number of prenex universal quantifiers in $M'$:

  \begin{itemize}

      \caseitem{$n = 0$ (base case).  Let $M = M'$}

        \begin{llproof}
          \conwfPf{\acontext}
            {Assumption}
          \congoesPf{\acontext}
            {\ccontext}
            {Assumption}
          \conwfPf{\ccontext}
            {Assumption}
          \wfpostypePf{\makedec{\acontext}}
            {P}
            {Subderivation}
          \wfpostypePf{\acontext}
            {P}
            {Since $\groundJudge{P}$, this reduces to}
            \trailingjust{$\FreeUV(P) \subseteq \ConUV(\acontext)$, which holds since}
            \trailingjust{$\subconunderscores$ preserves uvars and $\FreeUV(P) \subseteq \subcon{\ccontext} \acontext$}
            \trailingjust{(the latter holding by $\wfpostypeJudg{\makedec{\acontext}}{P}$).}

          \proofsep
          \wfnegtypePf{\acontext}
            {\forall \alpha \ldotp N}
            {Assumption}
          \wfnegtypePf{\acontext, \alpha}
            {N}
            {Inversion (\twfforall)}
          \wfnegtypePf{\acontext}
            {M}
            {Assumption}
          \NoSolvedVarsPf{\acontext}
            {\forall \alpha \ldotp N}
            {Assumption}
          \NoSolvedVarsPf{\acontext}
            {N}
            {\bydefsubcon}

          \proofsep
          \CnewJudgeNegPf{\acontext}
            {\subcon{P / \alpha} N}
            {M}
            {Subderivation}
          \Pf{\subcon{\ccontext, \guess{\alpha} = P} \acontext, \guess{\alpha} \entails \ \ \ \quad}
            {}
            {}
            {}
          \Pf{\subcon{\ccontext, \guess{\alpha} = P} \subcon{\guess{\alpha} / \alpha} N}
            {\negsubtype}
            {M}
            {Where $\guess{\alpha}$ is fresh}
          \conwfPf{\acontext, \guess{\alpha}}
            {By \cwfunsolvedguess}
          \wfnegtypePf{\acontext, \guess{\alpha}}
            {\subcon{\guess{\alpha} / \alpha} N}
            {Each application of \twfuvar involving $\alpha$}
            \trailingjust{becomes an application of \twfguess}
            \trailingjust{involving $\guess{\alpha}$, and $\guess{\alpha} \in \EV(\acontext, \guess{\alpha})$}
          \wfnegtypePf{\acontext, \guess{\alpha}}
            {M}
            {By \lemmaref{lemma:Term well-formedness weakening}}
          \congoesPf{\acontext, \guess{\alpha}}
            {\ccontext, \guess{\alpha} = P}
            {By \Csolveguess}
          \conwfPf{\ccontext, \guess{\alpha} = P}
            {By \cwfsolvedguess and}
            \trailingjust{\lemmaref{lemma:Context extension preserves term well-formedness}}
          \groundPf{M}
            {Assumption}
          \NoSolvedVarsPf{\acontext, \guess{\alpha}}
            {\subcon{\guess{\alpha} / \alpha} N}
            {$\acontext, \guess{\alpha}$ can not solve $\guess{\alpha}$ since $\acontext, \guess{\alpha} \conwf$, and $\subcon{\acontext} N = N$}
          \AJudgeNegPf{\acontext, \guess{\alpha}}
            {\subcon{\guess{\alpha} / \alpha} N}
            {M}
            {\acontext''}
            {By completeness i.h. (the type size of the ground}
          \trailingjust{side type in the declarative judgment is the same,}
          \trailingjust{but the total number of prenex quantifiers has}
          \trailingjust{decreased by 1)}
          \congoesPf{\acontext''}
            {\ccontext, \guess{\alpha} = P}
            {\ditto}

          \proofcomment{By inversion on $\acontext'' \congoes \ccontext, \guess{\alpha} = P$, have $\acontext' \congoes \ccontext$ and one of the following cases:}

          \proofcomment{Case $\acontext'' = \acontext', \guess{\alpha} = Q$ and $\MutualSubtypePosJudge{\makedec{\acontext'}} {Q} {P}$:}
          \AJudgeNegPf{\acontext, \guess{\alpha}}
            {\subcon{\guess{\alpha} / \alpha} N}
            {M}
            {\acontext', \guess{\alpha} = Q}
            {Substituting for $\acontext''$}
    \Hand \AJudgeNegPf{\acontext}
            {\forall \alpha \ldotp N}
            {M}
            {\acontext'}
            {By \aforalll}
    \Hand \congoesPf{\acontext'}
            {\ccontext}
            {Above}

          \proofcomment{Case $\acontext'' = \acontext', \guess{\alpha}$ and $\alpha \notin \FreeUV(N)$:}
          \AJudgeNegPf{\acontext, \guess{\alpha}}
            {\subcon{\guess{\alpha} / \alpha} N}
            {M}
            {\acontext', \guess{\alpha} = \times}
            {Where $\times$ represents \enquote{not solved}}
    \Hand \AJudgeNegPf{\acontext}
            {\forall \alpha \ldotp N}
            {M}
            {\acontext'}
            {By \aforalll}
    \Hand \congoesPf{\acontext'}
            {\ccontext}
            {Above}
        \end{llproof}

      \caseitem{$M'$ has $n+1$ prenex universal quantifiers, \ie $M' = \forall \beta \ldotp M$ where $M$ has $n$ prenex universal quantifiers}
        
        \begin{llproof}
          \CnewJudgeNegPf{\acontext, \beta}
            {\forall \alpha \ldotp N}
            {M}
            {Inversion (\dforallr)}
          \CnewCustomJudgeNegPf{\ccontext, \beta}
            {\acontext, \beta}
            {\forall \alpha \ldotp N}
            {M}
            {\bydefsubcon}
          \conwfPf{\acontext, \beta}
            {By \cwfuvar}
          \wfnegtypePf{\acontext, \beta}
            {\forall \alpha \ldotp N}
            {By \lemmaref{lemma:Term well-formedness weakening}}
          \wfnegtypePf{\acontext, \beta}
            {M}
            {Inversion (\twfforall)}
          \congoesPf{\acontext, \beta}
            {\ccontext, \beta}
            {By \Cuvar}
          \conwfPf{\ccontext, \beta}
            {By \cwfuvar}
          \groundPf{M}
            {\bydefground}
          \NoSolvedVarsPf{\acontext, \beta}
            {\forall \alpha \ldotp N}
            {\bydefsubcon}
          \AJudgeNegPf{\acontext, \beta}
            {\forall \alpha \ldotp N}
            {M}
            {\acontext''}
            {By i.h. of induction over prenex universal quantifiers}
          \congoesPf{\acontext''}
            {\ccontext, \beta}
            {\ditto}
          \eqPf{\acontext''}
            {\acontext', \beta}
            {Inversion (\Cuvar)}
    \Hand \congoesPf{\acontext'}
            {\ccontext}
            {\ditto}
          \AJudgeNegPf{\acontext, \beta}
            {\forall \alpha \ldotp N}
            {M}
            {\acontext', \beta}
            {Substituting for $\acontext''$}
    \Hand \AJudgeNegPf{\acontext}
            {\forall \alpha \ldotp N}
            {\forall \beta \ldotp M}
            {\acontext'}
            {By \aforallr}
        \end{llproof}

    \end{itemize}

    \DerivationProofCase{\darrow}
      {\CnewPosJudge{\acontext} {Q} {P} \\ \CnewNegJudge{\acontext} {N} {M}}
      {\CnewNegJudge{\acontext} {(P \funarrow N)} {(Q \funarrow M)}}

    \begin{llproof}
      \groundPf{Q \funarrow M}
        {Assumption}
      \NoSolvedVarsPf{\acontext}
        {(P \funarrow N)}
        {Assumption}
      \wfpostypePf{\acontext}
        {P \funarrow N}
        {Assumption}
      \wfpostypePf{\acontext}
        {Q \funarrow M}
        {Assumption}

      \proofsep
      \CnewJudgePosPf{\acontext}
        {Q}
        {P}
        {Subderivation}
      \conwfPf{\acontext}
        {Assumption}
      \wfpostypePf{\acontext}
        {Q}
        {Inversion (\twfarrow)}
      \wfpostypePf{\acontext}
        {P}
        {\ditto}
      \congoesPf{\acontext}
        {\ccontext}
        {Assumption}
      \conwfPf{\ccontext}
        {Assumption}
      \groundPf{Q}
        {By definition of ground}
      \NoSolvedVarsPf{\acontext}
        {P}
        {By definition of $\subconunderscores$}
      \AJudgePosPf{\acontext}
        {Q}
        {P}
        {\acontext'}
        {By i.h. (the type size of the ground side type of the}
        \trailingjust{declarative judgment has decreased)}
      \congoesPf{\acontext'}
        {\ccontext}
        {\ditto}
      \congoesPf{\acontext}
        {\acontext'}
        {By \lemmaref{lemma:equal declarative contexts}}
      \conwfPf{\acontext'}
        {\ditto}

      \proofsep
      \CnewJudgeNegPf{\acontext}
        {N}
        {M}
        {Subderivation}
      \CnewJudgeNegPf{\acontext'}
        {N}
        {M}
        {By \lemmaref{lemma:equal declarative contexts}}
      \wfnegtypePf{\acontext'}
        {N}
        {By inversion (\twfarrow) and}
        \trailingjust{\lemmaref{lemma:Context extension preserves term well-formedness}}
      \MutualJudgeNegPf{\makedec{\acontext'}}
        {\subcon{\ccontext} \subcon{\acontext'} N}
        {\subcon{\ccontext} N}
        {By \lemmaref{lemma:Context extension leads to isomorphic types}}

      \proofsep
      \wfnegtypePf{\acontext'}
        {\subcon{\acontext'} N}
        {By \lemmaref{lemma:Well-formed context substitution preserves term well-formedness}}
      \wfnegtypePf{\makedec{\acontext'}}
        {\subcon{\ccontext} \subcon{\acontext'} N}
        {By \lemmaref{lemma:Completing a context and type preserves well-formedness}}
      \wfnegtypePf{\makedec{\acontext'}}
        {\subcon{\ccontext} N}
        {By \lemmaref{lemma:Completing a context and type preserves well-formedness}}
      \wfnegtypePf{\acontext}
        {M}
        {Inversion (\twfarrow)}
      \wfnegtypePf{\acontext'}
        {M}
        {By \lemmaref{lemma:Context extension preserves term well-formedness}}
      \wfnegtypePf{\makedec{\acontext'}}
        {M}
        {By \lemmaref{lemma:Completing a context preserves well-formedness}}

      \proofsep
      \DJudgeNegPf{\makedec{\acontext'}}
        {\subcon{\ccontext} \subcon{\acontext'} N}
        {M}
        {By \lemmaref{lemma:Transitivity of declarative pnsubtype}}
      \conwfPf{\acontext'}
        {Above}
      \wfnegtypePf{\acontext'}
        {\subcon{\acontext'} N}
        {Above}
      \wfnegtypePf{\acontext'}
        {M}
        {Above}
      \congoesPf{\acontext'}
        {\ccontext}
        {Above}
      \conwfPf{\ccontext}
        {Above}
      \groundPf{M}
        {Since $Q \funarrow M$ ground}
      \NoSolvedVarsPf{\acontext'}
        {\subcon{\acontext'} N}
        {By \lemmaref{lemma:Context substitution idempotence}}
      \AJudgeNegPf{\acontext'}
        {\subcon{\acontext'} N}
        {M}
        {\acontext''}
        {By i.h. (the type size of the ground side type of the}
        \trailingjust{declarative judgment has decreased)}
\Hand \congoesPf{\acontext''}
        {\ccontext}
        {\ditto}
\Hand \AJudgeNegPf{\acontext}
        {P \funarrow N}
        {Q \funarrow M}
        {\acontext''}
        {By \aarrow}
    \end{llproof}

    \DerivationProofCase{\dshiftup}
      {\CnewPosJudge{\acontext} {Q} {P} \\ \DPosSubtypeJudge{\makedec{\acontext}} {\subcon{\ccontext} P} {Q}}
      {\CnewNegJudge{\acontext} {\shiftup P} {\shiftup Q}}

    \begin{llproof}
      \wfnegtypePf{\acontext}
        {\shiftup P}
        {Assumption}
      \wfnegtypePf{\acontext}
        {\shiftup Q}
        {Assumption}
      \groundPf{\shiftup Q}
        {Assumption}
      \NoSolvedVarsPf{\acontext}
        {\shiftup P}
        {Assumption}
      
      \proofsep
      \CnewJudgePosPf{\acontext}
        {Q}
        {P}
        {Subderivation}
      \conwfPf{\acontext}
        {Assumption}
      \wfpostypePf{\acontext}
        {Q}
        {Since $\wfpostypeJudg{\acontext} {\shiftup Q}$}
        \wfpostypePf{\acontext}
        {P}
        {Since $\wfpostypeJudg{\acontext} {\shiftup P}$}
      \congoesPf{\acontext}
        {\ccontext}
        {Assumption}
      \conwfPf{\ccontext}
        {Assumption}
      \groundPf{Q}
        {By definition of ground}
      \NoSolvedVarsPf{\acontext}
        {P}
        {By definition of $\subconunderscores$}
      \AJudgePosPf{\acontext}
        {Q}
        {P}
        {\acontext'}
        {By i.h.}
      \congoesPf{\acontext'}
        {\ccontext}
        {\ditto}
      \conwfPf{\acontext'}
        {By \lemmaref{lemma:well-formedness-pnsubtype}}
      \congoesPf{\acontext}
        {\acontext'}
        {\ditto}
      \groundPf{\subcon{\acontext'} P}
        {\ditto}

      \proofsep
      \DJudgePosPf{\makedec{\acontext}}
        {\subcon{\ccontext} P}
        {Q}
        {Subderivation}
      \DJudgePosPf{\makedec{\acontext'}}
        {\subcon{\ccontext} P}
        {Q}
        {By \lemmaref{lemma:equal declarative contexts}}
      \wfpostypePf{\acontext'}
        {P}
        {By \lemmaref{lemma:Context extension preserves term well-formedness}}
      \MutualJudgePosPf{\makedec{\acontext'}}
        {\subcon{\acontext'} P}
        {\subcon{\ccontext} P}
        {By \lemmaref{lemma:Context extension leads to isomorphic types (ground)}}
      
      \proofsep
      \wfpostypePf{\acontext'}
        {\subcon{\acontext'} P}
        {By \lemmaref{lemma:Well-formed context substitution preserves term well-formedness}}
      \wfpostypePf{\makedec{\acontext'}}
        {\subcon{\acontext'} P}
        {By \lemmaref{lemma:Completing a context preserves well-formedness}}
      \wfpostypePf{\makedec{\acontext'}}
        {\subcon{\ccontext} P}
        {By \lemmaref{lemma:Completing a context and type preserves well-formedness}}
      \wfpostypePf{\acontext'}
        {Q}
        {By \lemmaref{lemma:Context extension preserves term well-formedness}}
      \wfpostypePf{\makedec{\acontext'}}
        {Q}
        {By \lemmaref{lemma:Completing a context preserves well-formedness}}
      \DJudgePosPf{\makedec{\acontext'}}
        {\subcon{\acontext'} P}
        {Q}
        {By \lemmaref{lemma:Transitivity of declarative pnsubtype}}

      \proofsep
      \DJudgePosPf{\makedec{\acontext'}}
        {\subcon{\acontext'} P}
        {\subcon{\ccontext} Q}
        {By \lemmaref{lemma:Context substitution on ground terms}}
      \conwfPf{\acontext'}
        {Above}
      \wfpostypePf{\acontext'}
        {\subcon{\acontext'} P}
        {By \lemmaref{lemma:Well-formed context substitution preserves term well-formedness}}
      \wfpostypePf{\acontext'}
        {Q}
        {By \lemmaref{lemma:Context extension preserves term well-formedness}}
      \congoesPf{\acontext'}
        {\ccontext}
        {Above}
      \conwfPf{\ccontext}
        {Above}
      \groundPf{\subcon{\acontext'} P}
        {Above}
      \NoSolvedVarsPf{\acontext'}
        {Q}
        {By \lemmaref{lemma:Context substitution on ground terms}}

      \proofsep
      \termsizeEqPf{\subcon{\acontext'} P}
        {\subcon{\ccontext} P}
        {By \lemmaref{lemma:context extension ground substitution size lemma}}
      \continueeqPf{\termsize{\subcon{\ccontext} Q}}
        {By \lemmaref{lemma:Isomorphic types are the same size}}
      \Pf{}
        {<}
        {\termsize{\subcon{\ccontext} \shiftup Q}}
        {\bydeftermsize}
      \AJudgePosPf{\acontext'}
        {\subcon{\acontext'} P}
        {Q}
        {\acontext''}
        {By i.h. (the type size of the ground side type of the declarative}
        \trailingjust{judgment has decreased)}
\Hand \congoesPf{\acontext''}
        {\ccontext}
        {\ditto}
\Hand \AJudgeNegPf{\acontext}
        {\shiftup P}
        {\shiftup Q}
        {\acontext''}
        {By \ashiftup}
    \end{llproof}
  \end{itemize}
\end{proof}

\section{Determinism of subtyping}

\RestateSubtypingDeterministic*

\begin{proof}
  By rule induction on the first hypothesis.

  \begin{itemize}

  \DerivationProofCase{\arefl}
    {}
    {\APosSubtypeJudg{\aconwithhole{\alpha}}{\alpha}{\alpha}{\aconwithhole{\alpha}}}

    \begin{llproof}
      \AJudgePosPf{\aconwithhole{\alpha}}{\alpha}{\alpha}{\aconwithhole{\alpha}}{Assumption}
      \AJudgePosPf{\aconwithhole{\alpha}}{\alpha}{\alpha}{\acontext'_2}{Assumption}
      \proofsep

\Hand \eqPf{\aconwithhole{\alpha}}{\acontext'_2}{By the structure of $\alpha$, the instantiation above}
      \trailingjust{is the only possible instantiation of $\possubtype$}
    \end{llproof}

  \DerivationProofCase{\ainst}
    {\wfpostypeJudg{\acontext'}{P} \\ \groundJudge{P}}
    {\APosSubtypeJudg{\aconwithhole{\guess{\alpha}}}{P}{\guess{\alpha}}{\aconwithhole{\guess{\alpha} = P}}}

    \begin{llproof}
      \AJudgePosPf{\aconwithhole{\guess{\alpha}}}{P}{\guess{\alpha}}{\aconwithhole{\guess{\alpha} = P}}{Assumption}
      \AJudgePosPf{\aconwithhole{\guess{\alpha}}}{P}{\guess{\alpha}}{\acontext'_2}{Assumption}
      \proofsep

\Hand \eqPf{\aconwithhole{\guess{\alpha} = P}}{\acontext'_2}{By the structure of $\guess{\alpha}$, the instantiation above}
      \trailingjust{is the only possible instantiation of $\possubtype$}
    \end{llproof}

  \DerivationProofCase{\ashiftdown}
    {\ANegSubtypeJudg{\acontext}{M}{N}{\acontext'} \\ \ANegSubtypeJudg{\acontext'}{N}{\subterm{\acontext'} M}{\acontext''}}
    {\APosSubtypeJudg{\acontext}{\shiftd{N}}{\shiftd{M}}{\acontext''}}
    
    \begin{llproof}
      \AJudgePosPf{\acontext}{\shiftd{N}}{\shiftd{M}}{\acontext''}{Assumption}
      \AJudgePosPf{\acontext}{\shiftd{N}}{\shiftd{M}}{\acontext_2''}{Assumption}

      \proofcomment{By the structure of $\shiftd{N}$, the derivation of the second hypothesis must end with an application of the $\ashiftdown$ rule.}

      \AJudgeNegPf{\acontext}{M}{N}{\acontext'}{Subderivation}
      \AJudgeNegPf{\acontext}{M}{N}{\acontext'_2}{Subderivation}
      \eqPf{\acontext'}{\acontext'_2}{\byih}
      \proofsep

      \AJudgeNegPf{\acontext'}{N}{\subterm{\acontext'} M}{\acontext''}{Subderivation}
      \AJudgeNegPf{\acontext'_2}{N}{\subterm{\acontext'_2} M}{\acontext''_2}{Subderivation}
      \AJudgeNegPf{\acontext'}{N}{\subterm{\acontext'} M}{\acontext''_2}{Using $\acontext' = \acontext'_2$}
\Hand \eqPf{\acontext''}{\acontext_2''}{\byih}
    \end{llproof}

  \DerivationProofCase{\aforalll}
    {\ANegSubtypeJudg{\acontext, \guess{\alpha}}{\subterm{\guess{\alpha}/\alpha} N}{M}{\acontext', \guess{\alpha}\, [= P]} \\ M \neq \forall \beta \ldotp M'}
    {\ANegSubtypeJudg{\acontext}{\forall \alpha \ldotp N}{M}{\acontext'}}

    \begin{llproof}
      \AJudgeNegPf{\acontext}{\forall \alpha \ldotp N}{M}{\acontext'}{Assumption}
      \AJudgeNegPf{\acontext}{\forall \alpha \ldotp N}{M}{\acontext_2'}{Assumption}

      \proofcomment{By the structure of $\forall \alpha \ldotp N$, and since $M \neq \forall \beta \ldotp M'$, the derivation of the second hypothesis must conclude with an application of $\aforalll$.}

      \AJudgeNegPf{\acontext, \guess{\alpha}}{[\guess{\alpha}/\alpha]N}{M}{\acontext', \guess{\alpha}\, [= P]}{Subderivation}
      \AJudgeNegPf{\acontext, \guess{\alpha}_2}{[\guess{\alpha}_2/\alpha]N}{M}{\acontext_2', \guess{\alpha}_2\, [= P_2]}{Subderivation}
      \AJudgeNegPf{\acontext, \guess{\alpha}}{[\guess{\alpha}/\alpha]N}{M}{\acontext_2', \guess{\alpha}\, [= P_2]}{Renaming the free existential variable}
      \eqPf{\acontext', \guess{\alpha} [= P]}{\acontext_2', \guess{\alpha}\, [= P_2]}{\byih}
\Hand \eqPf{\acontext', \guess{\alpha} [= P]}{\acontext_2', \guess{\alpha}_2\, [= P_2]}{Substituting back the original name ($\guess{\alpha} = \guess{\alpha}_2$)}
    \end{llproof}

  \DerivationProofCase{\aforallr}
    {\ANegSubtypeJudg{\acontext, \alpha}{N}{M}{\acontext', \alpha}}
    {\ANegSubtypeJudg{\acontext}{N}{\forall \alpha \ldotp M}{\acontext'}}
    
    \begin{llproof}
      \AJudgeNegPf{\acontext}{N}{\forall \alpha \ldotp M}{\acontext'}{Assumption}
      \AJudgeNegPf{\acontext}{N}{\forall \alpha \ldotp M}{\acontext_2'}{Assumption}

      \proofcomment{By the structure of $\forall \alpha \ldotp M$, the derivation of the second hypothesis must conclude with an application of $\aforallr$.}

      \AJudgeNegPf{\acontext, \alpha}{N}{M}{\acontext', \alpha}{Subderivation}
      \AJudgeNegPf{\acontext, \alpha}{N}{M}{\acontext'_2, \alpha}{Subderivation}
      \eqPf{\acontext', \alpha}{\acontext_2', \alpha}{\byih}
\Hand \eqPf{\acontext'}{\acontext_2'}{By above}
    \end{llproof}

  \DerivationProofCase{\aarrow}
    {\APosSubtypeJudg{\acontext}{Q}{P}{\acontext'} \\ \ANegSubtypeJudg{\acontext'}{\subterm{\acontext'} N}{M}{\acontext''}}
    {\ANegSubtypeJudg{\acontext}{P \funarrow N}{Q \funarrow M}{\acontext''}}
    
    \begin{llproof}
      \AJudgeNegPf{\acontext}{P \funarrow N}{Q \funarrow M}{\acontext''}{Assumption}
      \AJudgeNegPf{\acontext}{P \funarrow N}{Q \funarrow M}{\acontext_2''}{Assumption}

      \proofcomment{By the structure of $\arrowtype{P}{N}$ and $\arrowtype{Q}{M}$, the derivation of the second hypothesis must conclude with an application of $\aarrow$.}

      \AJudgePosPf{\acontext}{Q}{P}{\acontext'}{Subderivation}
      \AJudgePosPf{\acontext}{Q}{P}{\acontext_2'}{Subderivation}
      \eqPf{\acontext'}{\acontext_2'}{\byih}
      \proofsep

      \AJudgeNegPf{\acontext'}{\subterm{\acontext'} N}{M}{\acontext''}{Subderivation}
      \AJudgeNegPf{\acontext'_2}{\subterm{\acontext'} N}{M}{\acontext_2''}{Subderivation}
      \AJudgeNegPf{\acontext'}{\subterm{\acontext'} N}{M}{\acontext_2''}{Using $\acontext' = \acontext'_2$}
\Hand \eqPf{\acontext''}{\acontext_2''}{\byih}
    \end{llproof}

  \DerivationProofCase{\ashiftup}
    {\APosSubtypeJudg{\acontext}{Q}{P}{\acontext'} \\ \APosSubtypeJudg{\acontext'}{\subterm{\acontext'} P}{Q}{\acontext''}}
    {\ANegSubtypeJudg{\acontext}{\shiftu{P}}{\shiftu{Q}}{\acontext''}}
    
    \begin{llproof}
      \AJudgeNegPf{\acontext}{\shiftu{P}}{\shiftu{Q}}{\acontext''}{Assumption}
      \AJudgeNegPf{\acontext}{\shiftu{P}}{\shiftu{Q}}{\acontext_2''}{Assumption}

      \proofcomment{By the structure of $\shiftu{P}$ and $\shiftu{Q}$, the derivation of the second hypothesis must conclude with an application of $\ashiftup$.}

      \AJudgePosPf{\acontext}{Q}{P}{\acontext'}{Subderivation}
      \AJudgePosPf{\acontext}{Q}{P}{\acontext_2'}{Subderivation}
      \eqPf{\acontext'}{\acontext_2'}{\byih}
      \proofsep

      \AJudgePosPf{\acontext'}{\subterm{\acontext'} P}{Q}{\acontext''}{Subderivation}
      \AJudgePosPf{\acontext'_2}{\subterm{\acontext'_2} P}{Q}{\acontext_2''}{Subderivation}
      \AJudgeNegPf{\acontext'}{\subterm{\acontext'} P}{Q}{\acontext_2''}{Using $\acontext' = \acontext'_2$}
\Hand \eqPf{\acontext''}{\acontext_2''}{\byih}
    \end{llproof}
\end{itemize}
\end{proof}

\section{Decidability of subtyping}

\subsection{Lemmas for decidability}

\RestateCompletedNonGroundSizeBounded*

\begin{proof}
  Proof sketch by rule induction on algorithmic subtyping judgment.
  The justification here for using the i.h. omits the reasoning for why the premises of well-formedness must hold for the subderivations if we know that they hold for the conclusion.
  This reasoning should be identical to that in \lemmaref{lemma:well-formedness-pnsubtype}.

  \begin{itemize}
    \DerivationProofCase{\arefl}
    {}
    {\aconwithhole{\alpha} \entails \alpha \possubtype \alpha \prodcon \aconwithhole{\alpha}}

    \begin{llproof}
      \eqPf{\subcon{\aconwithhole{\alpha}} \alpha}
        {\alpha}
        {\bydefsubcon}
\Hand \termsizeLeqPf{\subcon{\aconwithhole{\alpha}} \alpha}
        {\alpha}
        {Since the types are equal}
      
    \end{llproof}

    \DerivationProofCase{\ainst}
      {\acontext_L \entails P \postype \\ \groundJudge{P}}
      {\aconwithhole{\guess{\alpha}} \entails P \possubtype \guess{\alpha} \prodcon \aconwithhole{\guess{\alpha} = P}}
    
    \begin{llproof}
      \eqPf{\subcon{\aconwithhole{\guess{\alpha} = P}} \guess{\alpha}}
        {P}
        {\bydefsubcon}
\Hand \termsizeLeqPf{\subcon{\aconwithhole{\guess{\alpha} = P}} \guess{\alpha}}
        {P}
        {Since the types are equal}
    \end{llproof}

    \DerivationProofCase{\ashiftdown}
      {\acontext \entails M \negsubtype N \prodcon \acontext' \\
        \acontext' \entails N \negsubtype \subcon{\acontext'} M \prodcon \acontext''}
      {\acontext \entails \shiftdown N \possubtype \shiftdown M \prodcon \acontext''}

    \begin{llproof}
      \groundPf{\subcon{\acontext'} M}
        {By well-formedness on the first premise}
      \conwfPf{\acontext'}
        {\ditto}
      \congoesPf{\acontext'}{\acontext''}
        {By well-formedness on the first and second premises}
      \conwfPf{\acontext''}
        {\ditto}
      \proofsep

      \conwfPf{\acontext'}
        {Above}
      \groundPf{\subcon{\acontext'} M}
        {Above}
      \congoesPf{\acontext'}{\acontext''}{Above}
      \conwfPf{\acontext''}
        {Above}
      \termsizeEqPf{\subcon{\acontext'} M}{\subcon{\acontext''} M}
        {By \lemmaref{lemma:context extension ground substitution size lemma}}
      \proofsep

      \AJudgeNegPf{\acontext}{M}{N}{\acontext'}{Subderivation}
      \termsizeLeqPf{\subcon{\acontext''} M}
        {\subcon{\acontext'} M}
        {Since the sizes are equal}
      \continueTermsizeLeqPf
        {N}
        {\byih}
\Hand \termsizeLeqPf{\subcon{\acontext''} \shiftdown M}
        {\shiftdown N}
        {\bydeftermsize}
    \end{llproof}

    \DerivationProofCase{\aforallr}
      {\acontext, \alpha \entails N \negsubtype M \prodcon \acontext', \alpha}
      {\acontext \entails N \negsubtype \forall \alpha \ldotp M \prodcon \acontext'}
  
    \begin{llproof}
      \AJudgeNegPf{\acontext, \alpha}{N}{M}{\acontext', \alpha}
        {Subderivation}
      \termsizeLeqPf{\subcon{\acontext', \alpha} N}{M}{\byih}
      \eqPf{\subcon{\acontext', \alpha} N}
        {\subcon{\acontext'} N}
        {\bydefsubcon}
      \termsizeEqPf{M}{\forall \alpha \ldotp M}
        {\bydeftermsize}
\Hand \termsizeLeqPf{\subcon{\acontext'} N}
          {\forall \alpha \ldotp M}
          {Substituting above}
    \end{llproof}

    \DerivationProofCase{\aforalll}
      { \ANegSubtypeJudg{\acontext, \guess{\alpha}}
          {\subcon{\guess{\alpha} / \alpha} N}
          {M}
          {\acontext', \guess{\alpha} \,[= P]} \\
        M \neq \forall \alpha \ldotp M'
      }
      {\acontext \entails \forall \alpha \ldotp N \negsubtype M \prodcon \acontext'}
  
    \begin{llproof}
      \AJudgeNegPf{\acontext, \guess{\alpha}}
        {\subcon{\guess{\alpha} / \alpha} N}
        {M}
        {\acontext', \guess{\alpha} \,[= P]}
        {Subderivation}
      \termsizeLeqPf{\subcon{\acontext', \guess{\alpha} \,[= P]} \subcon{\guess{\alpha} / \alpha} N}
        {M}
        {\byih}
      
      \proofcomment{Case $\alpha \notin \FreeUV(N)$:}
      \eqPf{\subcon{\acontext', \guess{\alpha} \,[= P]} \subcon{\guess{\alpha} / \alpha} N}
        {\subcon{\acontext'} N}
        {\bydefsubcon}
      \termsizeLeqPf{\subcon{\acontext'} N}
        {M}
        {Substituting above}
\Hand \termsizeLeqPf{\subcon{\acontext'} \forall \alpha \ldotp N}
        {M}
        {\bydeftermsize}

      \proofcomment{Case $\alpha \in \FreeUV(N)$:}
      \conwfPf{\acontext', \guess{\alpha} \,[= P]}{By \lemmaref{lemma:well-formedness-pnsubtype}}
      \groundPf{\subcon{\acontext', \guess{\alpha} \,[= P]} \subterm{\guess{\alpha} / \alpha} N}
        {\ditto}
      \eqPf{(\acontext', \guess{\alpha} \,[= P])}
        {(\acontext', \guess{\alpha} = P)}
        {$\alpha \in \FreeUV(N)$ so $\guess{\alpha} \in \FreeEV(\subterm{\guess{\alpha} / \alpha} N)$. Since $\guess{\alpha}$ is}
      \trailingjust{not ground, the context must solve $\guess{\alpha}$ to make}
      \trailingjust{$\subcon{\acontext', \guess{\alpha} \,[= P]} \subterm{\guess{\alpha} / \alpha} N$ ground.}
      \groundPf{P}
        {Inversion (\cwfsolvedguess)}
      \eqPf{\subcon{\acontext', \guess{\alpha} = P} \subcon{\guess{\alpha} / \alpha} N}
        {\subcon{\acontext'} \subcon{P / \guess{\alpha}} \subcon{\guess{\alpha} / \alpha} N}
        {\bydefsubcon}
      \continueeqPf{\subcon{\acontext'} \subcon{P / \alpha} N}
        {\bydefsubcon}
      \continueeqPf{\subcon{\subcon{\acontext'} P / \alpha} \subcon{\acontext'} N}
        {Since the type being replaced is a universal variable}
      \continueeqPf{\subcon{P / \alpha} \subcon{\acontext'} N}
        {Since $P$ is ground}
      \termsizeEqPf{\subcon{\acontext'} \forall \alpha \ldotp N}
        {\subcon{\acontext'} N}
        {\bydeftermsize}
      \continueTermsizeLeqPf
        {\subcon{P / \alpha} \subcon{\acontext'} N}
        {The additional substitution cannot decrease}
      \trailingjust{the size of the type}
      \continueeqPf{\termsize{\subcon{\acontext', \guess{\alpha} = P} \subcon{\guess{\alpha} / \alpha} N}}
        {Above}
      \continueTermsizeLeqPf
        {M}
        {Above}
\Hand \termsizeLeqPf{\subcon{\acontext'} \forall \alpha \ldotp N}
        {M}
        {By transitivity of $\leq$}
    \end{llproof}

    \DerivationProofCase{\aarrow}
      {\acontext \entails Q \possubtype P \prodcon \acontext' \\
        \acontext' \entails \subcon{\acontext'} N \negsubtype M \prodcon \acontext''}
      {\acontext \entails P \funarrow N \negsubtype Q \funarrow M \prodcon \acontext''}
  
    \begin{llproof}
      \AJudgePosPf{\acontext}{Q}{P}{\acontext'}{Subderivation}
      \termsizeLeqPf{\subcon{\acontext'} P}{Q}{\byih}
      \groundPf{\subcon{\acontext'} P}
        {By w.f. applied to first subderivation}
      \congoesPf{\acontext'}{\acontext''}
        {By w.f. applied to second subderivation}
      \termsizeEqPf{\subcon{\acontext'} P}{\subcon{\acontext''} P}
        {By \lemmaref{lemma:context extension ground substitution size lemma}}
      \proofsep

      \AJudgeNegPf{\acontext}{\subcon{\acontext'} N}{M}{\acontext''}{Subderivation}
      \termsizeLeqPf{\subcon{\acontext''} \subcon{\acontext'} N}{M}{\byih}
      \termsizeEqPf{\subcon{\acontext''} \subcon{\acontext'} N}{\subcon{\acontext''} N}
        {By \lemmaref{lemma:context extension substitution size lemma}}
      \proofsep

      \eqPf{\termsize{\subcon{\acontext''} (P \funarrow N)}}
        {\termsize{\subcon{\acontext''} P} + \termsize{\subcon{\acontext''} N} + 1}
        {\bydeftermsize}
      \continueeqPf{\termsize{\subcon{\acontext'} P} + \termsize{\subcon{\acontext''} \subcon{\acontext'} N} + 1}
        {Substituting above}
      \continueLeqPf{\termsize{Q} + \termsize{M} + 1}
        {Using above inequalities}
\Hand \termsizeLeqPf{\subcon{\acontext''} (P \funarrow N)}
        {Q \funarrow M}
        {\bydeftermsize}
    \end{llproof}

    \DerivationProofCase{\ashiftup}
      {\acontext \entails Q \possubtype P \prodcon \acontext' \\
        \acontext' \entails \subcon{\acontext'} P \possubtype Q \prodcon \acontext''}
      {\acontext \entails \shiftup P \negsubtype \shiftup Q \prodcon \acontext''}

    Symmetric to \ashiftdown case.
  \end{itemize}
\end{proof}

\subsection{Statement}

\RestateDecidabilityAlgorithmicSubtyping*

\begin{proof}
  The ordering is the same lexicographic ordering we used earlier in \lemmaref{lemma:Transitivity of declarative pnsubtype} and \lemmaref{theorem:Completeness of algorithmic subtyping}:

  \begin{itemize}
  \item $(\termsize{P}, \numprenex{P} + \numprenex{Q})$ for positive judgments $\APosSubtypeJudg{\acontext}{P}{Q}{\acontext'}$
  \item $(\termsize{M}, \numprenex{M} + \numprenex{N})$ for negative judgments $\ANegSubtypeJudg{\acontext}{N}{M}{\acontext'}$
  \end{itemize}

  In this ordering, $\numprenex{A}$ is the number of prenex universal quantifiers in the type $A$ and $\termsize{A}$ is the size of the algorithmic type $A$ defined in \lemmaref{lemma:Completed non-ground size bounded} (N.B. universal quantifiers do not contribute to this size).

  Sketch of this proof: We will prove by rule induction that each subderivation compares less than each conclusion for each derivation of the algorithmic subtyping judgment.
  We will assume the following additional statements about the judgment being proved in the rule induction (the same assumptions used in \lemmaref{lemma:well-formedness-pnsubtype}):

  \begin{itemize}
    \begin{minipage}[t]{0.5\linewidth}
      \item For positive subderivations \\ $\APosSubtypeJudg{\acontext}{P}{Q}{\acontext'}$:
        \begin{enumerate}[noitemsep]
          \item $\acontext \conwf$
          \item $P$ ground
          \item $\subcon{\acontext} Q = Q$
        \end{enumerate}
    \end{minipage}
    \begin{minipage}[t]{0.5\linewidth}
      \item For negative subderivations \\ $\ANegSubtypeJudg{\acontext}{N}{M}{\acontext'}$:
        \begin{enumerate}[noitemsep]
          \item $\acontext \conwf$
          \item $M$ ground
          \item $\subcon{\acontext} N = N$
        \end{enumerate}
    \end{minipage}
  \end{itemize}

  The subtyping algorithm should first check that these well-formedness assumptions hold for the judgment in question.
  By the same argument as used in \lemmaref{lemma:well-formedness-pnsubtype}, we can show that they are preserved by the algorithmic subtyping rules from conclusion to subderivations.

  \begin{itemize}
  \item Checking the well-formedness of a type is decidable:
    \begin{itemize}
    \item Typing contexts are finite, so checking $\ConUV$ and $\EV$ is decidable.
    \item There is exactly one type well-formedness rule to apply for each type.
    \item The application of each rule reduces the natural size of the type (same as $\termsize{\_}$ except universal quantification contributes to this size).
    \end{itemize}
  \item Checking whether a type is ground is decidable, since types are finite.
  \item Checking context well-formedness is decidable.
  \begin{itemize}
    \item Checking type well-formedness is decidable.
    \item Checking whether a type is ground is decidable.
    \item There is exactly one rule to apply for each context.
    \item The application of each rule for each non-empty context reduces the number of items in the context by 1.
  \end{itemize}
  \item Applying a context as a substitution to a type is decidable since each rule decreases the lexicographic order (number of free existential variables, number of items in the context).
    This follows from the requirement that solutions to existential variables are ground.
  \end{itemize}

  The subtyping algorithm should then proceed to try and apply algorithmic subtyping rules based on the structure of the types until there are no more subderivations to prove.
  The structure of the types dictate a single rule to apply at each stage.
  We now sketch a proof that each derivation of an algorithmic subtyping judgment is finite.
  As with \lemmaref{lemma:Completed non-ground size bounded}, we skip justifications for why the same assumptions used in \lemmaref{lemma:well-formedness-pnsubtype} continue to hold.

  The key idea is that at each step in the proof, the subderivation either fails or the algorithm determines that it is derivable.
  If the first subderivation fails, the algorithm should terminate in a failure state, and therefore we do not need to prove anything about the second subderivation.
  This allows us to use the first subderivations of the shift rules in the proof that the second subderivations are smaller than the conclusions.
  We have omitted stating this reasoning in each of the proof cases.

  \begin{itemize}
    \DerivationProofCase{\arefl}
      {}
      {\aconwithhole{\alpha} \entails \alpha \possubtype \alpha \prodcon \aconwithhole{\alpha}}

      No algorithmic subtyping subderivations.

    \DerivationProofCase{\ainst}
      {\acontext_L \entails P \postype \\ \groundJudge{P}}
      {\aconwithhole{\guess{\alpha}} \entails P \possubtype \guess{\alpha} \prodcon \aconwithhole{\guess{\alpha} = P}}
      
      No algorithmic subtyping subderivations.

    \DerivationProofCase{\ashiftdown}
      {\acontext \entails M \negsubtype N \prodcon \acontext' \\
        \acontext' \entails N \negsubtype \subcon{\acontext'} M \prodcon \acontext''}
      {\acontext \entails \shiftdown N \possubtype \shiftdown M \prodcon \acontext''}

      \begin{llproof}
        \eqPf{\termsize{N}}{\termsize{\shiftdown N} - 1}{\bydeftermsize}
        \continueLtPf{\termsize{\shiftdown N}}{}

        \proofcomment{Therefore the first subderivation compares less than the conclusion.}
      
        \AJudgeNegPf{\acontext}{M}{N}{\acontext'}{Subderivation}
        \termsizeLeqPf{\subcon{\acontext'} M}{N}{By \lemmaref{lemma:Completed non-ground size bounded}}
        \continueLtPf{\termsize{\shiftdown N}}{\bydeftermsize}

        \proofcomment{Therefore the second subderivation compares less than the conclusion.}
      \end{llproof}

    \DerivationProofCase{\aforallr}
      {\acontext, \alpha \entails N \negsubtype M \prodcon \acontext', \alpha}
      {\acontext \entails N \negsubtype \forall \alpha \ldotp M \prodcon \acontext'}
    
      \begin{llproof}
        \termsizeEqPf{M}{\forall \alpha \ldotp M}{\bydeftermsize}
        \ltPf{\numprenex{M}}{\numprenex{\forall \alpha \ldotp M}}{The LHS has one fewer prenex quantifier}
        \ltPf{\numprenex{N} + \numprenex{M}}
          {\numprenex{N} + \numprenex{\forall \alpha \ldotp M}}
          {}

        \proofcomment{Therefore the subderivation compares less than the conclusion.}
      \end{llproof}

    \DerivationProofCase{\aforalll}
      { \ANegSubtypeJudg{\acontext, \guess{\alpha}}
          {\subcon{\guess{\alpha} / \alpha} N}
          {M}
          {\acontext', \guess{\alpha} \,[= P]} \\
        M \neq \forall \alpha \ldotp M'
      }
      {\acontext \entails \forall \alpha \ldotp N \negsubtype M \prodcon \acontext'}
    
      \begin{llproof}
        \termsizeEqPf{M}{M}{}
        \ltPf{\numprenex{\subterm{\guess{\alpha} / \alpha} N}}{\numprenex{\forall \alpha \ldotp N}}{The LHS has one fewer prenex quantifier}

        \proofcomment{Therefore the subderivation compares less than the conclusion.}
      \end{llproof}

    \DerivationProofCase{\aarrow}
      {\acontext \entails Q \possubtype P \prodcon \acontext' \\
        \acontext' \entails \subcon{\acontext'} N \negsubtype M \prodcon \acontext''}
      {\acontext \entails P \funarrow N \negsubtype Q \funarrow M \prodcon \acontext''}
    
      \begin{llproof}
        \eqPf{\termsize{Q}}{\termsize{Q \funarrow M} - \termsize{M} - 1}{\bydeftermsize}
        \continueLtPf{\termsize{Q \funarrow M}}{}
      
        \proofcomment{Therefore the first subderivation compares less than the conclusion.}

        \eqPf{\termsize{M}}{\termsize{Q \funarrow M} - \termsize{Q} - 1}{\bydeftermsize}
        \continueLtPf{\termsize{Q \funarrow M}}{}

        \proofcomment{Therefore the second subderivation compares less than the conclusion.}
      \end{llproof}

    \DerivationProofCase{\ashiftup}
      {\acontext \entails Q \possubtype P \prodcon \acontext' \\
        \acontext' \entails \subcon{\acontext'} P \possubtype Q \prodcon \acontext''}
      {\acontext \entails \shiftup P \negsubtype \shiftup Q \prodcon \acontext''}
  
      \begin{llproof}
        \eqPf{\termsize{Q}}{\termsize{\shiftup Q} - 1}{\bydeftermsize}
        \continueLtPf{\termsize{\shiftup Q}}{}

        \proofcomment{Therefore the first subderivation compares less than the conclusion.}

        \AJudgePosPf{\acontext'}{Q}{P}{\acontext'}{Subderivation}
        \termsizeLeqPf{\subcon{\acontext'} P}{Q}{By \lemmaref{lemma:Completed non-ground size bounded}}
        \continueLtPf{\termsize{\shiftup Q}}{\bydeftermsize}

        \proofcomment{Therefore the second subderivation compares less than the conclusion.}
      \end{llproof}
  \end{itemize}
\end{proof}

\section{Isomorphic types}

\IsomorphicTypeTypeSameExpressions*

\begin{proof}
  By mutual induction on the checking, synthesis, and spine judgments.

  We first define notions of the sizes of terms and spines:

  \vspace{1em}

  \judgboxx{|e|}{The size of the term $e$}
\judgboxx{|s|}{The size of the spine $s$}
\begin{align*}
    |x| = 1 & & |\{t\}| = |t| + 1 \\ 
    |\lambda x. t| = |t| + 1 & & |\gen{\alpha}{t}| = |t| + 1 \\
    |\text{\textsf{return }} v| = |v| + 1 \\
    |\letanno{x}{P}{f}{s}{t}| = |f| + |s| + |t| + 1 & & |\letplain{x}{f}{s}{t}| = |f| + |s| + |t| + 1 \\
    |\epsilon| = 1 & & |s, v| = |s| + |v| + 1
\end{align*}

  We perform the induction using the following metric on judgments:

  \vspace{1em}

  \DeclarePairedDelimiter\judgesize{\lvert}{\rvert}

\judgbox{\judgesize{J}}{The size of the judgment $J$}
\begin{align*}
    \judgesize{\algosynjudg{\acontext; \typeenv}{v}{P}{\acontext'}} &= (|f|, 0) \\
    \judgesize{\algosynjudg{\acontext; \typeenv}{t}{N}{\acontext'}} &= (|t|, 0) \\
    \judgesize{\algospinejudg{\acontext; \typeenv}{t}{\spine{N}{M}}{\acontext'}} &= (|t|, \numprenex{N}) \\
\end{align*}

  \begin{itemize}
    \DerivationProofCase{\Dvar}
      {x : P \in \Gamma}
      {\declsynjudg{\Theta; \Gamma}{x}{P}}

    \begin{llproof}
      \inPf{x : P}{\typeenv}{Premise}
      \declisovarctxjudgPf{\dcontext}{\typeenv}{\typeenv'}{Assumption}
      \Label{1}   \inPf{x : P'}{\typeenv'}{Inversion (\Eisovar)}
\Hand \MutualJudgePosPf{\dcontext}{P}{P'}{\ditto}
\Hand \declsynjudgPf{\Theta; \Gamma'}{x}{P'}{By \Dvar and (1)}
    \end{llproof}

    \DerivationProofCase{\Dfunabs}
      {\declsynjudg{\Theta; \Gamma, x : P}{t}{N}}
      {\declsynjudg{\Theta; \Gamma}{\lambda x : P \ldotp t}{P \to N}}

    \begin{llproof}
      \declisovarctxjudgPf{\dcontext}{\typeenv}{\typeenv'}{Assumption}
      \MutualJudgePosPf{\dcontext}{P}{P}{By \lemmaref{lemma:Reflexivity of declarative pnsubtype}}
      \declisovarctxjudgPf{\dcontext}{\typeenv, x : P}{\typeenv', x : P}{By \Eisovar}
      \declsynjudgPf{\Theta; \Gamma, x : P}{t}{N}{Subderivation}
      \MutualJudgeNegPf{\dcontext}{N}{N'}{\byih (term size has decreased)}
      \declsynjudgPf{\Theta; \Gamma', x : P}{t}{N'}{\ditto}
      \proofsep

\Hand \MutualJudgeNegPf{\dcontext}{\arrowtype{P}{N}}{\arrowtype{P}{N'}}{By \darrow}
\Hand \declsynjudgPf{\Theta; \Gamma'}{\lambda x : P \ldotp t}{P \to N'}{By \Dfunabs}
    \end{llproof}

    \DerivationProofCase{\Dtypeabs}
      {\declsynjudg{\Theta, \alpha; \Gamma}{t}{N}}
      {\declsynjudg{\Theta; \Gamma}{\gen{\alpha}{t}}{\forall\alpha\ldotp N}}

    \begin{llproof}
      \declisovarctxjudgPf{\dcontext}{\typeenv}{\typeenv'}{Assumption}
      \declsynjudgPf{\Theta, \alpha; \Gamma}{t}{N}{Subderivation}
      \MutualJudgeNegPf{\dcontext}{N}{N'}{\byih (term size has decreased)}
      \declsynjudgPf{\Theta, \alpha; \Gamma'}{t}{N'}{\ditto}
      \proofsep

\Hand \MutualJudgePosPf{\dcontext}{\forall\alpha\ldotp N}{\forall\alpha\ldotp N'}{By \dforalll (using $P = \alpha$) and \dforallr}
\Hand \declsynjudgPf{\Theta; \Gamma'}{\gen{\alpha}{t}}{\forall\alpha\ldotp N'}{By {\Dtypeabs}}
    \end{llproof}

    \DerivationProofCase{\Dthunk}
      {\declsynjudg{\Theta; \Gamma}{t}{N}}
      {\declsynjudg{\Theta; \Gamma}{\thunk{t}}{\shiftd{N}}}

    \begin{llproof}
      \declisovarctxjudgPf{\dcontext}{\typeenv}{\typeenv'}{Assumption}
      \declsynjudgPf{\dcontext; \typeenv}{t}{N}{Subderivation}
      \declnisotypejudgPf{\dcontext}{N}{N'}{\byih (term size has decreased)}
      \declsynjudgPf{\dcontext; \typeenv'}{t}{N'}{\ditto}
      \proofsep
\Hand \declnisotypejudgPf{\dcontext}{\shiftd{N}}{\shiftd{N'}}{By {\dshiftdown}}
\Hand \declsynjudgPf{\dcontext; \typeenv'}{\thunk{t}}{\shiftd{N'}}{By \Dthunk}
    \end{llproof}

    \DerivationProofCase{\Dreturn}
      {\declsynjudg{\Theta; \Gamma}{v}{P}}
      {\declsynjudg{\Theta; \Gamma}{\return{v}}{\shiftu{P}}}

    \begin{llproof}
      \declisovarctxjudgPf{\dcontext}{\typeenv}{\typeenv'}{Assumption}
      \declsynjudgPf{\dcontext; \typeenv}{v}{P}{Subderivation}
      \declpisotypejudgPf{\dcontext}{P}{P'}{\byih (term size has decreased)}
      \declsynjudgPf{\dcontext; \typeenv'}{v}{P'}{\ditto}
      \proofsep
\Hand \declnisotypejudgPf{\dcontext}{\shiftu{P}}{\shiftu{P'}}{By {\dshiftup}}
\Hand \declsynjudgPf{\dcontext; \typeenv'}{\return{v}}{\shiftu{P'}}{By \Dreturn}
    \end{llproof}

    \DerivationProofCase{\Dambiguouslet}
      {\declsynjudg{\dcontext; \typeenv}{v}{\shiftd M} \\
        \declspinejudg{\Theta; \Gamma}{s}{\spine{M}{\shiftu{Q}}} \\
        \DNegSubtypeJudge{\Theta}{\shiftu{P}}{\shiftu{Q}} \\
        \declsynjudg{\Theta; \Gamma, x : P}{t}{N}}
      {\declsynjudg{\Theta; \Gamma}{\letanno{x}{P}{v}{s}{t}}{N}}

    \begin{llproof}
      \declisovarctxjudgPf{\dcontext}{\typeenv}{\typeenv'}{Assumption}
      \proofsep

      \declsynjudgPf{\dcontext; \typeenv}{v}{\shiftd M'}{Subderivation}
      \MutualJudgePosPf{\dcontext}{\shiftd M}{\shiftd M'}{\byih (term size has decreased)}
\Label{1} \declsynjudgPf{\dcontext; \typeenv'}{v}{\shiftd M'}{\ditto}
      \proofsep

      \declspinejudgPf{\Theta; \Gamma}{s}{\spine{M}{\shiftu{Q}}}{Subderivation}
      \MutualJudgeNegPf{\dcontext}{M}{M'}{Inversion (\dshiftdown)}
      \declnisotypejudgPf{\dcontext}{\shiftu{Q}}{\shiftu{Q'}}{\byih (term size has decreased)}
\Label{2} \declspinejudgPf{\dcontext; \typeenv'}{s}{\spine{M'}{\shiftu{Q'}}}{\ditto}
      \proofsep
      \DJudgeNegPf{\dcontext}{\shiftu{P}}{\shiftu{Q}}{Premise}
\Label{3} \DJudgeNegPf{\dcontext}{\shiftu{P}}{\shiftu{Q'}}{By \lemmaref{lemma:Transitivity of declarative pnsubtype}}
      \proofsep

      \declisovarctxjudgPf{\dcontext}{\typeenv, x : P}{\typeenv', x : P}{By \Eisovar}
      \declsynjudgPf{\Theta; \Gamma, x : P}{t}{N}{Subderivation}
\Hand \declnisotypejudgPf{\dcontext}{N}{N'}{\byih (term size has decreased)}
\Label{4} \declsynjudgPf{\dcontext; \typeenv', x : P}{t}{N'}{\ditto}
      \proofsep
\Hand \declsynjudgPf{\dcontext; \typeenv'}{\letanno{x}{P}{v}{s}{t}}{N'}{By {\Dambiguouslet} and (1--4)}
    \end{llproof}

    \DerivationProofCase{\Dunambiguouslet}
      {\declsynjudg{\dcontext; \typeenv}{v}{\shiftd M} \\
        \declspinejudg{\Theta; \Gamma}{s}{\spine{M}{\shiftu{Q}}} \\
        \declsynjudg{\Theta; \Gamma, x : Q}{t}{N} \\
        \forall P \ldotp \text{if } \declspinejudg{\Theta;\Gamma}{s}{\spine{M}{\shiftu{P}}} \text{ then } \declpisotypejudg{\Theta}{Q}{P}}
      {\declsynjudg{\Theta; \Gamma}{\letplain{x}{v}{s}{t}}{N}}

    \begin{llproof}
      \declisovarctxjudgPf{\dcontext}{\typeenv}{\typeenv'}{Assumption}
      \proofsep

      \declsynjudgPf{\dcontext; \typeenv}{v}{\shiftd M}{Subderivation}
      \MutualJudgePosPf{\dcontext}{\shiftd M}{\shiftd M'}{\byih (term size has decreased)}
\Label{1} \declsynjudgPf{\dcontext; \typeenv'}{v}{\shiftd M'}{\ditto}
      \proofsep

      \declspinejudgPf{\Theta; \Gamma}{s}{\spine{M}{\shiftu{Q}}}{Subderivation}
      \MutualJudgeNegPf{\dcontext}{M}{M'}{Inversion (\dshiftdown)}
      \declnisotypejudgPf{\dcontext}{\shiftu{Q}}{\shiftu{Q'}}{\byih (term size has decreased)}
\Label{2} \declspinejudgPf{\dcontext; \typeenv'}{s}{\spine{M'}{\shiftu{Q'}}}{\ditto}

      \proofsep
      \MutualJudgePosPf{\dcontext}{Q}{Q'}{Inversion (\dshiftup)}
      \declisovarctxjudgPf{\dcontext}{\typeenv, x : Q}{\typeenv', x : Q'}{By \Eisovar}
\Hand \declnisotypejudgPf{\dcontext}{N}{N'}{\byih (term size has decreased)}
\Label{3} \declsynjudgPf{\dcontext; \typeenv', x : Q'}{t}{N'}{\ditto}

      \proofcomment{To show the final premise of the \Dunambiguouslet rule, let $P$ be arbitrary and assume $\declspinejudg{\dcontext; \typeenv'}{s}{\spine{M'}{\shiftu{P}}}$.
      Now show that $\declnisotypejudg{\dcontext}{Q'}{P}$:}

      \declspinejudgPf{\dcontext; \typeenv'}{s}{\spine{M'}{\shiftu{P}}}{Assumption}
      \declisovarctxjudgPf{\dcontext}{\typeenv}{\typeenv'}{Above}
      \declnisotypejudgPf{\dcontext}{M}{M'}{Above}
      \declspinejudgPf{\dcontext; \typeenv}{s}{\spine{M}{\shiftu{P'}}}{\byih (term size has decreased)}
      \declnisotypejudgPf{\dcontext}{\shiftu{P}}{\shiftu{P'}}{\ditto}
      \proofsep

      \declnisotypejudgPf{\dcontext}{P}{P'}{Inversion (\dshiftup)}
      \declnisotypejudgPf{\dcontext}{Q}{P'}{Applying subderivation}
      \declnisotypejudgPf{\dcontext}{Q}{Q'}{Above}
      \declnisotypejudgPf{\dcontext}{Q'}{P}{By \lemmaref{lemma:Transitivity of declarative pnsubtype}}

      \proofcomment{We have now shown the final premise of \Dunambiguouslet, so apply it to give the required typing judgment:}

\Hand \declsynjudgPf{\dcontext; \typeenv'}{\letanno{x}{P}{v}{s}{t}}{N'}{By {\Dunambiguouslet} and (1--3)}
    \end{llproof}

    \DerivationProofCase{\Dspinenil}
    {}
    {\declspinejudg{\dcontext; \typeenv}{\epsilon}{\spine{N}{N}}}

    \begin{llproof}
      \declisovarctxjudgPf{\dcontext}{\typeenv}{\typeenv'}{Assumption}
\Hand \declnisotypejudgPf{\dcontext}{N}{N'}{Assumption}

\Hand \declspinejudgPf{\dcontext; \typeenv'}{\epsilon}{\spine{N'}{N'}}{By {\Dspinenil}}

    \end{llproof}

    \DerivationProofCase{\Dspinecons}
      {\declsynjudg{\Theta; \Gamma}{v}{P} \\ \DPosSubtypeJudge{\Theta}{P}{Q} \\
        \declspinejudg{\Theta; \Gamma}{s}{\spine{N}{M}}}
      {\declspinejudg{\Theta; \Gamma}{v, s}{\spine{(Q \to N)}{M}}}

    \begin{llproof}
      \declisovarctxjudgPf{\dcontext}{\typeenv}{\typeenv'}{Assumption}
      \declnisotypejudgPf{\dcontext}{Q \funarrow N}{T}{Assumption}
    \end{llproof}

    By inversion, $T = \forall\alpha\cdots\forall\beta.P' \funarrow N'$. Therefore perform induction over the number of prenex universal quantifiers, $n$:

    \begin{itemize}
      \caseitem{$n = 0$}

      \begin{llproof}
        \eqPf{N}{Q' \funarrow N'}{By inversion}
        \declnisotypejudgPf{\dcontext}{Q \funarrow N}{Q' \funarrow N'}{Assumption}
        \MutualJudgePosPf{\dcontext}{Q}{Q'}{By inversion}
        \declnisotypejudgPf{\dcontext}{N}{N'}{By inversion}
        \proofsep
        
        \MutualJudgePosPf{\dcontext}{P}{P'}{By outer i.h. (term size has decreased)}
\Label{1} \declsynjudgPf{\dcontext; \typeenv'}{v}{P'}{\ditto}
        \proofsep

        \DJudgePosPf{\Theta}{P}{Q}{Subderivation}
\Label{2} \DJudgePosPf{\Theta}{P'}{Q'}{By \lemmaref{lemma:Transitivity of declarative pnsubtype}}
        \proofsep

        \declspinejudgPf{\Theta; \Gamma}{s}{\spine{N}{M}}{Subderivation}
\Hand   \declnisotypejudgPf{\dcontext}{M}{M'}{By outer i.h. (term size has decreased)}
\Label{3} \declspinejudgPf{\dcontext; \typeenv'}{s}{\spine{N'}{M'}}{\ditto}
        \proofsep
\Hand   \declspinejudgPf{\dcontext; \typeenv'}{v, s}{\spine{Q' \funarrow N'}{M'}}{By \Dspinecons and (1--3)}

      \end{llproof}

      \item \textbf{Case: }$n = k + 1$

      \begin{llproof}
        \eqPf{T}{\forall\alpha\ldotp T'}{By inversion}
        \MutualJudgePosPf{\dcontext}{P \funarrow N}{[P / \alpha]T'}{By inversion (\dforalll, \dforallr), for $\wfpostypeJudg{\dcontext}{P}$}

        \declspinejudgPf{\dcontext; \typeenv'}{v, s}{\spine{[P/\alpha]T'}{M'}}{By inner i.h.}
        \Hand     \declnisotypejudgPf{\dcontext}{M}{M'}{\ditto}

        \proofsep
        \declspinejudgPf{\dcontext; \typeenv'}{v, s}{\spine{(\forall\alpha\ldotp T')}{M'}}{By \Dspinetypeabs}
        \Hand     \declspinejudgPf{\dcontext; \typeenv'}{v, s}{\spine{T}{M'}}{By equality}
      \end{llproof}
    \end{itemize}

    \DerivationProofCase{\Dspinetypeabs}
      {\wfpostypeJudg{\Theta}{P} \\ \declspinejudg{\Theta; \Gamma}{s}{\spine{[P/\alpha]N}{M}}}
      {\declspinejudg{\Theta; \Gamma}{s}{\spine{(\forall\alpha\ldotp N)}{M}}}

    \begin{llproof}
      \declisovarctxjudgPf{\dcontext}{\typeenv}{\typeenv'}{Assumption}
      \declnisotypejudgPf{\dcontext}{\forall\alpha\ldotp N}{N''}{Assumption}
      \proofsep

      \eqPf{N''}{\forall \beta \ldotp N'}{Inversion (\dforalll / \dforallr)}
      \declnisotypejudgPf{\dcontext}{[P/\alpha]N}{[R/\beta] N'}{\ditto, for $\wfpostypeJudg{\dcontext}{R}$}
      \Hand \declnisotypejudgPf{\dcontext}{M}{M'}{\byih (term size is the same and the}
      \trailingjust{number of prenex quantifiers has decreased)}
      \declspinejudgPf{\dcontext; \typeenv'}{s}{\spine{[R/\beta] N'}{M'}}{\ditto}
      \proofsep

      \declspinejudgPf{\dcontext; \typeenv'}{s}{\spine{\forall \beta \ldotp N'}{M'}}{By \Dspinetypeabs}
\Hand \declspinejudgPf{\dcontext; \typeenv'}{s}{\spine{N''}{M'}}{By definition of $N''$}
    \end{llproof}
  \end{itemize}
\end{proof}

\section{Well-formedness of typing}

\RestrictedContextWf*

\begin{proof}
  By rule induction on the $\congoesWeakJudg{\acontext}{\acontext'}$ judgment.

  \begin{itemize}
    \DerivationProofCase{\Wcempty}
      {}
      {\congoesWeakJudg{\emptyacontext}{\emptyacontext}}

    \begin{llproof}
      \eqPf{\restrictcontext{\emptyacontext}{\emptyacontext}}{\emptyacontext}{By \Restrictempty}
      \Hand   \conwfPf{\emptyacontext}{By {\cwfempty}}
\Hand \congoesPf{\emptyacontext}{\emptyacontext}{By {\Cempty}}
\Hand \congoesWeakPf{\emptyacontext}{\emptyacontext}{By {\Wcempty}}
    \end{llproof}

    \DerivationProofCase{\Wcuvar}
      {\congoesWeakJudg{\Theta}{\Theta'}}
      {\congoesWeakJudg{\Theta, \alpha}{\Theta', \alpha}}

    \begin{llproof}
      \conwfPf{\acontext, \alpha}{Assumption}
      \conwfPf{\Theta}{Inversion (\cwfuvar)}
      \conwfPf{\acontext', \alpha}{Assumption}
      \conwfPf{\Theta'}{Inversion (\cwfuvar)}
      \congoesWeakPf{\acontext}{\acontext'}{Subderivation}
      \proofsep
      \congoesPf{\Theta}{\restrictcontext{\Theta'}{\Theta}}{\byih}
      \conwfPf{\restrictcontext{\Theta'}{\Theta}}{\ditto}
      \congoesWeakPf{\restrictcontext{\Theta'}{\Theta}}{\Theta'}{\ditto}
      \proofsep
      \eqPf{\restrictcontext{(\Theta', \alpha)}{(\Theta, \alpha)}}{(\restrictcontext{\Theta'}{(\Theta, \alpha)}), \alpha}{By \Restrictuvar}
      \proofsep
\Hand \conwfPf{(\restrictcontext{\Theta'}{\Theta}), \alpha}{By \cwfuvar}
\Hand \congoesPf{\Theta, \alpha}{(\restrictcontext{\Theta'}{\Theta}), \alpha}{By \Cuvar}
\Hand \congoesWeakPf{(\restrictcontext{\Theta'}{\Theta}), \alpha}{\Theta', \alpha}{By \Wcuvar}
    \end{llproof}

    \DerivationProofCase{\Wcunsolvedguess}
      {\congoesWeakJudg{\Theta}{\Theta'}}
      {\congoesWeakJudg{\Theta, \guess{\alpha}}{\Theta', \guess{\alpha}}}

    \begin{llproof}
      \conwfPf{\acontext, \guess{\alpha}}{Assumption}
      \conwfPf{\Theta}{Inversion (\cwfunsolvedguess)}
      \conwfPf{\acontext', \guess{\alpha}}{Assumption}
      \conwfPf{\Theta'}{Inversion (\cwfunsolvedguess)}
      \congoesWeakPf{\acontext}{\acontext'}{Subderivation}
      \proofsep
      \conwfPf{\restrictcontext{\acontext'}{\acontext}}{\byih}
      \congoesPf{\Theta}{\restrictcontext{\acontext'}{\acontext}}{\ditto}
      \congoesWeakPf{\restrictcontext{\acontext'}{\acontext}}{\Theta'}{\ditto}
      \proofsep
      \eqPf{\restrictcontext{(\acontext', \guess{\alpha})}{(\acontext, \guess{\alpha})}}{(\restrictcontext{\acontext'}{\acontext}), \guess{\alpha}}{By \Restrictguessin}
      \proofsep
\Hand \conwfPf{(\restrictcontext{\acontext'}{\acontext}), \guess{\alpha}}{By \cwfunsolvedguess}
\Hand \congoesPf{\Theta, \guess{\alpha}}{(\restrictcontext{\acontext'}{\acontext}), \guess{\alpha}}{By \Cunsolvedguess}
\Hand \congoesWeakPf{(\restrictcontext{\acontext'}{\acontext}), \guess{\alpha}}{\Theta', \guess{\alpha}}{By \Wcunsolvedguess}
    \end{llproof}

    \DerivationProofCase{\Wcsolveguess}
      {\congoesWeakJudg{\Theta}{\Theta'}}
      {\congoesWeakJudg{\Theta, \guess{\alpha}}{\Theta', \guess{\alpha} = P}}

    \begin{llproof}
      \conwfPf{\acontext, \guess{\alpha}}{Assumption}
      \conwfPf{\Theta}{Inversion (\cwfunsolvedguess)}
      \conwfPf{\acontext', \guess{\alpha} = P}{Assumption}
      \conwfPf{\Theta'}{Inversion (\cwfsolvedguess)}
      \congoesWeakPf{\acontext}{\acontext'}{Subderivation}
      \proofsep
      \conwfPf{\restrictcontext{\acontext'}{\acontext}}{\byih}
      \congoesPf{\Theta}{\restrictcontext{\acontext'}{\acontext}}{\ditto}
      \congoesWeakPf{\restrictcontext{\acontext'}{\acontext}}{\Theta'}{\ditto}
      \proofsep
      \eqPf{\restrictcontext{(\acontext', \guess{\alpha} = P)}{(\acontext, \guess{\alpha})}}{(\restrictcontext{\acontext'}{\acontext}), \guess{\alpha} = P}{By \Restrictguessin}
      \proofsep
\Hand \conwfPf{(\restrictcontext{\acontext'}{\acontext}), \guess{\alpha} = P}{By \cwfsolvedguess}
\Hand \congoesPf{\Theta, \guess{\alpha}}{(\restrictcontext{\acontext'}{\acontext}), \guess{\alpha} = P}{By \Csolveguess}
\Hand \congoesWeakPf{(\restrictcontext{\acontext'}{\acontext}), \guess{\alpha} = P}{\Theta', \guess{\alpha} = P}{By \Wcsolvedguess}
    \end{llproof}

    \DerivationProofCase{\Wcsolvedguess}
    {
      \congoesWeakJudg{\acontext}{\acontext'} \\
      \MutualSubtypePosJudge{\makedec{\acontext}} {P} {Q}
    }
    {\congoesWeakJudg{\acontext, \guess{\alpha} = P}{\acontext', \guess{\alpha} = Q}}

    \begin{llproof}
      \conwfPf{\acontext, \guess{\alpha} = P}{Assumption}
      \conwfPf{\Theta}{Inversion (\cwfsolvedguess)}
      \conwfPf{\acontext', \guess{\alpha} = Q}{Assumption}
      \conwfPf{\Theta'}{Inversion (\cwfsolvedguess)}
      \congoesWeakPf{\acontext}{\acontext'}{Subderivation}
      \proofsep
      \conwfPf{\restrictcontext{\acontext'}{\acontext}}{\byih}
      \congoesPf{\Theta}{\restrictcontext{\acontext'}{\acontext}}{\ditto}
      \congoesWeakPf{\restrictcontext{\acontext'}{\acontext}}{\Theta'}{\ditto}
      \proofsep
      \eqPf{\restrictcontext{(\acontext', \guess{\alpha} = Q)}{(\acontext, \guess{\alpha} = P)}}{(\restrictcontext{\acontext'}{\acontext}), \guess{\alpha} = Q}{By \Restrictguessin}
      \proofsep
\Hand \conwfPf{(\restrictcontext{\acontext'}{\acontext}), \guess{\alpha} = Q}{By \cwfsolvedguess}
      \MutualJudgePosPf{\makedec{\acontext}} {P} {Q} {Premise}
\Hand \congoesPf{\Theta, \guess{\alpha} = P}{(\restrictcontext{\acontext'}{\acontext}), \guess{\alpha} = Q}{By \Csolvedguess}
\Hand \congoesWeakPf{(\restrictcontext{\acontext'}{\acontext}), \guess{\alpha} = Q}{\Theta', \guess{\alpha} = Q}{By \Wcsolvedguess}
    \end{llproof}

    \DerivationProofCase{\Wcunsolvedextend}
    {\congoesWeakJudg{\Theta}{\Theta'}}
    {\congoesWeakJudg{\Theta}{\Theta', \guess{\alpha}}}

    \begin{llproof}
      \notinPf{\guess{\alpha}\,[= P]}{\acontext}{Since $\guess{\alpha}$ fresh}
      \eqPf{\restrictcontext{(\acontext', \guess{\alpha})}{\acontext}}{\restrictcontext{\acontext'}{\acontext}}{By \Restrictguessnotin}
      \proofsep
      \conwfPf{\acontext}{Assumption}
      \conwfPf{\acontext', \guess{\alpha}}{Assumption}
      \conwfPf{\Theta'}{Inversion (\cwfunsolvedguess)}
      \congoesWeakPf{\acontext}{\acontext'}{Subderivation}
      \proofsep
\Hand \conwfPf{\restrictcontext{\acontext'}{\acontext}}{\byih}
\Hand \congoesPf{\Theta}{\restrictcontext{\acontext'}{\acontext}}{\ditto}
      \congoesWeakPf{\restrictcontext{\acontext'}{\acontext}}{\Theta'}{\ditto}
\Hand \congoesWeakPf{\restrictcontext{\acontext'}{\acontext}}{\Theta', \guess{\alpha}}{By \Wcunsolvedextend}
    \end{llproof}

    \DerivationProofCase{\Wcsolvedextend}
    {\congoesWeakJudg{\Theta}{\Theta'}}
    {\congoesWeakJudg{\Theta}{\Theta', \guess{\alpha}} = P}

    \begin{llproof}
      \notinPf{\guess{\alpha}\,[= Q]}{\acontext}{Since $\guess{\alpha}$ fresh}
      \eqPf{\restrictcontext{(\acontext', \guess{\alpha} = P)}{\acontext}}{\restrictcontext{\acontext'}{\acontext}}{By \Restrictguessnotin}
      \proofsep
      \conwfPf{\acontext}{Assumption}
      \conwfPf{\acontext', \guess{\alpha} = P}{Assumption}
      \conwfPf{\Theta'}{Inversion (\cwfsolvedguess)}
      \congoesWeakPf{\acontext}{\acontext'}{Subderivation}
      \proofsep
\Hand \conwfPf{\restrictcontext{\acontext'}{\acontext}}{\byih}
\Hand \congoesPf{\Theta}{\restrictcontext{\acontext'}{\acontext}}{\ditto}
      \congoesWeakPf{\restrictcontext{\acontext'}{\acontext}}{\Theta'}{\ditto}
\Hand \congoesWeakPf{\restrictcontext{\acontext'}{\acontext}}{\Theta', \guess{\alpha} = P}{By \Wcsolvedextend}
    \end{llproof}
  \end{itemize}
\end{proof}

\begin{center}
  \AlphaNotInTypeWellFormed*
\end{center}

\begin{proof}
  By rule induction over the definition of well-formed types.

  \begin{itemize}
    \DerivationProofCase{\twfuvar}
      {\beta \in \FreeUV(\Theta_L, \alpha, \Theta_R)}
      {\wfpostypeJudg{\Theta_L, \alpha, \Theta_R}{\beta}}

    \begin{llproof}
      \notinPf{\alpha}{\FreeUV(\beta)}{Assumption}
      \neqPf{\beta}{\alpha}{By above}
      \wfpostypePf{\Theta_L, \alpha, \Theta_R}{\beta}{Assumption}
      \inPf{\beta}{\FreeUV(\Theta_L, \Theta_R)}{By above two statements}
\Hand \wfpostypePf{\Theta_L, \Theta_R}{\beta}{By {\twfuvar}}
    \end{llproof}

    \DerivationProofCase{\twfguess}
      {\guess{\alpha} \in \FreeEV(\Theta_L, \alpha, \Theta_R)}
      {\wfpostypeJudg{\Theta_L, \alpha, \Theta_R}{\guess{\alpha}}}

    \begin{llproof}
      \wfpostypePf{\Theta_L, \alpha, \Theta_R}{\guess{\alpha}}{Assumption}
      \inPf{\guess{\alpha}}{\FreeEV(\Theta_L, \Theta_R)}{By above}
\Hand \wfpostypePf{\Theta_L, \Theta_R}{\guess{\alpha}}{By {\twfguess}}
    \end{llproof}

    \DerivationProofCase{\twfshiftdown}
      {\wfnegtypeJudg{\Theta_L, \alpha, \Theta_R}{N}}
      {\wfpostypeJudg{\Theta_L, \alpha, \Theta_R}{\shiftd{N}}}

    \begin{llproof}
      \wfnegtypePf{\Theta_L, \alpha, \Theta_R}{N}{Premise}
      \wfnegtypePf{\Theta_L, \Theta_R}{N}{\byih}
\Hand \wfpostypePf{\Theta_L, \Theta_R}{\shiftd{N}}{By {\twfshiftdown}}
    \end{llproof}

    \DerivationProofCase{\twfforall}
      {\wfnegtypeJudg{\Theta_L, \alpha, \Theta_R, \beta}{N}}
      {\wfnegtypeJudg{\Theta_L, \alpha, \Theta_R}{\forall\beta. N}}

    \begin{llproof}
      \neqPf{\beta}{\alpha}{$\beta$ is fresh}
      \wfnegtypePf{\Theta_L, \alpha, \Theta_R, \beta}{N}{Premise}
      \wfnegtypePf{\Theta_L, \Theta_R, \beta}{N}{\byih}
\Hand \wfnegtypePf{\Theta_L, \Theta_R}{\forall\beta. N}{By {\twfforall}}
    \end{llproof}

    \DerivationProofCase{\twfarrow}
      {\wfpostypeJudg{\Theta_L, \alpha, \Theta_R}{P} \\ \wfnegtypeJudg{\Theta_L, \alpha, \Theta_R}{N}}
      {\wfnegtypeJudg{\Theta_L, \alpha, \Theta_R}{P \to N}}

    \begin{llproof}
      \wfpostypePf{\Theta_L, \alpha, \Theta_R}{P}{Premise}
      \wfpostypePf{\Theta_L, \Theta_R}{P}{\byih}
      \wfnegtypePf{\Theta_L, \alpha, \Theta_R}{N}{Premise}
      \wfnegtypePf{\Theta_L, \Theta_R}{N}{\byih}
\Hand \wfnegtypePf{\Theta_L, \Theta_R}{P \to N}{By {\twfarrow}}
    \end{llproof}

    \DerivationProofCase{\twfshiftup}
      {\wfpostypeJudg{\Theta_L, \alpha, \Theta_R}{P}}
      {\wfnegtypeJudg{\Theta_L, \alpha, \Theta_R}{\shiftu{P}}}

    \begin{llproof}
      \wfpostypePf{\Theta_L, \alpha, \Theta_R}{P}{Premise}
      \wfpostypePf{\Theta_L, \Theta_R}{P}{\byih}
\Hand \wfnegtypePf{\Theta_L, \Theta_R}{\shiftu{P}}{By {\twfshiftup}}
    \end{llproof}

  \end{itemize}
\end{proof}

\SubstitutionPreservesWFTypes*

\begin{proof}
  By rule induction over the definition of well-formed types.

  \begin{itemize}
    \DerivationProofCase{\twfuvar}
    {\beta \in \FreeUV(\Theta_L, \beta, \Theta_R)}
    {\wfpostypeJudg{\Theta_L, \alpha, \Theta_R}{\alpha}}

    Take cases on whether $\beta = \alpha$
    \begin{itemize}
      \caseitem{$\beta = \alpha$}

      \begin{llproof}
        \wfpostypePf{\Theta_L, \alpha, \Theta_R}{\alpha}{Assumption}

        \eqPf{[\guess{\alpha} / \alpha]\alpha}{\guess{\alpha}}{By definition}

        \wfpostypePf{\Theta_L, \guess{\alpha}, \Theta_R}{\guess{\alpha}}{By {\twfguess}}
  \Hand \wfpostypePf{\Theta_L, \guess{\alpha}, \Theta_R}{[\guess{\alpha}/\alpha]\beta}{By equality}
      \end{llproof}

      \caseitem{$\beta \neq \alpha$}

      \begin{llproof}
        \wfpostypePf{\Theta_L, \alpha, \Theta_R}{\beta}{Assumption}

        \eqPf{[\guess{\alpha} / \alpha]\beta}{\beta}{By definition}
      \end{llproof}

      Therefore, $\beta \in \FreeUV(\Theta_L)$ or $\beta \in \FreeUV(\Theta_R)$

      \begin{llproof}
        \inPf{\beta}{\FreeUV(\Theta_L, \guess{\alpha}, \Theta_R)}{}
        \wfpostypePf{\Theta_L, \guess{\alpha}, \Theta_R}{\beta}{By {\twfuvar}}
  \Hand \wfpostypePf{\Theta_L, \guess{\alpha}, \Theta_R}{[\guess{\alpha}/\alpha]\beta}{By equality}
      \end{llproof}
    \end{itemize}

    \DerivationProofCase{\twfguess}
    {\guess{\alpha} \in \FreeEV(\Theta_L,\alpha, \Theta_R)}
    {\wfpostypeJudg{\Theta_L, \alpha, \Theta_R}{\guess{\alpha}}}

    \begin{llproof}
      \wfpostypePf{\Theta_L, \alpha, \Theta_R}{\guess{\beta}}{Assumption}

      \eqPf{[\guess{\alpha} / \alpha]\guess{\beta}}{\guess{\beta}}{By definition}
    \end{llproof}

    Therefore, $\guess{\beta} \in \FreeEV(\Theta_L)$ or $\guess{\beta} \in \FreeEV(\Theta_R)$

    \begin{llproof}
      \inPf{\guess{\beta}}{\FreeEV(\Theta_L, \guess{\alpha}, \Theta_R)}{}
      \wfpostypePf{\Theta_L, \guess{\alpha}, \Theta_R}{\guess{\beta}}{By {\twfguess}}
      \Hand           \wfpostypePf{\Theta_L, \guess{\alpha}, \Theta_R}{[\guess{\alpha}/\alpha]\guess{\beta}}{By equality}
    \end{llproof}

    \DerivationProofCase{\twfshiftdown}
    {\wfnegtypeJudg{\Theta_L, \alpha, \Theta_R}{N}}
    {\wfpostypeJudg{\Theta_L, \alpha, \Theta_R}{\shiftd{N}}}

    \begin{llproof}
      \wfpostypePf{\Theta_L, \alpha, \Theta_R}{\shiftd{N}}{Assumption}
      \wfnegtypePf{\Theta_L, \alpha, \Theta_R}{N}{Premise}
      \wfnegtypePf{\Theta_L, \guess{\alpha}, \Theta_R}{[\guess{\alpha}/\alpha]N}{\byih}
      \wfpostypePf{\Theta_L, \guess{\alpha}, \Theta_R}{\shiftd{[\guess{\alpha}/\alpha]N}}{By {\twfshiftdown}}
\Hand \wfpostypePf{\Theta_L, \guess{\alpha}, \Theta_R}{[\guess{\alpha}/\alpha]\shiftd{N}}{By definition of substitution}
    \end{llproof}

    \DerivationProofCase{\twfforall}
    {\wfnegtypeJudg{\Theta_L, \alpha, \Theta_R, \beta}{N}}
    {\wfnegtypeJudg{\Theta_L, \alpha, \Theta_R}{\forall\beta. N}}

    $\beta$ fresh, and therefore $\beta \neq \alpha$.

    \begin{llproof}
      \wfnegtypePf{\Theta_L, \alpha, \Theta_R}{\forall\beta. N}{Assumption}
      \wfnegtypePf{\Theta_L, \alpha, (\Theta_R, \beta)}{N}{Premise}
      \wfnegtypePf{\Theta_L, \guess{\alpha}, (\Theta_R, \beta)}{[\guess{\alpha}/\alpha]N}{\byih}
      \wfnegtypePf{\Theta_L, \guess{\alpha}, \Theta_R}{\forall\beta. [\guess{\alpha}/ \alpha]N}{By {\twfforall}}
\Hand \wfnegtypePf{\Theta_L, \guess{\alpha}, \Theta_R}{[\guess{\alpha}/ \alpha](\forall\beta. N)}{By definition of substitution}
    \end{llproof}

    \DerivationProofCase{\twfarrow}
    {\wfpostypeJudg{\Theta_L, \alpha, \Theta_R}{P} \\ \wfnegtypeJudg{\Theta_L, \alpha, \Theta_R}{N}}
    {\wfnegtypeJudg{\Theta_L, \alpha, \Theta_R}{P \funarrow N}}

    \begin{llproof}
      \wfnegtypePf{\Theta_L, \alpha, \Theta_R}{P \funarrow N}{Assumption}
      \proofsep
      \wfpostypePf{\Theta_L, \alpha, \Theta_R}{P}{Assumption}
      \wfpostypePf{\Theta_L, \guess{\alpha}, \Theta_R}{[\guess{\alpha}/\alpha]P}{\byih}
      \proofsep
      \wfnegtypePf{\Theta_L, \alpha, \Theta_R}{N}{Assumption}
      \wfnegtypePf{\Theta_L, \guess{\alpha}, \Theta_R}{[\guess{\alpha}/\alpha]N}{\byih}
      \proofsep
      \wfnegtypePf{\Theta_L, \guess{\alpha}, \Theta_R}{([\guess{\alpha}/\alpha]P) \to ([\guess{\alpha}/\alpha]N)}{By {\twfarrow}}
\Hand \wfnegtypePf{\Theta_L, \guess{\alpha}, \Theta_R}{[\guess{\alpha}/\alpha](P \funarrow N)}{By definition of substitution}
    \end{llproof}

    \DerivationProofCase{\twfshiftup}
    {\wfpostypeJudg{\Theta_L, \alpha, \Theta_R}{P}}
    {\wfnegtypeJudg{\Theta_L, \alpha, \Theta_R}{\shiftu{P}}}

    \begin{llproof}
      \wfnegtypePf{\Theta_L, \alpha, \Theta_R}{\shiftu{P}}{Assumption}
      \wfpostypePf{\Theta_L, \alpha, \Theta_R}{P}{Premise}
      \wfpostypePf{\Theta_L, \guess{\alpha}, \Theta_R}{[\guess{\alpha}/\alpha]P}{\byih}
      \wfnegtypePf{\Theta_L, \guess{\alpha}, \Theta_R}{\shiftu{[\guess{\alpha}/\alpha]P}}{By {\twfshiftup}}
\Hand \wfnegtypePf{\Theta_L, \guess{\alpha}, \Theta_R}{[\guess{\alpha}/\alpha]\shiftu{P}}{By definition of substitution}
    \end{llproof}
  \end{itemize}
\end{proof}

\ContextExtensionVarSame*

\begin{proof}
  All rules ensure that the \lhs and \rhs contexts have the same set of free universal variables and the same set of existential variables.
\end{proof}

\AlgorithmicTypingWellFormed*

\begin{proof}
  By mutual rule induction over the algorithmic synthesis and spine judgments.

  \begin{itemize}
    \DerivationProofCase{\Avar}
      {x : P \in \Gamma}
      {\algosynjudg{\acontext; \Gamma}{x}{P}{\acontext}}

      \begin{llproof}
        \Hand   \conwfPf{\Theta}{Assumption}
\Hand   \congoesPf{\Theta}{\Theta}{By \lemmaref{lemma:context extension reflexive}}
        \envwfPf{\Theta}{\typeenv}{Assumption}
\Hand   \wfpostypePf{\Theta}{P}{Inversion (\Ewfvar)}
\Hand   \groundPf{P}{\ditto}
      \end{llproof}

    \DerivationProofCase{\Afunabs}
      {\algosynjudg{\acontext; \Gamma, x : P}{t}{N}{\acontext'}}
      {\algosynjudg{\acontext; \Gamma}{\lamterm{x}{P}{t}}{P \to N}{\acontext'}}

      \begin{llproof}
        \conwfPf{\Theta}{Assumption}
        \envwfPf{\Theta}{\Gamma} {Assumption}
        \wfpostypePf{\Theta}{P}{$P$ annotation}
        \groundPf{P}{\ditto}
        \envwfPf{\Theta}{\Gamma, x : P} {By {\Ewfvar}}
        \algosynjudgPf{\acontext; \Gamma, x : P}{t}{N}{\acontext'}{Subderivation}
        \proofsep

\Hand   \conwfPf{\Theta'}{\byih}
\Hand   \congoesPf{\Theta}{\Theta'}{\ditto}
        \wfnegtypePf{\Theta'}{N}{\ditto}
        \groundPf{N}{\ditto}
        \proofsep

\Hand \wfnegtypePf{\Theta'}{P \to N}{By \twfarrow}
\Hand \groundPf{P \to N} {By definition of ground}
      \end{llproof}

    \DerivationProofCase{\Atypeabs}
      {\algosynjudg{\acontext, \alpha; \Gamma}{t}{N}{\acontext', \alpha}}
      {\algosynjudg{\acontext; \Gamma}{\gen{\alpha}{t}}{\forall\alpha\ldotp N}{\acontext'}}

      \begin{llproof}
        \conwfPf{\Theta}{Assumption}
        \conwfPf{\Theta, \alpha}{By {\cwfuvar}}
        \envwfPf{\Theta}{\Gamma} {Assumption}
        \envwfPf{\Theta, \alpha}{\Gamma}{By weakening}
        \algosynjudgPf{\acontext, \alpha; \Gamma}{t}{N}{\acontext', \alpha}{Subderivation}
        \proofsep
        \conwfPf{\Theta', \alpha}{\byih}
        \congoesPf{\Theta, \alpha}{\Theta', \alpha}{\ditto}
        \wfnegtypePf{\Theta, \alpha}{N}{\ditto}
        \groundPf{\subcon{\acontext'} N} {\ditto}
        \proofsep

\Hand   \conwfPf{\Theta'}{Inversion ({\cwfuvar})}
\Hand   \congoesPf{\Theta}{\Theta'}{Inversion ({\Cuvar})}
\Hand   \wfnegtypePf{\Theta}{\forall\alpha\ldotp N}{By \twfforall}
\Hand \groundPf{\subcon{\acontext'} (\forall\alpha\ldotp N)} {By definition of ground}
      \end{llproof}

    \DerivationProofCase{\Athunk}
      {\algosynjudg{\acontext; \Gamma}{t}{N}{\acontext'}}
      {\algosynjudg{\acontext; \Gamma}{\thunk{t}}{\shiftd{N}}{\acontext'}}

      \begin{llproof}
        \conwfPf{\Theta}{Assumption}
        \envwfPf{\Theta}{\Gamma} {Assumption}
        \envwfPf{\Theta}{\Gamma} {By {\Ewfvar}}
        \algosynjudgPf{\acontext; \Gamma}{t}{N}{\acontext'}{Subderivation}
        \proofsep

\Hand   \conwfPf{\Theta'}{\byih}
\Hand   \congoesPf{\Theta}{\Theta'}{\ditto}
        \wfnegtypePf{\Theta'}{N}{\ditto}
        \groundPf{\subcon{\acontext'} N}{\ditto}
        \proofsep

\Hand \wfnegtypePf{\Theta'}{\shiftd{N}}{By \dshiftdown}
\Hand \groundPf{\subcon{\acontext'} \shiftd{N}} {By definition of ground}
      \end{llproof}

    \DerivationProofCase{\Areturn}
      {\algosynjudg{\acontext; \Gamma}{v}{P}{\acontext'}}
      {\algosynjudg{\acontext; \Gamma}{\return{v}}{\shiftu{P}}{\acontext'}}

      Symmetrical to \Athunk.

    \DerivationProofCase{\Aambiguouslet}
      {\algosynjudg{\acontext; \typeenv}{v}{\shiftd M}{\acontext'} \\
      \algospinejudg{\acontext'; \Gamma}{s}{\spine{M}{\shiftu{Q}}}{\acontext''} \\
      \APosSubtypeJudg{\acontext''}{P}{Q}{\acontext'''} \\
      \APosSubtypeJudg{\acontext'''}{\subcon{\acontext'''} Q}{P}{\acontext^{(4)}} \\
      \acontext^{(5)} = \restrictcontext{\acontext^{(4)}}{\acontext} \\
      \algosynjudg{\acontext^{(5)}; \Gamma, x : P}{t}{N}{\acontext^{(6)}}}
      {\algosynjudg{\acontext; \Gamma}{\letanno{x}{P}{v}{s}{t}}{N}{\acontext^{(6)}}}

    \begin{llproof}
      \proofcomment{Apply the induction hypothesis to the first premise:}

      \conwfPf{\acontext}{Assumption}
      \envwfPf{\Theta}{\typeenv}{Assumption}
      \algosynjudgPf{\acontext; \typeenv}{v}{\shiftdown M}{\acontext'}{Subderivation}
      \conwfPf{\acontext'}{\byih}
\Label{1} \congoesPf{\acontext}{\acontext'}{\ditto}
      \wfpostypePf{\acontext'}{\shiftdown{M}}{\ditto}
      \groundPf{\shiftdown{M}}{\ditto}

      \proofcomment{Apply the induction hypothesis again, this time to the second premise:}

      \conwfPf{\acontext}{Assumption}
      \congoesWeakPf{\acontext}{\acontext'}{By \lemmaref{lemma:weak context extension subsumes normal}}
      \envwfPf{\acontext'}{\typeenv}{By \lemmaref{lemma:context extension preserves wf envs}}
      \algospinejudgPf{\acontext'; \Gamma}{s}{\spine{M}{\shiftu{Q}}}{\acontext''}{Subderivation}
      \wfnegtypePf{\acontext'}{M}{Inversion ({\twfshiftdown})}
      \groundPf{M}{\bydefground and above}
      \eqPf{\subcon{\acontext'} M}{M}{By \lemmaref{lemma:Context substitution on ground terms}}
      \Label{2}   \congoesWeakPf{\acontext'}{\acontext''}{\byih}
      \conwfPf{\acontext''}{\ditto}
      \wfnegtypePf{\acontext''}{\shiftup{Q}}{\ditto}
      \eqPf{\subterm{\acontext''} \shiftup{Q}}{\shiftup{Q}}{\ditto}

      \proofcomment{Now apply the well-formedness of algorithmic subtyping to the third premise:}

      \AJudgePosPf{\acontext''}{P}{Q}{\acontext'''}{Subderivation}
      \conwfPf{\acontext''}{Above}
      \groundPf{P}{$P$ annotation}
      \NoSolvedVarsPf{\acontext''}{Q}{\bydefsubcon and above}
      \conwfPf{\acontext'''}{By \lemmaref{lemma:well-formedness-pnsubtype}}
\Label{3} \congoesPf{\acontext''}{\acontext'''}{\ditto}
      \groundPf{\subcon{\acontext'''} Q}{\ditto}

      \proofcomment{Apply it again to the fourth premise:}

      \AJudgePosPf{\acontext'''}{\subcon{\acontext'''} Q}{P}{\acontext^{(4)}}{Subderivation}
      \conwfPf{\acontext'''}{Above}
      \groundPf{\subcon{\acontext'''} Q}{Above}
      \NoSolvedVarsPf{\acontext'''}{P}{By \lemmaref{lemma:Context substitution on ground terms}}
      \conwfPf{\acontext^{(4)}}{By \lemmaref{lemma:well-formedness-pnsubtype}}
      \Label{4} \congoesPf{\acontext'''}{\acontext^{(4)}}{\ditto}

      \proofcomment{Make use of \lemmaref{lemma:restricted context wf} in the context of the fifth premise:}

      \congoesWeakPf{\acontext}{\acontext'}{Applying}
        \trailingjust{\lemmaref{lemma:weak context extension subsumes normal}}
        \trailingjust{to (1)}
      \congoesWeakPf{\acontext'}{\acontext''}{Above ((2))}
      \congoesWeakPf{\acontext''}{\acontext'''}{Applying}
      \trailingjust{\lemmaref{lemma:weak context extension subsumes normal}}
      \trailingjust{to (3)}
      \congoesWeakPf{\acontext'''}{\acontext^{(4)}}{Applying}
      \trailingjust{\lemmaref{lemma:weak context extension subsumes normal}}
      \trailingjust{to (4)}
      \proofsep

      \conwfPf{\acontext}{Above}
      \conwfPf{\acontext^{(4)}}{Above}
      \congoesWeakPf{\acontext}{\acontext^{(4)}}{By \lemmaref{lemma:weak context extension transitive}}
      \eqPf{\acontext^{(5)}}{\restrictcontext{\acontext^{(4)}}{\acontext}}{By premise}
\Label{5} \congoesPf{\acontext}{\acontext^{(5)}}{By \lemmaref{lemma:restricted context wf}}
      \conwfPf{\acontext^{(5)}}{\ditto}

      \proofcomment{Finally, apply the induction hypothesis to the last premise:}

      \conwfPf{\acontext^{(5)}}{Above}
      \congoesWeakPf{\acontext}{\acontext^{(5)}}{By \lemmaref{lemma:weak context extension subsumes normal}}
      \envwfPf{\acontext^{(5)}}{\typeenv}{By \lemmaref{lemma:context extension preserves wf envs}}
      \wfpostypePf{\Theta}{P}{$P$ is an annotation}
      \wfpostypePf{\acontext^{(5)}}{P}{By \lemmaref{lemma:Context extension preserves term well-formedness}}
      \groundPf{P}{$P$ is an annotation}
      \envwfPf{\acontext^{(5)}}{\Gamma, x : P}{By {\Ewfvar}}
      \algosynjudgPf{\acontext^{(5)}; \Gamma, x : P}{t}{N}{\acontext^{(6)}}{Subderivation}
\Hand \conwfPf{\acontext^{(6)}}{\byih}
\Label{6} \congoesPf{\acontext^{(5)}}{\acontext^{(6)}}{\ditto}
\Hand \wfnegtypePf{\acontext^{(6)}}{N}{\ditto}
\Hand \groundPf{N}{\ditto}
\Hand \congoesPf{\Theta}{\acontext^{(6)}}{Applying \lemmaref{lemma:context extension transitive} to (5) and (6)}
    \end{llproof}

    \DerivationProofCase{\Aunambiguouslet}
      {\algosynjudg{\acontext; \typeenv}{v}{\shiftd M}{\acontext'}  \\
      \algospinejudg{\acontext'; \Gamma}{s}{\spine{M}{\shiftu{Q}}}{\acontext''} \\
      \FreeEV(Q) = \emptyset \\
      \acontext''' = \restrictcontext{\acontext''}{\acontext} \\
      \algosynjudg{\acontext'''; \Gamma, x : Q}{t}{N}{\acontext^{(4)}}}
      {\algosynjudg{\acontext; \Gamma}{\letplain{x}{v}{s}{t}}{N}{\acontext^{(4)}}}

    \begin{llproof}
      \proofcomment{First apply the induction hypothesis to the first subderivation:}

      \conwfPf{\acontext}{Assumption}
      \envwfPf{\acontext}{\typeenv}{Assumption}
      \algosynjudgPf{\acontext; \typeenv}{v}{\shiftdown{M}}{\acontext'}{Subderivation}
      \proofsep

      \conwfPf{\acontext'}{\byih}
\Label{1} \congoesPf{\acontext}{\acontext'}{\ditto}
      \wfpostypePf{\acontext'}{\shiftdown{M}}{\ditto}
      \groundPf{\shiftdown{M}}{\ditto}

      \proofcomment{Now apply the induction hypothesis to the spine subderivation:}

      \conwfPf{\acontext'}{Above}
      \congoesWeakPf{\acontext}{\acontext'}{Applying \lemmaref{lemma:weak context extension subsumes normal} to (1)}
      \envwfPf{\acontext'}{\typeenv}{By \lemmaref{lemma:context extension preserves wf envs}}
      \algospinejudgPf{\acontext'; \typeenv}{s}{\spine{M}{\shiftu{Q}}}{\acontext''}{Subderivation}

      \groundPf{M}{By the definition of ground}
      \wfnegtypePf{\acontext'}{M}{Inversion ({\twfshiftdown})}
      \eqPf{\subcon{\acontext'} M}{M}{By \lemmaref{lemma:Context substitution on ground terms}}
      \proofsep

      \conwfPf{\acontext''}{\byih}
      \wfnegtypePf{\acontext''}{\shiftup{Q}}{\ditto}
      \congoesWeakPf{\acontext'}{\acontext''}{\ditto}

      \proofcomment{Produce a strong context extension judgment using \lemmaref{lemma:restricted context wf}:}

      \conwfPf{\acontext}{Above}
      \conwfPf{\acontext''}{Above}
      \congoesWeakPf{\acontext}{\acontext''}{By \lemmaref{lemma:weak context extension transitive}}
      \eqPf{\acontext'''}{\restrictcontext{\acontext''}{\acontext}}{Premise}
\Label{2} \congoesPf{\acontext}{\acontext'''}{By \lemmaref{lemma:restricted context wf}}
      \conwfPf{\acontext'''}{\ditto}

      \proofcomment{Finally, apply the induction hypothesis to the last premise:}

      \congoesWeakPf{\acontext}{\acontext'''}{Applying \lemmaref{lemma:weak context extension subsumes normal} to (2)}
      \envwfPf{\acontext'''}{\typeenv}{By \lemmaref{lemma:context extension preserves wf envs}}
      \wfpostypePf{\acontext''}{Q}{Inversion ({\twfshiftup})}
      \wfpostypePf{\acontext'''}{Q}{By \lemmaref{lemma:Context extension preserves term well-formedness}}
      \eqPf{\FreeEV(Q)}{\emptyset}{Premise}
      \groundPf{Q}{By definition of ground}
      \envwfPf{\acontext'''}{\typeenv, x : Q} {By {\Ewfvar}}
      \proofsep
      \conwfPf{\acontext'''}{Above}
      \envwfPf{\acontext'''}{\typeenv, x : Q} {Above}
      \algosynjudgPf{\acontext'''; \Gamma, x : Q}{t}{N}{\acontext^{(4)}}{Subderivation}
      \proofsep
\Hand \conwfPf{\acontext^{(4)}}{\byih}
\Label{3} \congoesPf{\acontext'''}{\acontext^{(4)}}{\ditto}
\Hand \wfnegtypePf{\acontext^{(4)}}{N}{\ditto}
\Hand \groundPf{N}{\ditto}
\Hand \congoesPf{\acontext}{\acontext^{(4)}}{Applying \lemmaref{lemma:context extension transitive} to (2) and (3)}
    \end{llproof}

    \DerivationProofCase{\Aspinenil}
      {}
      {\algospinejudg{\Theta; \Gamma}{\epsilon}{\spine{N}{N}}{\Theta}}

      \begin{llproof}
\Hand   \conwfPf{\Theta}{Assumption}
\Hand   \congoesWeakPf{\Theta}{\Theta}{By \lemmaref{lemma:weak context extension reflexive}}
\Hand   \wfnegtypePf{\Theta}{N}{Assumption}
\Hand   \NoSolvedVarsPf{\Theta}{N}{Assumption}
        \subseteqPf{\FreeEV(N)}{\FreeEV(N)}{By reflexivity of $\subseteq$}
\Hand   \subseteqPf{\FreeEV(N)}{\FreeEV(N) \cup (\FreeEV(\acontext') \setminus \FreeEV(\acontext))}{By definition of $\subseteq$}
      \end{llproof}

    \DerivationProofCase{\Aspinecons}
      {\algosynjudg{\acontext; \Gamma}{v}{P}{\acontext'} \\
      \APosSubtypeJudg{\acontext'}{P}{[\acontext']Q}{\acontext''} \\
      \algospinejudg{\acontext''; \Gamma}{s}{\spine{[\acontext'']N}{M}}{\acontext'''}}
      {\algospinejudg{\acontext; \Gamma}{v, s}{Q \to \spine{N}{M}}{\acontext'''}}

      \begin{llproof}
        \conwfPf{\acontext}{Assumption}
        \envwfPf{\acontext}{\typeenv}{Assumption}
        \algosynjudgPf{\acontext; \typeenv}{v}{P}{\acontext'}{Subderivation}
        \proofsep

        \conwfPf{\Theta'}{\byih}
        \congoesPf{\Theta}{\Theta'}{\ditto}
        \wfpostypePf{\acontext'}{P}{\ditto}
        \groundPf{P}{\ditto}
        \proofsep

        \AJudgePosPf{\acontext'}{P}{Q}{\acontext''}{Subderivation}
        \conwfPf{\acontext'}{Above}
        \groundPf{P}{Above}
        \NoSolvedVarsPf{\acontext'}{\subcon{\acontext'} Q}{By \lemmaref{lemma:Context substitution idempotence}}
        \conwfPf{\acontext''}{By \lemmaref{lemma:well-formedness-pnsubtype}}
        \congoesPf{\acontext'}{\acontext''}{\ditto}
        \groundPf{\subcon{\acontext''} \subcon{\acontext'} Q}{\ditto}
        \proofsep

        \conwfPf{\acontext''}{Above}
        \congoesPf{\acontext}{\acontext''}{By \lemmaref{lemma:Transitivity of declarative pnsubtype}}
        \congoesWeakPf{\Theta}{\Theta''}{By \lemmaref{lemma:weak context extension subsumes normal}}
        \envwfPf{\acontext''}{\typeenv}{By \lemmaref{lemma:context extension preserves wf envs}}
        \algospinejudgPf{\Theta''; \Gamma}{s}{\spine{\subterm{\Theta''} N}{\shiftup{Q}}}{\Theta'''}{Subderivation}
        \wfnegtypePf{\Theta}{N}{Inversion (\twfarrow)}
        \wfnegtypePf{\Theta''}{N}{By \lemmaref{lemma:Context extension preserves term well-formedness}}
        \wfnegtypePf{\Theta''}{\subterm{\Theta''} N}{By \lemmaref{lemma:Well-formed context substitution preserves term well-formedness}}
        \eqPf{\subterm{\Theta''}\subterm{\Theta''} N}{\subterm{\Theta''} N}{By \lemmaref{lemma:Context substitution idempotence}}
        \proofsep

\Hand   \conwfPf{\Theta'''}{\byih}
        \congoesWeakPf{\Theta''}{\Theta'''}{\ditto}
\Hand   \wfnegtypePf{\Theta'''}{M}{\ditto}
\Hand   \eqPf{[\Theta''']M}{M}{\ditto}
        \subseteqPf{\FreeEV(M)}{\FreeEV(N) \cup (\FreeEV(\acontext''') \setminus \FreeEV(\acontext''))}{\ditto}
        \proofsep

\Hand   \congoesWeakPf{\acontext}{\acontext'''}{By \lemmaref{lemma:weak context extension transitive}}
        \subseteqPf{\FreeEV(N)}{\FreeEV(Q \funarrow N)}{\bydeffev}
        \eqPf{\FreeEV(\Theta)}{\FreeEV(\Theta'')}{By \lemmaref{lemma:Context extension maintains variables}}
\Hand   \subseteqPf{\FreeEV(M)}{\FreeEV(Q \funarrow N) \cup (\FreeEV(\acontext''') \setminus \FreeEV(\acontext))}{Substituting above}
      \end{llproof}

    \DerivationProofCase{\Aspinetypeabsnotin}
      {\algospinejudg{\Theta; \Gamma}{s}{\spine{N}{M}}{\Theta'} \\ \alpha \notin \FreeUV(N)}
      {\algospinejudg{\Theta; \Gamma}{s}{\spine{(\forall\alpha\ldotp N)}{ M}}{\Theta'}}

      \begin{llproof}
        \conwfPf{\Theta}{Assumption}
        \envwfPf{\Theta}{\Gamma}{Assumption}
        \algospinejudgPf{\Theta; \Gamma}{s}{\spine{N}{M}}{\Theta'}{Subderivation}
        \wfnegtypePf{\Theta}{\forall\alpha\ldotp N}{Assumption}
        \wfnegtypePf{\Theta, \alpha}{N}{Inversion ({\twfforall})}
        \wfnegtypePf{\Theta}{N}{By \lemmaref{lemma:type well-formed with alpha removed}}
        \proofsep
        \eqPf{\subcon{\Theta} (\forall\alpha\ldotp N)}{\forall\alpha\ldotp N}{Assumption}
        \eqPf{\forall\alpha\ldotp \subcon{\Theta} N}{\forall\alpha \ldotp N}{\bydefsubcon}
        \eqPf{\subcon{\Theta} N}{N}{By equality}
        \proofsep
\Hand   \conwfPf{\Theta'}{\byih}
\Hand   \congoesWeakPf{\Theta}{\Theta'}{\ditto}
\Hand   \wfnegtypePf{\Theta'}{M}{\ditto}
\Hand   \eqPf{\subterm{\Theta'} M}{M}{\ditto}
        \subseteqPf{\FreeEV(M)}{\FreeEV(N) \cup (\FreeEV(\acontext') \setminus \FreeEV(\acontext))}{\ditto}
        \eqPf{\FreeEV(\forall \alpha \ldotp N)}{\FreeEV(N)}{\bydeffev}
\Hand   \subseteqPf{\FreeEV(M)}{\FreeEV(\forall \alpha \ldotp N) \cup (\FreeEV(\acontext') \setminus \FreeEV(\acontext))}{Substituting above}
      \end{llproof}

    \DerivationProofCase{\Aspinetypeabsin}
      {\algospinejudg{\Theta, \guess{\alpha}; \Gamma}{s}{\spine{\subterm{\guess{\alpha}/\alpha} N}{M}}{\Theta', \guess{\alpha}\, [= P]} \\ \alpha \in \FreeUV(N)}
      {\algospinejudg{\Theta; \Gamma}{s}{\spine{(\forall\alpha\ldotp  N)}{ M}}{\Theta', \guess{\alpha}\, [= P]}}

      \begin{llproof}
        \conwfPf{\Theta}{Assumption}
        \conwfPf{\Theta, \guess{\alpha}}{By {\cwfunsolvedguess}}
        \envwfPf{\acontext}{\typeenv}{Assumption}
        \congoesWeakPf{\acontext}{\acontext, \guess{\alpha}}{By \lemmaref{lemma:weak context extension reflexive}}
        \trailingjust{and \Wcunsolvedextend}
        \envwfPf{\acontext, \guess{\alpha}}{\typeenv}{By \lemmaref{lemma:context extension preserves wf envs}}
        \proofsep
        \algospinejudgPf{\Theta, \guess{\alpha}; \Gamma}{s}{\spine{\subterm{\guess{\alpha}/\alpha} N}{M}}{\Theta', \guess{\alpha}\, [= P]}{Subderivation}
        \wfnegtypePf{\Theta}{\forall\alpha\ldotp  N}{Assumption}
        \notinPf{\alpha}{\FreeUV(\Theta)}{$\alpha$ fresh}
        \wfnegtypePf{\Theta, \alpha}{N}{By {\twfforall}}
        \wfnegtypePf{\Theta, \guess{\alpha}}{\subterm{\guess{\alpha}/\alpha} N}{By \lemmaref{lemma:substituion preserves well-formedness of types}}
        \proofsep
        \eqPf{\subcon{\Theta} (\forall\alpha\ldotp N)}{\forall\alpha\ldotp N}{Assumption}
        \eqPf{\subcon{\Theta} N}{N}{\bydefsubcon}
        \eqPf{\subcon{\Theta} (\subterm{\guess{\alpha}/\alpha} N)}{\subterm{\guess{\alpha}/\alpha} N}{$\guess{\alpha}$ fresh}
        \eqPf{[\Theta,\guess{\alpha}](\subterm{\guess{\alpha}/\alpha} N)}{\subterm{\guess{\alpha}/\alpha} N}{\bydefsubcon}
      \end{llproof}
      \begin{llproof}
  \Hand \conwfPf{\Theta', \guess{\alpha}\, [= P]}{\byih}
          \congoesWeakPf{\Theta, \guess{\alpha}}{\Theta', \guess{\alpha}\, [= P]}{\ditto}
  \Hand \wfnegtypePf{\Theta', \guess{\alpha}\, [= P]}{M}{\ditto}
  \Hand \eqPf{[\Theta', \guess{\alpha}\, [= P]]M}{M}{\ditto}
        \subseteqPf{\FreeEV(M)}{\FreeEV(\subcon{\guess{\alpha} / \alpha} N) \cup (\FreeEV(\acontext', \guess{\alpha}\, [= P]) \setminus \FreeEV(\acontext, \guess{\alpha}))}{\ditto}
      \end{llproof}

      \begin{llproof}
        \congoesWeakPf{\Theta}{\Theta}{By \lemmaref{lemma:weak context extension reflexive}}
        \congoesWeakPf{\Theta}{\Theta, \guess{\alpha}}{By $\guess{\alpha}$ fresh \&  {\Wcunsolvedextend}}
  \Hand \congoesWeakPf{\Theta}{\Theta', \guess{\alpha}\, [= P]}{By \lemmaref{lemma:weak context extension transitive}}
      \end{llproof}
      \begin{llproof}
        \subseteqPf{\FreeEV(M)}{\FreeEV(\subcon{\guess{\alpha} / \alpha} N) \cup (\FreeEV(\acontext', \guess{\alpha}\, [= P]) \setminus \FreeEV(\acontext))}{\bydeffev}
        \subseteqPf{\FreeEV(\subcon{\guess{\alpha} / \alpha} N)}{\FreeEV(\forall \alpha \ldotp N) \union \{\guess{\alpha}\}}{\bydeffev}
        \subseteqPf{\{\guess{\alpha}\}}{\FreeEV(\acontext', \guess{\alpha}\, [= P]) \setminus \FreeEV(\acontext)}{\bydeffev}
  \Hand \subseteqPf{\FreeEV(M)}{\FreeEV(\forall \alpha \ldotp N) \cup (\FreeEV(\acontext', \guess{\alpha}\, [= P]) \setminus \FreeEV(\acontext))}{By above}
      \end{llproof}
  \end{itemize}
\end{proof}

\section{Determinism of typing}

\TypingDeterminacy*

\begin{proof}
  The algorithmic system is mostly syntax-oriented, with the only exceptions $\Aspinetypeabsnotin$ and $\Aspinetypeabsin$ (which have the same conclusion) being distinguished by whether $\alpha \in \FreeUV(N)$, a deterministic check.
  Therefore, determinacy of the system follows by a straightforward mutual rule induction over the algorithmic synthesis and spine judgments, making use of \lemmaref{lemma:subtyping determinacy}.
\end{proof}

\section{Decidability of typing}

\RestateDecidabilityAlgorithmicTyping*

\begin{proof}
  We use the same ordering of judgments as in \lemmaref{lemma:isomorphic types check expressions}.

  \begin{itemize}
    \DerivationProofCase{\Avar}
      {x : P \in \Gamma}
      {\algosynjudg{\acontext; \Gamma}{x}{P}{\acontext}}

    Testing membership of $\Gamma$ terminates since typing environments are finite.

    \DerivationProofCase{\Afunabs}
      {\algosynjudg{\acontext; \Gamma, x : P}{t}{N}{\acontext'}}
      {\algosynjudg{\acontext; \Gamma}{\lamterm{x}{P}{t}}{P \to N}{\acontext'}}

    \begin{llproof}
      \eqPf{|\lamterm{x}{P}{t}|}{|t| + 1}{By definition of $|\_|$}
      \continueGtPf{|t|}{}
\Hand \hugePf{(\algosynjudg{\Theta;\Gamma, x : P}{t}{N}{\Theta'})}{(\algosynjudg{\Theta; \Gamma}{\lamterm{x}{P}{t}}{P \funarrow N}{\Theta'})}{By definition of $\sqsubset$}
    \end{llproof}

    \DerivationProofCase{\Atypeabs}
      {\algosynjudg{\acontext, \alpha; \Gamma}{t}{N}{\acontext', \alpha}}
      {\algosynjudg{\acontext; \Gamma}{\gen{\alpha}{t}}{\forall\alpha\ldotp N}{\acontext'}}

    \begin{llproof}
      \eqPf{|\gen{\alpha}{t}|}{|t| + 1}{By definition of $|\_|$}
      \continueGtPf{|t|}{}
\Hand \hugePf{(\algosynjudg{\acontext, \alpha; \Gamma}{t}{N}{\acontext', \alpha})}{(\algosynjudg{\acontext; \Gamma}{\gen{\alpha}{t}}{\forall\alpha\ldotp N}{\acontext'})}{By definition of $\sqsubset$}
    \end{llproof}

    \DerivationProofCase{\Athunk}
      {\algosynjudg{\acontext; \Gamma}{t}{N}{\acontext'}}
      {\algosynjudg{\acontext; \Gamma}{\thunk{t}}{\shiftd{N}}{\acontext'}}

    \begin{llproof}
      \eqPf{|\thunk{t}|}{|t| + 1}{By definition of $|\_|$}
      \continueGtPf{|t|}{}
\Hand \hugePf{(\algosynjudg{\Theta; \Gamma}{t}{N}{\Theta'})}{(\algosynjudg{\acontext; \Gamma}{\thunk{t}}{\shiftd{N}}{\acontext'})}{By definition of $\sqsubset$}
    \end{llproof}

    \DerivationProofCase{\Areturn}
      {\algosynjudg{\acontext; \Gamma}{v}{P}{\acontext'}}
      {\algosynjudg{\acontext; \Gamma}{\return{v}}{\shiftu{P}}{\acontext'}}

    \begin{llproof}
      \eqPf{|\return{v}|}{|v| + 1}{By definition of $|\_|$}
      \continueGtPf{|v|}{}
\Hand \hugePf{(\algosynjudg{\acontext; \Gamma}{v}{P}{\acontext'})}{(\algosynjudg{\acontext; \Gamma}{\return{v}}{\shiftu{P}}{\acontext'})}{By definition of $\sqsubset$}
    \end{llproof}

    \DerivationProofCase{\Aambiguouslet}
      {\algosynjudg{\acontext; \typeenv}{v}{\shiftd M}{\acontext'} \\
      \algospinejudg{\acontext'; \Gamma}{s}{\spine{M}{\shiftu{Q}}}{\acontext''} \\
      \APosSubtypeJudg{\acontext''}{P}{Q}{\acontext'''} \\
      \APosSubtypeJudg{\acontext'''}{\subcon{\acontext'''} Q}{P}{\acontext^{(4)}} \\
      \acontext^{(5)} = \restrictcontext{\acontext^{(4)}}{\acontext} \\
      \algosynjudg{\acontext^{(5)}; \Gamma, x : P}{t}{N}{\acontext^{(6)}}}
      {\algosynjudg{\acontext; \Gamma}{\letanno{x}{P}{v}{s}{t}}{N}{\acontext^{(6)}}}

    The algorithmic subtyping judgments terminate per \lemmaref{lemma:Decidability of algorithmic subtyping}.

    \begin{llproof}
      \eqPf{|\letanno{x}{P}{v}{s}{t}|}{|v| + |s| + |t| + 1}{By definition of $|\_|$}
      \ltPf{|v|}{|\letanno{x}{P}{v}{s}{t}|}{}
\Hand \hugePf{(\algosynjudg{\acontext; \typeenv}{v}{\shiftd{M}}{\acontext'})}{(\algosynjudg{\Theta; \Gamma}{\letanno{x}{P}{v}{s}{t}}{N}{\Theta^{(6)}})}{By definition of $\sqsubset$}
      \proofsep

      \ltPf{|s|}{|\letanno{x}{P}{v}{s}{t}|}{}
\Hand \hugePf{(\algospinejudg{\Theta'; \Gamma}{s}{\spine{M}{\shiftu{Q}}}{\Theta''})}{(\algosynjudg{\Theta; \Gamma}{\letanno{x}{P}{v}{s}{t}}{N}{\Theta^{(6)}})}{By definition of $\sqsubset$}
      \proofsep

      \ltPf{|t|}{|\letanno{x}{P}{v}{s}{t}|}{}
\Hand \hugePf{(\algosynjudg{\Theta^{(5)}; \Gamma, x : P}{t}{N}{\Theta^{(6)}})}{(\algosynjudg{\Theta; \Gamma}{\letanno{x}{P}{v}{s}{t}}{N}{\Theta^{(6)}})}{By definition of $\sqsubset$}
    \end{llproof}

    \DerivationProofCase{\Aunambiguouslet}
      {\algosynjudg{\acontext; \typeenv}{v}{\shiftd M}{\acontext'}  \\
      \algospinejudg{\acontext'; \Gamma}{s}{\spine{M}{\shiftu{Q}}}{\acontext''} \\
      \FreeEV(Q) = \emptyset \\
      \acontext''' = \restrictcontext{\acontext''}{\acontext} \\
      \algosynjudg{\acontext'''; \Gamma, x : Q}{t}{N}{\acontext^{(4)}}}
      {\algosynjudg{\acontext; \Gamma}{\letplain{x}{v}{s}{t}}{N}{\acontext^{(4)}}}

    Determining the set of free universal variables of a finite type is terminating.

    \begin{llproof}
      \eqPf{|\letplain{x}{v}{s}{t}|}{|v| + |s| + |t| + 1}{By definition of $|\_|$}
      \ltPf{|v|}{|\letplain{x}{v}{s}{t}|}{}
\Hand \hugePf{(\algosynjudg{\acontext; \typeenv}{v}{\shiftd{M}}{\acontext'})}{(\algosynjudg{\Theta; \Gamma}{\letplain{x}{v}{s}{t}}{N}{\Theta^{(4)}})}{By definition of $\sqsubset$}
      \proofsep

      \ltPf{|s|}{|\letplain{x}{v}{s}{t}|}{}
\Hand \hugePf{(\algospinejudg{\Theta'; \Gamma}{s}{\spine{M}{\shiftu{Q}}}{\Theta''})}{(\algosynjudg{\Theta; \Gamma}{\letplain{x}{v}{s}{t}}{N}{\Theta^{(4)}})}{By definition of $\sqsubset$}
      \proofsep

      \ltPf{|t|}{|\letplain{x}{v}{s}{t}|}{}
\Hand \hugePf{(\algosynjudg{\Theta'''; \Gamma, x : P}{t}{N}{\Theta^{(4)}})}{(\algosynjudg{\Theta; \Gamma}{\letplain{x}{v}{s}{t}}{N}{\Theta^{(4)}})}{By definition of $\sqsubset$}
    \end{llproof}

    \DerivationProofCase{\Aspinenil}
    {}
    {\algospinejudg{\Theta; \Gamma}{\epsilon}{\spine{N}{N}}{\Theta}}

    Rule terminal.

    \DerivationProofCase{\Aspinecons}
      {\algosynjudg{\acontext; \Gamma}{v}{P}{\acontext'} \\
      \APosSubtypeJudg{\acontext'}{P}{[\acontext']Q}{\acontext''} \\
      \algospinejudg{\acontext''; \Gamma}{s}{\spine{[\acontext'']N}{M}}{\acontext'''}}
      {\algospinejudg{\acontext; \Gamma}{v, s}{Q \to \spine{N}{M}}{\acontext'''}}

      The algorithmic subtyping judgment terminates per \lemmaref{lemma:Decidability of algorithmic subtyping}.

    \begin{llproof}
      \eqPf{|v,s|}{|v| + |s| + 1}{By definition of $|\_|$}
      \continueGtPf{|v|}{}
\Hand \hugePf{(\algosynjudg{\Theta; \Gamma}{v}{P}{\Theta'})}{(\algospinejudg{\acontext; \Gamma}{v, s}{Q \to \spine{N}{M}}{\acontext'''})}{By definition of $\sqsubset$}
      \gtPf{|v,s|}{|s|}{}
\Hand \hugePf{(\algospinejudg{\acontext''; \Gamma}{s}{\spine{[\acontext'']N}{M}}{\acontext'''})}{\algospinejudg{\acontext; \Gamma}{v, s}{Q \to \spine{N}{M}}{\acontext'''}}{By definition of $\sqsubset$}
    \end{llproof}

    \DerivationProofCase{\Aspinetypeabsnotin}
    {\algospinejudg{\Theta; \Gamma}{s}{\spine{N}{M}}{\Theta'} \\ \alpha \notin \FreeUV(N)}
    {\algospinejudg{\Theta; \Gamma}{s}{\spine{(\forall\alpha\ldotp N)}{ M}}{\Theta'}}

    \begin{llproof}
      \eqPf{|s|}{|s|}{}
      \numPrenexGtPf{\forall \alpha \ldotp N}{N}{}
\Hand \hugePf{(\algospinejudg{\Theta; \Gamma}{s}{\spine{N}{M}}{\Theta'})}{(\algospinejudg{\Theta; \Gamma}{s}{\spine{(\forall\alpha\ldotp N)}{ M}}{\Theta'})}{By definition of $\sqsubset$}
    \end{llproof}

    We define types to be finite, therefore calculating $\FreeUV(N)$ is terminating.

    \DerivationProofCase{\Aspinetypeabsin}
    {\algospinejudg{\Theta, \hat{\alpha}; \Gamma}{s}{\spine{[\hat{\alpha}/\alpha]N}{M}}{\Theta', \hat{\alpha}\, [= P]} \\ \alpha \in \FreeUV(N)}
    {\algospinejudg{\Theta; \Gamma}{s}{\spine{(\forall\alpha\ldotp N)}{ M}}{\Theta', \hat{\alpha}\, [= P]}}

    \begin{llproof}
      \eqPf{|s|}{|s|}{}
      \numPrenexGtPf{\forall \alpha \ldotp N}{[\guess{\alpha}/\alpha]N}{Since $\alpha$ and $\guess{\alpha}$ are positive, the substitution}
      \trailingjust{cannot introduce any prenex quantifiers.}
\Hand \hugePf{(\algospinejudg{\Theta, \hat{\alpha}; \Gamma}{s}{\spine{[\hat{\alpha}/\alpha]N}{M}}{\Theta'}, \hat{\alpha} [= P])}{(\algospinejudg{\Theta; \Gamma}{s}{\spine{(\forall\alpha\ldotp N)}{ M}}{\Theta'})}{By definition of $\sqsubset$}
    \end{llproof}

    We define types to be finite, therefore calculating $\FreeUV(N)$ is terminating.
  \end{itemize}
\end{proof}

\section{Soundness of typing}

\subsection{Lemmas}

\ExtendedCompleteContext*

\begin{proof}
  We add the $\guess{\alpha}\,[= P]$ context items that newly appear in $\acontext'$ to the complete context.

  By rule induction on $\congoesWeakJudg{\acontext}{\acontext'}$:

  \begin{itemize}
    \DerivationProofCase{\Wcempty}
      {}
      {\congoesWeakJudg{\emptyacontext}{\emptyacontext}}

    The new context is $\emptyacontext$.

    \begin{llproof}
      \Hand \conwfPf{\emptyacontext}{By \cwfempty}
      \Hand \congoesPf{\emptyacontext}{\emptyacontext}{By \Cempty}
      \Hand \congoesWeakPf{\emptyacontext}{\emptyacontext}{By \Wcempty}
    \end{llproof}

    \DerivationProofCase{\Wcuvar}
      {\congoesWeakJudg{\acontext}{\acontext'}}
      {\congoesWeakJudg{\acontext, \alpha}{\acontext', \alpha}}

    We add the $\alpha$ context item onto the new context from the induction hypothesis.

    \begin{llproof}
      \conwfPf{\acontext', \alpha}{Assumption}
\Label{1} \conwfPf{\acontext'}{Inversion (\cwfuvar)}
      \congoesPf{\acontext, \alpha}{\ccontext}{Assumption}
      \eqPf{\ccontext}{\bar{\ccontext}, \alpha}{Inversion (\Cuvar)}
\Label{2} \congoesPf{\acontext}{\bar{\ccontext}}{\ditto}
      \conwfPf{\bar{\ccontext}, \alpha}{Assumption}
\Label{3} \conwfPf{\bar{\ccontext}}{Inversion (\cwfuvar)}
\Label{4} \congoesWeakPf{\acontext}{\acontext'}{Subderivation}
      \congoesPf{\restrictcontext{\acontext', \alpha}{\acontext, \alpha}}{\bar{\ccontext}, \alpha}{Assumption}
\Label{5} \congoesPf{\restrictcontext{\acontext'}{\acontext}}{\bar{\ccontext}}{Inversion (\Restrictuvar)}
      \proofsep
      \conwfPf{\bar{\ccontext}'}{\byih, using (1--5) and for some complete context $\bar{\ccontext}'$}
      \congoesPf{\acontext'}{\bar{\ccontext}'}{\ditto}
      \congoesWeakPf{\bar{\ccontext}}{\bar{\ccontext}'}{\ditto}
      \proofsep
\Hand \conwfPf{\bar{\ccontext}', \alpha}{By \cwfuvar}
\Hand \congoesPf{\acontext', \alpha}{\bar{\ccontext}', \alpha}{By \Cuvar}
\Hand \congoesWeakPf{\bar{\ccontext}, \alpha}{\bar{\ccontext}', \alpha}{By \Wcuvar}
    \end{llproof}

    \DerivationProofCase{\Wcunsolvedguess}
      {\congoesWeakJudg{\Theta}{\Theta'}}
      {\congoesWeakJudg{\Theta, \guess{\alpha}}{\Theta', \guess{\alpha}}}

      We add the $\guess{\alpha} = P$ context item to the new context from the induction hypothesis, where $P$ is the solution for $\guess{\alpha}$ in the complete context $\ccontext$.

    \begin{llproof}
      \conwfPf{\acontext', \guess{\alpha}}{Assumption}
\Label{1} \conwfPf{\acontext'}{Inversion (\cwfunsolvedguess)}
            \congoesPf{\acontext, \guess{\alpha}}{\ccontext}{Assumption}
      \congoesPf{\acontext, \guess{\alpha}}{\ccontext}{Assumption}
      \eqPf{\ccontext}{\bar{\ccontext}, \guess{\alpha} = P}{Inversion (\Csolveguess)}
\Label{2} \congoesPf{\acontext}{\bar{\ccontext}}{\ditto}
      \conwfPf{\bar{\ccontext}, \guess{\alpha} = P}{Assumption}
\Label{3} \conwfPf{\bar{\ccontext}}{Inversion (\cwfsolvedguess)}
      \wfpostypePf{\bar{\ccontext}}{P}{\ditto}
      \groundPf{P}{\ditto}
\Label{4} \congoesWeakPf{\acontext}{\acontext'}{Subderivation}
      \congoesPf{\restrictcontext{\acontext', \guess{\alpha}}{\acontext, \guess{\alpha}}}{\bar{\ccontext}, \guess{\alpha} = P}{Assumption}
      \congoesPf{(\restrictcontext{\acontext'}{\acontext}), \guess{\alpha}}{\bar{\ccontext}, \guess{\alpha} = P}{Inversion (\Restrictguessin)}
\Label{5} \congoesPf{\restrictcontext{\acontext'}{\acontext}}{\bar{\ccontext}}{Inversion (\Csolveguess)}

      \proofsep
      \conwfPf{\bar{\ccontext}'}{\byih, using (1--5) and for some complete context $\bar{\ccontext}'$}
      \congoesPf{\acontext'}{\bar{\ccontext}'}{\ditto}
      \congoesWeakPf{\bar{\ccontext}}{\bar{\ccontext}'}{\ditto}
      \proofsep
      \wfpostypePf{\bar{\ccontext}'}{P}{By \lemmaref{lemma:weak context extension preserves well-formedness}}
\Hand \conwfPf{\bar{\ccontext}', \guess{\alpha} = P}{By \cwfsolvedguess}
\Hand \congoesPf{\acontext', \guess{\alpha}}{\bar{\ccontext}', \guess{\alpha} = P}{By \Csolveguess}
      \MutualJudgePosPf{\subcon{\ccontext} \bar{\ccontext}}{P}{P}{By \lemmaref{lemma:Reflexivity of declarative pnsubtype}}
\Hand \congoesWeakPf{\bar{\ccontext}, \guess{\alpha} = P}{\bar{\ccontext}', \guess{\alpha} = P}{By \Wcsolvedguess}
    \end{llproof}

    \DerivationProofCase{\Wcsolveguess}
      {\congoesWeakJudg{\Theta}{\Theta'}}
      {\congoesWeakJudg{\Theta, \guess{\alpha}}{\Theta', \guess{\alpha} = P}}

    As before, we add the $\guess{\alpha} = Q$ context item to the new context from the induction hypothesis, where $Q$ is the solution for $\guess{\alpha}$ in the complete context $\ccontext$.

    \begin{llproof}
      \conwfPf{\acontext', \guess{\alpha} = P}{Assumption}
\Label{1} \conwfPf{\acontext'}{Inversion (\cwfsolvedguess)}
      \congoesPf{\acontext, \guess{\alpha}}{\ccontext}{Assumption}
      \eqPf{\ccontext}{\bar{\ccontext}, \guess{\alpha} = Q}{Inversion (\Csolveguess)}
\Label{2} \congoesPf{\acontext}{\bar{\ccontext}}{\ditto}
      \conwfPf{\bar{\ccontext}, \guess{\alpha} = Q}{Assumption}
\Label{3} \conwfPf{\bar{\ccontext}}{Inversion (\cwfsolvedguess)}
      \wfpostypePf{\bar{\ccontext}}{Q}{\ditto}
      \groundPf{Q}{\ditto}
\Label{4} \congoesWeakPf{\acontext}{\acontext'}{Subderivation}
      \congoesPf{\restrictcontext{\acontext', \guess{\alpha} = P}{\acontext, \guess{\alpha}}}{\bar{\ccontext}, \guess{\alpha} = Q}{Assumption}
      \congoesPf{(\restrictcontext{\acontext'}{\acontext}), \guess{\alpha} = P}{\bar{\ccontext}, \guess{\alpha} = Q}{Inversion (\Restrictguessin)}
\Label{5} \congoesPf{\restrictcontext{\acontext'}{\acontext}}{\bar{\ccontext}}{Inversion (\Csolvedguess)}
      \MutualJudgePosPf{\subcon{\ccontext} \restrictcontext{\acontext'}{\acontext}}{P}{Q}{\ditto}
      \proofsep

      \conwfPf{\bar{\ccontext}'}{\byih, using (1--5) and for some complete context $\bar{\ccontext}'$}
      \congoesPf{\acontext'}{\bar{\ccontext}'}{\ditto}
      \congoesWeakPf{\bar{\ccontext}}{\bar{\ccontext}'}{\ditto}
      \proofsep

      \wfpostypePf{\bar{\ccontext}'}{Q}{By \lemmaref{lemma:weak context extension preserves well-formedness}}
\Hand \conwfPf{\bar{\ccontext}', \guess{\alpha} = Q}{By \cwfsolvedguess}
      \MutualJudgePosPf{\makedec{\acontext'}}{P}{Q}{Since $\subcon{\ccontext} (\restrictcontext{\acontext'}{\acontext}) = \subcon{\ccontext} \acontext'$}
\Hand \congoesPf{\acontext', \guess{\alpha} = P}{\bar{\ccontext}', \guess{\alpha} = Q}{By \Csolvedguess}
      \MutualJudgePosPf{\subcon{\ccontext} \bar{\ccontext}}
        {Q}{Q}{By \lemmaref{lemma:Reflexivity of declarative pnsubtype}}
\Hand \congoesWeakPf{\bar{\ccontext}, \guess{\alpha} = Q}{\bar{\ccontext}', \guess{\alpha} = Q}{By \Wcsolvedguess}
    \end{llproof}

    \DerivationProofCase{\Wcsolvedguess}
      {\congoesWeakJudg{\Theta}{\Theta'} \\ \MutualSubtypePosJudge{\makedec{\acontext}}
      {P}{Q}}
      {\congoesWeakJudg{\Theta, \guess{\alpha} = P}{\Theta', \guess{\alpha} = Q}}

      We add the $\guess{\alpha} = R$ context item to the new context from the induction hypothesis, where $R$ is the solution for $\guess{\alpha}$ in the complete context $\ccontext$.

    \begin{llproof}
      \conwfPf{\acontext', \guess{\alpha} = Q}{Assumption}
\Label{1} \conwfPf{\acontext'}{Inversion (\cwfsolvedguess)}
      \congoesPf{\acontext, \guess{\alpha}}{\ccontext}{Assumption}
      \congoesPf{\acontext, \guess{\alpha} = P}{\ccontext}{Assumption}
      \eqPf{\ccontext}{\bar{\ccontext}, \guess{\alpha} = R}{Inversion (\Csolvedguess)}
\Label{2} \congoesPf{\acontext}{\bar{\ccontext}}{\ditto}
      \MutualJudgePosPf{\makedec{\acontext}}
        {P}{R}{\ditto}
      \conwfPf{\bar{\ccontext}, \guess{\alpha} = R}{Assumption}
\Label{3} \conwfPf{\bar{\ccontext}}{Inversion (\cwfsolvedguess)}
      \wfpostypePf{\bar{\ccontext}}{R}{\ditto}
      \groundPf{R}{\ditto}
\Label{4} \congoesWeakPf{\acontext}{\acontext'}{Subderivation}
      \congoesPf{\restrictcontext{\acontext', \guess{\alpha} = Q}{\acontext, \guess{\alpha} = P}}{\bar{\ccontext}, \guess{\alpha} = R}{Assumption}
      \congoesPf{(\restrictcontext{\acontext'}{\acontext}), \guess{\alpha} = Q}{\bar{\ccontext}, \guess{\alpha} = R}{Inversion (\Restrictguessin)}
\Label{5} \congoesPf{\restrictcontext{\acontext'}{\acontext}}{\bar{\ccontext}}{Inversion (\Csolvedguess)}
      \proofsep

      \conwfPf{\bar{\ccontext}'}{\byih, using (1--5) and for some complete context $\bar{\ccontext}'$}
      \congoesPf{\acontext'}{\bar{\ccontext}'}{\ditto}
      \congoesWeakPf{\bar{\ccontext}}{\bar{\ccontext}'}{\ditto}
      \proofsep

      \wfpostypePf{\bar{\ccontext}'}{R}{By \lemmaref{lemma:weak context extension preserves well-formedness}}
\Hand \conwfPf{\bar{\ccontext}', \guess{\alpha} = R}{By \cwfsolvedguess}
      \MutualJudgePosPf{\makedec{\acontext}}{P}{Q}{Premise}
      \MutualJudgePosPf{\makedec{\acontext}}
        {Q}{R}{By \lemmaref{lemma:Transitivity of declarative pnsubtype}}
      \MutualJudgePosPf{\makedec{\acontext'}}
        {Q}{R}{By \lemmaref{lemma:weak context extension equal declarative contexts}}
\Hand \congoesPf{\acontext', \guess{\alpha} = Q}{\bar{\ccontext}', \guess{\alpha} = R}{By \Csolvedguess}
      \MutualJudgePosPf{\subcon{\ccontext} \bar{\ccontext}}
        {P}{R}{By \lemmaref{lemma:equal declarative contexts}}
\Hand \congoesWeakPf{\bar{\ccontext}, \guess{\alpha} = P}{\bar{\ccontext}', \guess{\alpha} = R}{By \Wcsolvedguess}
    \end{llproof}

    \DerivationProofCase{\Wcunsolvedextend}
      {\congoesWeakJudg{\Theta}{\Theta'}}
      {\congoesWeakJudg{\Theta}{\Theta', \guess{\alpha}}}

    We add a solved context item for $\guess{\alpha}$ to the new complete context from the induction hypothesis.

\newcommand{\falsetype}[0]{\shiftd{\forall \alpha \ldotp \shiftu{\alpha}}}

    \begin{llproof}
      \conwfPf{\acontext', \alpha}{Assumption}
\Label{1} \conwfPf{\acontext'}{Inversion (\cwfuvar)}
\Label{2} \congoesPf{\acontext}{\ccontext}{Assumption}
\Label{3} \conwfPf{\ccontext}{Assumption}
\Label{4} \congoesWeakPf{\acontext}{\acontext'}{Subderivation}
      \congoesPf{\restrictcontext{(\acontext', \guess{\alpha})}{\acontext}}{\ccontext}{Assumption}
\Label{5} \congoesPf{\restrictcontext{\acontext'}{\acontext}}{\ccontext}{Inversion (\Restrictguessnotin)}
      \proofsep
      \conwfPf{\ccontext'}{\byih, using (1--5) and for some complete context $\ccontext'$}
      \congoesPf{\acontext'}{\ccontext'}{\ditto}
      \congoesWeakPf{\ccontext}{\ccontext'}{\ditto}
      \proofsep
\Hand \conwfPf{(\ccontext', \guess{\alpha} = \falsetype)}{By \cwfsolvedguess}
\Hand \congoesPf{\acontext', \guess{\alpha}}{(\ccontext', \guess{\alpha} = \falsetype)}{By \Csolveguess}
\Hand \congoesWeakPf{\ccontext}{(\ccontext', \guess{\alpha} = \falsetype)}{By \Wcunsolvedextend}
    \end{llproof}

    \DerivationProofCase{\Wcsolvedextend}
      {\congoesWeakJudg{\Theta}{\Theta'}}
      {\congoesWeakJudg{\Theta}{\Theta', \guess{\alpha}} = P}

    We add the $\guess{\alpha} = P$ context item onto the new context from the induction hypothesis.

    \begin{llproof}
      \conwfPf{\acontext', \guess{\alpha} = P}{Assumption}
\Label{1} \conwfPf{\acontext'}{Inversion (\cwfunsolvedguess)}
      \wfpostypePf{\acontext'}{P}{\ditto}
      \groundPf{P}{\ditto}
\Label{2} \congoesPf{\acontext}{\ccontext}{Assumption}
\Label{3} \conwfPf{\ccontext}{Assumption}
\Label{4} \congoesWeakPf{\acontext}{\acontext'}{Subderivation}
        \congoesPf{\restrictcontext{(\acontext', \guess{\alpha} = P)}{\acontext}}{\ccontext}{Assumption}
\Label{5} \congoesPf{\restrictcontext{\acontext'}{\acontext}}{\ccontext}{Inversion (\Restrictguessnotin)}
      \proofsep
      \conwfPf{\ccontext'}{\byih, using (1--5) and for some complete context $\ccontext'$}
      \congoesPf{\acontext'}{\ccontext'}{\ditto}
      \congoesWeakPf{\ccontext}{\ccontext'}{\ditto}
      \proofsep
      \wfpostypePf{\ccontext'}{P}{By \lemmaref{lemma:Context extension preserves term well-formedness}}
      \Hand \conwfPf{(\ccontext', \guess{\alpha} = P)}{By \cwfsolvedguess}
      \Hand \congoesPf{\acontext', \guess{\alpha}}{(\ccontext', \guess{\alpha} = P)}{By \Csolveguess}
      \Hand \congoesWeakPf{\ccontext}{(\ccontext', \guess{\alpha} = P)}{By \Wcunsolvedextend}
          \end{llproof}
  \end{itemize}
\end{proof}

\IdenticalRestrictedContexts*

\begin{proof}
  By rule induction on the $\restrictcontext{\acontext''}{\acontext}$ judgment.

  \begin{itemize}
    \DerivationProofCase{\Restrictempty}
      {}
      {\restrictcontext{\emptyacontext}{\emptyacontext} = \emptyacontext}

    \begin{llproof}
      \congoesPf{\emptyacontext}{\acontext'}{Assumption}
      \eqPf{\acontext'}{\emptyacontext}{Inversion (\Cempty)}
\Hand \eqPf{\restrictcontext{\emptyacontext}{\emptyacontext}}{\emptyacontext}{By \Restrictempty}
    \end{llproof}

    \DerivationProofCase{\Restrictuvar}
      {\restrictcontext{\acontext''}{\acontext} = \acontext'''}
      {\restrictcontext{\acontext'', \alpha}{\acontext, \alpha} = \acontext''', \alpha}

      \begin{llproof}
        \inPf{\alpha}{\acontext'}{By \lemmaref{lemma:Context extension maintains variables}}
        \eqPf{\restrictcontext{\acontext''}{\acontext'}}
          {\acontext'''}
          {\byih}
  \Hand \eqPf{\restrictcontext{\acontext'', \alpha}{\acontext'}}
          {\acontext''', \alpha}
          {By \Restrictuvar}
        \trailingjust{(the $\alpha$ context item must appear last in $\acontext'$ by well-formedness of $\acontext'$)}
      \end{llproof}

    \DerivationProofCase{\Restrictguessin}
      {\restrictcontext{\acontext''}{\acontext} = \acontext'''}
      {\restrictcontext{\acontext'', \guess{\alpha}\,[ = P ]}{\acontext, \guess{\alpha}\,[ = Q ]} = \acontext''', \guess{\alpha}\,[ = P ]}

      \begin{llproof}
        \inPf{\guess{\alpha}\,[ = R]}{\acontext'}{By \lemmaref{lemma:Context extension maintains variables}}
        \eqPf{\restrictcontext{\acontext''}{\acontext'}}
          {\acontext'''}
          {\byih}
  \Hand \eqPf{\restrictcontext{\acontext'', \guess{\alpha}\,[ = P ]}{\acontext'}}
          {\acontext''', \guess{\alpha}\,[ = P ]}
          {By \Restrictguessin}
        \trailingjust{(the $\guess{\alpha}\,[ = R]$ context item must appear in $\acontext'$ last by well-formedness of $\acontext'$)}
      \end{llproof}

    \DerivationProofCase{\Restrictguessnotin}
      {
        \restrictcontext{\acontext''}{\acontext} = \acontext''' \\
        \guess{\alpha}\,[ = Q ] \notin \acontext
      }
      {\restrictcontext{\acontext'', \guess{\alpha}\,[ = P ]}{\acontext} = \acontext'''}

      \begin{llproof}
        \notinPf{\guess{\alpha}\,[ = R]}{\acontext'}{By \lemmaref{lemma:Context extension maintains variables}}
        \eqPf{\restrictcontext{\acontext''}{\acontext'}}
          {\acontext'''}
          {\byih}
  \Hand \eqPf{\restrictcontext{\acontext'', \guess{\alpha}\,[ = P ]}{\acontext'}}
          {\acontext'''}
          {By \Restrictguessnotin}
      \end{llproof}
  \end{itemize}
\end{proof}

\subsection{Statement}

\begin{center}
  \Soundness*
\end{center}

\begin{proof}
  By mutual induction with \lemmaref{theorem:typing-completeness}, using the judgment ordering from \lemmaref{lemma:isomorphic types check expressions}.

  \begin{itemize}
    \DerivationProofCase{\Avar}
      {x : P \in \Gamma}
      {\algosynjudg{\acontext; \Gamma}{x}{P}{\acontext}}

    \begin{llproof}
      \inPf{x : P}{\Gamma}{Premise}
      \groundPf{P}{Typing environment only contains ground types}
      \inPf{x : \subcon{\ccontext}{P}}{\Gamma}{By \lemmaref{lemma:Context substitution on ground terms}}
\Hand \declsynjudgPf{\makedec{\acontext}; \Gamma}{x}{[\Omega]P} {By {\Dvar}}
    \end{llproof}

    \DerivationProofCase{\Afunabs}
      {\algosynjudg{\acontext; \Gamma, x : P}{t}{N}{\acontext'}}
      {\algosynjudg{\acontext; \Gamma}{\lamterm{x}{P}{t}}{P \to N}{\acontext'}}

    \begin{llproof}
      \conwfPf{\Theta}{Assumption}
      \envwfPf{\Theta}{\Gamma}{Assumption}
      \algosynjudgPf{\acontext; \Gamma}{\lamterm{x}{P}{t}}{P \to N}{\acontext'}{Assumption}
      \wfnegtypePf{\Theta}{P \funarrow N}{By \lemmaref{lemma:algorithmic-typing-well-formed}}
      \groundPf{P \funarrow N}{\ditto}
      \wfpostypePf{\Theta}{P}{Inversion (\twfarrow)}
      \groundPf{P \funarrow N}{Assumption}
      \groundPf{P}{By definition of ground}
      \envwfPf{\Theta}{\Gamma, x : P}{By {\Ewfvar}}
      \proofsep
      \declsynjudgPf{\makedec{\acontext}; \Gamma, x : P}{t}{[\Omega]N} {By i.h. (term size decreases)}
      \declsynjudgPf{\makedec{\acontext}; \Gamma, x : P}{t}{[\Omega]N} {\bydefsubcon}
      \declsynjudgPf{\makedec{\acontext}; \Gamma, x : \subcon{\ccontext}{P}}{t}{[\Omega]N} {By \lemmaref{lemma:Context substitution on ground terms}}
      \declsynjudgPf{\makedec{\acontext}; \Gamma}{\lambda x. t}{[\Omega]P \to [\Omega]N} {By {\Dfunabs}}
\Hand \declsynjudgPf{\makedec{\acontext}; \Gamma}{\lambda x. t}{[\Omega](P \funarrow N)} {\bydefsubcon}
    \end{llproof}

    \DerivationProofCase{\Atypeabs}
      {\algosynjudg{\acontext, \alpha; \Gamma}{t}{N}{\acontext', \alpha}}
      {\algosynjudg{\acontext; \Gamma}{\gen{\alpha}{t}}{\forall\alpha\ldotp N}{\acontext'}}

    \begin{llproof}
      \conwfPf{\Theta}{Assumption}
      \conwfPf{\Theta, \alpha}{By {\cwfuvar}}
      \envwfPf{\acontext}{\typeenv}{Assumption}
      \envwfPf{\acontext, \alpha}{\typeenv}{Weakening}
      \congoesPf{\acontext'}{\ccontext}{Assumption}
      \congoesPf{\acontext', \alpha}{\ccontext, \alpha}{By \Cuvar}
      \conwfPf{\ccontext}{Assumption}
      \conwfPf{\ccontext, \alpha}{By \cwfuvar}
      \proofsep
      \declsynjudgPf{\makedec{\Theta, \alpha}; \Gamma}{t}{[\Omega, \alpha]N} {\byih (term size decreases)}
      \declsynjudgPf{\makedec{\Theta}, \alpha; \Gamma}{t}{[\Omega]N} {By definitions of $\makedec{-}$ and $\subconunderscores$}
      \declsynjudgPf{\makedec{\acontext}; \Gamma}{\gen{\alpha}{t}}{\forall \alpha \ldotp ([\Omega]N)} {By {\Dtypeabs}}
\Hand \declsynjudgPf{\makedec{\acontext}; \Gamma}{\gen{\alpha}{t}}{[\Omega](\forall \alpha \ldotp N)} {\bydefsubcon}
    \end{llproof}

    \DerivationProofCase{\Athunk}
      {\algosynjudg{\acontext; \Gamma}{t}{N}{\acontext'}}
      {\algosynjudg{\acontext; \Gamma}{\thunk{t}}{\shiftd{N}}{\acontext'}}

    \begin{llproof}
      \algosynjudgPf{\acontext; \Gamma}{t}{N}{\acontext'}{Subderivation}
      \declsynjudgPf{\makedec{\acontext}; \Gamma}{t}{[\Omega]N} {\byih (term size decreases)}
      \declsynjudgPf{\makedec{\acontext}; \Gamma}{\{t\}}{\shiftd{[\Omega]N}} {By {\Dthunk}}
\Hand \declsynjudgPf{\makedec{\acontext}; \Gamma}{t}{[\Omega]\shiftd{N}} {\bydefsubcon}
    \end{llproof}

    \DerivationProofCase{\Areturn}
      {\algosynjudg{\acontext; \Gamma}{v}{P}{\acontext'}}
      {\algosynjudg{\acontext; \Gamma}{\return{v}}{\shiftu{P}}{\acontext'}}

    Symmetric to \Athunk case.

    \DerivationProofCase{\Aambiguouslet}
      {\algosynjudg{\acontext; \typeenv}{v}{\shiftd M}{\acontext'} \\
      \algospinejudg{\acontext'; \Gamma}{s}{\spine{M}{\shiftu{Q}}}{\acontext''} \\
      \APosSubtypeJudg{\acontext''}{P}{Q}{\acontext'''} \\
      \APosSubtypeJudg{\acontext'''}{\subcon{\acontext'''} Q}{P}{\acontext^{(4)}} \\
      \acontext^{(5)} = \restrictcontext{\acontext^{(4)}}{\acontext} \\
      \algosynjudg{\acontext^{(5)}; \Gamma, x : P}{t}{N}{\acontext^{(6)}}}
      {\algosynjudg{\acontext; \Gamma}{\letanno{x}{P}{v}{s}{t}}{N}{\acontext^{(6)}}}

    \begin{llproof}
      \proofcomment{Use well-formedness of the first premise:}

      \conwfPf{\acontext}{Assumption}
      \envwfPf{\acontext}{\typeenv}{Assumption}
      \algosynjudgPf{\acontext; \typeenv}{v}{\shiftdown{M}}{\acontext'}{Subderivation}
      \proofsep

      \congoesPf{\acontext}{\acontext'}{By \lemmaref{lemma:algorithmic-typing-well-formed}}
      \conwfPf{\acontext'}{\ditto}
      \wfpostypePf{\acontext'}{\shiftdown{M}}{\ditto}
      \groundPf{\shiftdown{M}}{\ditto}

      \proofcomment{Now use the well-formedness of $\algospinejudg{\Theta'; \typeenv}{s}{\spine{M}{\shiftu{Q}}}{\Theta''}$:}

      \conwfPf{\acontext'}{Above}
\Label{1} \congoesWeakPf{\acontext}{\acontext'}{By \lemmaref{lemma:weak context extension subsumes normal}}
      \envwfPf{\acontext'}{\typeenv}{By \lemmaref{lemma:context extension preserves wf envs}}
      \wfnegtypePf{\acontext'}{M}{Inversion (\twfshiftdown)}
      \groundPf{M}{By definition of ground}
      \NoSolvedVarsPf{\acontext'}{M}{By \lemmaref{lemma:Context substitution on ground terms}}
      \proofsep
\Label{2} \congoesWeakPf{\acontext'}{\acontext''}{By \lemmaref{lemma:algorithmic-typing-well-formed}}
      \conwfPf{\acontext''}{\ditto}
      \wfnegtypePf{\acontext''}{\shiftu{Q}}{\ditto}
      \NoSolvedVarsPf{\acontext''}{\shiftu{Q}}{\ditto}

      \proofcomment{Next use the well-formedness of $\APosSubtypeJudg{\acontext''}{P}{Q}{\acontext'''}$:}

      \conwfPf{\acontext''}{Above}
      \groundPf{P}{$P$ annotation}
      \NoSolvedVarsPf{\acontext''}{Q}{\bydefsubcon}
      \proofsep
      \conwfPf{\acontext'''}{By \lemmaref{lemma:algorithmic-typing-well-formed}}
      \congoesPf{\acontext''}{\acontext'''}{\ditto}
      \groundPf{[\acontext''']Q}{\ditto}

      \proofcomment{And the well-formedness of $\APosSubtypeJudg{\acontext'''}{\subcon{\acontext'''} Q}{P}{\acontext^{(4)}}$:}

      \conwfPf{\acontext'''}{Above}
      \groundPf{[\acontext''']Q}{Above}
      \NoSolvedVarsPf{\acontext'''}{P}{By \lemmaref{lemma:Context substitution on ground terms}}
      \proofsep
      \conwfPf{\acontext^{(4)}}{By \lemmaref{lemma:algorithmic-typing-well-formed}}
      \congoesPf{\acontext'''}{\acontext^{(4)}}{\ditto}

      \proofcomment{Use the well-formedness of the restricted context:}

      \conwfPf{\acontext}{Above}
      \conwfPf{\acontext^{(4)}}{Above}
      \congoesWeakPf{\acontext''}{\acontext'''}{By \lemmaref{lemma:weak context extension subsumes normal}}
      \congoesWeakPf{\acontext'''}{\acontext^{(4)}}{By \lemmaref{lemma:weak context extension subsumes normal}}
      \congoesWeakPf{\acontext}{\acontext^{(4)}}{Applying \lemmaref{lemma:weak context extension transitive}}
      \trailingjust{to (1), (2), and above}
      \eqPf{\acontext^{(5)}}{\restrictcontext{\acontext^{(4)}}{\acontext}}{Premise}
      \proofsep
      \conwfPf{\acontext^{(5)}}{By \lemmaref{lemma:restricted context wf}}
\Label{3} \congoesPf{\acontext}{\acontext^{(5)}}{\ditto}
\Label{4} \congoesWeakPf{\acontext^{(5)}}{\acontext^{(4)}}{\ditto}

      \proofcomment{Finally use the well-formedness of $\algosynjudg{\acontext^{(5)}; \typeenv, x : P}{t}{N}{\acontext''''}$:}

      \envwfPf{\acontext}{\typeenv}{Assumption}
      \congoesWeakPf{\acontext}{\acontext^{(5)}}{By \lemmaref{lemma:weak context extension subsumes normal}}
      \envwfPf{\acontext^{(5)}}{\typeenv}{By \lemmaref{lemma:context extension preserves wf envs}}
      \wfpostypePf{\acontext}{P}{$P$ annotation}
      \groundPf{P}{\ditto}
      \proofsep

      \conwfPf{\acontext^{(5)}}{Above}
      \envwfPf{\acontext^{(5)}}{\typeenv, x : P}{By \Ewfvar}
      \algosynjudgPf{\acontext^{(5)}; \typeenv; x : P}{t}{N}{\acontext^{(6)}}{Subderivation}
      \proofsep

      \conwfPf{\acontext^{(6)}}{By \lemmaref{lemma:algorithmic-typing-well-formed}}
      \congoesPf{\acontext^{(5)}}{\acontext^{(6)}}{\ditto}

      \proofcomment{Use \lemmaref{lemma:Extended complete context} to obtain a complete context for the second to fourth judgments:}

      \congoesPf{\acontext^{(5)}}{\acontext^{(6)}}{Above}
      \congoesPf{\acontext^{(6)}}{\ccontext}{Assumption}
      \congoesPf{\acontext^{(5)}}{\ccontext}{By \lemmaref{lemma:context extension transitive}}
      \proofsep

      \conwfPf{\acontext^{(4)}}{Above}
      \conwfPf{\ccontext}{Assumption}
      \congoesPf{\acontext^{(5)}}{\ccontext}{Above}
      \congoesWeakPf{\acontext^{(5)}}{\acontext^{(4)}}{Above}
      \eqPf{\acontext^{(5)}}{\restrictcontext{\acontext^{(4)}}{\acontext}}{Above}
      \congoesPf{\restrictcontext{\acontext^{(4)}}{\acontext}}{\ccontext}{Substituting using above}
      \congoesPf{\restrictcontext{\acontext^{(4)}}{\acontext^{(5)}}}{\ccontext}{By \lemmaref{lemma:Identical restricted contexts}}
      \proofsep

      \conwfPf{\ccontext'}{By \lemmaref{lemma:Extended complete context}}
      \congoesPf{\acontext^{(4)}}{\ccontext'}{\ditto}
      \congoesWeakPf{\ccontext}{\ccontext'}{\ditto}
      \proofsep

      \congoesPf{\acontext'''}{\acontext^{(4)}}{Above}
      \congoesPf{\acontext'''}{\ccontext'}{By \lemmaref{lemma:context extension transitive}}
      \proofsep

      \congoesPf{\acontext''}{\acontext'''}{Above}
      \congoesPf{\acontext''}{\ccontext'}{By \lemmaref{lemma:context extension transitive}}

      \proofcomment{Restrict $\ccontext'$ such that $\acontext'$ extends to it:}

      \congoesWeakPf{\acontext''}{\ccontext'}{By \lemmaref{lemma:weak context extension subsumes normal}}
      \congoesWeakPf{\acontext'}{\ccontext'}{By \lemmaref{lemma:weak context extension transitive}}
      \proofsep

      \LetPf{\ccontext''}{\restrictcontext{\ccontext'}{\acontext'}}{}
      \conwfPf{\ccontext''}{By \lemmaref{lemma:restricted context wf}}
      \congoesPf{\acontext'}{\ccontext''}{\ditto}
      \congoesWeakPf{\ccontext''}{\ccontext'}{\ditto}

      \proofcomment{Apply the induction hypothesis to the first premise:}

      \conwfPf{\acontext}{Above}
      \envwfPf{\acontext}{\typeenv}{Above}
      \congoesPf{\acontext'}{\ccontext''}{Above}
      \conwfPf{\ccontext''}{Above}
      \algosynjudgPf{\acontext; \typeenv}{v}{\shiftdown{M}}{\acontext'}{Subderivation}
      \declsynjudgPf{\makedec{\acontext}; \typeenv}{v}{\subcon{\ccontext''} \shiftdown{M}}{\byih (term size decreases)}
      \proofsep

      \wfnegtypePf{\ccontext''}{\shiftd{M}}{By \lemmaref{lemma:Context extension preserves term well-formedness}}
      \congoesWeakPf{\ccontext''}{\ccontext'}{Above}
      \groundPf{\subcon{\ccontext'} M}{By \lemmaref{lemma:Context substitution on ground terms}}
      \conwfPf{\ccontext''}{Above}
      \conwfPf{\ccontext'}{Above}
      \MutualJudgeNegPf{\makedec{\ccontext''}}{\subcon{\ccontext''} M}{\subcon{\ccontext'} M}{By \lemmaref{lemma:Weak context extension leads to isomorphic types (ground)}}
      \proofsep

      \MutualJudgeNegPf{\makedec{\acontext}}{\subcon{\ccontext''} M}{\subcon{\ccontext'} M}{By \lemmaref{lemma:weak context extension maintains variables} and}
      \trailingjust{\lemmaref{lemma:Context extension maintains variables}}

      \proofcomment{Next apply the induction hypothesis to the spine premise:}

      \conwfPf{\acontext'}{Above}
      \envwfPf{\acontext'}{\typeenv}{Above}
      \congoesPf{\acontext''}{\ccontext'}{Above}
      \conwfPf{\ccontext'}{Above}
      \algospinejudgPf{\acontext'; \typeenv}{s}{\spine{M}{\shiftu{Q}}}{\acontext''}{Subderivation}
      \wfnegtypePf{\acontext'}{M}{Above}
      \NoSolvedVarsPf{\acontext'}{M}{Above}
      \declspinejudgPf{\makedec{\acontext}; \typeenv}{s}{\spine{[\ccontext']M}{M'}} {\byih (term size decreases)}
      \MutualJudgeNegPf{\makedec{\acontext}}{[\ccontext']\shiftu{Q}}{M'}{\ditto}
      \eqPf{M'}{\shiftu{Q'}}{Declarative typing rules preserve shift structure}
      \declspinejudgPf{\makedec{\acontext}; \typeenv}{s}{\spine{[\ccontext']M}{\shiftu{Q'}}} {Substituting above}

      \proofsep
      \declspinejudgPf{\makedec{\acontext}; \typeenv}{s}{\spine{[\ccontext'']M}{\shiftu{Q''}}} {By \lemmaref{lemma:isomorphic types check expressions},}
      \trailingjust{using the fact that declarative typing rules preserve shift structure}
      \MutualJudgeNegPf{\makedec{\acontext}}{\shiftu{Q'}}{\shiftu{Q''}}{\ditto}

      \proofcomment{Show the third premise of \Dambiguouslet, first by establishing one direction of the isomorphism:}

\Label{5} \conwfPf{\acontext''}{Shown above}
\Label{6} \congoesPf{\acontext'''}{\ccontext'}{Above}
\Label{7} \wfpostypePf{\acontext''}{P}{Above}
\Label{8}  \groundPf{P}{Above}
      \wfnegtypePf{\acontext''}{\shiftu{Q}}{Above}
\Label{9}  \wfpostypePf{\acontext''}{Q}{By inversion of {\twfshiftup}}
      \eqPf{[\acontext'']\shiftu{Q}}{\shiftu{Q}}{By \lemmaref{lemma:algorithmic-typing-well-formed}}
\Label{10}  \eqPf{[\acontext'']Q}{Q}{\bydefsubcon}
      \proofsep
      \DJudgePosPf{\makedec{\acontext''}}{P}{[\ccontext']Q}{By (5 -- 10) \& \lemmaref{theorem:soundness-pnsubtype}}

      \proofcomment{Now establish the other direction:}

      \conwfPf{\acontext'''}{Shown above}
      \congoesPf{\acontext^{(4)}}{\ccontext'}{Above}
      \wfpostypePf{\acontext'''}{[\acontext''']Q}{By \lemmaref{lemma:Well-formed context substitution preserves term well-formedness}}
      \groundPf{[\acontext''']Q}{Above}
      \wfpostypePf{\acontext'''}{P}{By \lemmaref{lemma:Context extension preserves term well-formedness}}
      \eqPf{[\acontext''']P}{P}{By \lemmaref{lemma:Context substitution on ground terms}}
      \proofsep
      \DJudgePosPf{\makedec{\acontext'''}}{[\acontext''']Q}{[\ccontext']P}{By \lemmaref{theorem:soundness-pnsubtype}}

      \proofcomment{Show the fourth premise of \Dambiguouslet:}

      \DJudgePosPf{\makedec{\acontext'''}}{[\acontext''']Q}{P}{By \lemmaref{lemma:Context substitution on ground terms}}
      \MutualJudgePosPf{\makedec{\acontext'''}}{[\acontext'''] Q}{[\ccontext'] Q}{By \lemmaref{lemma:Context extension leads to isomorphic types (ground)}}
      \DJudgePosPf{\makedec{\acontext'''}}{[\ccontext'] Q}{P}{By \lemmaref{lemma:Transitivity of declarative pnsubtype}}
      \DJudgePosPf{\makedec{\acontext''}}{[\ccontext']Q}{P}{By \lemmaref{lemma:equal declarative contexts}}
      \DJudgePosPf{\makedec{\acontext}}{[\ccontext'] Q}{P}{By \lemmaref{lemma:weak context extension equal declarative contexts}}
      \proofsep

      \DJudgePosPf{\makedec{\acontext}}{P}{[\ccontext'] Q}{By \lemmaref{lemma:weak context extension equal declarative contexts}}
      \MutualJudgePosPf{\makedec{\acontext}}{P}{[\ccontext'] Q}{We have shown the subtyping in both directions}
      \proofsep

      \MutualJudgeNegPf{\makedec{\acontext}}{[\ccontext']\shiftu{Q}}{M'}{Above}
      \MutualJudgeNegPf{\makedec{\acontext}}{[\ccontext']\shiftu{Q}}{\shiftu{Q'}}{Substituting definition of $Q'$ into above}
      \MutualJudgePosPf{\makedec{\acontext}}{[\ccontext']Q}{Q'}{Inversion (\dshiftup)}
      \MutualJudgePosPf{\makedec{\acontext}}{P}{Q'}{By \lemmaref{lemma:Transitivity of declarative pnsubtype}}
      \DJudgeNegPf{\makedec{\acontext}}{\shiftu{P}}{\shiftu{Q'}}{By \dshiftup}
      \DJudgeNegPf{\makedec{\acontext}}{\shiftu{P}}{\shiftu{Q''}}{By \lemmaref{lemma:Transitivity of declarative pnsubtype}}

      \proofcomment{And now show the final premise of \Dambiguouslet:}

      \wfnegtypePf{\acontext}{N}{Assumption}
\Label{11}  \conwfPf{\acontext^{(5)}}{Above}
      \envwfPf{\acontext}{\typeenv}{Assumption}
      \envwfPf{\acontext^{(5)}}{\typeenv}{By \lemmaref{lemma:context extension preserves wf envs}}
      \wfpostypePf{\acontext^{(5)}}{P}{By \lemmaref{lemma:Context extension preserves term well-formedness}}
      \groundPf{P}{Above}
\Label{12}  \envwfPf{\acontext^{(5)}}{\typeenv, x : P}{By {\Ewfvar}}
\Label{13}  \congoesPf{\acontext^{(6)}}{\ccontext}{Assumption}
      \proofsep
      \declsynjudgPf{\makedec{\acontext}; \typeenv, x : P}{t}{[\ccontext]N}{By (11--13) \& i.h.}

      \proofcomment{Finally apply \Dambiguouslet:}

\Hand \declsynjudgPf{\makedec{\acontext}; \typeenv}{\letanno{x}{P}{v}{s}{t}}{[\ccontext]N}{By {\Dambiguouslet}}
    \end{llproof}

    \DerivationProofCase{\Aunambiguouslet}
      {\algosynjudg{\acontext; \typeenv}{v}{\shiftd M}{\acontext'}  \\
      \algospinejudg{\acontext'; \Gamma}{s}{\spine{M}{\shiftu{Q}}}{\acontext''} \\
      \FreeEV(Q) = \emptyset \\
      \acontext''' = \restrictcontext{\acontext''}{\acontext} \\
      \algosynjudg{\acontext'''; \Gamma, x : Q}{t}{N}{\acontext^{(4)}}}
      {\algosynjudg{\acontext; \Gamma}{\letplain{x}{v}{s}{t}}{N}{\acontext^{(4)}}}

    \begin{llproof}
      \proofcomment{As with the \Aambiguouslet case, first use the well-formedness of the first premise:}

      \conwfPf{\acontext}{Assumption}
      \envwfPf{\acontext}{\typeenv}{Assumption}
      \algosynjudgPf{\acontext; \typeenv}{v}{\shiftdown{M}}{\acontext'}{Subderivation}
      \proofsep

      \congoesPf{\acontext}{\acontext'}{By \lemmaref{lemma:algorithmic-typing-well-formed}}
      \conwfPf{\acontext'}{\ditto}
      \wfpostypePf{\acontext'}{\shiftdown{M}}{\ditto}
      \groundPf{\shiftdown{M}}{\ditto}

      \proofcomment{Now use the well-formedness of $\algospinejudg{\Theta'; \typeenv}{s}{\spine{M}{\shiftu{Q}}}{\Theta''}$:}

      \conwfPf{\acontext'}{Above}
\Label{1} \congoesWeakPf{\acontext}{\acontext'}{By \lemmaref{lemma:weak context extension subsumes normal}}
      \envwfPf{\acontext'}{\typeenv}{By \lemmaref{lemma:context extension preserves wf envs}}
      \wfnegtypePf{\acontext'}{M}{Inversion (\twfshiftdown)}
      \groundPf{M}{By definition of ground}
      \NoSolvedVarsPf{\acontext'}{M}{By \lemmaref{lemma:Context substitution on ground terms}}
      \proofsep
\Label{2} \congoesWeakPf{\acontext'}{\acontext''}{By \lemmaref{lemma:algorithmic-typing-well-formed}}
      \conwfPf{\acontext''}{\ditto}
      \wfnegtypePf{\acontext''}{\shiftu{Q}}{\ditto}

      \proofcomment{Use the well-formedness of the restricted context:}

      \conwfPf{\acontext}{Above}
      \conwfPf{\acontext''}{Above}
      \congoesWeakPf{\acontext}{\acontext''}{Applying \lemmaref{lemma:weak context extension transitive}}
      \trailingjust{to (1) and (2)}
      \eqPf{\acontext'''}{\restrictcontext{\acontext''}{\acontext}}{Premise}
      \proofsep
\Label{3} \conwfPf{\acontext'''}{By \lemmaref{lemma:restricted context wf}}
      \congoesPf{\acontext}{\acontext'''}{\ditto}
      \congoesWeakPf{\acontext'''}{\acontext''}{\ditto}

      \proofcomment{Finally use the well-formedness of $\algosynjudg{\acontext'''; \typeenv, x : P}{t}{N}{\acontext^{(4)}}$:}

      \envwfPf{\acontext}{\typeenv}{Assumption}
      \congoesWeakPf{\acontext}{\acontext'''}{By \lemmaref{lemma:weak context extension subsumes normal}}
      \envwfPf{\acontext'''}{\typeenv}{By \lemmaref{lemma:context extension preserves wf envs}}
      \wfpostypePf{\acontext}{P}{$P$ annotation}
      \groundPf{P}{\ditto}
      \proofsep

      \conwfPf{\acontext'''}{Above}
      \envwfPf{\acontext'''}{\typeenv, x : P}{By \Ewfvar}
      \algosynjudgPf{\acontext'''; \typeenv; x : P}{t}{N}{\acontext^{(4)}}{Subderivation}
      \proofsep
      \conwfPf{\acontext^{(4)}}{By \lemmaref{lemma:algorithmic-typing-well-formed}}
      \congoesPf{\acontext'''}{\acontext^{(4)}}{\ditto}

      \proofcomment{Use \lemmaref{lemma:Extended complete context} to obtain a complete context for the second judgment:}

      \congoesPf{\acontext'''}{\acontext^{(4)}}{Above}
      \congoesPf{\acontext^{(4)}}{\ccontext}{Assumption}
      \congoesPf{\acontext'''}{\ccontext}{By \lemmaref{lemma:context extension transitive}}
      \proofsep

      \conwfPf{\acontext''}{Above}
      \conwfPf{\ccontext}{Assumption}
      \congoesPf{\acontext'''}{\ccontext}{Above}
      \congoesWeakPf{\acontext'''}{\acontext''}{Above}
      \eqPf{\acontext'''}{\restrictcontext{\acontext''}{\acontext}}{Above}
      \congoesPf{\restrictcontext{\acontext''}{\acontext}}{\ccontext}{Substituting using above}
      \congoesPf{\restrictcontext{\acontext''}{\acontext'''}}{\ccontext}{By \lemmaref{lemma:Identical restricted contexts}}
      \proofsep

      \conwfPf{\ccontext'}{By \lemmaref{lemma:Extended complete context}}
      \congoesPf{\acontext''}{\ccontext'}{\ditto}
      \congoesWeakPf{\ccontext}{\ccontext'}

      \proofcomment{Restrict $\ccontext'$ such that $\acontext'$ extends to it:}

      \congoesWeakPf{\acontext''}{\ccontext'}{By \lemmaref{lemma:weak context extension subsumes normal}}
      \congoesWeakPf{\acontext'}{\ccontext'}{By \lemmaref{lemma:weak context extension transitive}}
      \proofsep

      \LetPf{\ccontext''}{\restrictcontext{\ccontext'}{\acontext'}}{}
      \conwfPf{\ccontext''}{By \lemmaref{lemma:restricted context wf}}
      \congoesPf{\acontext'}{\ccontext''}{\ditto}
      \congoesWeakPf{\ccontext''}{\ccontext'}{\ditto}

      \proofcomment{Apply the induction hypothesis to the first premise:}

      \conwfPf{\acontext}{Above}
      \envwfPf{\acontext}{\typeenv}{Above}
      \congoesPf{\acontext'}{\ccontext''}{Above}
      \conwfPf{\ccontext''}{Above}
      \algosynjudgPf{\acontext; \typeenv}{v}{\shiftdown{M}}{\acontext'}{Subderivation}
      \declsynjudgPf{\makedec{\acontext}; \typeenv}{v}{\subcon{\ccontext''} \shiftdown{M}}{\byih (term size decreases)}
      \proofsep

      \wfnegtypePf{\ccontext''}{\shiftd{M}}{By \lemmaref{lemma:Well-formed context substitution preserves term well-formedness}}
      \congoesWeakPf{\ccontext''}{\ccontext'}{Above}
      \groundPf{\subcon{\ccontext'} M}{By \lemmaref{lemma:Context substitution on ground terms}}
      \conwfPf{\ccontext''}{Above}
      \conwfPf{\ccontext'}{Above}
      \MutualJudgeNegPf{\makedec{\ccontext''}}{\subcon{\ccontext''} M}{\subcon{\ccontext'} M}{By \lemmaref{lemma:Weak context extension leads to isomorphic types (ground)}}
      \proofsep

      \MutualJudgeNegPf{\makedec{\acontext}}{\subcon{\ccontext''} M}{\subcon{\ccontext'} M}{By \lemmaref{lemma:weak context extension maintains variables} and}
      \trailingjust{\lemmaref{lemma:Context extension maintains variables}}

      \proofcomment{Next apply the induction hypothesis to the spine premise:}

      \conwfPf{\acontext'}{Above}
      \envwfPf{\acontext'}{\typeenv}{Above}
      \congoesPf{\acontext''}{\ccontext'}{Above}
      \conwfPf{\ccontext'}{Above}
      \algospinejudgPf{\acontext'; \typeenv}{s}{\spine{M}{\shiftu{Q}}}{\acontext''}{Subderivation}
      \wfnegtypePf{\acontext'}{M}{Above}
      \NoSolvedVarsPf{\acontext'}{M}{Above}
      \declspinejudgPf{\makedec{\acontext}; \typeenv}{s}{\spine{[\ccontext']M}{M'}} {\byih (term size decreases)}
      \MutualJudgeNegPf{\makedec{\acontext}}{[\ccontext']\shiftu{Q}}{M'}{\ditto}
      \eqPf{M'}{\shiftu{Q'}}{Declarative typing rules preserve shift structure}
      \declspinejudgPf{\makedec{\acontext}; \typeenv}{s}{\spine{[\ccontext']M}{\shiftu{Q'}}} {Substituting above equation}

      \proofsep
      \declspinejudgPf{\makedec{\acontext}; \typeenv}{s}{\spine{[\ccontext'']M}{\shiftu{Q''}}} {By \lemmaref{lemma:isomorphic types check expressions},}
      \trailingjust{using the fact that declarative typing rules preserve shift structure}
      \MutualJudgeNegPf{\makedec{\acontext}}{\shiftu{Q'}}{\shiftu{Q''}}{\ditto}

      \proofcomment{Next apply the induction hypothesis to the last premise:}

      \wfnegtypePf{\acontext}{N}{Assumption}
\Label{4}  \conwfPf{\acontext'''}{Above}
      \envwfPf{\acontext}{\typeenv}{Assumption}
      \envwfPf{\acontext'''}{\typeenv}{By \lemmaref{lemma:context extension preserves wf envs}}
      \wfpostypePf{\acontext'''}{Q}{By \lemmaref{lemma:Context extension preserves term well-formedness}}
      \eqPf{\FreeEV(Q)}{\emptyset}{Premise}
      \groundPf{Q}{\bydefground}
\Label{5}  \envwfPf{\acontext'''}{\typeenv, x : Q}{By {\Ewfvar}}
      \congoesPf{\acontext^{(4)}}{\ccontext}{Assumption}
      \congoesPf{\acontext'''}{\acontext^{(4)}}{Above}
\Label{6} \congoesPf{\acontext'''}{\ccontext}{By \lemmaref{lemma:context extension transitive}}
      \proofsep
      \declsynjudgPf{\makedec{\acontext}; \typeenv, x : Q}{t}{[\ccontext]N}{By (4--6) \& i.h.}

      \proofcomment{Rework the declarative judgment we got from the induction hypothesis to match the form we need to apply $\Dunambiguouslet$:}

      \MutualJudgeNegPf{\makedec{\acontext}}{[\ccontext']\shiftu{Q}}{M'}{Above}
      \MutualJudgeNegPf{\makedec{\acontext}}{\shiftu{Q}}{M'}{By \lemmaref{lemma:Context substitution on ground terms}}
      \MutualJudgeNegPf{\makedec{\acontext}}
      {\shiftu{Q}}{\shiftu{Q'}}{Substituting in the definition of $Q'$}
      \MutualJudgeNegPf{\makedec{\acontext}}
      {\shiftu{Q}}{\shiftu{Q''}}{By \lemmaref{lemma:Transitivity of declarative pnsubtype}}
      \MutualJudgePosPf{\makedec{\acontext}}{Q}{Q''}{Inversion ($\ashiftup$)}
      \declsynjudgPf{\makedec{\acontext}; \typeenv, x : Q''}{t}{[\ccontext]N}{Using \lemmaref{lemma:isomorphic types check expressions}}
      \trailingjust{to change the typing environment}

      \proofcomment{
        Now show that for all positive types $P$, if $\declspinejudg{\makedec{\acontext}; \Gamma}{s}{\spine{\subcon{\ccontext'} M}{\shiftu{P}}}$ then $\declpisotypejudg{\makedec{\acontext}}{Q''}{P}$.
        Let $P$ be an arbitrary positive type such that
        $\declspinejudg{\makedec{\acontext};\Gamma}{s}{\spine{\subcon{\ccontext'} M}{\shiftu{P}}}$.
      }

      \declspinejudgPf{\makedec{\Theta'};\Gamma}{s}{\spine{\subcon{\ccontext'} M}{\shiftu{P}}}{By \lemmaref{lemma:equal declarative contexts}}
      \algospinejudgPf{\Theta'; \Gamma}{s}{\spine{M}{\shiftu{R}}}{\hat{\Theta}''}{By \lemmaref{theorem:typing-completeness},}
      \trailingjust{for some $R$ and $\hat{\Theta}''$ (term size decreases)}
\Label{7} \eqPf{[\Omega']\shiftu{R}}{\shiftu{P}}{\ditto}
      \proofsep
      \algospinejudgPf{\Theta'; \Gamma}{s}{\spine{M}{\shiftu{Q}}}{\Theta''}{Subderivation}
      \eqPf{\shiftu{R}}{\shiftu{Q}}{By \lemmaref{lemma:subtyping determinacy}}
      \eqPf{[\Omega']\shiftu{Q}}{[\Omega']\shiftu{R}}{Applying $\Omega'$ to both sides}
      \eqPf{[\Omega']\shiftu{Q}}{\shiftu{P}}{Substituting using (7)}
\Label{9} \eqPf{[\Omega']Q}{P}{\bydefsubcon}
      \proofsep
      \DJudgePosPf{\makedec{\acontext}}{P}{P}{By \lemmaref{lemma:Reflexivity of declarative pnsubtype}}
      \MutualJudgePosPf{\makedec{\acontext}}{P}{P}{By definition of $\mutualSubtypePosNeg$}
      \declpisotypejudgPf{\makedec{\acontext}}{[\Omega']Q}{P}{Substituting using (8)}
      \MutualJudgePosPf{\makedec{\acontext}}{[\ccontext'] Q}{Q'}{Above}
      \MutualJudgePosPf{\makedec{\acontext}}{Q''}{P}{Applying \lemmaref{lemma:Transitivity of declarative pnsubtype} twice}
  
      \proofcomment{Finally apply \Dunambiguouslet:}
  
  \Hand \declsynjudgPf{\makedec{\acontext}; \typeenv}{\letplain{x}{v}{s}{t}}{[\ccontext]N}{By {\Dunambiguouslet}}
    \end{llproof}

    \DerivationProofCase{\Aspinenil}
      {}
      {\algospinejudg{\Theta; \Gamma}{\epsilon}{\spine{N}{N}}{\Theta}}

    \begin{llproof}
\Hand \declspinejudgPf{\makedec{\acontext}; \Gamma}{\epsilon}{\spine{[\Omega]N}{[\Omega]N}}{By {\Dspinenil}}
\Hand \MutualJudgeNegPf{\makedec{\acontext}}{\subcon{\ccontext} N}{\subcon{\ccontext} N}{By \lemmaref{lemma:Reflexivity of declarative pnsubtype}}
    \end{llproof}

    \DerivationProofCase{\Aspinecons}
      {\algosynjudg{\acontext; \Gamma}{v}{P}{\acontext'} \\
      \APosSubtypeJudg{\acontext'}{P}{[\acontext']Q}{\acontext''} \\
      \algospinejudg{\acontext''; \Gamma}{s}{\spine{[\acontext'']N}{M}}{\acontext'''}}
      {\algospinejudg{\acontext; \Gamma}{v, s}{Q \to \spine{N}{M}}{\acontext'''}}

    \begin{llproof}
      \wfnegtypePf{\Theta}{Q \funarrow N}{Assumption}
      \wfpostypePf{\Theta}{Q}{Inversion (\twfarrow)}
      \wfnegtypePf{\Theta}{N}{Inversion (\twfarrow)}
      \eqPf{[\Theta](Q \funarrow N)}{Q \funarrow N}{Assumption}
      \eqPf{[\Theta]Q \to [\Theta]N}{Q \funarrow N}{\bydefsubcon}
      \eqPf{[\Theta]Q}{Q}{By equality}
      \conwfPf{\Theta}{Assumption}
      \envwfPf{\Theta}{\Gamma}{Assumption}
      \conwfPf{\Omega}{Assumption}

      \proofcomment{Apply typing well-formedness to the first premise:}

      \congoesPf{\Theta}{\Theta'}{By \lemmaref{lemma:algorithmic-typing-well-formed}}
      \conwfPf{\Theta'}{\ditto}
      \wfpostypePf{\Theta'}{P}{\ditto}
      \groundPf{P}{\ditto}

      \proofcomment{Now apply the well-formedness of subtyping to the second premise:}

      \conwfPf{\Theta'}{Above}
      \groundPf{P}{Above}
      \NoSolvedVarsPf{\Theta'}{\subcon{\Theta'}{Q}}{By \lemmaref{lemma:Context substitution idempotence}}
      \proofsep
      \conwfPf{\Theta''}{By \lemmaref{lemma:well-formedness-pnsubtype}}
      \congoesPf{\Theta'}{\Theta''}{\ditto}
      \groundPf{[\Theta''][\Theta']Q}{\ditto}
      \groundPf{[\Theta'']Q}{By \lemmaref{lemma:extending-context-preserves-groundness}}

      \proofcomment{Apply typing well-formedness to the last premise:}

      \wfnegtypePf{\Theta''}{N}{By \lemmaref{lemma:Context extension preserves term well-formedness}}
      \wfnegtypePf{\Theta''}{\subcon{\acontext''} N}{By \lemmaref{lemma:Well-formed context substitution preserves term well-formedness}}
      \eqPf{[\Theta''][\Theta'']N}{[\Theta'']N}{By \lemmaref{lemma:Context substitution idempotence}}
      \envwfPf{\Theta}{\Gamma}{Assumption}
      \envwfPf{\Theta'''}{\Gamma}{By \lemmaref{lemma:context extension preserves wf envs}}
      \proofsep

      \congoesWeakPf{\Theta''}{\Theta'''}{By \lemmaref{lemma:algorithmic-typing-well-formed}}
      \conwfPf{\Theta'''}{\ditto}

      \proofcomment{
        Restrict $\ccontext$ such that $\acontext''$ extends to it:
      }

      \congoesPf{\Theta'''}{\Omega}{Assumption}
      \conwfPf{\Omega}{Assumption}
      \congoesWeakPf{\acontext'''}{\Omega}{By \lemmaref{lemma:weak context extension subsumes normal}}
      \congoesWeakPf{\acontext''}{\Omega}{By \lemmaref{lemma:weak context extension transitive}}
      \proofsep

      \conwfPf{\restrictcontext{\Omega}{\acontext''}}{By \lemmaref{lemma:restricted context wf}}
      \congoesPf{\acontext''}{\restrictcontext{\Omega}{\acontext''}}{\ditto}
      \congoesWeakPf{\restrictcontext{\Omega}{\acontext''}}{\Omega}{\ditto}
      \proofsep
      \congoesWeakPf{\acontext'}{\Omega}{By \lemmaref{lemma:weak context extension subsumes normal}}
      \trailingjust{and \lemmaref{lemma:weak context extension transitive}}
      \wfpostypePf{\ccontext}{P}{By \lemmaref{lemma:weak context extension preserves well-formedness}}
      \MutualJudgePosPf{\makedec{(\restrictcontext{\Omega}{\acontext''})}}{\subcon{\ccontext} P}{\subcon{(\restrictcontext{\Omega}{\acontext''})} P}{By \lemmaref{lemma:Weak context extension leads to isomorphic types (ground)}}
      \congoesPf{\acontext}{\restrictcontext{\Omega}{\acontext''}}{By \lemmaref{lemma:context extension transitive}}
      \MutualJudgePosPf{\makedec{\Theta}}{\subcon{\ccontext} P}{\subcon{(\restrictcontext{\Omega}{\acontext''})} P}{By \lemmaref{lemma:equal declarative contexts}}

      \proofcomment{Applying the induction hypothesis to the first premise:}

      \declsynjudgPf{\makedec{\acontext}; \Gamma}{v}{[(\restrictcontext{\Omega}{\acontext''})]P} {\byih (term size decreases)}
      \declsynjudgPf{\makedec{\acontext}; \Gamma}{v}{P} {By \lemmaref{lemma:Context substitution on ground terms}}

      \proofcomment{Applying soundness to the second premise:}

      \congoesPf{\acontext'}{\restrictcontext{\Omega}{\acontext''}}{By transitivity}
      \wfpostypePf{\acontext'}{\subcon{\acontext'} Q}{By \lemmaref{lemma:Context extension preserves term well-formedness}}
      \trailingjust{and \lemmaref{lemma:substituion preserves well-formedness of types}}
      \proofsep

      \DJudgePosPf{\makedec{\acontext'}}{P}{\subcon{\restrictcontext{\Omega}{\acontext''}} \subcon{\acontext'} Q}{By \lemmaref{theorem:soundness-pnsubtype}}
      \DJudgePosPf{\makedec{\acontext'}}{P}{\subcon{\restrictcontext{\Omega}{\acontext''}} Q}{By \lemmaref{lemma:Context extension leads to isomorphic types}}
      \DJudgePosPf{\makedec{\acontext}}{P}{\subcon{\ccontext} Q}{By \lemmaref{lemma:Weak context extension leads to isomorphic types (ground)}}
      \trailingjust{and \lemmaref{lemma:equal declarative contexts}}

      \proofcomment{Apply the induction hypothesis to the last premise:}

      \declspinejudgPf{\makedec{\acontext''}; \Gamma}{s}{\spine{[\Omega][\acontext''] N}{M'}} {\byih (term size decreases)}
      \MutualJudgeNegPf{\makedec{\acontext''}}{\subcon{\ccontext} M}{M'}{\ditto}

    \end{llproof}


    Reworking the spine declarative judgment:

    \begin{llproof}
      \MutualJudgeNegPf{\makedec{\acontext''}}{\subcon{\ccontext} \subcon{\acontext''} N}{\subcon{\ccontext} N}{By \lemmaref{lemma:Weak context extension leads to isomorphic types}}
      \declspinejudgPf{\makedec{\acontext''}; \Gamma}{s}{\spine{[\Omega] N}{M''}} {By \lemmaref{lemma:isomorphic types check expressions}}
      \MutualJudgeNegPf{\makedec{\acontext''}}{M'}{M''}{\ditto}
      \declspinejudgPf{\makedec{\acontext}; \Gamma}{s}{\spine{[\Omega] N}{M''}} {By \lemmaref{lemma:equal declarative contexts}}
 
      \proofcomment{Applying the declarative judgment, we have:}

      \declspinejudgPf{\makedec{\acontext}; \Gamma}{v, s}{\spine{[\Omega]Q \to [\Omega]N}{[\Omega]M}} {By {\Dspinecons}}
\Hand     \declspinejudgPf{\makedec{\acontext}; \Gamma}{v, s}{\spine{[\Omega](Q \funarrow N)}{M''}} {\bydefsubcon}
      \MutualJudgeNegPf{\makedec{\acontext''}}{\subcon{\ccontext} M}{M''}{By \lemmaref{lemma:Transitivity of declarative pnsubtype}}
\Hand \MutualJudgeNegPf{\makedec{\acontext}}{\subcon{\ccontext} M}{M''}{By \lemmaref{lemma:equal declarative contexts}}
    \end{llproof}

    \DerivationProofCase{\Aspinetypeabsnotin}
      {\algospinejudg{\acontext; \Gamma}{s}{\spine{N}{M}}{\acontext'} \\
      \alpha \notin \FreeUV(N)}
      {\algospinejudg{\acontext; \Gamma}{s}{\spine{(\forall\alpha\ldotp N)}{ M}}{\acontext'}}

    \begin{llproof}
      \wfnegtypePf{\Theta}{\forall\alpha\ldotp N}{Assumption}
      \wfnegtypePf{\Theta, \alpha}{N}{Inversion (\twfforall)}
\Label{1} \wfnegtypePf{\Theta}{N}{By \lemmaref{lemma:type well-formed with alpha removed}}
      \trailingjust{and $\alpha \notin \FreeUV(N)$}
      \eqPf{[\Theta](\forall\alpha \ldotp N)}{\forall\alpha \ldotp N}{Assumption}
      \eqPf{\forall\alpha \ldotp [\Theta]N}{\forall\alpha \ldotp N}{\bydefsubcon}
\Label{2} \eqPf{[\Theta]N}{N}{By equality}
\Label{3} \conwfPf{\Theta}{Assumption}
\Label{4} \envwfPf{\Theta}{\Gamma}{Assumption}
\Label{5} \congoesPf{\Theta'}{\Omega}{Assumption}
\Label{6} \conwfPf{\Omega}{Assumption}

      \proofcomment{Apply the induction hypothesis:}
      \declspinejudgPf{\makedec{\acontext}; \Gamma}{s}{\spine{[\Omega]N}{M'}}{\byih (term size stays the same and}
      \trailingjust{the number of prenex quantifiers decreases)}
\Hand \MutualJudgeNegPf{\makedec{\acontext}}{[\Omega]M}{M'}{\ditto}

      \proofcomment{Let $P$ be an arbitrary positive type, such that $\wfpostypeJudg{\Omega}{P}$:}

      \eqPf{[P / \alpha]N}{N}{As $\alpha \notin \FreeUV(N)$}
      \declspinejudgPf{\makedec{\acontext}; \Gamma}{s}{\spine{[\Omega][P / \alpha]N}{M'}}{By equality}
      \declspinejudgPf{\makedec{\acontext}; \Gamma}{s}{\spine{[P / \alpha][\Omega]N}{M'}}{\bydefsubcon}

      \proofcomment{Applying the declarative judgment:}

      \declspinejudgPf{\makedec{\acontext}; \Gamma}{s}{\spine{\forall \alpha \ldotp [\Omega]N}{M'}}{By {\Dspinetypeabs}}
\Hand \declspinejudgPf{\makedec{\acontext}; \Gamma}{s}{\spine{[\Omega](\forall \alpha \ldotp N)}{M'}}{\bydefsubcon}
    \end{llproof}

    \DerivationProofCase{\Aspinetypeabsin}
      {\algospinejudg{\acontext, \guess{\alpha}; \Gamma}{s}{\spine{[\guess{\alpha}/\alpha]N}{M}}{\acontext', \guess{\alpha}\, [= P]} \\
      \alpha \in \FreeUV(N)}
      {\algospinejudg{\acontext; \Gamma}{s}{\spine{(\forall\alpha\ldotp N)}{M}}{\acontext', \guess{\alpha}\, [= P]}}

    \begin{llproof}
      \wfnegtypePf{\Theta}{\forall \alpha \ldotp N}{Assumption}
      \wfnegtypePf{\Theta, \alpha}{N}{By {\twfforall}}
\Label{1} \wfnegtypePf{\Theta, \guess{\alpha}}{[\guess{\alpha}/\alpha]N}{By \lemmaref{lemma:substituion preserves well-formedness of types}}
      \proofsep
      \eqPf{[\Theta](\forall \alpha \ldotp N)}{\forall \alpha \ldotp N}{Assumption}
      \eqPf{[\Theta]N}{N}{\bydefsubcon}
      \eqPf{[\Theta][\guess{\alpha}/\alpha]N}{[\guess{\alpha}/\alpha]N}{$\guess{\alpha}$ fresh}
\Label{2} \eqPf{[\Theta,\guess{\alpha}][\guess{\alpha}/\alpha]N}{[\guess{\alpha}/\alpha]N}{\bydefsubcon}
      \proofsep
      \conwfPf{\Theta}{Assumption}
\Label{3} \conwfPf{\Theta, \guess{\alpha}}{By {\cwfunsolvedguess}}
      \envwfPf{\Theta}{\Gamma}{Assumption}
      \congoesWeakPf{\Theta}{\Theta, \guess{\alpha}}{By {\Wcunsolvedextend}}
\Label{4} \envwfPf{\Theta, \guess{\alpha}}{\Gamma}{By \lemmaref{lemma:context extension preserves wf envs}}
\Label{5} \conwfPf{\Omega}{Assumption}

      \proofcomment{Use well-formedness of the subderivation:}

      \congoesWeakPf{\acontext, \guess{\alpha}}{\acontext, \guess{\alpha} \, [= P]}{By \lemmaref{lemma:algorithmic-typing-well-formed}}

      \proofcomment{Obtain the solution to $\guess{\alpha}$ from the complete context:}

      \congoesPf{\Theta', \guess{\alpha}\, [= P]}{\Omega}{Assumption}
      \eqPf{\Omega}{\Omega', \guess{\alpha} = P'}{Inversion (\Csolveguess), since $\ccontext$ is a complete context}
      \conwfPf{\Omega', \guess{\alpha} = P'}{Above}
      \wfpostypePf{\Omega'}{P'}{Inversion (\cwfsolvedguess)}
      \groundPf{P'}{\ditto}
      \wfpostypePf{\Omega}{P'}{By \lemmaref{lemma:Term well-formedness weakening}}
      \congoesPf{\Theta', \guess{\alpha}\, [= P]}{\Omega}{Assumption}
      \wfpostypePf{\Theta', \guess{\alpha}\, [= P]}{P'}{By \lemmaref{lemma:Context extension maintains variables}}
      \wfpostypePf{\Theta, \guess{\alpha}}{P'}{By \lemmaref{lemma:weak context extension maintains variables}}
      \wfpostypePf{\makedec{\acontext}}{P'}{Since $P'$ ground}
    \end{llproof}

    Apply the induction hypothesis:

    \begin{llproof}
      \declspinejudgPf{\makedec{\acontext}; \Gamma}{s}{\spine{[\Omega][\guess{\alpha}/\alpha]N}{M'}}{\byih (Term size stays the same and the number}
      \trailingjust{of prenex universal quantifiers decreases. The}
      \trailingjust{substitution replaces a positive type by another}
      \trailingjust{positive type, so cannot add or remove prenex}
      \trailingjust{universal quantifiers.)}
\Hand \MutualJudgeNegPf{\makedec{\acontext}}{[\Omega]M}{M'}{\ditto}
      \declspinejudgPf{\makedec{\acontext}; \Gamma}{s}{\spine{[\ccontext', \guess{\alpha} = P'][\guess{\alpha}/\alpha]N}{M'}}{Substituting $\ccontext = \ccontext', \guess{\alpha} = P'$ from above}
      \declspinejudgPf{\makedec{\acontext}; \Gamma}{s}{\spine{[P' / \alpha][\ccontext', \guess{\alpha} = P']N}{M'}}{\bydefsubcon}  
      \declspinejudgPf{\makedec{\acontext}; \Gamma}{s}{\spine{[P' / \alpha][\Omega]N}{M'}}{Substituting $\ccontext = \ccontext', \guess{\alpha} = P'$ from above}
      \declspinejudgPf{\makedec{\acontext}; \Gamma}{s}{\spine{(\forall\alpha \ldotp  [\Omega]N)}{M'}} {By {\Dspinetypeabs}}
\Hand     \declspinejudgPf{\makedec{\acontext}; \Gamma}{s}{\spine{[\Omega](\forall\alpha \ldotp  N)}{M'}} {By definition of substitution}
    \end{llproof}
  \end{itemize}
\end{proof}

\section{Completeness of typing}

\subsection{Lemmas}

\begin{center}
  \WeakContextExtensionSameVars*
\end{center}

\begin{proof}
  All rules ensure the \lhs and \rhs contexts have the same set of free universal variables.
  \Wcempty, \Wcuvar, \Wcunsolvedguess, \Wcsolveguess, and \Wcsolvedguess ensure the \lhs and \rhs contexts have the same set of existential variables.
  The \rhs context in the \Wcunsolvedextend and \Wcunsolvedextend rules have a set of existential variables that is a superset of the set of existential variables on the \lhs context.
\end{proof}

\ReversingExtensionFromComplete*

\begin{proof}
  By rule induction on the $\congoesJudg{\ccontext}{\acontext}$ judgment:

  \begin{itemize}
    \DerivationProofCase{\Cempty}
      {}
      {\congoesJudg{\emptyacontext}{\emptyacontext}}

      \begin{llproof}
\Hand   \congoesPf{\emptyacontext}{\emptyacontext}{Assumption}
      \end{llproof}

    \DerivationProofCase{\Cuvar}
      {\congoesJudg{\ccontext}{\acontext}}
      {\congoesJudg{\ccontext, \alpha}{\acontext, \alpha}}

      \begin{llproof}
        \congoesPf{\ccontext}{\acontext}{Subderivation}
        \congoesPf{\acontext}{\ccontext}{\byih}
\Hand   \congoesPf{\acontext, \alpha}{\ccontext, \alpha}{By \Cuvar}
      \end{llproof}

    \DerivationProofCase{\Cunsolvedguess}
      {\congoesJudg{\ccontext}{\acontext}}
      {\congoesJudg{\ccontext, \guess{\alpha}}{\acontext, \guess{\alpha}}}

    Impossible since the LHS must be a complete context.

    \DerivationProofCase{\Csolveguess}
      {\congoesJudg{\ccontext}{\acontext}}
      {\congoesJudg{\ccontext, \guess{\alpha}}{\acontext, \guess{\alpha} = P}}

    Impossible since the LHS must be a complete context.

    \DerivationProofCase{\Csolvedguess}
      {\congoesJudg{\ccontext}{\acontext} \\ \MutualSubtypePosJudge{\makedec{\ccontext}} {P} {Q}}
      {\congoesJudg{\ccontext, \guess{\alpha} = P}{\acontext, \guess{\alpha} = Q}}

      \begin{llproof}
        \congoesPf{\ccontext}{\acontext}{Subderivation}
        \congoesPf{\acontext}{\ccontext}{\byih}
        \MutualJudgePosPf{\makedec{\ccontext}} {P} {Q} {Premise}
        \MutualJudgePosPf{\makedec{\acontext}} {P} {Q} {By \lemmaref{lemma:equal declarative contexts}}
        \MutualJudgePosPf{\makedec{\acontext}} {Q} {P} {By definition of $\MutualSubtypePosNegJudge{-}{-}{-}$}
\Hand   \congoesPf{\acontext, \guess{\alpha} = Q}{\ccontext, \guess{\alpha} = P}{By \Csolvedguess}
      \end{llproof}
  \end{itemize}
\end{proof}

\PullingBackRestrictedContexts*

\begin{proof}
  By rule induction on the $\congoesJudg{\acontext}{\acontext'}$ judgment:

  \begin{itemize}
    \DerivationProofCase{\Cempty}
      {}
      {\congoesJudg{\emptyacontext}{\emptyacontext}}

      \begin{llproof}
\Hand   \congoesPf{\restrictcontext{\emptyacontext}{\acontext''}}{\acontext'''}{Assumption}
      \end{llproof}

    \DerivationProofCase{\Cuvar}
      {\congoesJudg{\acontext}{\acontext'}}
      {\congoesJudg{\acontext, \alpha}{\acontext', \alpha}}

    \begin{llproof}
      \congoesPf{\restrictcontext{\acontext', \alpha}{\acontext''}}{\acontext'''}{Assumption}
      \congoesPf{\restrictcontext{\acontext'}{\bar{\acontext}''}}{\bar{\acontext}'''}{Inversion (\Restrictuvar)}
      \eqPf{\acontext''}{\bar{\acontext}'', \alpha}{\ditto}
      \eqPf{\acontext'''}{\bar{\acontext}''', \alpha}{\ditto}
      \congoesPf{\acontext}{\acontext'}{Subderivation}
      \proofsep

      \congoesPf{\restrictcontext{\acontext}{\bar{\acontext}''}}{\bar{\acontext}'''}{\byih}
      \congoesPf{(\restrictcontext{\acontext}{\bar{\acontext}''}), \alpha}{\bar{\acontext}''', \alpha}{By \Cuvar}
\Hand \congoesPf{\restrictcontext{\acontext, \alpha}{\acontext''}}{\acontext'''}{By \Restrictuvar}
    \end{llproof}

    \DerivationProofCase{\Cunsolvedguess}
      {\congoesJudg{\acontext}{\acontext'}}
      {\congoesJudg{\acontext, \guess{\alpha}}{\acontext', \guess{\alpha}}}

    \DerivationProofCase{\Csolvedguess}
      {\congoesJudg{\acontext}{\acontext'} \\ \MutualSubtypePosJudge{\makedec{\acontext}} {P} {Q}}
      {\congoesJudg{\acontext, \guess{\alpha} = P}{\acontext', \guess{\alpha} = Q}}

    Prove these two cases together.

    \begin{llproof}
      \congoesPf{\restrictcontext{\acontext', \guess{\alpha} \,[ = P ]}{\acontext''}}{\acontext'''}{Assumption}
    \end{llproof}

    Taking cases on whether $\guess{\alpha} \,[ = Q ] \in \acontext''$:

    \begin{itemize}
      \caseitem{$\guess{\alpha} \,[ = Q ] \in \acontext''$}

        \begin{llproof}
          \congoesPf{\restrictcontext{\acontext'}{\bar{\acontext}''}}{\bar{\acontext}'''}{Inversion (\Restrictguessin)}
          \eqPf{\acontext''}{\bar{\acontext}'', \guess{\alpha} \,[ = Q ]}{\ditto}
          \eqPf{\acontext'''}{\bar{\acontext}''', \guess{\alpha} \,[ = P ]}{\ditto}
          \proofsep

          \congoesPf{\restrictcontext{\acontext}{\bar{\acontext}''}}{\bar{\acontext}'''}{\byih}
          \congoesPf{(\restrictcontext{\acontext}{\bar{\acontext}''}), \guess{\alpha} \,[ = P ]}{\bar{\acontext}''', \guess{\alpha} \,[ = P ]}{By \Cunsolvedguess / \Csolvedguess}
\Hand     \congoesPf{\restrictcontext{\acontext, \guess{\alpha} \,[ = P ]}{\acontext''}}{\acontext'''}{By \Restrictguessin}
        \end{llproof}
      
      \caseitem{$\guess{\alpha} \,[ = Q ] \notin \acontext''$}

        \begin{llproof}
          \congoesPf{\restrictcontext{\acontext'}{\acontext''}}{\acontext'''}{Inversion (\Restrictguessnotin)}
          \congoesPf{\restrictcontext{\acontext}{\acontext''}}{\acontext'''}{\byih}
\Hand     \congoesPf{\restrictcontext{\acontext, \guess{\alpha} \,[ = P ]}{\acontext''}}{\acontext'''}{By \Restrictguessnotin}
        \end{llproof}
    \end{itemize}
  \end{itemize}
\end{proof}

\subsection{Statement}

\begin{center}
  \Completeness*
\end{center}

\begin{proof}
  By mutual induction with \lemmaref{theorem:soundness}, using the same judgment ordering as in \lemmaref{lemma:isomorphic types check expressions}.

  \begin{itemize}
    \DerivationProofCase{\Dvar}
      {x : P \in \Gamma}
      {\declsynjudg{\makedec{\acontext}; \Gamma}{x}{P}}

    \begin{llproof}
      \inPf{x : P}{\Gamma}{Premise}
\Hand \algosynjudgPf{\Theta; \Gamma}{x}{P}{\Theta}{By \Avar}
\Hand \congoesPf{\acontext}{\ccontext}{Assumption}
    \end{llproof}

    \DerivationProofCase{\Dfunabs}
      {\declsynjudg{\makedec{\acontext}; \Gamma, x : P}{t}{N}}
      {\declsynjudg{\makedec{\acontext}; \Gamma}{\lambda x : P \ldotp t}{P \to N}}

    \begin{llproof}
      \wfnegtypePf{\Theta}{P \funarrow N}{Assumption}
      \wfnegtypePf{\Theta}{N}{Inversion (\twfarrow)}
      \groundPf{P \funarrow N}{Assumption}
      \groundPf{N}{By definition of ground}

      \proofsep
      \conwfPf{\Theta}{Assumption}

      \proofsep
      \envwfPf{\Theta}{\Gamma}{Assumption}
      \wfnegtypePf{\Theta}{P \funarrow N}{Assumption}
      \wfpostypePf{\Theta}{P}{Inversion (\twfarrow)}
      \groundPf{P \funarrow N}{Assumption}
      \groundPf{P}{By definition of ground}
      \envwfPf{\Theta}{\Gamma, x : P}{By {\Ewfvar}}

      \proofsep
      \congoesPf{\Theta}{\Omega}{Assumption}
      \conwfPf{\Omega}{Assumption}

      \proofsep
      \algosynjudgPf{\Theta; \Gamma, x : P}{t}{N}{\Theta'}{\byih, for some context $\acontext'$ (term size decreases)}
\Hand \congoesPf{\Theta'}{\Omega}{\ditto}
\Hand \algosynjudgPf{\Theta; \Gamma}{\lambda x : P \ldotp t}{P \funarrow N}{\Theta'}{By \Afunabs}
    \end{llproof}

    \DerivationProofCase{\Dtypeabs}
      {\declsynjudg{\makedec{\Theta}, \alpha; \Gamma}{t}{N}}
      {\declsynjudg{\makedec{\Theta}; \Gamma}{\gen{\alpha}{t}}{\forall\alpha\ldotp N}}

    \begin{llproof}
      \declsynjudgPf{\makedec{\Theta, \alpha}; \Gamma}{t}{N}{\bydefmakedec}
      \proofsep
      \wfnegtypePf{\Theta}{\forall\alpha\ldotp N}{Assumption}
      \wfnegtypePf{\Theta, \alpha}{N}{By {\twfforall}}

      \proofsep
      \groundPf{\forall\alpha\ldotp N}{Assumption}
      \groundPf{N}{By definition of ground}

      \proofsep
      \conwfPf{\Theta}{Assumption}
      \conwfPf{\Theta, \alpha}{By {\cwfuvar}}

      \proofsep
      \envwfPf{\Theta}{\Gamma}{Assumption}

      \proofsep
      \congoesPf{\Theta}{\Omega}{Assumption}
      \congoesPf{\Theta, \alpha}{\Omega, \alpha}{By {\Cuvar}}

      \proofsep
      \conwfPf{\Omega}{Assumption}
      \conwfPf{\Omega, \alpha}{By {\cwfuvar}}

      \proofsep
      \algosynjudgPf{\Theta, \alpha; \Gamma}{t}{N}{\Theta'}{\byih, for some context $\acontext'$ (term size decreases)}
      \congoesPf{\Theta'}{\Omega, \alpha}{\ditto}
      \eqPf{\Theta'}{\Theta'', \alpha}{Inversion (\Cuvar)for some context $\acontext''$}
\Hand \congoesPf{\Theta''}{\Omega}{\ditto}

      \algosynjudgPf{\Theta, \alpha; \Gamma}{t}{N}{\Theta'', \alpha}{Substituting for $\acontext'$}
\Hand \algosynjudgPf{\Theta; \Gamma}{\gen{\alpha}{t}}{\forall\alpha\ldotp N}{\Theta''}{By \Atypeabs}
    \end{llproof}

    \DerivationProofCase{\Dthunk}
      {\declsynjudg{\makedec{\Theta}; \Gamma}{t}{N}}
      {\declsynjudg{\makedec{\Theta}; \Gamma}{\thunk{t}}{\shiftd{N}}}

    \begin{llproof}
      \conwfPf{\Theta}{Assumption}
      \envwfPf{\Theta}{\Gamma}{Assumption}
      \congoesPf{\Theta}{\Omega}{Assumption}
      \conwfPf{\Omega}{Assumption}

      \algosynjudgPf{\Theta; \Gamma}{t}{N}{\Theta'}{\byih (term size decreases)}
\Hand \congoesPf{\Theta'}{\Omega}{\ditto}
\Hand \algosynjudgPf{\Theta; \Gamma}{\thunk{t}}{\shiftd{N}}{\Theta'}{By {\Athunk}}
    \end{llproof}

    \DerivationProofCase{\Dreturn}
      {\declsynjudg{\makedec{\Theta}; \Gamma}{v}{P}}
      {\declsynjudg{\makedec{\Theta}; \Gamma}{\return{v}}{\shiftu{P}}}

      Symmetric to \Dthunk case.

    \DerivationProofCase{\Dambiguouslet}
      {\declsynjudg{\makedec{\acontext}; \typeenv}{v}{\shiftd M} \\
        \declspinejudg{\makedec{\acontext}; \Gamma}{s}{\spine{M}{\shiftu{Q}}} \\
        \DNegSubtypeJudge{\makedec{\acontext}}{\shiftu{P}}{\shiftu{Q}} \\
        \declsynjudg{\makedec{\acontext}; \Gamma, x : P}{t}{N}}
      {\declsynjudg{\makedec{\acontext}; \Gamma}{\letanno{x}{P}{v}{s}{t}}{N}}

    \begin{llproof}
      \conwfPf{\acontext}{Assumption}
      \envwfPf{\acontext}{\typeenv}{Assumption}
      \congoesPf{\acontext}{\ccontext}{Assumption}
      \conwfPf{\ccontext}{Assumption}
      \declsynjudgPf{\makedec{\acontext}; \typeenv}{v}{\shiftdown{M}}{Subderivation}

      \proofcomment{Apply the induction hypothesis to give a context $\acontext'$ such that:}

      \algosynjudgPf{\acontext; \typeenv}{v}{\shiftdown M}{\acontext'}{\byih (term size decreases)}
      \congoesPf{\acontext'}{\ccontext}{\ditto}

      \proofcomment{Applying well-formedness:}

      \congoesPf{\acontext}{\acontext'}{By \lemmaref{lemma:algorithmic-typing-well-formed}}
      \conwfPf{\acontext'}{\ditto}
      \wfpostypePf{\acontext'}{\shiftdown{M}}{\ditto}
      \groundPf{\shiftdown{M}}{\ditto}

      \proofcomment{Rework the second premise so we can apply the induction hypothesis:}

      \declspinejudgPf{\makedec{\acontext}; \Gamma}{s}{\spine{M}{\shiftu{Q}}}{Premise}
      \declspinejudgPf{\makedec{\acontext}; \Gamma}{s}{\spine{[\Omega]M}{\shiftu{Q}}}{By \lemmaref{lemma:Context substitution on ground terms}}
      \declspinejudgPf{\makedec{\acontext'}; \Gamma}{s}{\spine{[\Omega]M}{\shiftu{Q}}}{By \lemmaref{lemma:equal declarative contexts}}

      \proofcomment{Next show the antecedents of the second premise's induction hypothesis:}

      \conwfPf{\acontext'}{Above}
      \congoesWeakPf{\acontext}{\acontext'}{By \lemmaref{lemma:weak context extension subsumes normal}}
      \envwfPf{\acontext'}{\typeenv}{By \lemmaref{lemma:context extension preserves wf envs}}
      \congoesPf{\acontext'}{\ccontext}{Above}
      \conwfPf{\ccontext}{Above}
      \wfnegtypePf{\acontext'}{M}{Inversion (\twfshiftdown)}
      \NoSolvedVarsPf{\acontext'}{M}{By \lemmaref{lemma:Context substitution on ground terms}}

      \proofcomment{Apply the induction hypothesis to give a contexts $\acontext''$, $\ccontext'$ and a type $Q'$ such that:}

      \algospinejudgPf{\acontext'; \typeenv}{s}{\spine{M}{\shiftu{Q'}}}{\acontext''}{\byih (term size decreases)}
      \congoesWeakPf{\Omega}{\Omega'}{\ditto}
      \congoesPf{\acontext''}{\Omega'}{\ditto}
      \MutualJudgeNegPf{\makedec{\acontext'}}{\subcon{\ccontext'} \shiftu{Q'}}{\shiftu{Q}}{\ditto}
      \eqPf{[\acontext'']\shiftu{Q'}}{\shiftu{Q'}}{\ditto}
      \conwfPf{\Omega'}{\ditto}

      \proofcomment{Applying well-formedness:}

      \congoesWeakPf{\acontext'}{\acontext''}{By \lemmaref{lemma:algorithmic-typing-well-formed}}
      \conwfPf{\acontext''}{\ditto}
      \wfnegtypePf{\acontext''}{\shiftu{Q'}}{\ditto}
      \eqPf{[\acontext'']\shiftu{Q'}}{\shiftu{Q'}}{\ditto}

      \proofcomment{Now rework the third premise to match algorithmic rule. First show the third premise of the declarative rule:}

      \DJudgeNegPf{\makedec{\acontext}}{\shiftu{P}}{\shiftu{Q}}{Premise}
      \congoesWeakPf{\acontext}{\acontext''}{By \lemmaref{lemma:weak context extension transitive}}
      \DJudgeNegPf{\makedec{\acontext''}}{\shiftu{P}}{\shiftu{Q}}{By \lemmaref{lemma:weak context extension equal declarative contexts}}
      \MutualJudgeNegPf{\makedec{\acontext''}}{\subcon{\ccontext'} \shiftu{Q'}}{\shiftu{Q}}{By \lemmaref{lemma:weak context extension equal declarative contexts}}
      \DJudgeNegPf{\makedec{\acontext''}}{\shiftu{P}}{\shiftu{[\Omega']Q'}}{By \lemmaref{lemma:Transitivity of declarative pnsubtype}}
      \DJudgePosPf{\makedec{\acontext''}}{P}{[\Omega']Q'}{Inversion (\dshiftup)}

      \proofcomment{Show the antecedents of completeness:}

      \conwfPf{\acontext''}{Above}
      \congoesPf{\acontext''}{\ccontext'}{Above}
      \conwfPf{\ccontext'}{Above}

      \proofsep
      \wfpostypePf{\Theta}{P}{$P$ annotation}
      \wfpostypePf{\Theta''}{P}{By \lemmaref{lemma:weak context extension preserves well-formedness}}
      \wfnegtypePf{\Theta''}{\shiftu{Q'}}{Above}
      \wfpostypePf{\Theta''}{Q'}{Inversion (\twfshiftup)}

      \proofsep
      \groundPf{P}{$P$ annotation}
      \NoSolvedVarsPf{\acontext''}{\shiftu{Q'}}{Above}
      \NoSolvedVarsPf{\acontext''}{Q'}{\bydefsubcon}

      \proofcomment{Applying \lemmaref{theorem:Completeness of algorithmic subtyping}, we have the following for some context $\acontext'''$:}

      \AJudgePosPf{\acontext''}{P}{Q'}{\acontext'''}{By \lemmaref{theorem:Completeness of algorithmic subtyping}}
      \congoesPf{\acontext'''}{\Omega'}{\ditto}

      \proofcomment{Now appeal to the well-formedness of the algorithmic subtyping judgment:}

      \conwfPf{\acontext'''}{By \lemmaref{lemma:algorithmic-typing-well-formed}}
      \congoesPf{\acontext''}{\acontext'''}{\ditto}
      \groundPf{[\acontext''']Q'}{\ditto}

      \proofcomment{Now show the fourth premise of the declarative rule:}

      \DJudgeNegPf{\makedec{\acontext''}}{\shiftu{P}}{\shiftu{[\Omega']Q'}}{Above}
      \DJudgeNegPf{\makedec{\acontext'''}}{\shiftu{P}}{\shiftu{[\Omega']Q'}}{By \lemmaref{lemma:equal declarative contexts}}
      \DJudgePosPf{\makedec{\acontext'''}}{[\Omega']Q'}{P}{Inversion (\dshiftup)}
      \DJudgePosPf{\makedec{\acontext'''}}{[\ccontext'][\acontext''']Q'}{[\ccontext']Q'}{By \lemmaref{lemma:Context extension leads to isomorphic types}}
      \DJudgePosPf{\makedec{\acontext'''}}{[\ccontext'][\acontext''']Q'}{P}{By \lemmaref{lemma:Transitivity of declarative pnsubtype}}

      \proofcomment{Show the antecedents of completeness:}

      \conwfPf{\acontext'''}{Above}
      \congoesPf{\acontext'''}{\ccontext'}{Above}
      \conwfPf{\ccontext'}{Above}

      \proofsep
      \wfpostypePf{\Theta'''}{P}{By \lemmaref{lemma:Context extension preserves term well-formedness}}
      \wfpostypePf{\Theta'''}{Q'}{By \lemmaref{lemma:Context extension preserves term well-formedness}}
      \wfpostypePf{\Theta'''}{\subcon{\acontext'''} Q'}{By \lemmaref{lemma:Well-formed context substitution preserves term well-formedness}}

      \proofsep
      \groundPf{\subcon{\acontext'''} Q'}{Above}
      \NoSolvedVarsPf{\acontext'''}{P}{By \lemmaref{lemma:Context substitution on ground terms}}

      \proofcomment{Applying \lemmaref{theorem:Completeness of algorithmic subtyping}, we have the following for some context $\acontext'''$:}

      \AJudgePosPf{\acontext'''}{\subcon{\acontext'''} Q'}{P}{\acontext^{(4)}}{By \lemmaref{theorem:Completeness of algorithmic subtyping}}
      \congoesPf{\acontext^{(4)}}{\Omega'}{\ditto}

      \proofcomment{Now appeal to the well-formedness of the algorithmic subtyping judgment:}

      \conwfPf{\acontext^{(4)}}{By \lemmaref{lemma:algorithmic-typing-well-formed}}
      \congoesPf{\acontext'''}{\acontext^{(4)}}{\ditto}

      \proofcomment{Let the restricted context $\acontext^{(5)} = \restrictcontext{\acontext^{(4)}}{\acontext}$:}

      \congoesPf{\acontext}{\Theta^{(5)}}{By \lemmaref{lemma:restricted context wf}}
      \conwfPf{\Theta^{(5)}}{\ditto}
      \proofsep

      \conwfPf{\ccontext}{Above}
      \conwfPf{\ccontext'}{Above}
      \congoesWeakPf{\ccontext}{\ccontext'}{Above}
      \congoesPf{\ccontext}{\restrictcontext{\ccontext'}{\ccontext}}{By \lemmaref{lemma:restricted context wf}}
      \proofsep

      \congoesPf{\restrictcontext{\ccontext'}{\ccontext}}{\ccontext}{By \lemmaref{lemma:Reversing context extension from a complete context}}
      \congoesPf{\acontext^{(4)}}{\ccontext'}{Above}
      \congoesPf{\restrictcontext{\acontext^{(4)}}{\ccontext}}{\ccontext}{By \lemmaref{lemma:Pulling back restricted contexts}}
      \congoesPf{\acontext}{\ccontext}{Above}
      \congoesPf{\restrictcontext{\acontext^{(4)}}{\acontext}}{\ccontext}{By \lemmaref{lemma:Identical restricted contexts}}
      \congoesPf{\acontext^{(5)}}{\ccontext}{Substituting in definition of $\acontext^{(5)}$}

      \proofcomment{Rework the final premise to match the algorithmic rule:}

      \declsynjudgPf{\makedec{\acontext}; \Gamma, x : P}{t}{N}{Premise}
      \declsynjudgPf{\makedec{\Theta^{(5)}}; \Gamma, x : P}{t}{N}{By \lemmaref{lemma:equal declarative contexts}}

      \proofcomment{Show the antecedents of the final premise's induction hypothesis:}

      \congoesPf{\Theta}{\acontext^{(5)}}{Above}
      \wfnegtypePf{\Theta}{N}{Assumption}
      \wfnegtypePf{\acontext^{(5)}}{N}{By \lemmaref{lemma:Context extension preserves term well-formedness}}
      \groundPf{N}{Assumption}
      \conwfPf{\acontext^{(5)}}{As shown above}
      \congoesPf{\acontext^{(5)}}{\Omega}{Above}
      \conwfPf{\Omega}{Assumption}

      \proofsep
      \envwfPf{\Theta}{\typeenv}{Assumption}
      \congoesWeakPf{\acontext}{\acontext^{(5)}}{By \lemmaref{lemma:weak context extension subsumes normal}}
      \envwfPf{\acontext^{(5)}}{\typeenv}{By \lemmaref{lemma:context extension preserves wf envs}}
      \wfpostypePf{\Theta}{P}{$P$ an annotation}
      \wfpostypePf{\acontext^{(5)}}{P}{By \lemmaref{lemma:Context extension preserves term well-formedness}}
      \groundPf{P}{$P$ an annotation}
      \envwfPf{\acontext^{(5)}}{\typeenv, x : P}{By {\Ewfvar}}

      \proofcomment{Apply the induction hypothesis, to give a context $\acontext^{(6)}$, such that:}

      \Hand \congoesPf{\acontext^{(6)}}{\Omega}{\byih (term size decreases)}
      \algosynjudgPf{\Theta^{(5)}; \typeenv, x : P}{t}{N}{\acontext^{(6)}}{\ditto}
\Hand \algosynjudgPf{\Theta; \typeenv}{\letanno{x}{P}{v}{s}{t}}{N}{\Theta^{(6)}}{By {\Aambiguouslet}}
    \end{llproof}

    \DerivationProofCase{\Dunambiguouslet}
      {\declsynjudg{\makedec{\acontext}; \typeenv}{v}{\shiftd M} \\
        \declspinejudg{\makedec{\acontext}; \Gamma}{s}{\spine{M}{\shiftu{Q}}} \\
        \declsynjudg{\makedec{\acontext}; \Gamma, x : Q}{t}{N} \\
        \forall P \ldotp \text{if } \declspinejudg{\makedec{\acontext};\Gamma}{s}{\spine{M}{\shiftu{P}}} \text{ then } \declpisotypejudg{\makedec{\acontext}}{Q}{P}}
      {\declsynjudg{\makedec{\acontext}; \Gamma}{\letplain{x}{v}{s}{t}}{N}}

    \begin{llproof}
      \conwfPf{\acontext}{Assumption}
      \envwfPf{\acontext}{\typeenv}{Assumption}
      \congoesPf{\acontext}{\ccontext}{Assumption}
      \conwfPf{\ccontext}{Assumption}
      \declsynjudgPf{\makedec{\acontext}; \typeenv}{v}{\shiftdown{M}}{Subderivation}

      \proofcomment{Apply the induction hypothesis to give a context $\acontext'$ such that:}

\Label{1} \algosynjudgPf{\acontext; \typeenv}{v}{\shiftdown M}{\acontext'}{\byih (term size decreases)}
      \congoesPf{\acontext'}{\ccontext}{\ditto}

      \proofcomment{Applying well-formedness:}

      \congoesPf{\acontext}{\acontext'}{By \lemmaref{lemma:algorithmic-typing-well-formed}}
      \conwfPf{\acontext'}{\ditto}
      \wfpostypePf{\acontext'}{\shiftdown{M}}{\ditto}
      \groundPf{\shiftdown{M}}{\ditto}

      \proofcomment{Rework the second premise so we can apply the induction hypothesis:}

      \declspinejudgPf{\makedec{\acontext}; \Gamma}{s}{\spine{M}{\shiftu{Q}}}{Premise}
      \declspinejudgPf{\makedec{\acontext}; \Gamma}{s}{\spine{[\Omega]M}{\shiftu{Q}}}{By \lemmaref{lemma:Context substitution on ground terms}}
      \declspinejudgPf{\makedec{\acontext'}; \Gamma}{s}{\spine{[\Omega]M}{\shiftu{Q}}}{By \lemmaref{lemma:equal declarative contexts}}

      \proofcomment{Next show the antecedents of the second premise's induction hypothesis:}

      \conwfPf{\acontext'}{Above}
      \congoesWeakPf{\acontext}{\acontext'}{By \lemmaref{lemma:weak context extension subsumes normal}}
      \envwfPf{\acontext'}{\typeenv}{By \lemmaref{lemma:context extension preserves wf envs}}
      \congoesPf{\acontext'}{\ccontext}{Above}
      \conwfPf{\ccontext}{Above}
      \wfnegtypePf{\acontext'}{M}{Inversion (\twfshiftdown)}
      \NoSolvedVarsPf{\acontext'}{M}{By \lemmaref{lemma:Context substitution on ground terms}}

      \proofcomment{Apply the induction hypothesis to give a contexts $\acontext''$, $\ccontext'$ and a type $Q'$ such that:}

\Label{2} \algospinejudgPf{\acontext'; \typeenv}{s}{\spine{M}{\shiftu{Q'}}}{\acontext''}{\byih (term size decreases)}
      \congoesWeakPf{\Omega}{\Omega'}{\ditto}
      \congoesPf{\acontext''}{\Omega'}{\ditto}
\Label{3} \MutualJudgeNegPf{\makedec{\acontext'}}{\subcon{\ccontext'} \shiftu{Q'}}{\shiftu{Q}}{\ditto}
      \eqPf{[\acontext'']\shiftu{Q'}}{\shiftu{Q'}}{\ditto}
      \conwfPf{\Omega'}{\ditto}

      \proofcomment{Applying well-formedness:}

      \congoesWeakPf{\acontext'}{\acontext''}{By \lemmaref{lemma:algorithmic-typing-well-formed}}
      \subseteqPf{\FreeEV(\shiftu{Q'})}{\FreeEV(M) \cup (\FreeEV(\acontext'') \setminus \FreeEV(\acontext'))}{\ditto}
      \conwfPf{\acontext''}{\ditto}
      \wfnegtypePf{\acontext''}{\shiftu{Q'}}{\ditto}
      \eqPf{[\acontext'']\shiftu{Q'}}{\shiftu{Q'}}{\ditto}

      \proofcomment{From these conclusions we can deduce:}

      \subseteqPf{\FreeEV(Q')}{\FreeEV(M) \cup (\FreeEV(\acontext'') \setminus \FreeEV(\acontext'))}{\bydeffev and above}
      \eqPf{\FreeEV(M)}{\emptyset}{Since $\groundJudge{M}$}
      \subseteqPf{\FreeEV(Q')}{\FreeEV(\acontext'') \setminus \FreeEV(\acontext')}{Substituting above equations}
      \subseteqPf{\FreeEV(\acontext')}{\FreeEV(\acontext'')}{By \lemmaref{lemma:weak context extension maintains variables}}
      \eqPf{\FreeEV(Q') \cap \FreeEV(\acontext')}{\emptyset}{By definition of $\subseteq$}

      \proofcomment{
        Next prove by contradiction that we have $\groundJudge{Q'}$ from the induction hypothesis.
        Assume $\FreeEV(Q') \neq \emptyset$.
        Then:
      }

\Label{4} \inPf{\guess{\alpha}}{\FreeEV(Q')}{Necessarily true for some $\guess{\alpha}$, otherwise}
      \trailingjust{$Q'$ would not be ground}

      \proofcomment{Let $R = \subcon{\ccontext'} \guess{\alpha}$ and define $\ccontext''$ as the complete context obtained by taking $\ccontext'$ and substituting the $\guess{\alpha} = R$ context item with $\guess{\alpha} = \shiftdown{\shiftup{R}}$. Now apply soundness to the algorithmic judgment but using $\ccontext''$ as the complete context:}

      \conwfPf{\acontext'}{Above}
      \envwfPf{\acontext'}{\typeenv}{Above}
      \congoesPf{\acontext''}{\ccontext''}{Since (a) $\congoesJudg{\acontext''}{\ccontext'}$ and $\guess{\alpha} \in \FreeEV(Q')$ implies}
      \trailingjust{(b) $\guess{\alpha}$ is unsolved in $\acontext''$}
      \algospinejudgPf{\acontext'; \typeenv}{s}{\spine{M}{\shiftup{Q'}}}{\acontext''}{Above}
      \wfpostypePf{\ccontext'_L}{R}{Inversion (\cwfsolvedguess), for some prefix $\ccontext'_L$ of $\ccontext'$}
      \trailingjust{(this is also a prefix of $\ccontext''$ by definition of $\ccontext''$)}
      \groundPf{R}{\ditto}
      \wfpostypePf{\ccontext'_L}{\shiftdown{\shiftup{R}}}{By \twfshiftup and \twfshiftdown}
      \groundPf{\shiftdown{\shiftup{R}}}{\bydefground}
      \conwfPf{\ccontext''}{By $\conwfJudg{\ccontext'}$ and above two statements}
      \wfnegtypePf{\acontext'}{M}{Above}
      \NoSolvedVarsPf{\acontext'}{M}{Above}
      \proofsep

      \declspinejudgPf{\makedec{\acontext'}; \typeenv}{s}{\spine{\subcon{\ccontext''} M}{Q''}}{By \lemmaref{theorem:soundness}}
      \trailingjust{(term size decreases)}
      \MutualJudgePosPf{\makedec{\acontext'}}{Q''}{\subcon{\ccontext''} \shiftup{Q'}}{\ditto}
\Label{5} \MutualJudgePosPf{\makedec{\acontext}}{Q''}{\subcon{\ccontext''} \shiftup{Q'}}{By \lemmaref{lemma:equal declarative contexts}}
      \congoesWeakPf{\ccontext}{\ccontext'}{Above}
      \congoesWeakPf{\ccontext}{\ccontext''}{Replacing the instance of \Wcsolveguess}
      \declspinejudgPf{\makedec{\acontext}; \typeenv}{s}{\spine{\subcon{\ccontext''} M}{Q''}}{By \lemmaref{lemma:weak context extension equal declarative contexts}}
      \declspinejudgPf{\makedec{\acontext}; \typeenv}{s}{\spine{M}{Q''}}{By \lemmaref{lemma:Context substitution on ground terms}}

      \proofcomment{Now make use of the final premise:}

      \MutualJudgePosPf{\makedec{\acontext}}{Q}{Q''}{Instantiating final premise with $P = Q''$}
      \MutualJudgePosPf{\makedec{\acontext}}{Q}{\subcon{\ccontext''} \shiftup{Q'}}{Applying \lemmaref{lemma:Transitivity of declarative pnsubtype}}
      \trailingjust{to above and (5)}
      \MutualJudgeNegPf{\makedec{\acontext}}{\subcon{\ccontext'} \shiftu{Q'}}{\shiftu{Q}}{Applying \lemmaref{lemma:equal declarative contexts}}
      \trailingjust{to (3)}
      \MutualJudgePosPf{\makedec{\acontext}}{\subcon{\ccontext'} Q'}{Q}{Inversion (\dshiftup)}
      \MutualJudgePosPf{\makedec{\acontext}}{\subcon{\ccontext'} Q'}{\subcon{\ccontext''} Q'}{By \lemmaref{lemma:Transitivity of declarative pnsubtype}}

      \proofcomment{However:}

      \NotMutualJudgePosPf{\makedec{\acontext}}{R}{\shiftup{\shiftdown{R}}}{Must have the same number of shifts on both sides}
      \trailingjust{of the declarative subtyping judgment}
      \NotMutualJudgePosPf{\makedec{\acontext}}{\subcon{\ccontext'} Q'}{\subcon{\ccontext''} Q'}{Since $\guess{\alpha} \in \FreeEV(Q')$ by (4)}

      \proofcomment{This is a contradiction, hence $Q'$ must be ground:}

\Label{6} \eqPf{\FreeEV(Q')}{\emptyset}{Since $\groundJudge{Q'}$}

      \proofcomment{Next, restrict the output context of the spine judgment:}

\Label{7} \LetPf{\acontext'''}{\restrictcontext{\acontext''}{\acontext}}{}
      \conwfPf{\acontext}{Above}
      \conwfPf{\acontext''}{Above}
      \congoesWeakPf{\acontext}{\acontext''}{Above}
      \proofsep

      \congoesPf{\acontext}{\acontext'''}{By \lemmaref{lemma:restricted context wf}}
      \conwfPf{\acontext'''}{\ditto}
      \proofsep

      \conwfPf{\ccontext}{Above}
      \conwfPf{\ccontext'}{Above}
      \congoesWeakPf{\ccontext}{\ccontext'}{Above}
      \congoesPf{\ccontext}{\restrictcontext{\ccontext'}{\ccontext}}{By \lemmaref{lemma:restricted context wf}}
      \congoesPf{\restrictcontext{\ccontext'}{\ccontext}}{\ccontext}{By \lemmaref{lemma:Reversing context extension from a complete context}}
      \proofsep

      \congoesPf{\acontext''}{\ccontext'}{Above}
      \congoesPf{\restrictcontext{\acontext''}{\ccontext}}{\ccontext}{By \lemmaref{lemma:Pulling back restricted contexts}}
      \congoesPf{\acontext}{\ccontext}{Above}
      \congoesPf{\restrictcontext{\acontext''}{\acontext}}{\ccontext}{By \lemmaref{lemma:Identical restricted contexts}}
      \congoesPf{\acontext'''}{\ccontext}{Substituting in definition of $\acontext'''$}

      \proofcomment{Rework the third premise to match the algorithmic judgment:}

      \declsynjudgPf{\makedec{\acontext}; \Gamma, x : Q}{t}{N}{Premise}
      \MutualJudgeNegPf{\makedec{\acontext}}{\subcon{\ccontext'} \shiftu{Q'}}{\shiftu{Q}}{Above}
      \declsynjudgPf{\makedec{\acontext}; \Gamma, x : [\Omega']Q'}{t}{N}{By \lemmaref{lemma:isomorphic types check expressions}}
      \declsynjudgPf{\makedec{\acontext}; \Gamma, x : Q'}{t}{N}{By \lemmaref{lemma:Context substitution on ground terms}}
      \declsynjudgPf{\makedec{\acontext'''}; \Gamma, x : Q'}{t}{N}{By \lemmaref{lemma:equal declarative contexts}}

      \proofcomment{Next show the antecedents of the induction hypothesis:}

      \conwfPf{\acontext'''}{Above}
      \proofsep

      \envwfPf{\Theta}{\Gamma}{Above}
      \envwfPf{\Theta'''}{\Gamma}{By \lemmaref{lemma:context extension preserves wf envs}}
      \wfpostypePf{\Theta''}{Q'}{Inversion (\twfshiftup)}
      \groundPf{Q'}{Above}
      \wfpostypePf{\Theta'''}{Q'}{By definition of restricted context $\acontext'''$ and since $\groundJudge{Q'}$}
      \envwfPf{\Theta'''}{\Gamma, x : Q'}{By {\Ewfvar}}
      \proofsep

      \congoesPf{\acontext'''}{\ccontext}{Above}
      \conwfPf{\ccontext}{Above}
      \declsynjudgPf{\makedec{\acontext'''}; \Gamma, x : Q'}{t}{N}{Above}
      \wfnegtypePf{\Theta}{N}{Assumption}
      \wfnegtypePf{\Theta'''}{N}{By $\congoesJudg{\Theta}{\Theta'''}$ and \lemmaref{lemma:Context extension preserves term well-formedness}}
      \groundPf{N}{Assumption}

      \proofcomment{Applying the induction hypothesis, we have for a context $\acontext^{(4)}$:}

\Label{8} \algosynjudgPf{\Theta'''; \Gamma, x : Q'}{t}{N}{\acontext^{(4)}}{\byih (term size decreases)}
\Hand \congoesPf{\acontext^{(4)}}{\Omega}{\ditto}
\Hand \algosynjudgPf{\Theta; \Gamma}{\letplain{x}{v}{s}{t}}{N}{\acontext^{(4)}}{Applying \Aunambiguouslet to (1), (2), (6), (7), and (8)}
    \end{llproof}

    \DerivationProofCase{\Dspinenil}
      {}
      {\declspinejudg{\makedec{\acontext}; \typeenv}{\epsilon}{\spine{N}{N}}}

    \begin{llproof}
      \proofcomment{The output context will be $\Theta$, the complete context will be $\ccontext$, and the output type will be $M$.}

      \proofsep
\Hand \algospinejudgPf{\Theta; \Gamma}{\epsilon}{\spine{N}{N}}{\Theta}{By {\Aspinenil}}
\Hand \congoesWeakPf{\Omega}{\Omega}{By \lemmaref{lemma:weak context extension reflexive}}
\Hand \congoesPf{\Theta}{\Omega}{Assumption}
      \MutualJudgeNegPf{\makedec{\acontext}}{\subcon{\ccontext} \subcon{\acontext} N}{\subcon{\acontext} N}{By \lemmaref{lemma:Context extension leads to isomorphic types}}
\Hand \eqPf{[\Theta]N}{N}{Assumption}
\Hand \MutualJudgeNegPf{\makedec{\acontext}}{\subcon{\ccontext} N}{N}{Substituting in above equation}
\Hand \eqPf{[\Theta]N}{N}{Assumption}
\Hand \conwfPf{\Omega}{Assumption}
    \end{llproof}

    \DerivationProofCase{\Dspinecons}
      {\declsynjudg{\makedec{\acontext}; \Gamma}{v}{P} \\ \DPosSubtypeJudge{\makedec{\acontext}}{P}{\subcon{\ccontext} Q} \\
        \declspinejudg{\makedec{\acontext}; \Gamma}{s}{\spine{\subcon{\ccontext} N}{M}}}
      {\declspinejudg{\makedec{\acontext}; \Gamma}{v, s}{\spine{\subcon{\ccontext} (Q \to N)}{M}}}

    \begin{llproof}
      \conwfPf{\Theta}{Assumption}
      \envwfPf{\Theta}{\Gamma}{Assumption}
      \congoesPf{\Theta}{\Omega}{Assumption}
      \conwfPf{\Omega}{Assumption}

      \proofcomment{By the induction hypothesis for the first premise, there exists a context $\Theta'$, such that:}

      \algosynjudgPf{\Theta; \Gamma}{v}{P}{\Theta'}{\byih (term size decreases)}
      \congoesPf{\Theta'}{\Omega}{\ditto}

      \proofcomment{Apply well-formedness to this algorithmic judgment:}

      \congoesPf{\Theta}{\Theta'}{By \lemmaref{lemma:algorithmic-typing-well-formed}}
      \conwfPf{\Theta'}{\ditto}
      \wfpostypePf{\Theta}{P}{\ditto}
      \groundPf{P}{\ditto}

      \proofcomment{Now use completeness of subtyping:}

      \conwfPf{\acontext'}{Above}
      \congoesPf{\acontext'}{\ccontext}{Above}
      \conwfPf{\ccontext}{Above}
      \proofsep

      \DJudgePosPf{\makedec{\acontext}}{P}{\subcon{\ccontext} Q}{Subderivation}
      \DJudgePosPf{\makedec{\acontext'}}{P}{\subcon{\ccontext} Q}{By \lemmaref{lemma:equal declarative contexts}}
      \DJudgePosPf{\makedec{\acontext'}}{P}{\subcon{\ccontext} \subcon{\acontext'} Q}{By \lemmaref{lemma:Context extension leads to isomorphic types}}
      \proofsep

      \wfpostypePf{\Theta'}{P}{By \lemmaref{lemma:Context extension preserves term well-formedness}}
      \wfnegtypePf{\Theta}{Q \funarrow N}{Assumption}
      \wfpostypePf{\Theta}{Q}{By {\twfarrow}}
      \wfpostypePf{\Theta'}{Q}{By \lemmaref{lemma:Context extension preserves term well-formedness}}
      \wfpostypePf{\Theta'}{\subcon{\acontext'} Q}{By \lemmaref{lemma:Well-formed context substitution preserves term well-formedness}}
      \groundPf{P}{Above}
      \NoSolvedVarsPf{\acontext'}{[\acontext'] Q}{By \lemmaref{lemma:Context substitution idempotence}}

      \proofsep
      \AJudgePosPf{\acontext'}{P}{\subcon{\acontext'} Q}{\acontext''}{By \lemmaref{theorem:Completeness of algorithmic subtyping}}
      \congoesPf{\acontext''}{\ccontext}{\ditto}

      \proofcomment{Applying well-formedness:}

      \conwfPf{\acontext''}{By \lemmaref{lemma:well-formedness-pnsubtype}}
      \congoesPf{\acontext'}{\acontext''}{\ditto}
      \groundPf{\subcon{\acontext''} \subcon{\acontext'} Q}{\ditto}

      \proofcomment{Next rework the third premise to match the algorithmic rule:}

      \declspinejudgPf{\makedec{\acontext''}; \Gamma}{s}{\spine{[\Omega]N}{M}}{By \lemmaref{lemma:equal declarative contexts}}
      \declnisotypejudgPf{\makedec{\acontext''}}{[\Omega][\Theta'']N}{[\Omega]N}{By \lemmaref{lemma:Context extension leads to isomorphic types}}
      \declspinejudgPf{\makedec{\acontext''}; \Gamma}{s}{\spine{[\Omega][\Theta'']N}{M'}}{By \lemmaref{lemma:isomorphic types check expressions}}
      \MutualJudgeNegPf{\makedec{\acontext''}}{M}{M'}{\ditto}

      \proofcomment{Now show the antecedents of the third premise's induction hypothesis:}

      \conwfPf{\Theta''}{Above}
      \envwfPf{\Theta}{\Gamma}{Assumption}
      \envwfPf{\Theta''}{\Gamma}{By \lemmaref{lemma:context extension preserves wf envs}}
      \proofsep

      \wfnegtypePf{\Theta}{P \funarrow N}{Assumption}
      \wfnegtypePf{\Theta}{N}{By {\twfarrow}}
      \congoesPf{\acontext}{\acontext''}{By \lemmaref{lemma:context extension transitive}}
      \wfnegtypePf{\Theta''}{N}{By \lemmaref{lemma:Context extension preserves term well-formedness}}
      \wfnegtypePf{\Theta''}{[\Theta''] N}{By \lemmaref{lemma:Well-formed context substitution preserves term well-formedness}}

      \proofsep
      \eqPf{[\Theta''][\Theta'']N}{[\Theta'']N}{By \lemmaref{lemma:Context substitution idempotence}}

      \proofsep
      \congoesPf{\Theta''}{\Omega}{\byih (term size decreases)}

      \conwfPf{\Omega}{Assumption}

      \proofcomment{Now apply the induction hypothesis to this third premise. This gives the following for some contexts $\Theta'''$, $\Omega'$ and type $M''$:}

      \algospinejudgPf{\Theta''; \Gamma}{s}{\spine{[\Theta'']N}{M''}}{\Theta'''}{\byih (term size decreases)}
\Hand \congoesWeakPf{\Omega}{\Omega'}{\ditto}
\Hand \congoesPf{\Theta'''}{\Omega'}{\ditto}
      \MutualJudgeNegPf{\makedec{\acontext''}}{\subcon{\ccontext'} M''}{M'}{\ditto}
\Hand \eqPf{[\Theta''']M''}{M''}{\ditto}
\Hand \conwfPf{\Omega'}{\ditto}
      \proofsep

      \MutualJudgeNegPf{\makedec{\acontext'}}{\subcon{\ccontext} M''}{M}{By \lemmaref{lemma:Transitivity of declarative pnsubtype}}
\Hand \MutualJudgeNegPf{\makedec{\acontext}}{\subcon{\ccontext} M''}{M}{By \lemmaref{lemma:equal declarative contexts}}

      \proofcomment{Finally, apply the algorithmic judgment:}

\Hand \algospinejudgPf{\Theta; \Gamma}{v, s}{\spine{P \funarrow N}{M''}}{\Theta''}{By {\Aspinecons}}
    \end{llproof}

    \DerivationProofCase{\Dspinetypeabs}
    {\wfpostypeJudg{\makedec{\acontext}}{P} \\ \declspinejudg{\makedec{\acontext}; \Gamma}{s}{\spine{[P/\alpha]([\Omega]N)}{M}}}
    {\declspinejudg{\makedec{\acontext}; \Gamma}{s}{\spine{[\Omega](\forall\alpha\ldotp N)}{M}}}

    Take cases on whether $\alpha \in \FreeUV(N)$
    \begin{itemize}
      \caseitem{$\alpha \notin \FreeUV(N)$}

      \begin{llproof}
        \eqPf{[P/\alpha]([\Omega]N)}{[\Omega]N}{By $\alpha \notin \FreeUV(N)$}

        \proofsep
        \declspinejudgPf{\makedec{\acontext}; \Gamma}{s}{\spine{[\Omega]N}{M}}{By equality}

        \proofsep
        \wfnegtypePf{\Theta}{\forall\alpha\ldotp N}{Assumption}
        \wfnegtypePf{\Theta, \alpha}{N}{By {\twfforall}}
        \wfnegtypePf{\Theta}{N}{By $\alpha \notin \FreeUV(N)$ and \lemmaref{lemma:type well-formed with alpha removed}}

        \proofsep
        \eqPf{[\Theta](\forall\alpha\ldotp N)}{\forall\alpha\ldotp N}{Assumption}
        \eqPf{[\Theta]N}{N}{\bydefsubcon}

        \proofsep
        \conwfPf{\Theta}{Assumption}
        \envwfPf{\Theta}{\Gamma}{Assumption}
        \congoesPf{\Theta}{\Omega}{Assumption}
        \conwfPf{\Omega}{Assumption}
      
        \proofcomment{By the induction hypothesis we have contexts $\Theta'$, $\Omega'$ and type $M'$, such that:}

  \Hand     \congoesWeakPf{\Omega}{\Omega'}{\byih (Term size stays the same and the number of}
            \trailingjust{prenex universal quantifiers decreases. Since applying}
            \trailingjust{the context only replaces positive types by positive}
            \trailingjust{types, it cannot change the number of prenex}
            \trailingjust{universal quantifiers.)}
  \Hand     \congoesPf{\Theta'}{\Omega'}{\ditto}
  \Hand     \MutualJudgeNegPf{\makedec{\acontext}}{\subcon{\ccontext} M'}{M}{\ditto}
  \Hand     \eqPf{[\Theta']M'}{M'}{\ditto}
  \Hand     \conwfPf{\Omega'}{\ditto}
        \algospinejudgPf{\Theta; \Gamma}{s}{\spine{N}{M'}}{\Theta'}{\ditto}

        \proofcomment{Applying the algorithmic judgment:}

  \Hand     \algospinejudgPf{\Theta; \Gamma}{s}{\spine{\forall\alpha\ldotp N}{M'}}{\Theta'}{By {\Aspinetypeabsnotin}}
      \end{llproof}

      \caseitem{$\alpha \in \FreeUV(N)$}

      First rework the premise to match the algorithmic rule:

      \begin{llproof}
        \declspinejudgPf{\makedec{\acontext}; \Gamma}{s}{\spine{[P / \alpha]([\Omega]N)}{M}}{Premise}
        \declspinejudgPf{\makedec{\acontext}; \Gamma}{s}{\spine{[[\Omega]P / \alpha]([\Omega]N)}{M}}{$\groundJudge{P}$ and \lemmaref{lemma:Context substitution on ground terms}}
        \declspinejudgPf{\makedec{\acontext}; \Gamma}{s}{\spine{[\Omega]([P / \alpha]N)}{M}}{\bydefsubcon}
      \end{llproof}
      \begin{llproof}
        \declspinejudgPf{[\Omega, \guess{\alpha} = P](\Theta, \guess{\alpha}); [\Omega, \guess{\alpha} = P]\Gamma}{s}{\spine{[\Omega, \guess{\alpha} = P]([\guess{\alpha} / \alpha]N)}{M}}{For fresh $\guess{\alpha}$}
      \end{llproof}

      Now show the antecedents of the induction hypothesis:

      \begin{llproof}
        \wfnegtypePf{\Theta}{\forall\alpha\ldotp N}{Assumption}
        \wfnegtypePf{\Theta, \alpha}{N}{By {\twfforall}}
        \wfnegtypePf{\Theta, \guess{\alpha}}{[\guess{\alpha}/\alpha]N}{By \lemmaref{lemma:substituion preserves well-formedness of types}}

        \proofsep
        \eqPf{[\Theta](\forall\alpha\ldotp N)}{\forall\alpha\ldotp N}{Assumption}
        \eqPf{[\Theta]N}{N}{\bydefsubcon}
        \eqPf{[\guess{\alpha}/\alpha]([\Theta]N)}{[\guess{\alpha}/\alpha]N}{By equality}
        \eqPf{[\Theta]([\guess{\alpha}/\alpha]N)}{[\guess{\alpha}/\alpha]N}{$\guess{\alpha}$ fresh}
        \eqPf{[\Theta, \guess{\alpha}]([\guess{\alpha}/\alpha]N)}{[\guess{\alpha}/\alpha]N}{\bydefsubcon}

        \proofsep
        \conwfPf{\Theta}{Assumption}
        \conwfPf{\Theta, \guess{\alpha}}{By {\cwfunsolvedguess}}

        \proofsep
        \congoesWeakPf{\Theta}{\Theta}{By \lemmaref{lemma:weak context extension reflexive}}
        \congoesWeakPf{\Theta}{\Theta, \guess{\alpha}}{By {\Wcunsolvedextend}}
        \envwfPf{\Theta}{\Gamma}{Assumption}
        \envwfPf{\Theta, \guess{\alpha}}{\Gamma}{By $\congoesWeakJudg{\Theta}{\Theta, \guess{\alpha}}$ and}
        \trailingjust{\lemmaref{lemma:context extension preserves wf envs}}

        \proofsep
        \congoesPf{\Theta}{\Omega}{Assumption}
        \congoesPf{\Theta, \guess{\alpha}}{\Omega, \guess{\alpha} = P}{By {\Csolveguess}}

        \proofsep
        \conwfPf{\Omega}{Assumption}
        \groundPf{P}{$P$ declarative type}
        \wfpostypePf{\makedec{\acontext}}{P}{Premise}
        \wfpostypePf{\Omega}{P}{Since $\congoesJudg{\acontext}{\ccontext}$, and context extension cannot add}
        \trailingjust{or remove universal variables}
        \conwfPf{\Omega, \guess{\alpha} = P}{By {\cwfsolvedguess}}

        \proofcomment{Applying the induction hypothesis, we have contexts $\Theta'$, $\Omega'$ and a type $M'$, such that:}

        \congoesWeakPf{\Omega,\guess{\alpha} = P}{\Omega'}{\byih (Term size stays the same and the number of}
        \trailingjust{prenex universal quantifiers decreases. Since applying}
        \trailingjust{the context only replaces positive types by positive}
        \trailingjust{types, it cannot change the number of prenex}
        \trailingjust{universal quantifiers.)}
  \Hand     \congoesPf{\Theta'}{\Omega'}{\ditto}
  \Hand     \MutualJudgeNegPf{\makedec{\acontext}}{\subcon{\ccontext'} M'}{M}{\ditto}
  \Hand     \eqPf{[\Theta']M'}{M'}{\ditto}
  \Hand     \conwfPf{\Omega'}{\ditto}
        \algospinejudgPf{\Theta, \guess{\alpha}; \Gamma}{s}{\spine{[\guess{\alpha}/\alpha]N}{M'}}{\Theta'}{\ditto}

        \proofsep
        \congoesWeakPf{\Omega,\guess{\alpha} = P}{\Omega'}{Above}
        \congoesWeakPf{\Omega}{\Omega}{By \lemmaref{lemma:weak context extension reflexive}}
        \congoesWeakPf{\Omega}{\Omega,\guess{\alpha} = P}{By {\Wcsolvedextend}}
  \Hand     \congoesWeakPf{\Omega}{\Omega'}{By \lemmaref{lemma:weak context extension transitive}}

        \proofcomment{Finally, applying the algorithmic judgment:}

  \Hand     \algospinejudgPf{\Theta; \Gamma}{s}{\spine{(\forall\alpha\ldotp N)}{M'}}{\Theta'}{By {\Aspinetypeabsin}}
      \end{llproof}

    \end{itemize}
\end{itemize}
\end{proof}